\documentclass[rmp,aps,floatfix,eqsecnum]{revtex4}
\usepackage{amssymb}

\usepackage{epsfig}

\begin{document}

\title{Odd Triplet Superconductivity and Related Phenomena in
Superconductor-Ferromagnet Structures}
\author{F. S. Bergeret}
\affiliation{Departamento de F\'{i}sica Te\'{o}rica de la Materia Condensada\\
Universidad Aut\'{o}noma de Madrid, Spain}
\author{A. F. Volkov}
\affiliation{Theoretische Physik III,\\
 Ruhr-Universit\"{a}t Bochum, Germany.\\
Institute of Radioengineering and Electronics of \\
the Russian Academy of Sciences,Moscow, Russia.}
\author{K. B. Efetov}
\affiliation{Theoretische Physik III,\\ Ruhr-Universit\"{a}t
Bochum, Germany.\\
L.D. Landau Institute for Theoretical Physics,\\  Moscow, Russia.}


\begin{abstract}
We consider novel unusual effects in superconductor-ferromagnet (S/F)
structures. In particular we analyze the triplet component (TC) of the
condensate generated in those systems.This component is odd in frequency and
even in the momentum, which makes it insensitive to non-magnetic impurities.
If the exchange field is not homogeneous in the system the triplet component
is not destroyed even by a strong exchange field and can penetrate the
ferromagnet over long distances. Some other effects considered here and
caused by the proximity effect are: enhancement of the Josephson current due
to the presence of the ferromagnet, induction of a magnetic moment in
superconductors resulting in a screening of the magnetic moment, formation
of periodic magnetic structures due to the influence of the superconductor,
etc. We compare the theoretical predictions with existing experiments.
\end{abstract}

\maketitle
\tableofcontents

\section{Introduction}

\label{INT}

Although superconductivity has been discovered by H. Kammerlingh Onnes
almost one century ago {(1911), the interest in studying }this phenomenon is
far from declining. The great attention to superconductivity within the last
15 years is partly due to the discovery of the high temperature
superconductors (HTSC) \cite{mueller}, which promises important
technological applications. It is clear that issues such as the origin of
the high critical temperature superconductivity, effects of external fields
and impurities on HTCS, etc, will remain fields of interest for years to
come.

Due to the successful investigations of the HTSC and its possible
technological applications, the interest in studying properties of
traditional (low $T_{c}$) superconductors was not as broad.
Nevertheless this field has also undergone a tremendous
development. Technologically, the traditional superconductors are
often easier to manipulate than high $T_{c}$ cuprates. One of the
main achievements{\ }of the last decade is the
making of high quality contacts between superconductors and normal metals $%
(S/N)$, superconductors and ferromagnets $(S/F)$, superconductors and
insulators $(S/I)$, etc. All these heterostructures can be very small with
the characteristic sizes of submicrometers.

This has opened a new field of research. The small size of these structures
provides the coherence of superconducting correlations over the full length
of the $N$ region. The length of the condensate penetration into the $N$
region $\xi _{N}$\ is restricted by decoherence processes (inelastic or
spin-flip scattering). At low temperatures the characteristic length over
which these decoherence processes occur may be quite long (a few microns).
Superconducting coherent effects in $S/N$ nanostructures, such as
conductance oscillations in an external magnetic field, were studied
intensively during the last decade (see for example the review articles by %
\citet{beenakker_rev,lambert_rev}).

The interplay between a superconductor $(S)$ and a normal metal $(N)$ in
simpler types of $S/N$\ structures (for example, $S/N$\ bilayers) has been
under study for a long time and the main physics of this so called proximity
effect is well described in the review articles by \citet{de_gennes} and %
\citet{deutscher}. In these works it was noticed that not only the
superconductor changes the properties of the normal metal but also the
normal metal has a strong effect on the superconductor. It was shown that
near the $S/N$ interface the superconductivity is suppressed over the
correlation length $\xi _{S}$, which means that the order parameter $\Delta $
is reduced at the interface in comparison with its bulk value far away from
the interface. At the same time, the superconducting condensate penetrates
the normal metal over the length $\xi _{N}$, which at low temperatures may
be much larger than $\xi _{S}$. Due to the penetration of the condensate
into the normal metal over large distances the Josephson effect is possible
in $S/N/S$ junctions with the thicknesses of the $N$ regions of the order of
a few hundreds nanometers. The Josephson effects in $S/N/S$ junctions were
studied in many papers and a good overview, both experimental and
theoretical, is given by \citet{kulik_book}, \citet{likharev}, and %
\citet{barone_book}.

The situation described above is quite different if an insulating layer $I$
is placed between two superconductors. The thickness of the insulator in $%
S/I/S$ structures cannot be as large as of the normal metals because
electron wave functions decay in the insulator on atomic distances. As a
consequence, the Josephson current is extremely small in $S/I/S$ structures
with a thick insulating layer.

But what about $S/F/S$ heterojunctions, where $F$ denotes a ferromagnetic
metal? In principle, the electron wave function can extend in the
ferromagnet over a rather large distance without a considerable decay.
However, it is well known that electrons with different spins belong to
different energy bands. The energy shift of the two bands can be considered
as an effective exchange field acting on the spin of the electrons. The
condensate of conventional superconductors is strongly influenced by this
exchange field of the ferromagnets and usually this reduces drastically the
superconducting correlations.

The suppression of the superconducting correlations is a consequence of the
Pauli principle. In most superconductors the wave function of the Cooper
pairs is singlet so that the electrons of a pair have opposite spins. In
other words, both the electrons cannot be in the same state, which would
happen if they had the same spin. If the exchange field of the ferromagnet
is sufficiently strong, it tries to align the spins of the electrons of a
Cooper pair parallel to each other, thus destroying the superconductivity.
Regarding the $S/F$ interfaces and the penetration of the condensate into
the ferromagnet, these effects mean that the superconducting condensate
decays fast in the region of the ferromagnet. A rough estimate leads to the
conclusion that the ratio of the condensate penetration depth in
ferromagnets to the one in non-magnetic metals with a high impurity
concentration is of the order of $\sqrt{T_{c}/h}$, where $h$ is the exchange
energy and $T_{c}$ is the critical temperature of the superconducting
transition. The exchange energy in conventional ferromagnets like $Fe$ or $%
Co $ is several orders of magnitude higher than $T_{c}$ and therefore the
penetration depth in the ferromagnets is much smaller than that in the
normal metals.

Study of the proximity effect in the $S/F$ structures started not
long ago but it has already evolved into a very active field of
research (for a review see
\citet{proshin,golubov_rev,buzdin_rev,pokrovsky}). The effect of
the suppression of superconductivity by the ferromagnetism is
clearly seen experimentally and this corresponds to the simple
picture of the destruction of the singlet superconductivity by the
exchange field as discussed above.

At first glance, it seems that due to the strong suppression of
the superconductivity the proximity effect in $S/F$ structures is
less interesting than in the $S/N$ systems. However, this is not
so because the physics of the proximity effect in the $S/F$
structures is not exhausted by the suppression of the
superconductivity and new very interesting effects come into play.
Moreover, under some circumstances superconductivity is not
necessarily suppressed by the ferromagnets because the presence of
the latter may lead to a triplet superconducting pairing
\cite{BVE2,kadigrobov}. In some cases  not only the ferromagnetism
tends to destroy the superconductivity but also the
superconductivity may suppress the ferromagnetism
\cite{buzdin_crypto,BEL}. This may concern ``real'' strong
ferromagnets like iron or nickel with a Curie temperature much
larger than the transition temperature of the superconductor.

In all, it is becoming more and more evident from recent experimental and
theoretical studies that the variety of non-trivial effects in the $S/F$
structures exceeds considerably what one would have expected before. Taking
into account possible technological applications, there is no wonder that $%
S/F$ systems attract nowadays a lot of attention.

This review article is devoted to the study of new ``exotic'' phenomena in
the $S/F$ heterojunctions. By the word ``exotic'' we mean phenomena that
could not be expected from the simple picture of a superconductor in contact
with a homogeneous ferromagnet. Indeed, the most interesting effects should
occur when the exchange field is not homogeneous. These non-homogeneities
can be either intrinsic for the ferromagnetic material, like e.g. domain
walls, or arise as a result of experimental manipulations, such as
multilayered structures with different directions of the magnetization,
which can also be spoken of as a non-homogeneous alignment of the magnetic
moments.

Of course, we are far from saying that there is nothing interesting to be
seen when the exchange field is homogeneous. Although it is true that in
this case the penetration depth of the superconducting condensate into the
ferromagnet is short, the exponential decay of the condensate function into
ferromagnets is accompanied by oscillations in space. These oscillations
lead, for example, to oscillations of the critical superconducting
temperature $T_{c}$ and the critical Josephson current $I_{c}$ in $S/F$
structures as a function of the thickness $d_{F}$. Being predicted by %
\citet{buzdin_kupr_tc} and \citet{radovic2}, the observation of
such oscillatory behavior was first reported by \citet{jiang} on
$Gd/Nb$ structures. {Indications to a non-monotonic behavior of
$T_{c}$ as a
function of $d_{F}$ was also reported by %
\citet{wong,strunk,mercaldo,obi,velez,muehge_prl}.}

{However, in other experiments the dependence of }$T_{c}${\ on
}$d_{F}${\ was monotonic. For example in Ref. \cite{dynes02} the
critical temperature of the bilayer Pb/Ni decreased by increasing
the F
layer thickness }$d_{F}${\ in a monotonic way.} In the experiments by %
\citet{muehge} on $Fe/Nb/Fe$ structures and by \citet{aarts} on $V/Fe$
systems {both a monotonic and non-monotonic behavior of }$T_{c}$%
{\ has been observed. This different behavior was attributed to
changes of the transmittance of the }$S/F${\ interface.} {A
comprehensive analysis taking into account the samples quality was
made for different materials by \citet{chien_reich}.}

{More convincing results were found by measuring the Josephson critical
current in a }${S/F/S}${\ junction. Due to the oscillatory behavior of the
superconducting condensate in the }${F}${\ region the critical Josephson
current should change its sign in a $S/F/S$ junction ($\pi -$junction). This
phenomenon predicted long ago by \citet{bula_kuzi} has been confirmed
experimentally only recently \cite
{ryazanov,kontos01,kontos02,palevski,sellier,strunk04}.}

Experiments on transport properties of $S/F$ structures were also performed
in the last years. For example, \citet{petrashov} and \citet{pannetier}
observed an unexpected decrease of the resistance of a ferromagnetic wire
attached to a superconductor when the temperature is lowered below $T_{c}$.
In both of the experiments strong ferromagnets $Ni$ and $Co$, respectively,
were used. One would expect that the change of the resistance must be very
small due to the destruction of the superconductivity by the ferromagnets.
However, the observed drop was about $10\%$ and this can only be explained
by a long-range proximity effect.

This raises a natural question: how can such long range superconducting
effects occur in a ferromagnet with a strong exchange field? We will see in
the subsequent chapters that provided the exchange field is not homogenous a
long-range component of the condensate may be induced in the ferromagnet.
This component is in a triplet state and can penetrate the $F$ region over
distances comparable with $\xi _{N}$, as in the case of a normal metal.

We outline now the structure of the present review.

In Chapter \ref{PRO} we discuss the proximity effects in $S/N$ structures
and $S/F$ structures with a homogeneous magnetization. Chapter \ref{PRO} may
serve as an introduction into the field. The main results illustrated there
have been presented in other reviews and we discuss them here in order to
give the reader an impression about works done previously. Chapter \ref{PRO}
can also help in getting the basic knowledge about calculational methods
used in subsequent chapters. One can see from this discussion that already
homogeneous ferromagnets in contact to superconductors lead to new and
interesting physics.

Nevertheless, the non-homogeneities bring even more. We review below several
different effects arising in the non-homogeneous situation. It turns out
that a non-homogeneous alignment of the exchange field leads to a
complicated spin structure of the superconducting condensate. As a result,
not only the singlet component of the condensate exists but also a triplet
one with all possible projections of the total spin of the Cooper pair ($%
S_{z}=0,\pm 1$). In contrast to the singlet component, the spins of the
electrons in the triplet one with $S_{z}=\pm 1$ are parallel to each other.
The condensate (Gor'kov) function $f_{tr}$ of the triplet state is an odd
function of the Matsubara frequency\footnote{%
Superconductivity caused by the triplet odd in $\omega $ condensate is
called here odd superconductivity.}. The singlet part $f_{sng}$ is, as
usual, an even function of $\omega $ but it changes sign when interchanging
the spin indices. This is why the anticommutation relations for the
equal-time functions $f_{tr}(t,t)$ and $f_{sng}(t,t)$ remain valid; in
particular, $f_{tr}(t,t)=0$ and $f_{sng}(t,t)\neq 0$. Therefore the
superconductivity in the $S/F$ structures can be very unusual: alongside
with the usual BCS singlet part it may contain also the triplet part which
is symmetric in the momentum space (in the diffusive case) and odd in
frequency. Both components are insensitive to the scattering by non-magnetic
impurities and hence survive in the $S/F$ structures even if the mean free
path $l$ is short. When generated, the triplet component is not destroyed by
the exchange field and can penetrate the ferromagnet over long distances of
the order of $\xi _{N}=\sqrt{D_{F}/2\pi T}$.

In Chapter \ref{EXO} we analyze properties of this new type of
superconductivity that may arise in $S/F$ structures. We emphasize that this
triplet superconductivity is generated by the exchange field and, in the
absence of the field, one would have the conventional singlet pairing.

The superconductor-ferromagnet multilayers are a very interesting and
natural object for observation of Josephson effects. The thickness of both
the superconductor and ferromagnetic layers, as well as the transparency of
the interface, can be varied experimentally. This makes possible a detailed
study of many interesting physical quantities. As we have mentioned, an
interesting manifestation of the role played by the ferromagnetism is the
possibility of a $\pi $-junction.

However, this is not the only interesting effect and several new
ones have been recently proposed theoretically. As not so much
time has been passed, they have not been confirmed experimentally
unambiguously but there is no doubt that proper experiments will
have been performed soon. {In Chapter \ref{JOS} we discuss new
Josephson effects in multilayered $S/F$ structures taking into
account a possible change of the mutual direction of the
magnetization in the ferromagnetic layers.} We discuss a simple
situation when the directions of the magnetic moments in a
$SF/I/FS$ structure are collinear and the Josephson current flows
through an insulator ($I$) but not through the ferromagnets.
Naively, one could expect that the presence of the ferromagnets
leads to a reduction of the value of the critical current.
However, the situation is more interesting. The critical current
is larger when the magnetic moments of the $F$-layers are
antiparallel than when they are parallel. Moreover, it turns out
that the critical current for the antiparallel configuration is
even larger than the one in the absence of any ferromagnetic
layer. In other words, the ferromagnetism can enhance the critical
current \cite{BVE1}

Another setup is suggested in order to observe the odd triplet
superconductivity discussed in Chapter \ref{EXO}. Here the current
should flow through the ferromagnetic layers. Usually, one could
think that the critical current would just decay very fast with
increasing the thickness of the ferromagnetic layer. However,
another effect is possible. Changing the mutual direction of the
additional ferromagnetic layers one can generate the odd triplet
component of the superconducting condensate. This component can
penetrate the ferromagnetic layer as if it were a normal metal,
leading to large values of the critical current.

{Such structures can be of use for detecting and manipulating the
triplet component of the condensate in experiments.  In
particular, we will see  that in some S/F structures the type of
superconductivity is different in different directions: }in the
longitudinal direction (in-plane superconductivity) it is caused
mainly by the singlet component, whereas in the transversal
direction {the triplet component } mainly contributes to the
superconductivity. We discuss also possibilities of an
experimental observation of the triplet component.

Although the most pronounced effect of the interaction between the
superconductivity and ferromagnetism is the suppression of the
former by the latter, the opposite is also possible and this is
discussed in Chapter \ref {RED}. Of course, a weak ferromagnetism
should be strongly affected by the superconductivity and this
situation is realized in so called magnetic superconductors
\cite{bula_adv}. Less trivial is that the conventional strong
ferromagnets in the $S/F$ systems may also be considerably
affected by the superconductivity. This can happen provided the
thickness of the ferromagnetic layer is small enough. Then, it can
be energetically more profitable to enforce the magnetic moment to
rotate in space than to destroy the superconductivity. If the
period of such oscillations is smaller than the size of the Cooper
pairs $\xi _{S}$, the influence of the magnetism on the
superconductor becomes very small and the superconductivity is
preserved. In thick layers such an oscillating structure
(cryptoferromagnetic state) would cost much energy and the
destruction of the superconductivity is more favorable. Results of
several experiments have been interpreted in this way
{\cite{garifullin,muehge}. }

Another unexpected phenomenon, namely, the inverse proximity
effect is also presented in Chapter \ref{RED}. It turns out that
not only the superconducting condensate can penetrate the
ferromagnets but also a magnetic moment can be induced in a
superconductor that is in contact with a ferromagnet. This effect
has a very simple explanation. There is a probability that some of
the electrons of Cooper pairs enter the ferromagnet and its spin
tends to be parallel to the magnetic moment. At the same time, the
spin of the second electron of the Cooper pair should be opposite
to the first one (the singlet pairing or the triplet one with
$S_{z}=0$ is assumed). As a result, a magnetic moment with the
direction opposite to the magnetic moment in the ferromagnet is
induced in the superconductor over distances of the
superconducting coherence length $\xi _{S}$.

In principle, the total magnetic moment can be completely screened by the
superconductor. Formally, the appearance of the magnetic moment in the
superconductor is due the triplet component of the condensate that is
induced in the ferromagnet $F$ and penetrates into the superconductor $S$.
It is important to notice that this effect should disappear if the
superconductivity is destroyed by, e.g. heating, and this gives a
possibility of an observation of the effect. In addition to the Meissner
effect, this is one more mechanism of the screening of the magnetic field by
superconductivity. In contrast to the Meissner effect where the screening is
due to the orbital electron motion, this is a kind of spin screening.

Finally, in Chapter VI we discuss the results presented in the
review and try to anticipate future directions of the research.
The Appendix \ref{ApA} contains necessary information about the
quasiclassical approach in the theory of superconductivity.

We should mention that several review articles on $S/F$ related
topics have been published recently
\cite{proshin,golubov_rev,buzdin_rev,pokrovsky}. In these reviews
various properties of the $S/F$ structures are discussed for the
case of a homogeneous magnetization. In the review by
\citet{pokrovsky} the main attention is paid to effects caused by
a magnetic interaction between the ferromagnet and superconductor
(for example, a spontaneous creation of vortices in the
superconductor due to the magnetic interaction between the
magnetic moment of vortices and the magnetization in the
ferromagnet). We emphasize that, in contrast to these reviews, we
focus on the discussion of the triplet component with all possible
projections of the magnetic moment ($S_{z}=0,\pm 1$) arising only
in the case of a nonhomogeneous magnetization. In addition, we
discuss the inverse proximity effect, that is, the influence of
superconductivity on the magnetization $M$ of $S/F$ structures and
some other effects. Since the experimental study of the proximity
effects in the $S/F$ structures still remains in its infancy, we
hope that this review will help in understanding the conditions
under which one can observe the new type of superconductivity and
 other interesting effects and hereby will stimulate experimental
activity in this hot area.

\section{The proximity effect}

\label{PRO}

In this section we will review the basic features of the proximity effect in
different heterostructures. The first part is devoted to
superconductors-normal metals structures, while in the second part
superconductors in contact with homogeneous ferromagnets are considered.

\subsection{Superconductor-normal metal structures}

\begin{figure}[h]
\includegraphics[scale=0.35]{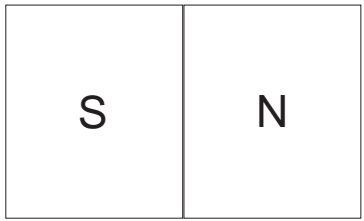}
\caption{S/N bilayer.}
\label{Fig.1}
\end{figure}

If a superconductor is brought in contact with a non-superconducting
material the physical properties of both materials may change. This
phenomenon called the \textit{proximity effect} has been studied for many
years. Both experiments and theory show that the properties of
superconducting layers in contact with insulating ($I$) materials remain
almost unchanged. For example, for superconducting films evaporated on glass
substrates, the critical temperature $T_{c}$ is very close to the bulk
value. However, physical properties of both metals of a normal
metal/superconductor ($N/S$, see FIG. \ref{Fig.1}) heterojunction with a
high $N/S$ interface conductance can change drastically.

Study of the proximity effect goes back to the beginning of the 1960's and
was reviewed in many publications (see, e.g. \citet{de_gennes} and %
\citet{deutscher}). It was found that the critical temperature of
the superconductor in a $S/N$ system decreased with increasing
$N$ layer thickness. This behavior can be interpreted as the
breaking down of some Cooper pairs due to the penetration of one
of the electrons of the pairs into the normal metal where they are
no longer attracted by the other electrons of the pairs.

\begin{figure}[h]
\includegraphics[scale=0.3]{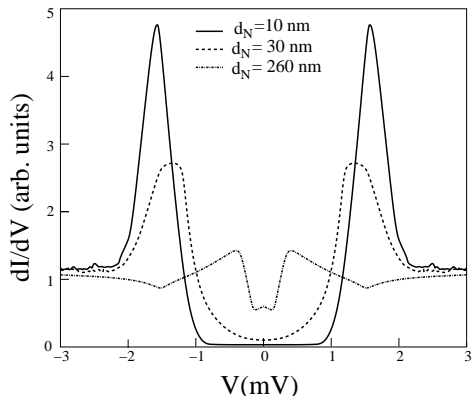}
\caption{Tunnelling density of states measured at 60 mK at the Au surface of
different Nb/Au bilayers samples with varying Au thickness $d_N$. Adapted
from \citet*{pannetier_dos_sn}. }
\label{Fig.2}
\end{figure}

At the same time, penetrating into the normal metal the Cooper
pairs induce superconducting correlations. For example, the
influence of the superconductivity on the physical properties of
the $N$ metal manifests itself in the suppression of the density
of states.  Experiments determining the density of states of S/N
bilayers with the help of tunneling spectroscopy were performed
many years ago  \cite{toplicar,adkins}. While spatially resolved
density of states were later measured by
\citet{pannetier_dos_sn,esteve96,esteve03} (see FIG. \ref{Fig.2}).
The simplest way to describe the proximity effect is to use the
Ginzburg-Landau (G-L) equation for the order parameter $\Delta $ \cite{GL}.
This equation is valid if the temperature is close to the critical
temperature of the superconducting transition $T_{c}$. In this case all
quantities can be expanded in the small parameter $\Delta /T_{c}$ and slow
variations of the order parameter $\Delta $ in space.

Using the G-L equation written as
\begin{equation}
\xi _{GL}\frac{\partial ^{2}\Delta \left( \mathbf{r}\right) }{\partial
\mathbf{r}^{2}}+\Delta \left( \mathbf{r}\right) (\mbox{${\rm sgn}$}%
(T_{c,N,S}-T)-\Delta ^{2}\left( \mathbf{r}\right) /\Delta _{0}^{2})=0\;.
\label{GL}
\end{equation}
one can describe the spatial distribution of the order parameter in any $N/S$
structure. Here $\xi _{GL}$ is the coherence length in the $N$ and $S$
regions at temperatures close to the critical temperatures $T_{cN,S}$. In
the diffusive limit this length is equal to
\begin{equation}
\xi _{GL}=\sqrt{\pi D_{N,S}/8|T-T_{cN,S}|}  \label{e1a}
\end{equation}
where $D_{N,S}$ is the diffusion coefficient in the $N$ and $S$ regions. The
quantity $\Delta _{0}$ is the bulk value of the order parameter in the
superconductor $S$. It vanishes when T reaches the transition temperature $%
T_{c}$.

It should be noticed though, that the region of the applicability of Eq. (%
\ref{GL}) for the description of the $S/N$ contacts is rather
restricted. Of course, the temperature must be close to the
transition temperature $T_{c}$ but this is not sufficient. The G-L
equation describes variations of the order parameters correctly
only if they are slow on the scales $v_{F}/T_{c}$ for the clean
case\ or $\sqrt{D_{N,S}/T_{c}}$ in the diffusive ``dirty'' case.
This can be achieved if the normal metal is a superconducting
material taken at a temperature exceeding its transition
temperature $T_{cN}$ and the transition temperatures $T_{cS}$ and
$T_{cN}$ are close to each other. If this condition is not
satisfied (e.g. $T_{cN}=0$) one should use more complicated
equations even at temperatures close to $T_{cS}$, {as we show
below.}

It follows from Eq. (\ref{GL}) that in the $S$ region, far from the $N/S$
interface, the order parameter $\Delta \left( \mathbf{r}\right) $ equals the
bulk value $\Delta _{0}$, whereas in the $N$ region $\Delta \left( \mathbf{r}%
\right) $ decays exponentially to zero on the length $\xi _{N}$.

The order parameter $\Delta \left( \mathbf{r}\right) $ is related to the
condensate function (or Gor'kov function)
\begin{equation}
f(t,t^{\prime })=<\psi _{\uparrow }(t)\psi _{\downarrow }(t^{\prime })>
\label{e2}
\end{equation}
via the self-consistency equation
\begin{equation}
\Delta _{N,S}(t)=\lambda _{N,S}f(t,t)\;,  \label{SelfCon}
\end{equation}
where $\lambda _{N,S}$ is the electron-electron coupling constant leading to
the formation of the superconducting condensate.

Eq.(\ref{GL}) describes actually a contact between two
superconductors with different critical temperatures $T_{cN,S}$,
when the temperature is chosen between $T_{cS}$ and $T_{cN}$. In
the case of a real normal metal the coupling constant $\lambda
_{N}$ is equal to zero and therefore $\Delta _{N}=0$. However,
this does not imply that the normal metal does not possess
superconducting properties in this case. The point is that many
important physical quantities are related not to the order
parameter $\Delta $ but to the condensate function $f$, Eq.
(\ref{e2}). For example, the non-dissipative condensate current
$j_{S}$ is expressed in terms of the function $f$ but not of
$\Delta $. If the contact between the $N$ and $S$ regions is good,
the condensate penetrates {the normal metal leading to a finite
value of $j_{S}\neq 0$ in this region.}

In the general case of an arbitrary $\lambda _{N}$ {it is
convenient to describe the penetration of the condensate (Cooper
pairs) into the }${N}${\ region in the diffusive limit by the
Usadel equation \cite{usadeleq} which is valid for all
temperatures and for  distances exceeding the mean free path
}$l${. This equation determines the so called quasiclassical
Green's functions (see Appendix \ref{ApA}) {which can be
conveniently used in problems involving length scales larger than
the Fermi wave length $\lambda_F$ and energies much smaller than
the Fermi energy.} Alternatively, one could try to find an exact
solution (the normal and anomalous electron Green's functions) for
the Gor'kov equations, {but this is in most of the cases a
difficult task.}}

{In order to illustrate the convenience of using the
quasiclassical method we calculate now the change of the
tunnelling density of states (DOS) in the normal metal due to the
proximity effect with the help of the Usadel equation. The DOS is
a very important  quantity that can be measured experimentally
and, at the same time, {can be calculated without difficulties}. }

We consider the $S/N$ structure shown in FIG. \ref{Fig.1} and
assume that the system is diffusive (i.e. the condition $\epsilon
\tau <<1$ is assumed to be fulfilled, where $\tau $ is the
momentum relaxation time and $\epsilon $ is the energy) and that
the transparency of the $S/N$ is low enough. In this case the
condensate Green's function $f(\epsilon )=\int dtf(t-t^{\prime
})\exp (i\epsilon (t-t^{\prime }))$ is small in the $N$ region and
the Usadel equation can be linearized (see Appendix \ref{ApA}).

Assuming that the boundary between the superconductor and normal metal is
flat and choosing the coordinate $x$ perpendicular to the boundary we reduce
the Usadel equation in the $N$ region to the form
\begin{equation}
D_{N}\partial ^{2}f/\partial x^{2}+2i\epsilon f=0\; ,  \label{NUs}
\end{equation}
where $D_{N}=v_{F}l/3$ is the classical diffusion coefficient.

The solution of this equation can be found easily and we write it as
\begin{equation}
f=f_{0}\exp \left( -x\sqrt{-2i\epsilon /D_{N}}\right) \; ,
\label{e3}
\end{equation}
where $f_{0}$ is a constant that is to be found from the boundary conditions.

We see that the solution for the condensate function $f$ decays in the $N$
region exponentially at distances inversely proportional to $\sqrt{\epsilon }
$. In many cases the main contribution to physical quantities comes from the
energies $\epsilon $ of the order of the temperature, $\epsilon \sim T$.
This means that the superconducting condensate penetrates the $N$ region
over distances of the order of $\xi _{N}=\sqrt{D_{N}/2\pi T}$. At low
temperatures this distance becomes very large, and if the thickness of the
normal metal layer is smaller than the inelastic relaxation length, the
condensate spreads throughout the entire $N$ region.

In order to calculate the DOS it is necessary to know the normal
Green's function $g$ which is related to the condensate function
$f$ via the normalization condition (see Appendix \ref{ApA})
\begin{equation}
g^{2}-f^{2}=1  \label{NorCon}
\end{equation}
Eqs. (\ref{NUs}) and (\ref{NorCon}) are written for the retarded
Green's function ($f=f^{R}$, see Appendix \ref{ApA}). They are
also valid for the advanced Green's functions provided $(\epsilon
+i0)$ is replaced by $(\epsilon -i0)$. The normalized
density-of-states (we normalize the DOS to the DOS of
non-interacting electrons) $\nu (\epsilon )$ is given by the
expression
\begin{equation}
\nu (\epsilon )=\mathrm{Re}g(\epsilon )  \label{DOS}
\end{equation}
As the condensate function $f$ is small, a correction $\delta \nu $ to the
DOS due to the proximity effect is also small. In the main approximation the
DOS $\nu $ is very close to its value in the absence of the superconductor, $%
\nu \approx 1$. Corrections to the DOS $\delta \nu $ are
determined by the condensate function $f$. From Eq. (\ref{NorCon})
one gets
\[
\delta \nu \approx f^{2}/2\; .
\]

Now we consider another case when the function $f$ is not small and the
correction\ $\delta \nu $ is of the order of unity. Then the linearized Eq. (%
\ref{NUs}) may no longer be used and we should write a more
general one. {For a S/N system} the general equation can be
written as (see Appendix \ref{ApA})
\begin{equation}
-iD_{S,N}\partial (\hat{g}\partial \hat{g}/\partial x)_{S,N}/\partial
x+\epsilon \lbrack \hat{\tau}_{3},\hat{g}_{S,N}]+[\hat{\Delta}_{S},\hat{g}%
_{S,N}]=0\;.  \label{Us1}
\end{equation}
This non-linear equation contains the quasiclassical matrix Green's function $%
\hat{g}$. Both normal $g$ and anomalous Green's functions $f$ enter as
elements of this matrix through the following relation (the phase in the
superconductor is set to zero)
\begin{equation}
\hat{g}_{N}=g_{N}\hat{\tau}_{3}+f_{N}i\hat{\tau}_{2}\;,  \label{g_N}
\end{equation}
where $\tau _{i}$, $i=1,2,3$ are Pauli matrices and $\left[ A,B\right]
=AB-BA $ is the commutator for any matrices $A$ and $B$.

We consider a flat $S/N$ interface normal to the $x$-axis. The
normal metal occupies the region $0<x<d_{N}.$ We assume that in
the normal metal $N$ there is no electron-electron interaction
($\lambda _{N}=0$, see Eq.(\ref {SelfCon})) so that in this
region the superconducting order parameter
vanishes, $\Delta _{N}=0.$ In the superconductor the matrix $\hat{\Delta}%
_{S} $ has the structure $\hat{\Delta}_{S}=\Delta i\hat{\tau}_{2}.$

At large distances from the $S/N$ interface the Green's functions $\hat{g}%
_{S}$ of the superconductor do not depend on coordinates and the first term
in Eq. (\ref{Us1}) can be neglected. Then we obtain a simpler equation
\begin{equation}
\epsilon \lbrack \hat{\tau}_{3},\hat{g}_{S}]+\Delta \lbrack i\hat{\tau}_{2},%
\hat{g}_{S}]=0\;.  \label{BCSeq}
\end{equation}
The solution for this equation satisfying the normalization condition (\ref
{NorCon}) is
\begin{equation}
g_{BCS}=\epsilon /\xi _{\epsilon };\text{ }f_{BCS}=\Delta /\xi _{\epsilon
}\;,  \label{BCS}
\end{equation}
where $\xi _{\epsilon }=\sqrt{\epsilon ^{2}-\Delta ^{2}}$. Eq.
(\ref{BCS}) is just the BCS solution {for a bulk superconductor.}

In order to find the matrix $\hat{g}(x)$ both in the $S$ and $N$ regions,
Eq.(\ref{Us1}) should be complemented by boundary conditions and this is a non-trivial problem. Starting from the initial Hamiltonian $\hat{H}%
_{tot}$, Eq. (\ref{e5}), one does not need boundary conditions at
the interface between the superconductor and the ferromagnet
because the interface can be described by introducing a proper
potential in the Hamiltonian. In this case the self-consistent
Gor'kov equations can be derived.

However, deriving the Usadel equation, Eq. (\ref{usadel}), we have
simplified the initial Gor'kov equations using the quasiclassical
approximation. {Possible spatial variation of the interface
potential on  a very small scale, due to the roughness of the
interface cannot be included in the quasiclassical equations. }
Nevertheless,  this problem is avoided deriving the quasiclassical
equations at distances from the interface exceeding the
wavelength. In the diffusive case one should go away from the
interface to distances larger than the mean free path $l$. In
order to match the solutions in the superconducting and
non-superconducting regions one should solve exact the  equations
near the interface and compare the asymptotic behavior of this
solution at large distances with the solutions of the Usadel
equation. This procedure is equivalent to solving the
quasiclassical equations with some boundary
conditions. These conditions were derived by \citet{zaitsev} and %
\citet{kuprianov} (see also Appendix \ref{ApA}, where these
conditions are discussed in more details). For the present case
they can be written as
\begin{equation}
2\gamma _{S,N}(\hat{g}\partial \hat{g}/\partial x)_{S,N}=[\hat{g}_{S},\hat{g}%
_{N}]|_{x=0}  \label{SNbounCon}
\end{equation}
where $\gamma _{S,N}=R_{b}\sigma _{S,N}$, $R_{b},$ measured in units $\Omega
cm^{2}$, is the $S/N$ interface resistance per unit area in the normal
state, and $\sigma _{S,N}$ are the conductivities of the $S$ and $N$ metals
in the normal state.

We assume that the thickness of the normal metal $d_{N}$ is smaller than the
characteristic penetration length $\xi _{N}(\epsilon )=\sqrt{D_{N}/\epsilon }
$ for a given energy $\epsilon $, that is\footnote{%
The quantity $E_{Th}=D_{N}/d_{N}^{2}$ is the so called Thouless energy} $%
\epsilon <<D_{N}/d_{N}^{2}=E_{T}$. Then the functions $g$ and $f$
remain almost constant over the thickness of the metal, and for
finding them, one can average the Usadel equation over the
thickness. In other words,  we assume that the thickness $d_{N}$
of the N layer satisfies the inequality
\begin{equation}
\;d_{N\text{ }}<<\sqrt{D_{N}/\epsilon },\text{ }\epsilon \sim \epsilon _{bN}
\label{Ineq1}
\end{equation}
($\epsilon _{bN}$ is a characteristic energy in the DOS of the $N$ layer)
and average Eq. (\ref{Us1}) over the thickness  $d_N$ considering $%
\hat{g}_{N}$ as a constant in the second term of this equation. Using the
boundary condition, Eq.(\ref{SNbounCon}), the first term in Eq. (\ref{Us1})
can be replaced after the integration by the commutator $[\hat{g}_{S},\hat{g}%
_{N}]|_{x=0}$. At $x=d_{N}$ the product $(\hat{g}\partial \hat{g}/\partial
x)_{N}$ is zero because the barrier resistance $R_{b}(d_{N})$ is infinite
(the current cannot flow into the vacuum). Finally, we obtain \cite{SubgapZ}
\begin{equation}
(\epsilon +i\epsilon _{b}g_{S}(0))[\hat{\tau}_{3},\hat{g}_{N}]+\epsilon
_{bN}if_{S}(0)[i\hat{\tau}_{2},\hat{g}_{N}]=0\;.  \label{EqN}
\end{equation}
where $\epsilon _{bN}=(D_{N}/2\gamma _{N}d_{N})$ is a new characteristic
energy that is determined by the $S/N$ interface resistance $R_{b}$. This
equation looks similar to Eq.(\ref{BCSeq}) after making the replacement $%
\hat{g}_{S}\rightarrow \hat{g}_{N}.$ The solution is similar to the solution
(\ref{BCS})
\begin{equation}
g_{N}=\tilde{\epsilon}/\tilde{\xi}_{\epsilon };\text{ }f_{N}=\tilde{\epsilon}%
_{bN}/\tilde{\xi}_{\epsilon }\;,  \label{gfN1}
\end{equation}
where $\tilde{\epsilon}=\epsilon +i\epsilon _{bN}g_{S}(0),$ $\tilde{\xi}%
_{\epsilon }=\sqrt{\tilde{\epsilon}^{2}-\tilde{\epsilon}_{bN}^{2}},$ $\tilde{%
\epsilon}_{bN}=\epsilon _{bN}if_{S}(0).$ Therefore the Green's functions in
the $N$ layer $g_{N}$ and $f_{N}$ are determined by the Green's functions on
the $S$ side of the $S/N$ interface $g_{S}(0)$ and $f_{S}(0).$ In order to
find the values of $g_{S}(0)$ and $f_{S}(0),$ one has to solve Eq. (\ref{Us1}%
) on the superconducting side ($x<0$). However, provided the inequality
\begin{equation}
\gamma _{N}/\gamma _{S}=\sigma _{N}/\sigma _{S}<<1  \label{Ineq2}
\end{equation}
is fulfilled one can easily show that, in the main approximation, the
solution in the $S$ region coincides with the solution for bulk
superconductors (\ref{BCS}). If the transparency of the $S/N$ interface is
not high, $\epsilon _{bN}<<\Delta $, the characteristic energies $\epsilon
\sim \epsilon _{bN}$ are much smaller than $\Delta $ and the functions $%
g_{S}(0)$ and $f_{S}(0)$ are equal to: $g_{S}(0)\approx g_{BCS}(0)\approx
\epsilon /i\Delta ,$ $f_{S}(0)\approx f_{BCS}(0)\approx 1/i.$ For these
energies the functions $g_{N}$ and $f_{N}$ have the same form as the BCS
functions $g_{BCS}$ and $f_{BCS}$ (\ref{BCS}) with the replacement $\Delta
\rightarrow \epsilon _{bN}$
\begin{equation}
g_{N}=\frac{\epsilon }{\sqrt{\epsilon ^{2}-\epsilon _{bN}^{2}}},\text{ \ }%
f_{N}=\frac{\epsilon _{bN}}{\sqrt{\epsilon ^{2}-\epsilon
_{bN}^{2}}}\; ,\label{gfN2}
\end{equation}
where $\epsilon _{bN}=D_{N}/(2R_{b}\sigma _{N}d_{N})$. The energy $\epsilon
_{bN}$\ can be represented in another form
\begin{equation}
\epsilon _{bN}=\frac{\pi ^{2}}{2}(\frac{R_{Q}}{R_{b}k_{F}^{2}})\hbar \frac{%
v_{F}}{d_{N}}=\hbar \frac{v_{F}}{d_{N}}(\frac{T_{b}}{4})\; .
\label{Subgap1}
\end{equation}
$R_{Q}=\hbar /e^{2}$\ is the resistance quantum, $v_{F}$\ and
$k_{F}$\ are the Fermi velocity and wave vector. When obtaining
the latter expression, we used a relation between the barrier
resistance $R_{b}$\ and an effective coefficient of transmission
$T_{b}$\ through the S/N interface
\cite{kuprianov,zaitsev}: $R_{b}\sigma _{n}=(2/3)(l/T_{b}),$\ where $%
l=v_{F}\tau $\ is the mean free path, $T_{b}=<T(\theta )\cos \theta
/(1-T(\theta )>,$\ $\theta $\ is the angle between the momentum of an
incoming electron and the vector normal to the S/N interface, and $T(\theta
) $\ is the angle dependent transmission coefficient. The angle brackets
mean an averaging over $\theta $.

An important result follows from Eq.(\ref{gfN2}): the DOS is zero at $%
|\epsilon |<\epsilon _{bN}$, \textit{i.e}., $\epsilon _{bN}$ is a minigap in
the excitation spectrum \cite{mcmillan}. Remarkably, in the considered limit
$\epsilon _{bN}<<\Delta $ the value of $\epsilon _{bN}$ does not depend on $%
\Delta $, but is determined by the interface transparency or, in
other words, by the interface resistance $R_{b}$. The appearance
of the minigap is related to Andreev reflections \cite{andreev}.

Eq. (\ref{Subgap1}) for the minigap is valid if the inequalities (\ref{Ineq1}) $%
\ $and $\epsilon _{bN}<\Delta $\ are fulfilled. Both inequalities can be
written as
\begin{equation}
(D_{N}/\Delta )/d_{b}<d_{N}<d_{b}
\end{equation}
where $d_{b}=2R_{b}\sigma _{N}$\ is a characteristic length. In the case of
a small interface resistance $R_{b}$ or a large thickness of the N layer,
that is, if the condition $\sqrt{\mathit{D}_{N}\mathit{/\Delta }},d_{b\text{
}}<d_{N}$\ is fulfilled, the value of the minigap in the N layer is given by
\cite{golubov_kupriyanov_96}
\begin{equation}
\mathit{\epsilon }_{bN}\mathit{=c}_{1}\frac{D_{N}}{d_{N}^{2}}\mathit{\;}
\label{Subgap2}
\end{equation}
where $c_{1}$\ is a factor of the order 1. This result has been
obtained from a numerical solution of the Usadel equation. {The
DOS for the case of  arbitrary
thickness }$d_{N}$ {\ and  interface transparency was calculated by %
\citet{bruder}.}

{The situation changes in the clean limit. Let us consider, for
example, a normal slab of a thickness }$d_{N}$ {in contact with an
infinite superconductor. If the Thouless energy }$E_{Th}=v_{F}/d_{N}${%
\ is less than }$\Delta ${, then discrete energy levels }$\epsilon
_{n}${\ appear \cite{sjames} in the N region due to Andreev
reflections \cite{andreev}. As a result, the DOS has sharp peaks at } $%
\epsilon =\epsilon _{n}$ {\ (for a recent review see %
\citet{deutscher_rev}).} If $E_{Th}$ {is much larger than }$\Delta ,$%
{\ the DOS }$\nu (\epsilon )$ {is zero at $\epsilon =0$ and
increases with increasing the energy }$\epsilon $ {(no gap).
However, this is true only for such a simple geometry. For samples
of more complicated shapes the behavior of the DOS }$\nu (\epsilon
)${\ depends on whether the electron dynamics in the N region is
chaotic or integrable
\cite{beenakker96,lodder,bruder,altland01,beenakker_rev}.}

Finally, it has been shown by \citet{altland} and
\citet{ostrovsky} that mesoscopic fluctuations smear out the
singularity in the DOS at $|\epsilon |=\epsilon _{bN}$ and the
DOS {in the diffusive limit} is finite, although small, for $%
|\epsilon |<\epsilon _{bN}.$ The minigap discussed above has been observed
on a Nb/Si bilayer system and on a Pb/Ag granular system by %
\citet{heslinga,kouh}, respectively.

{\ From this analysis we see that the proximity effect changes the DOS of
the normal metal which acquires superconducting properties.} In the next
section we will focus our attention on the case that the normal metal is a
ferromagnet. We will see that new interesting physics will arise from the
mutual interaction of superconductivity and magnetism.

\subsection{Superconductor-ferromagnet structures with an uniform
magnetization}

\label{PRO-SF}

In this section we consider the proximity effect between a superconductor $S$
and a ferromagnet $F$. We assume that the ferromagnet is a metal and has a
conduction band. In addition, there is an exchange field due to spins of
electrons of other bands.

As has been already mentioned, the effective exchange field acts on spins of
the conduction electrons in the ferromagnet, and an additional term $\hat{H}%
_{ex}$ describing this action appears in the total Hamiltonian
(for more details see Appendix \ref{ApA})
\begin{equation}
\hat{H}_{tot}=\hat{H}+\hat{H}_{ex}  \label{e5}
\end{equation}
\begin{equation}
\hat{H}_{ex}=-\int d^{3}\mathbf{r}\psi _{\alpha }^{+}\left( \mathbf{r}%
\right) \left( \mathbf{h}\left( \mathbf{r}\right) \sigma _{\alpha \beta
}\right) \psi \left( \mathbf{r}\right) d\mathbf{r}  \label{e6}
\end{equation}
\ where $\psi ^{+}\left( \psi \right) $ are creation and destruction
operators, $\mathbf{h}$ is the exchange field, $\mathbf{\sigma }_{\alpha
\beta }$ are Pauli matrices, and $\alpha ,\beta $ are spin indices. The
Hamiltonian $\hat{H}$ stands for a non-magnetic part of the Hamiltonian. It
includes the kinetic energy, impurities, external potentials, etc. and is
sufficient to describe all properties of the system in the absence of the
exchange field.

The energy of the spin-up electrons differs from the energy of the
spin-down
electrons by the Zeeman energy $2h$. Due to the presence of the term $\hat{H}%
_{ex}$ describing the exchange  interaction  all functions,
including the condensate Green's function $f$,  are generally
speaking non-trivial matrices in the spin space with non-zero
diagonal and off-diagonal elements.

The situation is  simpler if the direction of the exchange field
does not depend on coordinates. In this case, choosing the
$z$-axis along the direction of $\mathbf{h}$ one can consider
electrons with spin ``up'' and ``down'' separately. In this
Section we concentrate on this case. This can help the reader to
understand several interesting effects and get an intuition about
what one can expect from the presence of the exchange field. The
results of this section will also help in understanding which
effects in the superconductor-ferromagnet structures can be
considered as rather usual and what kind of behavior is
``exotic''. We will see that the exotic phenomena occur in cases
when the exchange field is not homogeneous and therefore postpone
their discussion until the next chapters.

If the exchange field $h$ is homogeneous the matrix $\hat{f}$ describing the
condensate $\hat{f}$ is diagonal and can be represented in the form
\begin{equation}
\hat{f}=f_{3}\hat{\sigma}_{3}+f_{0}\hat{\sigma}_{0}\;  \label{F1}
\end{equation}
where $f_{3}$ is the amplitude of the singlet component and $f_{0}$ is the
amplitude of the triplet component with zero projection of the magnetic
moment of Cooper pairs on the $z$ axis ($S_{z}=0$). {Note that in the case
of a }${S/N}${\ structure the condensate function has a singlet structure
only, i.e. it is proportional to $\mbox{$\hat{\sigma}_3$}$. The presence of
the exchange field leads to the appearance of the triplet term proportional
to $\mbox{$\hat{\sigma}_0$}$}

The amplitudes of the singlet and triplet components are related
to the correlation functions $\left\langle \psi _{\alpha }\psi
_{\beta }\right\rangle $ as follows \cite{legget,woelfle_book}
\begin{eqnarray}
f_{3}(t) &\sim &\left\langle \psi _{\uparrow }(t)\psi _{\downarrow
}(0)\right\rangle -\left\langle \psi _{\downarrow }(t)\psi _{\uparrow
}(0)\right\rangle \;,  \nonumber \\
f_{0}(t) &\sim &\left\langle \psi _{\uparrow }(t)\psi _{\downarrow
}(0)\right\rangle +\left\langle \psi _{\downarrow }(t)\psi _{\uparrow
}(0)\right\rangle \;,
\end{eqnarray}
One can see that a permutation of spins does not change the function $%
f_{3}(0),$ whereas such a permutation leads to a change of the sign of $%
f_{0}\left( 0\right) $. {This means that the amplitude of the
triplet component taken at equal times is zero in agreement with
the Pauli exclusion principle.} Later we will see that in the case
of a non-homogeneous magnetization all triplet components
including $\left\langle \psi _{\uparrow }(t)\psi _{\uparrow
}(0)\right\rangle $ and $\left\langle \psi _{\downarrow }(t)\psi
_{\downarrow }(0)\right\rangle $ differ from zero.

{ Once one determines the condensate function, Eq. (\ref{F1}), one
is able to determine physical quantities ,as DOS, the critical
temperature $T_C$, or the Josephson critical current through a
S/F/S junction.}

Next paragraphs are devoted to a discussion of these physical
properties in $F/S$ systems with  homogeneous magnetization.


\subsubsection{Density of states (DOS)}

{In this section we discuss the difference between the DOS in
}${S/N}${\ and }${S/F}${\ structures.} General equations for the
quasiclassical Green's functions describing the system can be
written  but they are rather complicated (see Appendix \ref{ApA}).
In order to simplify the problem and, at the same time, give the
basic idea about the effects it is sufficient to consider some
limiting cases. This will be done in the present section leaving
the general equations for the Appendix \ref{ApA}.

In the case of a weak proximity effect, the condensate function
$\hat{f}$ is small outside the $S$ region. We consider again the
diffusive limit. Then, the general Eq. (\ref{usadel}) can be
linearized and one obtains an equation for the matrix $\hat{f}$ $\
$similar to Eq.(\ref{NUs}) but containing an extra term due to the
exchange field $h$
\begin{equation}
D_{F}\partial ^{2}\hat{f}_{F}/\partial x^{2}+2i(\epsilon \hat{\sigma}_{0}+h%
\hat{\sigma}_{3})\hat{f}_{F}=0\;.  \label{2.B.1}
\end{equation}
 The subscript $F$ stands for the $F$ region.

In the absence of the exchange field $h$, Eq. (\ref{2.B.1}) reduces to Eq. (%
\ref{NUs}). It is important to emphasize that Eq. (\ref{2.B.1}) is valid for
a homogeneous $h$ only. Any variation of $h$ in space makes the equation
much more complicated.

Eq. (\ref{2.B.1}) should be complemented by boundary conditions
which take the form (see Appendix \ref{ApA})
\begin{equation}
\gamma _{F}\partial \hat{f}_{F}/\partial x=-\hat{f}_{S}  \label{2.B.2}
\end{equation}
where $\gamma _{F}=R_{b}\sigma _{F}$, $R_{b}$ is the boundary resistance per
unit area, $\sigma _{F}$ is the conductivity of the $F$ region, $\hat{f}%
_{F,S}$ are the condensate matrix functions in the $F$ and $S$ regions.
Since we assume a weak proximity effect, a deviation of the $\hat{f}_{S}$
from its $BCS$ value $\hat{f}_{BCS}=\hat{\sigma}_{3}f_{BCS}$ is small.
Therefore on the right-hand side of Eq.(\ref{2.B.2}) one can write $\hat{f}%
_{S}\approx \hat{\sigma}_{3}f_{BCS},$ where $f_{BCS}$ is defined in Eq.(\ref
{Us1}). At the ferromagnet/vacuum interface the boundary condition is given
by the usual expression $\partial _{x}\hat{f}_{F}=0$, which follows from the
condition $R_{b}\rightarrow \infty $.

Using Eq. (\ref{2.B.2}), one can easily solve Eq. (%
\ref{2.B.1}). We assume, as in the previous section, that the normal metal
(ferromagnet) is in a contact with the superconductor at $x=0$ ($x$ is the
coordinate perpendicular to the interface). The other boundary of the
ferromagnet is located at $x=d_{F}$ and the space at $x>d_{F}$ is empty.

The proper solution for the diagonal matrix elements $f_{\pm }\equiv
f_{11(22)}$ can be written as
\begin{equation}
f_{\pm }(x) =\left\{
\begin{array}{cc}
\pm \frac{f_{BCS}}{\kappa _{\epsilon \pm }\gamma _{F}}\frac{\cosh
(\kappa _{\epsilon \pm
}(x-d_{F}))}{\sinh (\kappa _{\epsilon \pm }d_{F})} & 0<x<d_{F} \\
\text{\ }0\text{ } & x>d_{F}
\end{array}
\right. \text{ }\;.  \label{2.B.3}
\end{equation}
Here $\kappa _{\epsilon \pm }=\sqrt{-2i(\epsilon \pm h)/D_{F}}$ is a
characteristic wave vector that determines the inverse penetration depth of
the condensate functions $f_{0,3}$ into the ferromagnet.

Usually, the exchange energy $h$ is much larger than the energy $\epsilon $ (%
$\epsilon \propto \max \{\Delta ,T\}$). This means that the
condensate penetration depth $\xi _{F}=\sqrt{D_{F}/h}$ is much
shorter than the penetration depth into a normal (non-magnetic)
metal $\xi _{N}$. The strong suppression of the condensate in the
ferromagnet is caused by the exchange interaction that tries to
align the spins of \ electrons parallel to the magnetization. This
effect destroys the Cooper pairs with zero total magnetic moment.

It is worth mentioning that the condensate function $f_{\pm }$ experiences
oscillations in space. Indeed, for a thick $F$ layer \ ($d_{F}>>\xi _{F}$)\
\ we obtain from Eq. (\ref{2.B.3}).
\begin{equation}
f_{\pm }=\pm \frac{\Delta}{E_{\epsilon }\kappa _{F\pm }\gamma
_{F}}\exp (-x/\xi _{F})[\cos (x/\xi _{F})\pm i\sin (x/\xi _{F})].
\label{2.B.4}
\end{equation}
where $E_{\epsilon }=\sqrt{\epsilon ^{2}-\Delta ^{2}},\kappa _{F\pm }=\kappa
_{\epsilon \pm }(\epsilon )$ at $\epsilon =0$. The damped oscillations of $%
\;f_{\pm }$ lead to many interesting effects and, in particular, to a
non-monotonic dependence of the critical temperature on the thickness $d_{F}$
of a $F/S$ bilayer which will be discussed in the next section.

\begin{figure}[h]
\includegraphics[scale=0.35]{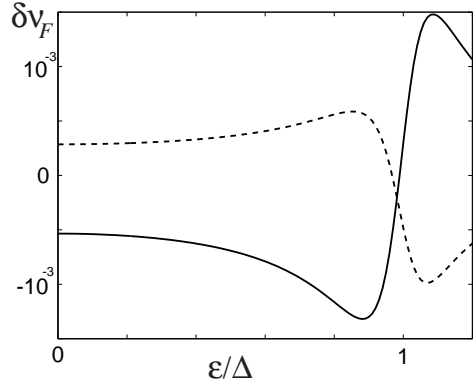}
\caption{Calculated change of the local density of states for a S/F bilayer
at the outer F interface. The solid line corresponds to a F thickness $%
d_{F}=0.5\protect\xi _{0}$, where $\protect\xi _{0}=\protect\sqrt{%
D_{F}/\Delta }$, while the dashed one corresponds to $d_{F}=0.8\protect\xi
_{0}$. The latter curve is multiplied by a factor of 10. }
\label{Fig.3}
\end{figure}

{In order to calculate the DOS we have to use the normalization condition,
Eq. (\ref{NorCon}), which} is also valid for the matrix elements $f_{\pm }$
and $g_{\pm }$. Thus, for $g_{\pm }$ we obtain $g_{\pm }=\sqrt{1+f_{\pm }^{2}%
}$, which can be written for small $f_{\pm }$ as $g_{\pm }\approx
1+f_{\pm }^{2}/2$. Then the correction to the normalized DOS in
the $F$ region $\delta \nu _{F}=\nu _{F}-1$ takes the form
\begin{equation}
\delta \nu _{F}(x)=\mathrm{Re}(f_{+}^{2}+f_{-}^{2})/4\;.  \label{FDOSvar}
\end{equation}
Substituting Eq. (\ref{2.B.3}) into Eq.(\ref{FDOSvar}), we obtain finally
the DOS variation at the edge of the F film
\begin{eqnarray}
\delta \nu _{F}(d_{F})&=&(1/4)\mathrm{Re}\left\{(\frac{f_{BCS}}{\gamma _{F}}%
)^{2}\left[(\kappa _{\epsilon +}\sinh (\kappa _{\epsilon
+}d_{F}))^{-2}+\right.\right.\nonumber\\
& & \left.\left.+(\kappa _{\epsilon -}\sinh (\kappa _{\epsilon
-}d_{F}))^{-2}\right]\right\}\;.
\end{eqnarray}
In FIG. \ref{Fig.3} we plot the function $\delta \nu _{F}(\epsilon )$ for
different thicknesses $d_{F}$ and $h/\Delta =20$. It is seen that at zero
energy $\epsilon =0$ the correction to DOS $\delta \nu _{F}$ is positive for
$F$ films with $d_{F}=0.8\xi _{0}$ while it is negative for films with $%
d_{F}=0.5\xi _{0}$ where $\xi _{0}=\sqrt{D_{F}/\Delta }$.
{Such a behavior of the DOS, which is typical for S/F systems, has
been observed experimentally by \citet{kontos01} in a bilayer
consisting of a thin PdNi film (}$5nm<d_{F}<7.5nm${) on the top of
a thick superconductor. The DOS was determined by tunnelling
spectroscopy.} This type
of dependence of $\delta \nu _{N}$ on $d_{N}$ can also be obtained in $%
N/S$ contacts but for finite energies $\epsilon $. In the $F/S$
contacts the energy $\epsilon $ is shifted, $\epsilon \rightarrow
\epsilon \pm h$ (time-reversal symmetry breaking) and this leads
to a non-monotonic dependence of $\nu _{F}$ on the thickness
$d_{F}$ even at zero energy. {On the other hand, a non oscillatory
behavior of the\ DOS }$\nu (\epsilon )${\ has been found recently
in experiments on Nb/CoFe bilayers \cite{beasley05}. The
discrepancy between the existing theory and the experimental data
may be due to the small thicknesses of the ferromagnetic layer
(}$0.5nm<d_{F}<2.5nm${) which is comparable with the Fermi wave
length $\lambda _{F}\approx 0.3nm$. Strictly speaking, in this
case the Usadel equation cannot be applied.}

The DOS in $F/S$ structures was studied theoretically in many papers. %
\citet{halterman} studied the DOS variation numerically for ballistic $F/S$
structures. The DOS in quasiballistic $F/S$ structures was investigated by %
\citet{baladie}, \citet{BVE_dos} and \citet{nazarov_dos} and for
dirty $F/S$ structures by \citet{fazio} and \citet{buzdin}. {The
subgap in a
dirty S/F/N structure was investigated in a recent publication by %
\citet{golubov05}}.

It is interesting to note that in the ballistic case ($\tau h>>1$, $\tau $
is the momentum relaxation time) the DOS in the $F$ layer is constant in the
main approximation in the parameter $1/(\tau h)$ while in the diffusive case
($\tau h<<1$) it\ experiences the damped oscillations. The reason for the
constant DOS in the ballistic case is that both the parts of $f$ , the
symmetric and antisymmetric in the momentum space, contribute to the DOS.
Each of them oscillates in space. However, while in the diffusive case the
antisymmetric part is small, in the ballistic case the contributions of both
parts to the DOS are equal to each other, but opposite in sign, thus
compensating each other.

Finally, we would like to emphasize that both, the singlet and
triplet components, contribute to the DOS. As it is seen from
Eq.(\ref{FDOSvar}), the changes of the DOS can be represented in
the form $\delta \nu _{F}={\rm Re}(f_{0}^{2}+f_{3}^{2})/4$, which
demonstrates explicitly this fact.

\subsubsection{Transition temperature}

\label{PRO-SF-TC}

As we have seen previously, the exchange field affects greatly the singlet
pairing in conventional superconductors. Therefore the critical temperature
of the superconducting transition $T_{c}$ is considerably reduced in $S/F$
structures with a high interface transparency.

\begin{figure}[h]
\includegraphics[scale=0.35]{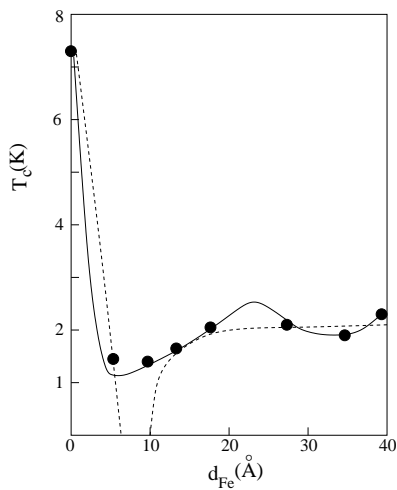}
\caption{Dependence of the superconducting transition temperature on the
thickness of the Fe layer as determined by resistivity measurements. The
dashed line is a fit assuming a perfect interface transparency while the
solid line corresponds to a non-perfect interface. Adapted from \citet*{lazar%
}. }
\label{Fig.4}
\end{figure}

{The critical temperature for $S/F$ bilayer and multilayered
structures was calculated in many works \cite
{radovic2,demler,khusainov,tagirov_C,baladie_tc,golubov_tc,bagrets,golubov_triplet,proshin98, proshin99,proshin01,buzdin_kupr1,buzdin05,you,buzdin_wr}%
.} Experimental studies of the $T_{c}$ were also reported in many
publications \cite{jiang,muehge,aarts,lazar,bader}. A good
agreement between
theory and experiment has been achieved in some cases (see FIG. \ref{Fig.4}%
). 

One has to mention that, despite of many papers published on this subject,
the problem of the transition temperature $T_{c}$ in the $S/F$ structures is
not completely clear. {For example, \citet{jiang} and \citet{ogrin} claimed
that the non-monotonic dependence of $T_{c}$ on the thickness of the
ferromagnet observed on }${Gd/Nb}${\ samples was due to the oscillatory
behavior of the condensate function in }${F}${. However, \citet{aarts} in an
other experiment on }${V/FeV}${\ have shown that the interface transparency
plays a crucial role in the interpretation of the experimental data {%
that showed both non-monotonic and monotonic dependence } of $T_{c}$ on $%
(d_{F})$. {In other experiments \cite{dynes02} the critical
temperature of the bilayer Pb/Ni decreases with increasing
}$d_{F}${\ in a monotonic way.}}

From the theoretical point of view the $T_{c}$ problem in a
general case cannot be solved exactly. In most papers it is
assumed that the transition
to the superconducting state is of second order, i.e. the order parameter $%
\Delta $ varies continuously from zero to a finite value with decreasing the
temperature $T$. However, generally this is not so.

Let us consider, for example, a thin $S/F$ bilayer with thicknesses obeying
the condition: $d_{F}<\xi _{F}$, $d_{S}<\xi _{S}$, where $d_{F,S}$ are the
thicknesses of the $F(S)$ layer. In this case the Usadel equation can be
averaged over the thickness (see for instance, \citet{BVE1}) and reduced to
an equation describing an uniform magnetic superconductor with an effective
exchange energy $\tilde{h}$ and order parameter $\tilde{\Delta}$.

This problem can easily be solved. The Green's functions $g_{\pm }$ and $%
f_{\pm }$ are given by
\begin{equation}
g_{\pm }=\frac{\epsilon \pm \tilde{h}}{E_{\epsilon }},\text{ \ }f_{\pm }=%
\frac{\tilde{\Delta}}{E_{\epsilon }}\;,
\end{equation}
where $E_{\epsilon }=\sqrt{(\epsilon \pm \tilde{h})^{2}-\tilde{\Delta}^{2}},$
$\tilde{h}=r_{F}h$, $\tilde{\Delta}=r_{S}\Delta $, $r_{F}=1-r_{S}=\nu
_{F}d_{F}/(\nu _{F}d_{F}+\nu _{S}d_{S})$. In this case the Green's functions
are uniform in space and have the same form as in a magnetic superconductors
or in a superconducting film in a parallel magnetic field acting on the
spins of electrons.

The difference between the $S/F$ bilayer system and a magnetic
superconductors is that the effective exchange energy $\tilde{h}$ depends on
the thickness of the $F$ layer and may be significantly reduced in
comparison with its value in a bulk ferromagnet. A thin superconducting film
in a strong magnetic field $H=\tilde{h}/\mu _{B}$ ($\mu _{B}$ is an
effective Bohr magneton) is described by the same Green's functions. The
behavior of these systems and, in particular, the critical temperature of
the superconducting transition $T_{c}$, was studied long ago by %
\citet{lo_state,fulde_state,sarma,maki}. It was established that
both  first  and second order phase transitions may occur in these
systems if $\tilde{h}$ is less or of the order of
$\tilde{\Delta}$. If the effective exchange field $\tilde{h}$
exceeds the value $\tilde{\Delta}/\sqrt{2}\approx
0.707\tilde{\Delta},$ the system remains in the normal state {(the %
\citet{clogston} and \citet{chandra} limit). Independently from each other %
\citet{lo_state} and \citet{fulde_state} found that in a clean
system and in a narrow interval of $\tilde{h}$ the homogeneous
state is unstable and an inhomogeneous state with the order
parameter varying in space is established in the system. This
state, denoted as the Fulde-Ferrel-Larkin-Ovchinnikov (LOFF)
state.}  has not been observed yet in bulk superconductors. In
bilayered $S/F$ systems such a state cannot be realized because of
a short mean free path.

In the case of a first order phase transition from the
superconducting to the normal state the order parameter $\Delta $
drops from a finite value to zero. {The study of this transition
requires the use of nonlinear equations for $\Delta $. It was
shown by \citet{tollis} that under some assumptions both the first
and second order phase transitions may occur in a S/F/S
structure.}

In the case of a second order phase transition one can linearize the
corresponding equations (the Eilenberger or Usadel equation) for the order
parameter and use the Ginzburg-Landau expression for the free energy
assuming that the temperature $T$ is close to the critical temperature $%
T_{c} $. Just this case was considered in most papers on this topic. The
critical temperature of an $S/F$ structure can be found from an equation
which is obtained from the self-consistency condition Eq. (\ref{SelfCon}).
In the Matsubara representation it has the form
\begin{equation}
\ln \frac{T_{c}}{T_{c}^{\ast }}=(\pi T_{c}^{\ast })\sum_{\omega }(\frac{1}{%
|\omega _{n}|}-if_{\omega }/\Delta ),  \label{Tc}
\end{equation}
where $T_{c}$ is the critical temperature in the absence of the proximity
effect and $T_{c}^{\ast }$ is the critical temperature with taking into
account the proximity effect.

The function $f_{\omega }$ is the condensate (Gor'kov) function in the
superconductor; it is related to the function $f_{S3}(\epsilon )$ as $%
f_{S3}(i\omega _{n})=f_{\omega },$ where $\omega _{n}=\pi (2n+1)$ is the
Matsubara frequency. Strictly speaking, Eq.(\ref{Tc}) is valid for a
superconducting film with a thickness smaller than the coherence length $\xi
_{S}$ because in this case $f_{\omega }$ is almost constant in space.

The quasiclassical Green's function $f_{\omega }$ obeys the Usadel
equation (in the diffusive case) or the more general Eilenberger
equation. One of
these equations has to be solved by using the boundary conditions at the $%
S/F $ interface (or $S/F$ interfaces in case of multilayered structures).
This problem was solved in different situations in many works where an
oscillation of $T_{c}$ as a function of the F thickness was predicted (see
FIG. \ref{Fig.4}). In most of these papers it was assumed that magnetization
vectors $\mathbf{M}$ in different $F$ layers are collinear. Only %
\citet{golubov_triplet} considered the case of an arbitrary angle $\alpha $
between the $\mathbf{M}$ vectors in two $F$ layers separated by a
superconducting layer.

As mentioned previously, in this case the triplet components with all
projections of the spin $S$ of the Coopers pair arise in the $F/S/F$
structure. It was shown that $T_{c}$ depends on $\alpha $ decreasing from a
maximum value $T_{c\max }$ at $\alpha =0$ to a minimum value $T_{c\min }$ at
$\alpha =\pi .$ We will not discuss the problem of $T_{c}$ for $S/F$
structures in detail because this problem is discussed in other review
articles \cite{proshin, buzdin_rev}.

\subsubsection{The Josephson effect in SFS junctions}

\label{PRO-SF-IC} 
{The oscillations of the condensate function in the ferromagnet (see Eq.(\ref
{2.B.4})) lead to interesting peculiarities not only in the dependence $%
T_{c}(d_{F})$ but also in the Josephson effect in the }${S/F/S}${\
junctions. Although, as it has been mentioned in the previous section, the
experimental results concerning the dependence $T_{c}(d_{F})$ are still
controversial, there is a more clear evidence for these oscillations in
experiments on the Josephson current measurements that we will discuss here.}

It turns out that under certain conditions the Josephson critical current $%
I_{c}$ changes its sign and becomes negative. In this case the energy of the
Josephson coupling $E_{J}=(\hbar I_{c}/e)[1-\cos \varphi ]$ has a minimum in
the ground state when the phase difference $\varphi $ is equal not to $0$,
as in ordinary Josephson junctions, but to $\pi $ (the so called $\pi -$%
junction).

This effect was predicted for the first time by \citet{bula_kuzi}. The
authors considered a Josephson junction consisting of two superconductors {%
separated by a region containing magnetic impurities}. The
Josephson current  through a $S/F/S$ junction was calculated for
the first time by \citet{buzdin_sfs}. Different aspects of the
Josephson effect in S/F/S structures has been studied in many
subsequent papers
\cite[e.g]{buzdin_kupr1,radovic03,fogel00,barash,golubov_phi,schoen,nazarov_sfs,zyuzin}.
Recent experiments confirmed the $0$-$\pi$ transition of the
critical current in S/F/S junctions
\cite{ryazanov,kontos02,palevski,strunk04,sellier}.

In the experiments of \citet{ryazanov} and \citet{palevski}, $Nb$ is used as
a superconductor and a $Cu_{x}Ni_{1-x}$ alloy as a ferromagnet. %
\citet{kontos02} used a more complicated $S_{1}/F/I/S$ structure, where $%
S_{1}$ is a $Nb/Al$ bilayer, $S$ is $Nb$, $I$ is the insulating $Al_{2}O_{3}$
layer and $F$ is a thin ($40$\AA $<d_{F}<150$\AA ) magnetic layer of a $PdNi$
alloy. All these structures exhibit oscillations of the critical current $%
I_{c}$. 
\begin{figure}[h]
\includegraphics[scale=0.3]{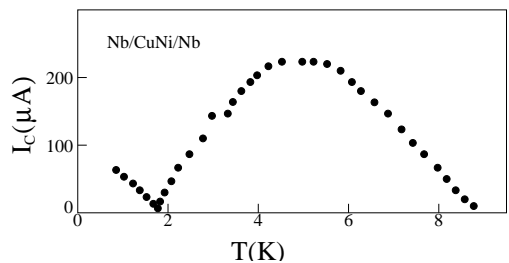}
\caption{Measurement of the critical current $I_c$ as a function of the
temperature for a Nb/Cu$_{0.48}$Ni$_{0.52}$/Nb junction. The thickness of
the CuNi layer is $d_F=22$nm. Adapted from \citet*{ryazanov} }
\label{Fig.5}
\end{figure}
In FIG. \ref{Fig.5} the temperature dependence of $I_{c}$ measured by %
\citet{ryazanov} is shown. It is seen that the critical current in the
junction with $d_{F}=27nm$ turns to zero at $T\approx 2$ $K$, rises again
with increasing temperature and reaches a maximum at $T\approx 5.5$ $K$. If
temperature increases further, $I_{c}$ decreases. In FIG. \ref{Fig.6} we
also show the dependence of $I_{c}$ on the thickness $d_{F}$ measured by %
\citet {palevski}. The measured oscillatory dependence is well fitted with
the theoretical dependence calculated by \citet{buzdin_sfs} and %
\citet{BVE_josephson}. 
\begin{figure}[h]
\includegraphics[scale=0.3]{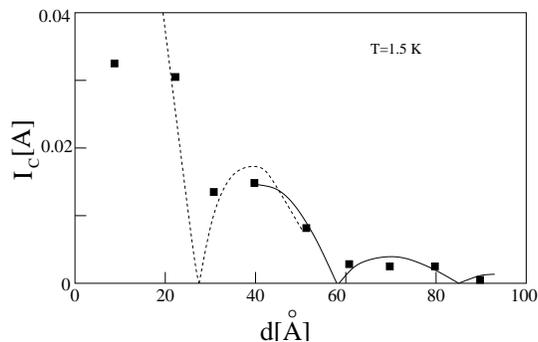}
\caption{Critical current of a Nb/Cu/Ni/Cu/Nb junction as a function of the
Ni layer thickness $d$. The squares are the measured points. The theoretical
fits are done according to \citet{buzdin_sfs} (dashed line) and
\citet{BVE_josephson} (solid line). Adapted from \citet*{palevski} }
\label{Fig.6}
\end{figure}
The $\pi $-state in a Josephson junction leads to some observable phenomena.
As was shown by \citet{bula_kuzi}, a spontaneous supercurrent may arise in a
superconducting loop with a ferromagnetic $\pi $-junction. This current has
been measured by \citet{strunk04}. Note also that the fractional Shapiro
steps in a ferromagnetic $\pi $-junction were observed by \citet{sellier} at
temperatures at which the critical current $I_{c}$\ turns to zero.

Oscillations of the Josephson critical current $I_{c}$ are related
to the oscillatory behavior of the condensate function $f$ in
space (see Eq.(\ref {2.B.4})). The critical current $I_{c}$ in a
$S/F/S$ junction can easily be obtained once the condensate
function in the $F$ region is known. We use the following formula
for the superconducting current $I_{S}$ in the diffusive limit
which follows in the equilibrium case from a general expression
(see Appendix \ref{ApA})
\begin{equation}
\;I_{S}=L_{y}L_{z}\sigma _{F}(i\pi T/4e)\sum_{\omega }Tr(\hat{\tau}_{3}[\;%
\check{f}_{+}\partial \;\check{f}_{+}/\partial x+\;\check{f}_{-}\partial \;%
\check{f}_{-}/\partial x]),  \label{I_S}
\end{equation}
where $L_{y}L_{z}$ is the area of the interface and $\sigma _{F}$ is the
conductivity of the $F$ layer.

In the considered case of a non-zero phase difference the condensate
functions $f_{\pm }$ are matrices in the particle-hole space. If in Eq.(\ref
{I_S}) instead of $f_{\pm }$ we write a 4$\times $4 matrix for $\check{f}$,
then $\Delta $ is given by $\hat{\Delta}=\Delta (i\hat{\tau}_{2}\cos
(\varphi /2)\mp i\hat{\tau}_{1}\sin (\varphi /2))\hat{\sigma}_{3}.$ We set
the phase of the right (left) superconductor equal to $\pm \varphi /2$. For
simplicity we assume that the overlap between the condensate functions $%
f_{\pm }$ induced in the $F$ region by each superconductor is small. This
assumption is correct in the case $d_{F}>>\xi _{F}$. Under this assumption
the condensate function may be written in the form of two independently
induced $f$ functions
\begin{eqnarray}
\;\hat{f}_{\pm }(x) &=&(1/\xi _{\epsilon }\kappa _{F\pm }\gamma _{F})i\hat{%
\tau}_{2}[\hat{\Delta}_{l}\exp (-\kappa _{\epsilon \pm }(x+d_{F}/2)
\nonumber \\
&+&\hat{\Delta}_{r}\exp (-\kappa _{\epsilon \pm }(-x+d_{F}/2)].
\label{FfJos}
\end{eqnarray}
Here $\hat{\Delta}_{r(l)}$ is the order parameter in the right
(left) superconductor. Substituting Eq.(\ref{FfJos}) into
Eq.(\ref{I_S}), we get
\begin{eqnarray}
\;I_{S}\equiv I_{c}\sin (\varphi ) &=&4\pi T(L_{y}L_{z})\sigma _{F}/(\kappa
_{F}\gamma _{F}^{2})\exp (-d_{F}/\xi _{F})  \nonumber \\
&&\cos (d_{F}/\xi _{F})\sum_{\omega }\frac{\Delta ^{2}}{\Delta ^{2}+\omega
^{2}}\sin \varphi .  \label{I_S1}
\end{eqnarray}
When deriving Eq.(\ref{I_S1}), it was assumed that the exchange energy $h$
is much larger than both $T$ and $\Delta .$

Calculating the sum in Eq. (\ref{I_S1}), we come to the final formula for
the critical current
\begin{equation}
\;I_{c}=\Delta \tanh (\Delta /2T)\sigma _{F}/(\kappa _{F}\gamma
_{F}^{2})\exp (-d_{F}/\xi _{F})\cos (d_{F}/\xi _{F}).  \label{I_c}
\end{equation}
As expected, according to Eq.(\ref{I_c}) {\ the critical current oscillates
with varying the thickness of the ferromagnet $d_{F}$. The period of these
oscillations gives the value of $\xi _{F}$ and therefore the value of the
exchange energy $h.$ For example, according to the experiments on Nb/CuNi
performed by \citet{palevski} $h\approx 110meV$, which is a quite reasonable
value for CuNi.}

The non-monotonic dependence of the critical current on temperature observed
by \citet{ryazanov} can be obtained only in the case of an exchange energy $%
h $ comparable with $\Delta $ (at least, the ratio $h/\Delta $ should not be
too large). If the exchange energy were not too large, the effective
penetration length $\xi _{F,eff}$ would be temperature dependent. According
to estimates presented by \citeauthor{ryazanov} $h\approx 30K$, which means
that the exchange energy in this experiment was much smaller than in the one
performed by \citeauthor{palevski} and by \citeauthor{kontos02} (in the last
reference $h\approx 35meV$).

The conditions under which the $\pi -$ state is realized in $S/F/S$
Josephson junctions of different types were studied theoretically in many
papers \cite{buzdin_kupr1,buzdin_kupr2,nazarov_sfs,krivoruchko1,li,buzdin_pi}%
. In all these papers it was assumed that the ferromagnet consisted of a
single domain with a magnetization $M$ fixed in space. The case of a $S/F/S$
Josephson junction with a two domain ferromagnet was analyzed by %
\citet{blanter}. The Josephson critical current $I_{c}$ was calculated for
parallel and anti-parallel magnetization orientations in both ballistic and
diffusive limits. It turns out that in such a junction the current $I_{c}$
is larger for the anti-parallel orientation.

A similar effect arises in a $S/F/S$ junction with a rotating in space
magnetization, {as it was shown by \citet {BVE_josephson}.} In this case not
only the singlet and triplet component with projection $S_{z}=0$, but also
the triplet component with $S_{z}=\pm 1$ arises in the ferromagnet. The last
component penetrates the ferromagnet over a large length of the order of $%
\xi _{N}$ and contributes to the Josephson current.
\begin{figure}[h]
\includegraphics[scale=0.4]{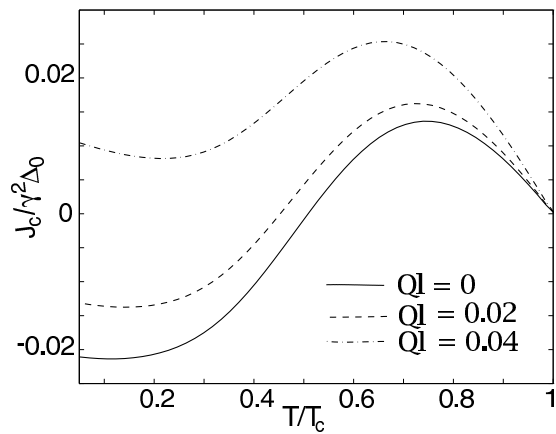}
\caption{Dependence of the critical current on T for $h\protect\tau =0.06$,$%
\Delta _{0}\protect\tau =0.03$, $d/l=\protect\pi $ and different values of $%
Ql$. Here $\protect\tau $ is the momentum relaxation time. }
\label{Fig.7}
\end{figure}
In FIG. \ref{Fig.7} the temperature dependence of the critical current is
presented for different values of $Ql,$ where $Q=2\pi /L_{m}$, $L_{m}$ is
the period of the spatial rotation of the magnetization and $l$ is the mean
free path. It is seen that at $Q=0$ (homogeneous ferromagnet) and low
temperatures $T$ the critical current $I_{c}$ is negative ($\pi -$ state$)$,
whereas with increasing temperature, $I_{c}$ becomes positive ($0-$ state).
If $Q$ increases, the interval of negative $I_{c}$ gets narrower and
disappears completely at $Ql\approx 0.04$, that is, the $S/F/S$ structure
with a non-homogeneous $M$ is an ordinary Josephson junction with a positive
critical current.

It is interesting to note that the $\pi $-type Josephson coupling may also
be realized in $S/N/S$ junctions provided the distribution function of
quasiparticles $n(\epsilon )$ in the $N$ region deviates significantly from
the equilibrium. This deviation may be achieved with the aid of a
non-equilibrium quasiparticles injection through an additional electrode in
a multiterminal $S/N/S$ junction. The Josephson current in such a junction
is again determined by Eq.(\ref{I_S}) in which one has to put $h=0,$ $%
f_{+}=f_{-}$ and replace $\tanh (\epsilon \beta )=(1-2n(\epsilon ))$ by $%
(1/2)[\tanh ((\epsilon +eV)\beta )+\tanh ((\epsilon -eV)\beta )],$ where $V$
is a voltage difference between $N$ and $S$ electrodes.

At a certain value of $V$ the critical current changes sign. {Thus, there is
some analogy between the sign reversal effect in a }${S/F/S}${\ junction and
the one in a multiterminal }${S/N/S}${\ junction under non-equilibrium
conditions. }

{Indeed, when calculating $I_{S}$ in a multiterminal }${S/N/S}${\ junction
one can shift the energy by $eV$ or $-eV$. Then the function $(1/2)[\tanh
((\epsilon +eV)\beta )+\tanh ((\epsilon -eV)\beta )]$ is transformed into $%
\tanh (\epsilon \beta )$ while in the other functions one performs the
substitution $\epsilon \rightarrow \epsilon \pm eV$. So, we see that $eV$ is
analogous to the exchange energy $h$ that appears in the case of a }${S/F/S}$%
{\ junction.}

The sign reversal effect in a multiterminal $S/N/S$ junction under
non-equilibrium conditions has been observed by \citet{vanWees} and studied
theoretically by \citet{volkov,yip,wilhelm_schon}. Later \citet{schoen}
studied theoretically a combined effect of a non-equilibrium quasiparticle
distribution on the current $I_{c}$ in a $S/F/S$ Josephson junction.

{\ Concluding this Section we note that the experimental results by %
\citet{ryazanov}, \citet{kontos02}, \citet{palevski} and \citet{strunk} seem
to confirm the theoretical prediction of an oscillating condensate function
in the ferromagnet and the possibility of switching between the 0 and the $%
\pi $-state. }

\section{ Odd triplet superconductivity in S/F structures}

\label{EXO} 

\subsection{Conventional and unconventional superconductivity}

\label{EXO-CON} 
Since the development of the BCS theory of \ superconductivity by \citet*{BCS%
}, and over many years only one type of superconductivity was observed in
experiments. This type is characterized by the $s$-wave pairing between the
electrons with opposite spin orientations due to the electron-phonon
interaction. It can be called conventional since it is observed in most
superconductors with critical temperature below $20$ $K$ (the so-called low
temperature superconductors).

{\ \citet*{mueller} discovered that a }${La}_{2-x}{Sr}_{x}{CuO}_{4}${\
compound is\ a superconductor with a critical temperature of }${30K}${. This
was the first known high-$T_{c}$ copper-oxide (cuprate) superconductor.
Nowadays many cuprates have been discovered with critical temperatures above
the temperature of liquid nitrogen. These superconductors (the so called
high $T_{c}$ superconductors) show in general a }${d}${-wave symmetry} and,
like the conventional superconductors, are in a singlet state. That is, the
order parameter $\Delta _{\alpha \beta }$ is represented in the form: $%
\Delta _{\alpha \beta }=\Delta \cdot (i\sigma _{3})_{\alpha \beta }$, where $%
\sigma _{3}$ is the Pauli matrix in the spin space. The difference between
the $s$ and $d$ pairing is due to a different dependence of the order
parameter $\Delta $ on the Fermi momentum $\mathbf{p}_{F}=\hbar \mathbf{k}%
_{F}$. In isotropic conventional superconductors $\Delta $ is a $\mathbf{k}$%
-(almost) independent quantity. In anisotropic conventional superconductors $%
\Delta $ depends on the $\mathbf{k_{F}}$ direction but it does not change
sign as a function of the momentum $\mathbf{k_{F}}$ orientation in space. In
high $T_{c}$ superconductors where the $d$-wave pairing occurs, the order
parameter $\Delta (\mathbf{k_{F}})$ changes sign at certain points at the
Fermi surface.

On the other hand, the Pauli principle requires the function $\Delta (%
\mathbf{k_{F}})$ to be an even function of \ $\mathbf{k_{F}}$, which imposes
certain restrictions for the dependence of the order parameter on the
Fermi-momentum. For example, for $d$-pairing the order parameter is given by
$\Delta (\mathbf{k_{F}})=\Delta (0)(k_{x}^{2}-k_{y}^{2})$, where $k_{x,y}$
are the components of the $\mathbf{k_{F}}$ vector in the $Cu-O$ plane. This
means that the order parameter may have either positive or negative sign
depending on the direction.

The change of the sign of the order parameter leads to different physical
effects. For example, if a Josephson junction consists of two high $T_{c}$
superconductors with properly chosen crystallographic orientations, the
ground state of the system may correspond to the phase difference $\varphi
=\pi $ ($\pi -$junction). In some high $T_{c}$ superconductors the order
parameter may consist of a mixture of $s$- and $d$-wave components \cite
{kirtley}.

Another type of pairing, the spin-triplet superconductivity, has been
discovered in materials with strong electronic correlations. The triplet
superconductivity has been found in heavy fermion intermetallic compounds
and also in organic materials (for a review see \citet{mineev}). Recently a
lot of work has been carried out to study the superconducting properties of
strontium ruthenate $Sr_{2}RuO_{4}$. Convincing experimental data have been
obtained in favor of the triplet, $p$-wave superconductivity. For more
details we refer the reader to the review articles by \citet{maeno} and %
\citet{annett}.

Due to the fact that the condensate function $<\psi _{\alpha }(r,t)\psi
_{\beta }(r^{\prime },t^{\prime })>$ must be an odd function with respect to
the permutations $\alpha \leftrightarrow \beta ,$ $r\leftrightarrow
r^{\prime }$ (for equal times, $t=t^{\prime }$), the wave function of a
triplet Cooper pair has to be an odd function of the orbital momentum, that
is, the orbital angular momentum $L$ is an odd number: $L=1$ ($p$-wave), $3$
etc. Thus, the superconducting condensate is sensitive to the presence of
impurities. Only the $s$-wave $(L=0)$ singlet condensate is not sensitive to
the scattering by nonmagnetic impurities (Anderson theorem). In contrast,
the $p$-wave condensate in an impure material is suppressed by impurities
and therefore the order parameter $\Delta _{\alpha \beta }=\sum_{k}\Delta
_{\alpha \beta }(\mathbf{k_{F}})\sim $ $\sum_{k}<\psi _{\alpha }(r,t)\psi
_{\beta }(r^{\prime },t)>_{k}$ is also suppressed \cite{larkin_triplet}.
That is why the superconductivity in impure $Sr_{2}RuO_{4}$ samples has not
been observed. In order to observe the triplet $p$-wave superconductivity
(or another orbital order parameter with higher odd $L$), one needs to use
clean samples of appropriate materials.

At first glance one cannot avoid this fact and there is no hope to see a
non-conventional superconductivity in impure materials. However, another
nontrivial possibility for the triplet pairing exists. The Pauli principle
imposes restrictions on the correlation function $<\psi _{\alpha }(r,t)\psi
_{\beta }(r^{\prime },t)>_{k}$ for equal times. In the Matsubara
representation this means that the sum $\sum_{\omega }<\psi _{\alpha
}(r,\tau )\psi _{\beta }(r^{\prime },\tau ^{\prime })>_{k,\omega }$ must
change sign under the permutation $r\leftrightarrow r^{\prime }$ (for the
triplet pairing the diagonal matrix elements ($\alpha =\beta $) of these
correlation functions are not zero). This implies that the sum $\sum_{\omega
}<\psi _{\alpha }(r,\tau )\psi _{\beta }(r^{\prime },\tau ^{\prime
})>_{k,\omega }$ has to be either an odd function of $k$ or just turn to
zero. The latter possibility does not mean that the pairing must vanish. It
can remain finite if the average $<\psi _{\alpha }(r,\tau )\psi _{\beta
}(r^{\prime },\tau ^{\prime })>_{k,\omega }$ is an odd function of the
Matsubara frequency $\omega $ (in this case it must be an even function of $%
k).$ Then the sum over all frequencies is zero and therefore the Pauli
principle for the equal-time correlation functions is not violated.

This type of pairing was first suggested by \citet{berezin} as a possible
mechanism of superfluidity in $^{3}He$. He assumed that the order parameter $%
\Delta (\omega )\propto \sum_{\omega ,k}<\psi _{\alpha }(r,\tau )\psi
_{\beta }(r^{\prime },\tau ^{\prime })>_{k,\omega }$ is an odd function of $%
\omega :$ $\ \Delta (\omega )=-\Delta (-\omega )$. However, experiments on
superfluid $^{3}He$ have shown that the Berezinskii state is only a
hypothetical state and the $p$-pairing in $^{3}He$ has different symmetries.
As it is known nowadays, the condensate in $^{3}He$ is antisymmetric in the
momentum space and symmetric (triplet) in the spin space. Thus, the
Berezinskii hypothetical pairing mechanism remained unrealized for few
decades.

However, in recent theoretical works it was found that a superconducting
state similar to the one suggested by Berezinskii might be induced in
conventional $S/F$ systems due to the proximity effect \cite
{BVE2,BVE_manifestation}. {In the next sections we will analyze this new
type of the superconductivity with the triplet pairing that is odd in the
frequency and even in the momentum. This pairing is possible not only in the
clean limit but also in samples with a high impurity concentration}.

It is important to note that, in spite of the similarity, there is a
difference between this new superconducting state in the $S/F$ structures
and that proposed by Berezinskii. In the $S/F$ structures both the singlet
and triplet types of the condensate $f$ coexist. However, the order
parameter $\Delta $ is not equal to zero only in the $S$ region (we assume
that the superconducting coupling in the $F$ region is zero) and is
determined there by the singlet part of the condensate only. This contrasts
the Berezinskii state where the order parameter $\Delta $ should contains a
triplet component.

{Note that attempts to find conditions for the existence of the
odd superconductivity were undertaken in several papers in
connections with the pairing mechanism in high }$T_{c}${\
superconductors
\cite{kirkpatrick91,kirkpatrick92,balatsky92,coleman93,coleman94,coleman95,balatsky95,abrahams,hashimoto}.
In these papers a singlet pairing odd in frequency and in the
momentum was considered. }

We would like to emphasize that, while theories of unconventional
superconductivity are often based on the presence of strong correlations
where one has to use a phenomenology, the triplet state induced in the $S/F$
structures can be studied within the framework of the BCS theory, which is
valid in the weak-coupling limit. This fact drastically simplifies the
problem not only from the theoretical, but also from experimental point of
view since well known superconductors grown in a controlled way may be used
in order to detect the triplet component.

We summarize the properties of this new type of superconductivity which we
speak of as \textit{triplet odd superconductivity}:

\begin{itemize}
\item  {It contains the triplet component. In particular the
components with projection $S_z=\pm 1$ on the direction of the
field are insensitive to the presence of an exchange field and
therefore long-range proximity effects arise in }${S/F}${\
structures.}

\item  {In the dirty limit it has a }${s}${--wave symmetry. The condensate
function is even in the momentum $\mathbf{p}$ and therefore, contrary to
other unconventional superconductors, is not destroyed by the presence of
non-magnetic impurities.}

\item  {The triplet condensate function is \textit{odd} in frequency}.
\end{itemize}

Before we turn to a quantitative analysis let us make the last remark: we
assume that in the ferromagnetic regions no attractive electron-electron
interaction exists, and therefore $\Delta =0$ in the $F$-regions. The
superconducting condensate arises in the ferromagnet only due to the
proximity effect. This will become more clear later.

Another type of triplet superconductivity in the $S/F$ structures that
differs from the one considered in this review was analyzed by \citet{Edel03}%
. The author assumed that spin-orbit interaction takes place at the $S/F$
interface due to a strong electric field which exists over interatomic
distances (the so-called Rashba term in the Hamiltonian \cite{rashba}). It
was also assumed that electron-electron interaction is not zero not only in
the $s$-wave singlet channel but also in the $p$-wave triplet channel. The
spin-orbit interaction mixes both the triplet and singlet components. Then,
the triplet component can penetrate into the $F$ region over a large
distance.

However, in contrast to odd superconductivity, the triplet
component analyzed by \citeauthor{Edel03} is odd in the momentum
and therefore must be destroyed by scattering on ordinary
nonmagnetic impurities. This type of triplet component was also
studied in $2$-dimensional systems and in $S/N$ structures in the
presence of the Rashba-type spin-orbit interaction \cite
{Edel89,GorkovRashba,Edel03}{.}

\subsection{Odd triplet component (homogeneous magnetization)}

\label{EXO-HOM} 
As we have mentioned in section \ref{PRO-SF}, even in the case of a
homogeneous magnetization the triplet component with the zero projection $%
S_z=0$ of the total spin on the direction of the magnetic field
appears in the $S/F$ structure. Unlike the singlet component it is
an odd function of the Matsubara frequency $\omega $. In order to
see this, we look for a solution of the Usadel equation in the
Matsubara representation. In this representation {the linearized
Usadel equation} for the ferromagnet takes the form
\begin{equation}
D_{F}\partial ^{2}\hat{f}_{F}/\partial x^{2}-2(|\omega |\hat{\sigma}%
_{0}-ih_{\omega }\hat{\sigma}_{3})\hat{f}_{F}=0\;,  \label{3.1}
\end{equation}
where $\omega =\pi T(2n+1)$ is the Matsubara frequency and $h_{\omega }=%
\mathrm{sgn}(\omega )h$.

The solution of Eq. (\ref{3.1}) corresponding to Eq.(\ref{2.B.4}) can be
written as
\begin{equation}
\;f_{\pm }(\omega )=\pm (\Delta /i\xi _{\omega }\kappa _{\pm }(\omega
)\gamma _{F})\exp (-\kappa _{\pm }(\omega )x)\;.  \label{3.2}
\end{equation}
where
\begin{equation}
\kappa _{\pm }(\omega )=\sqrt{2(|\omega |\mp ih_{\omega })/D_{F}}
\label{3.2a}
\end{equation}
and $\xi _{\omega }=\sqrt{\omega ^{2}+\Delta ^{2}}$.

For the amplitudes of the triplet $(f_{0}=(f_{+}+f_{-})/2)$ and singlet $%
(f_{3}=(f_{+}-f_{-})/2)$ components we get  in the ferromagnet
\begin{equation}
\;f_{3,0}(\omega ,x)=(\Delta /2i\xi _{\omega }\gamma _{F})[\frac{\exp
(-\kappa _{+}(\omega )x)}{\kappa _{+}(\omega )}\pm \frac{\exp (-\kappa
_{-}(\omega )x)}{\kappa _{-}(\omega )}]\;.  \label{3.3}
\end{equation}
Eqs. (\ref{3.2}) and (\ref{3.3}) show that both the singlet and
the triplet component with $S_z=0$ of the condensate functions
decay in the ferromagnet on the scale of $\mathrm{Re}\kappa _{+}$
having oscillations with $\mathrm{Im}\kappa _{+}$. Taking into
account that $\kappa _{+}(\omega )=\kappa _{-}(-\omega )$, we see
that $f_{3}(\omega )$ is an even function of $\omega ,$ whereas
the
amplitude of the triplet component, $f_{0}(\omega )$, is an odd function of $%
\omega $. {The mixing between the triplet and singlet components is due to
the term proportional to $h_{\omega }\hat{\sigma}_{3}$ in Eq.(\ref{3.1}).
This term breaks the time-reversal symmetry. }

{Due to the proximity effect the triplet component $f_{0}$ penetrates also
into the superconductor and the characteristic length of the decay is the
coherence length $\xi _{S}$. The spatial dependence of this component inside
the superconductor can be found provided} the Usadel equation is linearized
with respect to a deviation of the $\hat{f}_{S}$ matrix from its bulk BCS
form $\hat{f}_{BCS}$.{\ In the presence of an exchange field the Green's
functions} $\check{g}$ are $4\times 4$ matrices in the particle-hole and
spin space. In the case of a homogeneous magnetization they can be
represented as a sum of two terms (the $\hat{\tau}$ matrices operate in the
particle-hole space)
\begin{equation}
\check{g}=\hat{g}\hat{\tau}_{3}+\hat{f}i\hat{\tau}_{2}\;,  \label{3.4}
\end{equation}
where $\hat{g}$ and $\hat{f}$ are matrices in the spin space.

In a bulk superconductor these matrices are equal to
\begin{equation}
\hat{g}_{BCS}=g_{BCS}(\omega )\hat{\sigma}_{0};\text{ }\hat{f}%
_{BCS}=f_{BCS}(\omega )\hat{\sigma}_{3}\;,  \label{3.5}
\end{equation}
where
\begin{equation}
g_{BCS}(\omega )=\omega /\xi _{\omega };\text{ }f_{BCS}(\omega )=\Delta
/i\xi _{\omega }.  \label{3.6}
\end{equation}
and $\xi _{\omega }=\sqrt{\omega ^{2}+\Delta ^{2}}$. {\ }

{We linearize now} the Usadel equation with respect to a small deviation $%
\delta \check{g}_{S}\equiv \delta \hat{g}_{S}\hat{\tau}_{3}+\delta \hat{f}%
_{S}i\hat{\tau}_{2}=\check{g}_{S}-\check{g}_{BCS}$ and obtain for the
condensate function $\delta \hat{f}_{S}$ in the superconductor the following
equation
\begin{equation}
(\partial ^{2}/\partial x^{2})\delta \hat{f}_{S}-\kappa _{S}^{2}\delta \hat{f%
}_{S}=2i(\delta \Delta /D_{S})g_{BCS}^{2}\hat{\sigma}_{3},  \label{3.7}
\end{equation}
where $\kappa _{S}^{2}=2\sqrt{(\omega ^{2}+\Delta ^{2})}/D_{S}$ and $\delta
\Delta \left( x\right) $ is a deviation of the superconducting order
parameter from its BCS value in the bulk.

A solution for Eq.(\ref{3.7}) determines the triplet component $\delta
f_{S0} $ and a correction $\delta f_{S3}$ to the singlet component. To find
the component $\delta f_{S3}$ is a much more difficult task than to find $%
\delta f_{S0}$ because $\delta \Delta (x)$ is a function of $x$ and, in its
turn, is determined by the amplitude $\delta f_{S3}.$ Therefore, the singlet
component $\delta f_{S3}$ obeys a non-linear integro-differential equation.
That is why the critical temperature $T_{c}$ can be calculated only
approximately \cite
{buzdin_kupr_tc,radovic2,demler,proshin,tagirov_C,baladie_tc,bagrets}. %
\citet{golubov_tc} proposed an analytical trick that reduces the $T_{c}$
problem to a form allowing a simple numerical solution.

On the contrary, {the triplet component $\delta f_{S0}$ proportional to $%
\mbox{$\hat{\sigma}_0$}$ can be found exactly (in the linear approximation).}
The solution for $\delta f_{S0}\left( 0\right) $ takes the form
\begin{equation}
\;\delta f_{S0}(x)=\delta f_{S0}(0)\exp (-\kappa _{S}(\omega )x).
\label{3.8}
\end{equation}
The constant $\delta f_{S0}(0)$ can be found from the boundary
condition (see Appendix \ref{ApA})
\begin{equation}
\partial \delta f_{S0}(x)/\partial x|_{x=0}=f_{F0}(0)/\gamma_S\;.
\label{3.9}
\end{equation}
As follows from this equation, the triplet component in the superconductor $%
\delta f_{S0}$ has the same symmetry as the component $f_{F0},$ that is, it
is odd in frequency. So, the triplet component of the condensate is
inevitably generated by the exchange field both in the ferromagnet and
superconductor. Both the singlet component and the triplet component with $%
S=0$ decay fast in the ferromagnet because the exchange field $h$ is usually
very large (see Eq. (\ref{3.2a})). At the same time, the triplet component
decays much slower in the superconductor because the inverse characteristic
length of the decay $k_{S}$ is much smaller.

To illustrate some consequences of the presence of the triplet component in
the superconductor, we use the fact that {the normalization condition $%
\mbox{$\check{g}$}^{2}=1$ results in the relation}
\begin{equation}
\;g_{0}g_{3}=f_{3}f_{0}  \label{3.10}
\end{equation}
The function $g_{0}$ entering Eq. (\ref{3.10}) determines the change of the
local DOS
\begin{equation}
\;\nu (\epsilon )=\mathrm{Re}g_{0}(\epsilon )  \label{3.11}
\end{equation}
while the function $g_{3}$ determines the {magnetic moment $M_{z}$
of the itinerant electrons (see Appendix \ref{ApA})}
\begin{equation}
\;M_{z}=\mu _{B}\nu i\pi T\sum_{\omega }g_{3}(\omega )  \label{3.12}
\end{equation}
{We see that the appearance of the triplet component in the superconductor
leads to a finite magnetic moment in the }${S}$-region, which can be spoken
of as{\ an inverse proximity effect. This problem will be discussed in more
detail in section \ref{RED-INV}.}

{Thus,} even in the case of a homogeneous magnetization, the
triplet component with $S_z=0$ arises in the $S/F$ structure. This
fact was
overlooked in many papers and has been noticed for the first time by %
\citet{BVE_manifestation}. This component, as well as the singlet one,
penetrates {the ferromagnet} over a short length $\xi _{F}$ because it {%
consists of} averages of two operators with opposite spins $<\psi
_{\uparrow }\psi _{\downarrow }>$ \ and is strongly suppressed by
the exchange field. The triplet component with projections
$S_z=\pm 1$ on the direction of the field results in more
interesting properties of the system {since it} is not suppressed
by the exchange interaction. It can be generated by a
non-homogeneous magnetization as we will discuss in the next
section.

\subsection{Triplet odd superconductivity ( inhomogeneous magnetization)}

\label{EXO-INH} 
{According to the results of the last section the presence of an
exchange field leads to the formation of the triplet component of
the condensate function. In a homogeneous exchange field, only the
component with the projection $S_z=0$ is induced. }

{A natural question arises: Can the other components with $S_z=\pm
1$ be induced? If they could, this would lead to a long range
penetration of the superconducting correlations into the
ferromagnet because these components correspond\ to the
correlations of the type }$<\psi _{\uparrow }\psi _{\uparrow }>$
with parallel spins and {are not as sensitive to the exchange
field as the other ones.}

In what follows we analyze some examples of $S/F$ structures in which all
projections of the triplet component are induced. The common feature of
these structures is that the magnetization is nonhomogeneous.

In order to determine the structure of the condensate we will use as before
the method of quasiclassical Green's functions. This allows us to
investigate all interesting phenomena except those that are related to
quantum interference effects.

The method of the quasiclassical Green's functions can be used at spatial
scales much longer than the Fermi wave length \footnote{%
Note that, as was shown by \citet{zaikin_b,shelankov}, in the
ballistic case and in the presence of several potential barriers
some effects similar to the quantum interference effects may be
important. We do not consider purely ballistic systems assuming
that the impurity scattering is important. In this case the
quasiclassical approach is applicable. The applicability of the
quasiclassical approximation was discussed long ago by Larkin and
Ovchinnikov \cite{larkin_ovch}.}. As we have mentioned already, in
order to describe the $S/F$ structures the Green's functions have
to be $4\times 4$ matrices in the particle-hole and spin space.
Such $4\times 4$ matrix Green's functions (not necessarily in the
quasiclassical form) have been used long ago by
\cite{vaks,maki_book}. Equations for the quasiclassical Green's
functions in the presence of the exchange field similar to the
Eilenberger and Usadel equations can be derived in the same way as
the one used in the non-magnetic case (see Appendix \ref{ApA}).
For example, a generalization of the Eilenberger equation was
presented by \citet*{BEL} and applied to the study of
cryptoferromagnetism.

\subsubsection{F/S/F trilayer structure}

\label{EXO-INH-FSF} 
\begin{figure}[h]
\includegraphics[scale=0.4]{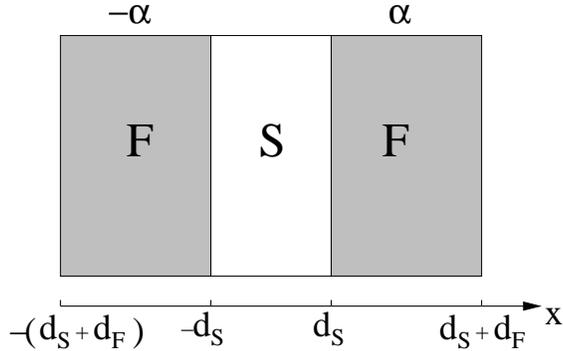}
\caption{Trilayer geometry. The magnetization of the left (right) F layer
makes an angle $\protect\alpha$ ($-\protect\alpha$) with the $z$-axis.}
\label{Fig.8}
\end{figure}

We start the analysis of the non-homogeneous case by considering the $F/S/F$
system shown in FIG. \ref{Fig.8}. The structure consists of one $S$ layer
and two $F$ layers with magnetizations inclined at the angle $\pm \alpha $
with respect to the $z$-axis (in the $yz$ plane).

We want to demonstrate now that the triplet component with $S_{z}=\pm 1$
inevitably arises due to the overlap of the triplet components generated by
the ferromagnetic layers in the $S$ layer. It is not difficult to understand
why it should be so.

As we have seen in the previous section, each of the layers generates the
triplet component with the zero total projection of the spin, $S_{z}=0,$ on
the direction of the exchange field. If the magnetic moments of the layers
are collinear to each other (parallel or antiparallel), the total projection
remains zero. However, if the moments of the ferromagnetic layers are not
collinear the superposition of the triplet components coming from the
different layers should have all possible projections of the total spin.

From this  argument we can expect  the generation of the triplet
component with all projections of the total spin provided the
thickness of the $S$ layer is not too large. Since the only
relevant length in the superconductors is  $\xi _{S}\approx
\sqrt{D_{S}/\pi T_{c}}$, we assume that the thickness of the
superconducting layer $S$ does not exceed this length.

Now we perform explicit calculations that support the qualitative conclusion
about the generation of the triplet component with all projections of the
total spin. We consider the diffusive case when the Usadel equation is
applicable. This means that the condition
\begin{equation}
h\tau <<1  \label{3.13}
\end{equation}
is assumed to be fulfilled ($\tau $ is the elastic scattering time).

The linearized Usadel equation in the $F$ region takes the form
(see Appendix \ref{ApA})
\begin{equation}
\partial ^{2}_{xx}\check{f}-\kappa _{\omega }^{2}\check{f}%
+\frac{i\kappa _{h}^{2}}{2}\left\{\hat{\tau} _{0}[\hat{\sigma}_{3},\check{f}%
]_{+}\cos \alpha \pm \hat{\tau}
_{3}[\hat{\sigma}_{2},\check{f}]\sin \alpha \right\}=0\;,
\label{3.14}
\end{equation}
where $\check{f}$ is a $4\times 4$ matrix (condensate function) which is
assumed to be small and $[\hat{\sigma}_{3},\check{f}]_{+}=\hat{\sigma}_{3}.%
\check{f}+\check{f}.\hat{\sigma}_{3}$. The wave vectors $\kappa _{\omega }$
and $\kappa _{h}$ entering Eq. (\ref{3.14}) have the form
\begin{equation}
\kappa _{\omega }^{2}=2|\omega |/D_{F}  \label{3.14a}
\end{equation}
and
\begin{equation}
\kappa _{h}^{2}=2h\mathrm{sgn}\omega /D_{F}  \label{3.14b}
\end{equation}

The magnetization vector {\ }$\mathbf{M}$ lies in the
($y,z$)-plane and has the components: $\mathbf{M}=M\{0,\pm \sin
\alpha ,\cos \alpha \}$. {The sign ``$+$'' (``$-$'') corresponds
to the right (left) }${F}${\ film.} We consider here the simplest
case of a highly transparent $S/F$ interface and temperatures
close to the critical temperature of the superconducting
transition $T_{c}$. In this case the {\ function $\check{f}$,
being small, obeys a linear equation similar to Eq.(\ref{3.7})}
\begin{equation}
(\partial ^{2}\check{f}/\partial x^{2})-\kappa _{S}^{2}\check{f}=2i(\delta
\breve{\Delta}/D_{S})g_{BCS}^{2}\;,  \label{3.15}
\end{equation}
where $\kappa _{S}^{2}=2|\omega |/D_{S}$.

The boundary conditions at the $S/F$ interfaces are
\begin{eqnarray}
\check{f}_{x=d_{S}+0} &=&\check{f}_{x=d_{S}-0}  \label{3.16} \\
\gamma (\partial \check{f}/\partial x)|_{F} &=&(\partial \check{f}/\partial
x)|_{S}\;.  \label{3.17}
\end{eqnarray}
where $\gamma =\sigma _{F}/\sigma _{S}$ and $\sigma _{F}\left( \sigma
_{S}\right) $ is the conductivity in the ferromagnet (superconductor).

{The first condition, Eq. (\ref{3.16}), corresponds to the continuity of the
condensate function at the }${S/F}${\ interface with a high transparency,
whereas Eq (\ref{3.17}) ensures the continuity of the current across the }${%
S/F}${\ interface} \cite{BVE3}.

A solution for Eqs.(\ref{3.14}-\ref{3.15}) with the boundary conditions (\ref
{3.16}-\ref{3.17}) can easily be found. The matrix $\check{f}$ can be
represented as
\begin{equation}
\check{f}=i\hat{\tau}_{2}\otimes \hat{f}_{2}+i\hat{\tau}_{1}\otimes \hat{f}%
_{1}\;,  \label{3.18}
\end{equation}
where
\begin{equation}
\hat{f}_{1}=b_{1}(x)\hat{\sigma}_{1},\text{ \ }\hat{f}_{2}=b_{3}(x)\hat{%
\sigma}_{3}+b_{0}(x)\hat{\sigma}_{0},  \label{3.19}
\end{equation}
\ In the left $F$ layer the functions $b_{k}(x)$ are to be replaced by $\bar{%
b}_{k}(x).$ For simplicity we assume that the thickness of the $F$ films $%
d_{F}$ exceeds $\xi _{F}$ (the case of an arbitrary $d_{F}$ was analyzed by %
\citet{BVE_manifestation}). Using the representation, Eqs. (\ref{3.18}-\ref
{3.19}), we find the functions $b_{i}(x)$ and $\bar{b}_{i}(x)$. They are
decaying exponential functions and can be written as
\begin{equation}
b_{k}(x)=b_{k}\exp (-\kappa (x-d_{S})),\text{ }\bar{b}_{k}(x)\text{ }=\bar{b}%
_{k}\exp (\kappa (x+d_{S}))  \label{3.20}
\end{equation}
Substituting Eq.(\ref{3.20}) into Eqs.(\ref{3.14})-(\ref{3.15}), we obtain a
set of linear equations for the coefficients $b_{k}.$ The condition for the
existence of non-trivial solutions yields an equation for the eigenvalues $%
\kappa .$ This equation reads
\begin{equation}
(\kappa ^{2}-\kappa _{\omega }^{2})[(\kappa ^{2}-\kappa _{\omega
}^{2})^{2}+\kappa _{h}^{4}]=0  \label{3.21}
\end{equation}
Eq. (\ref{3.21}) is of the sixth order and therefore has six solutions.
Three of these solutions should be discarded because the corresponding to $%
b_{k}\left( x\right) $ grow when going away from the interface. The
remaining three solutions of Eq. (\ref{3.21}){\ give} three different
physical values of $\kappa $.

If the exchange energy $h$ is sufficiently large ($h>>\{T,\Delta \}$), the
eigenvalues are
\begin{eqnarray}
\kappa &=&\kappa _{\omega }  \label{3.22} \\
\kappa _{\pm } &\approx &(1\pm i)\kappa _{h}  \label{3.23}
\end{eqnarray}
We see that these solutions are completely different. {The roots $\kappa
_{\pm }$ proportional to }$\kappa _{h}$ (\textit{cf.} Eq. (\ref{3.14b})),
are very large and therefore the corresponding solutions $b_{k}\left(
s\right) $ decay very fast (similar to the singlet component). This is the
solution that exists for a homogeneous magnetization (collinear
magnetization vectors).

In contrast, the value for $\kappa $ given by Eq. (\ref{3.22}), is
much smaller (see Eq. (\ref{3.14a})) and corresponds to a slow
decay of the superconducting correlations. The solutions
corresponding to the root given by Eq.(\ref{3.22}) describe a
long-range penetration of the triplet component into the
ferromagnetic region. For each root one can easily obtain
relations between the coefficients $b_{k}(x)$. As a result, we
obtain
\begin{eqnarray}
b_{1}(x) &=&b_{\omega }e^{-\kappa _{\omega }(x-d_{S})} \nonumber\\
& &\!\!\!\!\!\!\!\!\!\!\!\!-\sin \alpha \left[ b_{3+}e^{ -\kappa
_{+}(x-d_{S})}-b_{3-}e^{ -\kappa _{-}(x-d_{S})}\right]\;
\label{3.26}
\end{eqnarray}
\begin{eqnarray}
b_{0}(x) &=&-\tan \alpha \cdot b_{\omega }e^{-\kappa _{\omega
}(x-d_{S})}
\nonumber \\
&&\!\!\!\!\!\!\!\!\!\!\!\!-\cos \alpha \left[ b_{3+}e^{ -\kappa
_{+}(x-d_{S})}-b_{3-}e^{-\kappa _{-}(x-d_{S})}\right]
\end{eqnarray}
and
\begin{equation}
b_{3}(x)=b_{3+}\exp (-\kappa _{+}(x-d_{S}))+b_{3+}\exp (-\kappa
_{-}(x-d_{S}))  \label{3.27}
\end{equation}
The function $b_{1}(x)$ is the amplitude of the triplet component
penetrating into the $F$ region over a long distance of the order of $\kappa
_{\omega }^{-1}\sim \xi _{N}.$ Its value as well as the values of the other
functions $b_{k}(x)$ is found from the boundary conditions (\ref{3.16}-\ref
{3.17}) at the $S/F$ interfaces.

What remains to be done is to match the solutions for the superconductor and
the ferromagnets at the interfaces between them. The solution for the
superconductor satisfies Eq.(\ref{3.15}) and can be written as
\begin{eqnarray}
f_{3}(x) &=&\Delta /iE_{\omega }+a_{3}\cosh (\kappa _{S}x)  \label{3.28} \\
f_{0}(x) &=&a_{0}\cosh (\kappa _{S}x) \\
f_{1}(x) &=&a_{1}\sinh (\kappa _{S}x)\;,
\end{eqnarray}
where $E_{\omega }=\sqrt{\omega ^{2}+\Delta ^{2}}$.

Matching these solutions with Eqs. (\ref{3.26}-\ref{3.27}) at the $S/F$
interfaces we obtain the coefficients $b_{k}$ and $\bar{b}_{k}$ as well as $%
a_{k}$. Note that $b_{3\pm }=$ $\bar{b}_{3\pm }$ and $b_{\omega }=$ $-\bar{b}%
_{\omega }.$ Although the solution can be found for arbitrary parameters
entering the equations, we present here for brevity the expressions for $%
b_{3\pm }$ and $b_{\omega }$ in some limiting cases only.

Let us\ consider first the case when the parameter $\gamma \kappa
_{h}/\kappa _{S}$\ \ is small, that is, we assume the condition
\begin{equation}
\gamma \kappa _{h}/\kappa _{S}\approx \frac{\nu _{F}}{\nu _{S}}\sqrt{\frac{%
D_{F}h}{D_{S}\pi T_{c}}}<<1  \label{3.29}
\end{equation}
to be fulfilled.

Here $\nu _{F,S}$ is the density of states in the ferromagnet and
superconductor, respectively (in the quasiclassical approximation the DOS
for electrons with spin up and spin down is nearly the same: $h<<\epsilon
_{F}$). The condition, Eq. (\ref{3.29}), can be fulfilled in the limit $%
D_{F}<<D_{S}$. Taking, for example, $\nu _{F}\approx \nu _{S},$ $%
l_{F}\approx 30$ \AA\ and $l_{S}\approx 300$ \AA , we find that $h$ should
be smaller than $30T_{c}$.

In this limit the coefficients $b_{1,3\pm }$ and $a_{1}$ can be written in a
rather simple form
\begin{equation}
b_{\omega }\approx -\frac{2\Delta }{E_{\omega }}(\frac{\gamma \kappa _{h}}{%
\kappa _{S}})\frac{\sin \alpha \cos ^{2}\alpha }{\sinh (2\Theta _{S})},
\label{3.31}
\end{equation}
\begin{equation}
b_{3+}\approx b_{3-}\approx \frac{\Delta }{2i\xi _{\omega }},  \label{3.32}
\end{equation}
\begin{equation}
a_{3}=-\frac{\Delta }{iE_{\omega }}\frac{\gamma \kappa _{h}}{\kappa _{S}}%
\frac{1}{\sinh (2\Theta _{S})},  \label{3.33}
\end{equation}
where {\ }$\Theta =\kappa _{S}d_{S}.$

As follows from the first of these equations, Eq. (\ref{3.32}), the
correction to the bulk BCS solution for the singlet component is small in
this approximation and this justifies our approach.

At the $S/F$ interface the amplitude of the triplet component $b_{\omega }$
is small in comparison with the magnitude of the singlet one $b_{3+}$ .
However the triplet component decays over a long distance $\xi _{N}$ while
the singlet one vanishes at distances exceeding the short length $\xi _{F}$.
The amplitudes $b_{\omega }$ and $b_{3\pm }$ become comparable if the
parameter $\gamma \kappa _{h}/\kappa _{S}$ is of the order of unity.

It follows also from Eq.(\ref{3.31}) that the amplitude of the triplet
component $b_{\omega }$ is zero in the case of collinear vectors of
magnetization, i.e. at $\alpha =0$ or $\alpha =\pi /2$. It reaches the
maximum at the angle $\alpha _{m}$ for which $\sin \alpha _{m}=1/\sqrt{3}$.
Therefore the maximum angle-dependent factor in Eq.(\ref{3.31}) is $\sin
\alpha _{m}\cos ^{2}\alpha _{m}=2/3\sqrt{3}\approx 0.385.$

One can see from Eq.(\ref{3.31}) that $b_{\omega }$ becomes exponentially
small if the thickness $d_{S\text{ }}$of the $S$ films significantly exceeds
the coherence length $\xi _{S}\approx \sqrt{D_{S}/\pi T_{c}}$. This means
that in order to have a considerable penetration of the superconducting
condensate into the ferromagnet one should not make the superconducting
layer too thick.

On the other hand, if the thickness $d_{S}$ {is too small} the critical
temperature $T_{c}$ is suppressed. In order to avoid this suppression one
has to use, for instance, an $F/S/F$ structure with a small width of the $F$
films. Similar systems were considered by \citet{Beck}, where non-local
effects of Andreev reflections in a $S/F$ nanostructure were studied .

Another limiting case that allows a comparatively simple solution is the
limit of small angles $\alpha $ \cite{BVE3} but an arbitrary parameter $%
\gamma \kappa _{h}/\kappa _{S}$, Eq. (\ref{3.29}). At small angles $\alpha $%
\ the amplitudes of the triplet and singlet component are given by the
following formulae
\begin{widetext}
\begin{equation}
b_{\omega }\approx -\frac{\Delta }{E_{\omega }}\frac{\sin \alpha (\gamma
\kappa _{h}/\kappa _{S})\tanh \Theta _{S}}{\cosh ^{2}\Theta _{S}|\tanh
\Theta _{S}+(\gamma \kappa _{h}/\kappa _{S})|^{2}(1+(\gamma \kappa
_{h}/\kappa _{S})\tanh \Theta _{S})},  \label{3.34}
\end{equation}
\begin{equation}
b_{3\pm }\approx \frac{\Delta }{2iE_{\omega }}\frac{1}{1+(\gamma \kappa
_{\pm }/\kappa _{S})\tanh \Theta _{S}},  \label{3.35}
\end{equation}
\end{widetext}
One can see from Eqs. (\ref{3.34}-\ref{3.35}) that, provided the parameter
given by Eq. (\ref{3.29}) is not small and $\alpha ,|\Theta _{S}|\sim 1$,
the amplitudes $b_{\omega }$ and $b_{3\pm }$ are again{\ }comparable with
each other.

The amplitudes of the triplet and singlet components were calculated by %
\citet{BVE_manifestation} in a more general case of an arbitrary $S/F$
interface transparency and a finite thickness of the $F$ films.
\begin{figure}[h]
\includegraphics[scale=0.4]{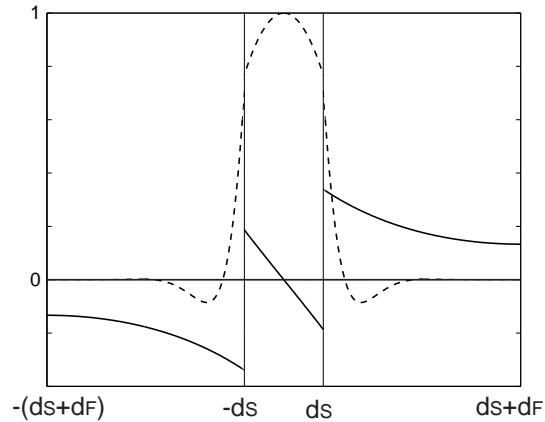}
\caption{ The spatial dependence of $\mathrm{Im}$(SC) (dashed line) and the
long-range part of $\mathrm{Re}$(TC) (solid line). We have chosen $\protect%
\sigma_F/\protect\sigma_S=0.2$, $h/T_C=50$, $\protect\sigma_F R_b/\protect\xi%
_F=0.05$, $d_F\protect\sqrt{T_C/D_S}=2$, $d_S\protect\sqrt{T_C/D_S}=0.4$ and
$\protect\alpha=\protect\pi/4$. The discontinuity of the TC at the S/F
interface is because the short-range part is not shown in this figure. Taken
from \citet*{BVE_manifestation} }
\label{Fig.9}
\end{figure}

In FIG. \ref{Fig.9} we plot the spatial dependence of the triplet (TC) and
singlet (SC) components in a $F/S/F$ structure. It is seen from this figure
that, as expected, the triplet component decays slowly, whereas the singlet
component decays fast over the short length $\xi _{h}$. For this reason, in
a multilayered $S/F$ structure with a varying direction of the magnetization
vector $\mathbf{M}$ and thick $F$ layers ($\xi _{h}<<d_{F}$) a
Josephson-like coupling between neighboring layers {can be} realized via the
odd triplet component. In this case the in-plane superconductivity is caused
by both triplet and singlet components. Properties of such S/F multilayered
structures will be discussed in the next chapter.

Let us mention\ an important fact. The quasiclassical Green's function $\check{%
g}(\vartheta )$ in the diffusive case can be expanded in spherical
harmonics. In the present approach, only the first two terms of this
expansion are taken into account such that
\begin{equation}
\;\check{g}=\check{g}_{sym}+\check{g}_{as}\cos \vartheta  \label{3.38}
\end{equation}
where $\vartheta $ is the angle between the momentum $p$ and the
$x$-axis, and $\check{g}_{as}=-l\check{g}_{sym}\partial
\check{g}_{sym}/\partial x$ is the antisymmetric part of
$\check{g}(\vartheta )$ and $\check{g}_{sym}$ is the isotropic
part  of $\check{g}(\vartheta )$,  which does not   depend on
$\vartheta $. {The antisymmetric part of $\mbox{$\check{g}$}$
determines the electric current in the system. }

{Higher order terms in the expansion of $\check{g}$ are small in the
diffusive limit and can be neglected.} In the case of a weak proximity
effect the antisymmetric part of the condensate function in the $F$ region
can be written as
\begin{equation}
\;\check{f}_{as}\cos \vartheta \approx -l\hat{\tau}_{3}\otimes \hat{\sigma}%
_{0}\mbox{${\rm sgn}$}\omega \partial \check{f}_{sym}/\partial x\cos
\vartheta \;.  \label{3.39}
\end{equation}
{This expression follows from the fact that $\check{g}_{0}\approx -\hat{\tau}%
_{3}\otimes \hat{\sigma}_{0}\mbox{${\rm sgn}$}\omega $ (corrections to $%
\check{g}_{0}$ are proportional to $\;\check{f}_{0}^{2}$). Eq. (\ref{3.39})}
holds for both the singlet and triplet components.

As we have clarified previously, the symmetric part $\check{f}_{0}$ is an
odd function of $\omega $. Thus, according to Eq.(\ref{3.35}) the
antisymmetric part is an even function of $\omega $ so that the total
condensate function $\check{f}=\check{f}_{0}+\check{f}_{1}\cos \vartheta $
is neither odd nor even function of $\omega $. {However, in the diffusive
limit it is still legitimate to speak about the odd superconductivity since
the symmetric part is much larger than antisymmetric part of $\check{f}$.}

If the parameter $h\tau $ is not small, i.e. the system is not diffusive,
the symmetric and antisymmetric parts are comparable, and one cannot speak
of the odd superconductivity. {All this} distinguishes the superconductivity
in $S/F$ structures from the odd superconductivity suggested by %
\citet{berezin} who assumed that the order parameter $\Delta (\omega )$ was
an odd function of $\omega $. In our discussion it is assumed that the order
parameter $\Delta $ is an $\omega -$ independent quantity and it is
determined by the singlet component of the condensate function $\check{f}%
_{0} $.

\subsubsection{Domain wall at the S/F interface}

\label{EXO-INH-DW} 
In the previous section we have seen how the generation of the
triplet component takes place. The appearance of this component
leads to long range effects in a structure where the angle between
the directions of magnetization in the different layers can be
changed experimentally. This is an example of a situation when the
long range triplet component of the superconducting condensate can
be produced under artificial experimental conditions.

In this section we show  that the conditions under which the
triplet long range superconducting correlations occur are
considerably more general. It is well known that the magnetization
of any ferromagnet can be quite inhomogeneous due to the presence
of domain walls. They are especially probable near interfaces
between the ferromagnets and other materials. Therefore, making an
interface between the ferromagnets and superconductors one
produces almost inevitably domain walls, and one should take
special care to get rid off them.
\begin{figure}[h]
\includegraphics[scale=0.35]{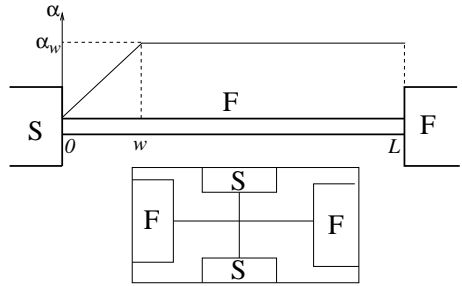}
\caption{S/F structure with a domain wall in the region $0<x<w$.
In this region $\protect\alpha=Qx$, where $Q$ is the wave vector
which describes the spiral structure of the domain wall. For $x>w$
it is assumed that the
magnetization is homogeneous, i.e. $\protect\alpha_w\equiv\protect\alpha%
(x>w)=Qw$. }
\label{Fig.10}
\end{figure}
In this section we consider a domain wall like structure and show that it
will also lead to the triplet long range correlations. This structure is
shown schematically in FIG. \ref{Fig.10}. It consists of a $S/F$ bilayer
with a non-homogeneous magnetization in the $F$ layer. We assume for
simplicity that the magnetization vector $\mathbf{M}=M(0,\sin \alpha(x)
,\cos \alpha(x) )$ rotates in the $F$ film starting from the $S/F$ interface
($x=0$) and the rotation angle has a simple piece-wise $x$-dependence
\begin{equation}
\alpha (x)=\left\{
\begin{array}{l}
Qx,\text{ \ }0<x<w \\
\alpha_w=Qw,\text{ \ }w<x
\end{array}
\right.  \label{3.2.1}
\end{equation}
{This }form{\ {\ }means} that the $\mathbf{M}$ vector is aligned parallel to
the $z$-axis at the $S/F$ interface and rotates by the angle $\alpha_w$($%
=\alpha(w)$) over the length $w$ ($w$ may be the width of a domain wall). At
$x>w$ the orientation of the vector $\mathbf{M}$ is fixed.

We calculate the condensate function in the $F$ region and show that it
contains the long range triplet component (LRTC). As in the preceding
section, we assume that the condensate function in the $F$ region is small.
The smallness of $\check{f}$ in this case is either due to a mismatch of the
Fermi velocities in the superconductor and ferromagnet or due to a possible
potential barrier at the $S/F$ interface. In such cases the transparency of
the interface is small and only a small portion of the superconducting
electrons penetrates the ferromagnet.

Due to the smallness of the transparency of the interface {\ }the function $%
\check{f}$ can experience a jump at the $S/F$ {interface, which contrasts
the preceding case. The boundary condition for the $4\times 4$ matrix $%
\check{f}$ has the same form as in Eq.(\ref{2.B.2}) }
\begin{equation}
\;\gamma _{F}\partial_x\check{f}=-\check{f}_{S}  \label{3.2.2}
\end{equation}
The function $\check{f}_{S}$ on the right-had side is the
condensate matrix Green's function in the superconductor that, in
the limit considered here, should be close to the {bulk solution}
\begin{equation}
\;\check{f}_{S}\;=f_{BCS}i\hat{\tau}_{2}\otimes \hat{\sigma}_{3}
\label{3.2.3}
\end{equation}
We have to solve again Eq.(\ref{3.14}) with the boundary conditions (\ref
{3.2.2}). Therefore we assume that the domain wall thickness $w$ is larger
than the mean free path $l\ $and the condition, Eq. ({\ref{3.13}) is
fulfilled }(dirty limit). This case was analyzed by \citet{BVE1}. Another
case of a thin domain wall ($w<l$) was considered by \citet{kadigrobov}.

The problem of finding the condensate functions in the case of the
magnetization varying continuously in space is more difficult than the
previous one because the angle $\alpha $ depends now on $x$. However, one
can use a trick that helps to solve the problem, namely{,} we exclude the
dependence $\alpha (x)$ introducing a new matrix $\check{f}_{n}$ related to $%
\check{f}$ via an unitary transformation (a rotation in the particle-hole
and spin-space)
\begin{equation}
\;\check{f}\;=\check{U}.\check{f}_{n}.\check{U}^{+}  \label{3.2.4}
\end{equation}
where $\check{U}=\exp (i\hat{\tau}_{3}\otimes \hat{\sigma}_{1}\alpha (x)/2).$

Performing this transformation we obtain instead of Eq.(\ref{3.14}) a new
equation
\begin{eqnarray}
\;(\partial ^{2}_{xx}-Q^{2}/2)\check{f}_{n}&-&\kappa _{\omega }^{2}%
\check{f}_{n}\;+i\kappa _{h}^{2}[\hat{\sigma}_{3},\check{f}%
_{n}]_{+}\nonumber\\
-\frac{Q^{2}}{2}(\hat{\sigma}_{1}\check{f}_{n}\hat{\sigma}%
_{1})&+&iQ\hat{\tau}_{3}[\hat{\sigma}_{1},\partial_x
\check{f}_{n}]_{+}=0  \label{3.2.5}
\end{eqnarray}
{Correspondingly, the boundary condition, Eq. (\ref{3.2.2}), takes the form
\begin{equation}
\;\gamma _{F}\{(Q/2)i\hat{\tau}_{3}[\hat{\sigma}_{1,}\check{f}%
_{n}]_{+}+\partial \check{f}_{n}\;/\partial x\}=-\check{f}_{s}  \label{3.2.6}
\end{equation}
Eq.(\ref{3.2.5}) complemented by this boundary condition} has to be solved
in the region $0<x<w$. In the region $w<x$ one needs to solve Eq.(\ref{3.14}%
) with $Q=0$. Both the solutions should be matched at $x=w$ under the
assumption {that there is no barrier }at this point.{\ Therefore,} the
matrix $\check{f}_{n}$ and its ''generalized'' derivative should be
continuous at $x=w$%
\begin{eqnarray}
\check{f}_{n}\mid _{x=w-0} &=&\check{f}_{n}\mid _{x=w+0} \\
\frac{Q}{2}i\hat{\tau}_{3}[\hat{\sigma}_{1,}\check{f}_{n}]_{+}&+&\partial_x \check{f}%
_{n}\mid _{x=w-0} =\partial _{x}\check{f}_{n}\mid _{x=w+0}
\label{3.2.7}
\end{eqnarray}
In this case the solution has the same structure as Eq.(\ref{3.18}) {\ }but
small changes should be done. {The eigenvalues }$\kappa $ {\ obey the
equation}
\begin{equation}
\lbrack (\kappa ^{2}-Q^{2}-\kappa _{\omega }^{2})^{2}+4Q^{2}\kappa _{\omega
}^{2}](\kappa ^{2}-\kappa _{\omega }^{2})+\kappa _{h}^{4}[\kappa
^{2}-Q^{2}-\kappa _{\omega }^{2}]=0  \label{3.2.8A}
\end{equation}
{where }$\kappa _{\omega ,h}^{2}${\ are determined in Eqs.(\ref{3.14a},\ref
{3.14b}).}{\ }The eigenvalue given by Eq. (\ref{3.22}) changes. Now it is
equal to
\begin{equation}
\;\kappa _{Q}^{2}=Q^{2}+\kappa _{\omega }^{2}\;,  \label{3.2.8}
\end{equation}
{while the eigenvalues $\kappa _{\pm },$ Eq. (\ref{3.23}),} remain unchanged
provided the condition
\begin{equation}
\;Q,\kappa _{\omega }<<\kappa _{h}  \label{3.2.9}
\end{equation}
is fulfilled.

In the opposite limit of large $Q>>\kappa _{h}$, the eigenvalues $\kappa
_{\pm }$ take the form
\begin{equation}
\;\kappa _{\pm }=\pm iQ[1\mp i\kappa _{h}^{2}/\sqrt{2}Q^{2}],  \label{3.2.10}
\end{equation}
Thus, in this limit $\kappa _{\pm }$\ is imaginary in the main
approximation, which means that the function\ $\check{f}_{n}(x)$\ oscillates
fast in space with the period $2\pi /Q$. In this case the eigenvalues (\ref
{3.2.8}) change also and have the form
\begin{equation}
\kappa ^{2}=\kappa _{\omega }^{2}+\frac{\kappa _{h}^{4}}{Q^{2}}  \label{e11}
\end{equation}
Therefore the limit of a very fast rotating magnetization ($\kappa
_{h}/Q\rightarrow 0$) is analogous to the case of a normal metal, \textit{%
i.e.} when the condensate penetrates the ferromagnet over the length $\kappa
_{\omega }^{-1}\sim \sqrt{D_{F}/2\pi T}$\ which is the characteristic
penetration length of the condensate in a S/N system.

More interesting and realistic is the opposite limit when the condition (\ref
{3.2.9}) is fulfilled and the long-range penetration of the triplet
component into the ferromagnet becomes possible.

In the limit of large $\kappa _{h}$, (Eq. (\ref{3.2.9})), the singlet
component penetrates the ferromagnet over a short length of the order $\xi
_{h}=1/\kappa _{h}$ while the LRTC penetrates over the length $\sim 1/\kappa
_{Q}$. As follows from Eq. (\ref{3.2.8}), this penetration length is about $%
1/Q$ (provided $w/\alpha _{w}$ is smaller than the length $\xi _{N}$).

Now let us find the amplitude of the LRTC. The solution for Eq.(\ref{3.2.5})
in the interval $0<x<w$ is determined by Eqs.(\ref{3.18}, \ref{3.19}) with
the functions $b_{i}\left( x\right) $, $i=0,1,3$ given by the following
formulae
\begin{equation}
b_{1}(x)=b_{Q}\exp (\kappa _{Q}x)+\bar{b}_{Q}\exp (-\kappa _{Q}x)\;
\label{3.2.11}
\end{equation}
\begin{equation}
b_{0}(x)=-b_{3+}\exp (-\kappa _{+}x)+b_{3-}\exp (-\kappa _{-}x)\;
\label{3.2.12}
\end{equation}
and
\begin{equation}
b_{3}(x)=b_{3+}\exp (-\kappa _{+}x)+b_{3-}\exp (-\kappa _{-}x)\;
\label{3.2.13}
\end{equation}
In the region $w<x$ the solution for the condensate function $\check{f}_{n}$
takes the form
\begin{equation}
\;\;\check{f}_{n}\;=i\hat{\tau}_{1}\otimes \hat{\sigma}_{1}c_{\omega }\exp
(-\kappa _{\omega }(x-w))  \label{3.2.14}
\end{equation}
where $c_{\omega }$ is a coefficient that has to be found by matching the
solutions {\ }at {\ }$x=w$.

{Terms of the order of $Q/\kappa _{h}$ are small and they are omitted now.}

Then we find from the matching conditions at the $S/F$ interface, Eq. (\ref
{3.2.7}), the following relations for the coefficients
\begin{equation}
b_{3\pm }=\frac{f_{BCS}}{2\gamma _{F}\kappa _{\pm }}  \label{3.2.15}
\end{equation}
and
\begin{equation}
b_{Q}=-\bar{b}_{Q}=(Q/\kappa _{Q})(b_{3+}-b_{3-})  \label{3.2.16}
\end{equation}
(the parameter $\gamma _{F}$ given by Eq. (\ref{3.2.2}))

One can see from the above equations {\ }that the condensate function $|%
\check{f}|$ is small provided parameter $|\gamma _{F}\kappa _{\pm }|$ is
large. It follows from Eq.(\ref{3.2.16}) that the amplitude of the LRTC, $%
b_{Q}$, is not zero only if the magnetization is nonhomogeneous, i.e., $%
Q\neq 0.$

Matching the solutions (\ref{3.2.11}-\ref{3.2.14}) at $x=w$, we find for the
amplitude of the LRTC
\begin{equation}
\;\;c_{\omega }=-\frac{if_{BCS}}{2\gamma _{F}}[\frac{Q}{\kappa _{Q}\sinh
\alpha _{w}+\kappa _{\omega }\cosh \alpha _{w}}](\frac{h\mathrm{sgn}\omega
/D_{F}}{|\kappa _{+}|^{2}\mathrm{Re}\kappa _{+}})  \label{3.2.17}
\end{equation}
where $\alpha _{w}=Qw$ is the total angle of the magnetization rotation. As
it has been mentioned, the amplitude of the LRTC is an odd function of $%
\omega $.

{As one can see from the last expression} the amplitude $c_{\omega
}$ increases from zero when increasing $Q$, reaches a maximum at
$Q_{max}$ corresponding a certain angle {$\alpha _{max}$ and then
exponentially decreases at $\alpha_w >>$ $\alpha _{max}.$}

{The maximum of }$c_{\omega }$ {\ is achieved at }
\begin{equation}
\;\alpha _{max}=(w\kappa _{\omega })\sqrt{\sqrt{5}-1}/\sqrt{2}\approx
0.786(w\kappa _{\omega })\;,  \label{3.2.18}
\end{equation}
{At }$\alpha _{w}=\alpha _{max}$ {\ the ratio in the square brackets in Eq.(%
\ref{3.2.17}) is equal to }$\approx 0.68$ {. }This means that the amplitude
of the LRTC is of the order of the singlet component at the $S/F$ interface.
The width $w$ should not be too small because in deriving the expression for
$c_{Q}$ we assumed the condition $w>>\xi _{h}.$

\begin{figure}[h]
\includegraphics[scale=0.35]{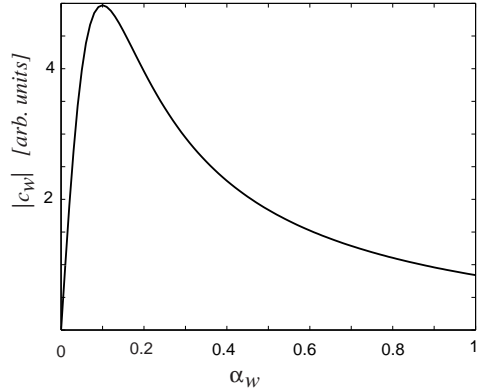}
\caption{Dependence of the amplitude of the triplet component on $\protect%
\alpha _{w}=Qw$. We have chosen $w\protect\kappa _{\protect\omega }=0.01$. }
\label{Fig.11}
\end{figure}

\begin{figure}[h]
\includegraphics[scale=0.3]{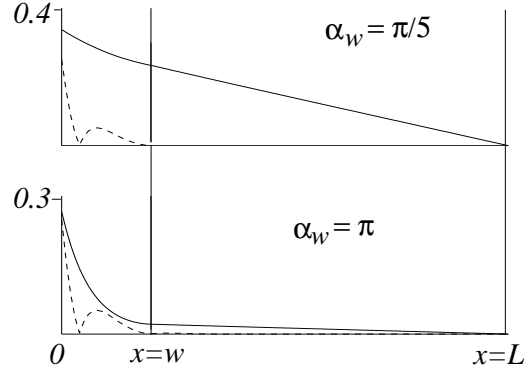}
\caption{Spatial dependence of the amplitudes of the singlet (dashed line)
and triplet (solid line) components of the condensate function in the F wire
for different values of $\protect\alpha_w$. Here $w=L/5$, $\protect\epsilon%
=E_T $ and $h/E_T=400$. $E_T=D/L^2$ is the Thouless energy (From \citet*{BVE2%
}). }
\label{Fig.12}
\end{figure}

In FIG. \ref{Fig.11} we represented the dependence of $\left| c_{\omega
}\right| $ on $\alpha _{w}$ for a fixed $w$. The spatial dependence of the
LRTC and the singlet component is shown in FIG. \ref{Fig.12}. It is seen
that for the parameters chosen the LRTC is larger than the singlet component
and decays much slower with increasing the distance $x$.

If the magnetization vector $\mathbf{M}$ rotates by the angle $\pi $ (a
domain wall) over a small length $w$ so that $Q\sim \pi /w>>\kappa _{w}$,
then the ratio in brackets in Eq.(\ref{3.2.17}) is equal to
\begin{equation}
\;\;(\frac{Q}{\kappa _{Q}\sinh \alpha _{w}+\kappa _{\omega }\cosh \alpha _{w}%
})\approx Q/(Q\sinh \pi )\approx 0.087  \label{3B.19}
\end{equation}
which shows that the amplitude of the LRTC in this case is smaller than the
amplitude of the singlet component.

We can conclude from {this analysis }that in order to get a large LRTC, a
small total angle of the rotation of the magnetization vector is more
preferable.

The amplitude of the condensate function calculated here enters different
physical quantities. In section \ref{EXO-LRPE} we discuss how the long-range
penetration of the triplet component into the ferromagnet affects transport
properties of $F/S$ structures.

{It is interesting to note that the type of magnetic structure
discussed in this section differs drastically from the one in the
case of an
in-plane rotating magnetization. The latter was considered recently by %
\citet{eschrig_pp,eschrig05}. It was assumed that the magnetization vector }$M_{F}$%
{\ was parallel to the S/F interface and rotates; that is, it has
the form }$M_{F}=M_{0}\{0,\sin (Qy),\cos (Qy)\}${\ (the x-axis is
normal to the S/F interface plane). As shown by \citet{eschrig05},
the odd triplet component arises also in this case but it
penetrates into the ferromagnetic region over a short distance of
the order of }$\xi _{h}$.

\subsubsection{Spin-active Interfaces}

\label{EXO-INH-SAI} In almost all papers containing discussions of the $S/F$
structures it is assumed that the $S/F$ interface is spin-inactive, i.e. the
spin of an electron does not change when the electron goes through the
interface.

Although in many cases it is really so {,} one can imagine another situation
when the spin of an electron passing through the interface changes. One can
consider a region with a domain wall at the interface also as a ``
spin-active interface'' provided the width $w$ of the domain wall is very
small but the product $Qw$ is of the order unity. As we have seen in section
\ref{EXO-INH-DW}, at such type of interfaces the triplet condensate arises.

Boundary conditions at spin-active $S/F$ interfaces for the quasiclassical
Green's functions were derived in a number of publications \cite{millis,kopu}
and were used in studying different problems. \citet{kulic_endres} employed
these boundary conditions in the study of a system similar to the one shown
in FIG. \ref{Fig.8}. Contrary to \citet{BVE_manifestation}, they assumed
that the ferromagnets $F$ are insulators so that the condensate does not
penetrate them. Nevertheless, the calculated critical temperature $T_{c}$ of
the superconducting transition depends on the mutual orientation of the
magnetization $M_{F}$ in the ferromagnets. In accordance with %
\citet{baladie_tc,tagirov_C,golubov_tc} where metallic ferromagnets were
considered in a $F/S/F$ structure, Kulic and Endres found that the critical
temperature $T_{c}$ was maximal for the antiparallel magnetization
orientation. If the directions of magnetization vector $M_{F}$ are
perpendicular to each other, a triplet component also arises in the
superconductor. The authors considered a clean case only, so that the
influence of impurity scattering on the triplet component remained unclear.

{According to \citet{nazarov02} a spin-active N/F interface plays an
important role in the absolute spin-valve effect which can take place in a
S/N/F mesoscopic structure. The authors considered a structure with a thin
normal metal layer (N) and a ferromagnetic insulator F. The DOS variation in
a conventional superconductor which is in contact with a ferromagnetic
insulator was analyzed by \citet{tokuyasu88}.}

{\citet{schoen_half} considered a clean }$S/F/S$ {\ Josephson
junction in which the ferromagnet }$F$ {\ was a half metal so that
the electrons with only one spin orientation (say the spin-up
}$\uparrow ${\ electrons) existed in the ferromagnet. In this case
only the triplet component corresponding to the condensate
function }$<\psi _{\uparrow }\psi _{\uparrow }>$ {\ may penetrate
the ferromagnet. Assuming the p-wave triplet condensate function,
the authors have calculated the critical Josephson current
}$I_{c}$ {. They showed that the }$\pi - $ {\ state (negative
critical current }$I_{c}$ {) is possible in this junction. The dc
Josephson effect in a junction consisting of two superconductors
and a spin-active interface between them was analyzed by
\citet{fogel00}.}

It would be of interest to analyze the influence of impurities on the
critical current in such type of Josephson junctions because, as we noted,
in a clean case the singlet component can penetrate the ferromagnet (not a
half metal) over a large distance.

\subsection{Long-range proximity effect}

\label{EXO-LRPE} 
In the last decade transport properties of mesoscopic superconductor/normal
metal $S/N$ structures were intensively studied (see for example the review
articles by \citet{beenakker_rev,lambert_rev} and references therein). In
the course of these studies many interesting phenomena have been discovered.
Among them is a non-monotonic voltage and temperature dependence of the
conductance in $S/N$ mesoscopic structures, i.e. structures whose dimensions
are less than the phase coherence length $L_{\varphi }$ and the inelastic
scattering length $l_{\varepsilon }$. This means that the resistance $R$ of
a $S/N$ structure changes non-monotonically when the temperature decreases
below the critical temperature $T_{c}$.

This complicated behavior is due to the fact that there are two
contributions to the resistance in such systems: the one coming from the $%
S/N $ interface resistance and the resistance of the normal wire itself. The
experimentally observed changes of the resistance can be both positive ($%
\delta R>0$) and negative ($\delta R<0$) \cite{Sanquer02,shapira}. The
increase or decrease of the resistance $R$ depends, in particular, on the
interface resistance $R_{S/N}$. If the latter is very small, the resistance
of the $S/N$ structure is determined mainly by the resistance of the $N$
wire $R_{N}$. This resistance decreases with decreasing the temperature $T$,
reaches a minimum at a temperature of the order of the Thouless energy $%
D_{N}/L_{N}^{2}$, and increases again returning to the value in the normal
state $R_{N}(T_{c})$ at low $T$, where $D_{N}$ is the diffusion coefficient
and $L_{N}$ is the length of the $N$ wire. This is the so called re-entrant
behavior observed in many experiments \cite
{gubankov,pothier,vanWees_95,petrashov_sn,volkov_pannetier,chien1,shapira}.

Theoretical explanations for the non-monotonic behavior of the resistance
variation as a function of the temperature $T$ or voltage $V$ in $S/N$
structures have been presented by %
\citet{artemenko,VZK,nazarov_stoof,volkov_allsopp,golubov_zaikin,shapira}.{\
Such a variation of the resistance of the normal metal wire can be explained
in terms of the proximity effect that leads to the penetration of the
condensate into the }${N}${\ wire.} Due to this penetration there are two
types of contributions to the conductance $G_{N}$ \cite
{volkov_pavlo,golubov_zaikin}. One of them reduces the DOS in the $N$ wire
and therefore reduces the conductance $G_{N}.$ The {\ }other term, similar
to the Maki-Thompson term \cite{golubov_zaikin,volkov_pavlo}, leads to an
increase of the conductance of the $N$ wire.

In principle, the magnitude of the conductance variation $\delta G_{N}$ may
be comparable with the conductance $G_{N}$. So, there are no doubts that the
proximity effect plays a very important role in many experiments on $S/N$
structures.

Recently, similar investigations have been carried out also on mesoscopic $%
F/S$ structures in which ferromagnets ($F$) were used instead of normal
(nonmagnetic) metals. According to our previous discussion, the depth of the
condensate penetration into an impure ferromagnet equals $\xi _{F}=\sqrt{%
\hbar D/h}$. This length is extremely short ($5-50$\AA ) for strong
ferromagnets like $Fe$ or $Ni$. Therefore one might expect that the
influence of the proximity effect on the transport properties of such
structures should be negligibly small.

It was a great surprise that experiments carried out recently on $F/S$
structures showed that the resistance variation $\delta R$ were quite
visible (varying from about $1$ to $10\%$) when decreasing the temperature
below $T_{c}$ \cite{Giordano96,Giordano99, petrashov,chandrasekhar,pannetier}%
. For example in the experiments by \citet{Giordano96,Giordano99}, where an $%
Sn/Ni$ structure was studied, the effective condensate penetration length
estimated from the measured resistance was about $400$\AA . This quantity
exceeds $\xi _{F}$ by order of magnitude. Similar results have been obtained
by \citet{pannetier} on $Co/Al$ structures, by \citet{petrashov} on a $Ni/Al$
structures and by \citet {chandrasekhar} on $Ni/Al$ structures.

It is worth mentioning that the change of the {\ }resistance was both
positive and negative. In some experiments the variation $\delta R_{F}$ was
related to a change of the interface resistance \cite{chandrasekhar},
whereas in others \cite{Giordano96,Giordano99,petrashov,pannetier} to the
resistance variation of the ferromagnetic wire $\delta R_{F}$.

\begin{figure}[h]
\includegraphics[scale=0.3]{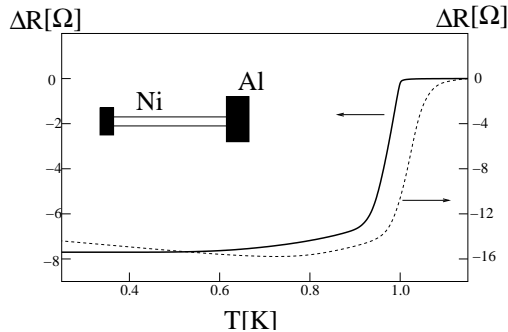}
\caption{Reduction of the resistance of a Ni wire attached to a
superconductor(Al). Adapted from \citet{petrashov}. }
\label{Fig.13}
\end{figure}

In FIG. \ref{Fig.13} we show the temperature dependence of the resistance of
a $Ni$ wire attached to an $Al$ bank measured by \citet{petrashov}.
According to estimates of $\xi _{F}$ performed in this experiments, the
observed $\delta R_{F}$ is by two orders of magnitude larger than it might
be expected from the conventional theory of {\ }$S/F$ the contacts.
Therefore these results cannot be explained in terms of the penetration of
the singlet component.

\begin{figure}[h]
\includegraphics[scale=0.5]{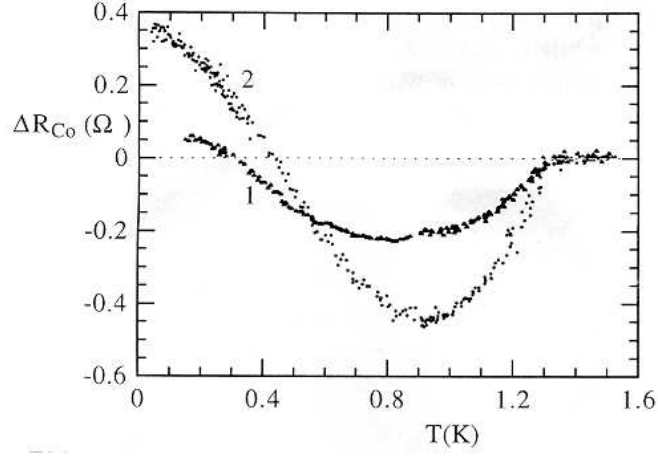}
\caption{Temperature dependence of the resistance of a Co wire attached to a
superconductor (Al) measured by \citet{pannetier}. Note that at low
temperatures the authors observed a reentrance behavior. }
\label{Fig.14}
\end{figure}

{In FIG. \ref{Fig.14} {\ we show similar }data from the{\ {\ }experiment on }%
${Co/Al}${\ structures performed by \citet{pannetier}. In this experiments a
reentrance behavior of $\delta R$ was observed. In the limit of very low
temperatures $T\rightarrow 0$ the resistance was even larger than in the
normal state. } }

The final {explanation of this effect remains until now unclear.}
However, the long range proximity effects considered in the
previous sections may {definitely contribute to the conductance
variation.} In order to support this point of view we analyze
qualitatively the changes of the conductance due to the LRTC
penetration into the ferromagnet and demonstrate that the LRTC may
lead to the conductance variation comparable with that observed in
the experiments.

However, before presenting these calculations it is reasonable to understand
if one can explain the experiments in a more simple way. Actually, the
resistance of the $S/F$ structures has been analyzed in many theoretical
works. For example, \citet{dejong,golubov,nazarov_belzig} analyzed a
ballistic $S/F$ contact. It was shown that at zero exchange field ($h=0$),
the contact conductance $G_{F/S}$ is twice as large as its conductance $%
G_{F/N}$ in the normal state (above $T_{c}$), as it should be. This agrees
with a conductance in a $N/S$ ballistic contact according to theoretical
predictions. At the same time, it drops to zero at $h=E_{F},$ where $E_{F}$
is the Fermi energy.

The conductance of a diffusive point contact $G_{F/S}$ has been calculated
by \citet{golubov} who showed that $G_{F/S}$ was always smaller than the
conductance $G_{F/N}$ in the normal state. In the case of a mixed
conductivity mechanism (partly diffusive and partly ballistic) the
conductance $G_{F/S}$ has been calculated by \citet{nazarov_belzig}.
According to their calculations it can be both larger or smaller than the
conductance in the normal state $G_{F/N}$.

The resistance $R_{F}$ of a ferromagnetic wire attached to a superconductor
was calculated by \citet{falko_volkov1,vanWees_99,BPVE} and let us shortly
describe what happens in such a system.

The proximity effect was neglected in these works but a difference in the
conductivities $\sigma _{\uparrow \downarrow }$ for spin-up and down
electrons was taken into account. The change of the conductance (or
resistance) $\delta G_{F}$ is caused by a different form of the distribution
functions below and above $T_{c}$ because of Andreev reflections.

The conductance $G_{F}(T_{c})$ of the $F$ wire in the normal state ($T>T_{c}$%
) is given by the simple expression
\begin{equation}
\;\;G_{F}(T_{c})=G_{\uparrow }+G_{\downarrow }\;,  \label{4.1}
\end{equation}
where $G_{\uparrow \downarrow }=\sigma _{\uparrow \downarrow }L_{F}A,$ $%
L_{F} $ and $A$ are the length and cross-section area of the $F$ wire.

This means that the total conductance is the sum of the conductances of the
spin-up and down channels. In this case not only the electric current but
also the spin current is not zero. It turns out that {\ }below $T_{c}$ ($%
T<T_{c}$) the conductance decreases and at zero temperature it is equal to
\begin{equation}
\;\;G_{F}(0)=4G_{\uparrow }G_{\downarrow }/(G_{\uparrow }+G_{\downarrow })
\label{4.2}
\end{equation}
Eq. (\ref{4.2}) shows {\ }that the zero-temperature conductance $G_{F}(0)$
for the system considered is smaller than the normal state conductance $%
G_{F}(T_{c}).$

It is possible to obtain the explicit formulae not only in the limiting
cases, Eq. (\ref{4.1}, \ref{4.2}), but also to describe the system at
arbitrary temperatures. The general formula for the conductance of the $F$
wire can be written as
\begin{equation}
\;\;G_{F}(T)=G_{F}(0)\tanh (\Delta /2T)+G_{F}(T_{c})(1-\tanh (\Delta /2T))
\label{4.3}
\end{equation}
Eqs. (\ref{4.1}) and (\ref{4.2}) can be obtained from Eq. (\ref{4.3}) by
putting $\Delta $\ or $T$\ to zero. Eqs.(\ref{4.1}-\ref{4.3}) are valid
provided the length $L_{F}$ satisfies the condition
\begin{equation}
l_{\uparrow \downarrow }<L_{F}<L_{SO},L_{in}\;,  \label{4.4}
\end{equation}
where $\;l_{\uparrow \downarrow }$ is the mean free path of spin-up and
spin-down electrons,while $L_{SO}$ and $L_{in}$ are the spin-orbit and
inelastic relaxation length, respectively.

{The resistance of multiterminal }$S/F$ {\ structures was calculated by %
\citet{melin01,melin03,melin04_2} on the basis of the tunnel Hamiltonian
method. The influence of superconducting contacts on giant magnetoresistance
in multilayered structures was studied by \citet{lambert01}. \citet{falko02}
studied an enhancement of Andreev reflection at the S/F interface due to
inelastic magnon-assisted scattering}.

\begin{figure}[h]
\includegraphics[scale=0.25]{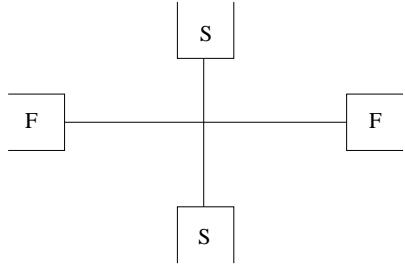}
\caption{The cross geometry used to measured the changes of the resistance
of a F wire due to the proximity effect. }
\label{Fig.15}
\end{figure}

One can conclude from the works listed above that neglecting the penetration
of the LRTC into the $F$ wire an increase in the conductance $G_{F}$ cannot
be explained. Therefore, {\ }let us discuss the consequences of the LRTC
penetration into the ferromagnetic wire. In order to avoid the consideration
of the $S/F$ interface contribution to the total resistance, we consider a
cross geometry (see FIG. \ref{Fig.15}) and assume that the resistance of the
interface between the $F$ wire and $F$ reservoirs is negligible. Such a
geometry was used, for example, in the experiments by \citet{petrashov_sn}.
The structure under consideration consists of two $F$ wires attached to the $%
F$ and $S$ reservoirs. We assume that there is a significant mismatch
between parameters of the superconductor and ferromagnet so that the
condensate amplitude induced in $F$ is small and is determined by Eqs.(\ref
{3.34}) or (\ref{3.2.17}).

According to our results obtained previously the long range proximity effect
is possible provided there is a domain wall near the interface between the
superconductor and ferromagnet and we assume that this is the case for the
setup shown in FIG. \ref{Fig.15}. Another possibility to generate the
triplet condensate would be to attach to the superconductor {\ } an
additional ferromagnet with a non-collinear magnetization.

The conductance can be found on the basis of a general formula for
the current (see for example the book by \citet{kopnin} and
Appendix \ref{ApA})
\begin{equation}
I=(1/16e)(L_{y}L_{z})\sigma _{F}\mathrm{Tr}\hat{\sigma}_{0}\otimes \hat{\tau}%
_{3}\circ \int \mathrm{d}\epsilon \lbrack \check{g}^{R}\partial _{x}\check{g}%
^{K}+\check{g}^{K}\partial _{x}\check{g}^{A}]  \label{4.5}
\end{equation}
where $\sigma _{F}$ is the conductivity of the $F$ wire in the normal state.

The matrix Green's function $\check{g}^{K}=\check{g}^{R}\check{F}-\check{F}$ $%
\check{g}^{A}$ is the Keldysh function related to a matrix distribution
function $\check{F}$. The distribution function consists of two parts,{\ }%
namely, one of them is symmetric with respect to the energy $\epsilon $, the
{\ }other one is antisymmetric in $\epsilon $ and determines the dissipative
current.

In the limit of a weak proximity effect the retarded (advanced)
Green's function has the form
\begin{equation}
\check{g}^{R(A)}\approx \pm \hat{\tau}_{3}\otimes \hat{\sigma}_{0}+\;\check{f%
}^{R(A)}\;,  \label{4.6}
\end{equation}
where $\check{f}^{R(A)}$ is given by Eqs.(\ref{3.34}) or (\ref{3.2.17}).

We have to find the conductance of the vertical $F$ wire in FIG. \ref{Fig.15}%
. In the main approximation the distribution function in this $F$ wire is
equal to
\begin{equation}
\check{F}=F_{0}\cdot \hat{\tau}_{0}\otimes \hat{\sigma}_{0}+F_{3}\cdot \hat{%
\tau}_{3}\otimes \hat{\sigma}_{0}\;,  \label{4.7}
\end{equation}
where $F_{0,3}=[\tanh ((\epsilon +V)/2T)\pm \tanh ((\epsilon -V)/2T)].$

The distribution function $F_{3}$ symmetric in $\epsilon $ determines the
current $I.$ The differential conductance $G_{d}=dI/dV$ {can} be represented
as
\begin{equation}
G_{d}=G_{0}+\delta G\;,  \label{4.8}
\end{equation}
where $G_{0}=\sigma _{F}L_{F}A$ is the conductance in the normal state (here
we neglect for simplicity the difference between $\sigma _{\uparrow }$ and $%
\sigma _{\downarrow }$).

The {normalized} correction to the conductance due to the
proximity effect $\delta S(T)\equiv \delta G/G_{0}${\ can be found
using a general formula }\cite{BVE2}
\begin{equation}
\delta S(T)=(32T)^{-1}\mathrm{Tr}\hat{\sigma}_{0}\int d\epsilon <(\hat{f}%
^{R}-\hat{f}^{A})^{2}>F_{V}^{\prime }(\epsilon )
\end{equation}
{where }

$F_{V}^{\prime }(\epsilon )=[${\ }$\cosh ^{-2}((\epsilon
+eV)/2T)+\cosh ^{-2}((\epsilon -eV)/2T)]/2${. }

{The angle brackets }$<...>${\ denote the average over the length
of the ferromagnetic wire between the F (or N) reservoirs. The
functions }$\hat{f}^{R(A)}${\ are given by expressions similar to Eq.(%
\ref{3.2.17}). This formula shows that if }$T<D_{F}/L^{2}$, {\ on
the
order of magnitude }$\delta S(T)\sim |f_{tr}|^{2}${,where }$L${%
\ is the length of the ferromagnetic wire and }$|f_{tr}|${\ is the
amplitude of the triplet component at the S/F interface at a
characteristic
energy }$\epsilon _{ch}\sim \min \{T,D_{F}/L\}.${\ According to Eq.(%
\ref{3.2.17}) the amplitude of the triplet component is of the order of }$%
c_{1}(\rho \xi _{h}/R_{b})${\ where }$\rho ${\ is the resistivity
of the ferromagnet and }$c_{1}${\ is determined by the factor in
the square brackets, that is, by the
characteristics of the domain wall. In principle the amplitude }$|f_{tr}|$%
{\ may be of the order of 1. }

Strictly speaking, both the singlet and triplet components contribute to the
conductance. However if the length $L_{F}$ much exceeds the short length $%
\xi _{F}$ only the contribution of the LRTC is essential.

\begin{figure}[h]
\includegraphics[scale=0.3]{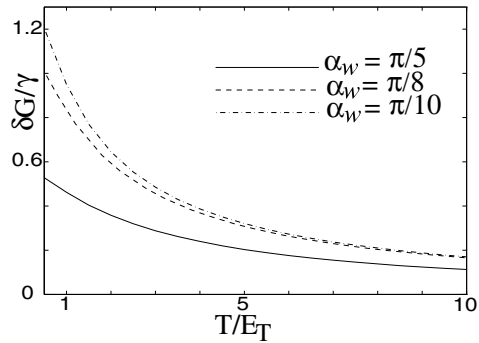}
\caption{The $\protect\delta G(T)$ dependence. Here $\protect\gamma =\protect%
\rho \protect\xi _{h}/R_{b}$. $\Delta /E _{T}\gg 1$ and $w/L=0.05$ (from
\citet*{BVE2}). }
\label{Fig.16}
\end{figure}

In FIG. \ref{Fig.16} we present the temperature dependence of the correction
to the conductance {\ }$\delta G(T).$ It is seen that with increasing the
temperature $\delta G_{F}(T)$ decreases in a monotonous way. This dependence
differs from the re-entrant behavior {discussed above} that occurs in the $%
S/N$ structures. The reason for this difference is that the time-reversal
symmetry in $S/F$ structures is broken and this leads to a difference in
transport properties. In a $S/N$ system, a relation$\hat{f}^{R}(\epsilon )=%
\hat{f}^{A}(\epsilon )|_{\epsilon =0}$ holds and this equality is a
consequence the time-reversal symmetry. That is why $\delta G(0)=\delta
G(T_{c})=0$ in $S/N$ structures, whereas in a $S/F$ structure $\hat{f}%
^{R}(\epsilon )\neq \hat{f}^{A}(\epsilon )|_{\epsilon =0}$ and
{that is why }$\delta S(T)_{T=0}\neq 0.$

\bigskip {Although the LRTC may be the reason for the enhancement of
the conductivity in the }$S/F${\ structures (this possibility was
also pointed out in the work by \citet{giroud03}),\ }our
understanding is based on the assumption that the magnetic moment
is fixed and does not change with the temperature. \citet{Geim}
suggested another mechanism based on an assumption about a domain
redistribution when the temperature drops below $T_{c}.$ The
ferromagnetic wires (or strips) used in different experiments may
consist of many domains. Their form and number depend on the
sample geometry and parameters of the system. When the temperature
decreases below $T_{c}$, stray magnetic fields excite the Meissner
currents in the superconductor attached to the $F$ wire. Therefore
the demagnetizing factors change, which may lead{\ to a new}
domain structure. At the same time, {\ }the total conductance (or
resistance) $G_{F}$ depends on the form and the number of
domains. So, one might expect that{\ }the conductance $G_{F}(T)$ below $%
T_{c} $ would differ from $G_{F}$ in the normal state. This idea was
supported by measurements carried out by \citet{Geim}. In this work a
structure consisting of a two-dimensional electron gas and five Hall probes
was used. An $F/S$ system ($Ni$+$Al$ disks) was placed on top of this
structure. Measuring the Hall voltage, the authors were able to probe local
magnetic fields around the ferromagnetic disks. They found that these fields
really changed when the temperature dropped below $T_{c}$.

On the other hand, the Meissner currents and, hence, the effect of the
redistribution of the domain walls may be considerably reduced in wires, as
discussed previously. Changing the thickness of the superconducting wires in
a controlled way and measuring the conductance could help to distinguish
experimentally between the contribution to the conductivity of the triplet
condensate and the effects of the redistribution of the domain walls.

\bigskip

{An experiment in which the domain redistribution was excluded has
been performed by \citet{petrashov04}. The authors measured the
resistance variation of a ferromagnetic wire
(Ni}$_{1-x}${Cu}$_{x}${)
lowering the temperature }$T${\ below the critical temperature }$%
T_{c} ${\ of the superconductor (Al or Pb), which was attached to
the middle part of the ferromagnetic wire. A magnetic field,
strong enough to align all domains in the ferromagnet in one
direction but not too strong to suppress the superconductivity,
was applied to the system. Under these conditions a small increase
in the resistance (}$\delta R/R\approx 3\cdot 10^{-3}${) was
observed when the temperature }$T${\ drops below }$T_{c}${. The
analysis presented above shows that the triplet component leads to
an increase of the conductance but not in the resistance of the
ferromagnetic wire. Therefore this particular experiment can
hardly be explained in terms of the long-range proximity effect.\
Perhaps the small
increase in the resistance of the ferromagnetic wire observed in %
\citet{petrashov04} was related to the ''kinetic'' mechanism
discussed above (see Eq. (\ref{4.3})) or to weak localization
corrections caused by the triplet Cooperons \cite{mccann00}.
According to \citet{mccann00} the change of the resistance of the
ferromagnetic wire is positive (contrary to the contribution of
the LRTC) and its  order of magnitude } is $(e^{2}/\hbar )R_{F}$,
{where }$R_{F}$ {is the resistance of the F wire in the normal
state. In order to clarify the role of the LRTC in the transport
properties of S/F structures, further theoretical and experimental
investigations are needed. Note that strong ferromagnets like Fe
are not suitable materials for observing the contribution of the
LRTC into the conductance variation because of the strong exchange
field }$h$.{ In this case, according to Eq.(\ref {3.31}) and
Eq.(\ref{3.2.17}), the amplitude of the LRTC is small because it
contains }$h${\ in the denominator.}

\section{Josephson effect in S/F systems (inhomogeneous magnetization)}

\label{JOS} 
As we have mentioned above, one of the most interesting issues in the $S/F$
structures is the possibility of switching between the so called 0- and $\pi$%
-states in Josephson $S/F/S$ junctions. The $\pi $-state denotes the state
for which the Josephson critical current $I_{c}$\ becomes negative. This
occurs for a certain thickness $d_F$ and temperature $T$. In this state the
minimum of the Josephson coupling energy $E_{J}=(\hbar I_{c}/e)(1-\cos \phi
) $\ corresponds to a phase difference of $\phi=\pi $ but not to $\phi =0$
as in conventional Josephson junctions.

The reason for the sign reversal of $I_{c}$\ is the oscillatory dependence
of the condensate functions $\hat{f}$\ on the thickness $d_{F}$\ (see Eq.(%
\ref{I_c})). Since the critical current $I_{c}$\ is sensitive to the phase
of the condensate function at the boundary, the $\pi $-state is a rather
natural consequence of the oscillations.

The possibility of the $\pi -$ state was predicted by \citet{bula_kuzi} and %
\citet{buzdin_sfs}, and studied later in many other works %
\citep[e.g.][]{radovic2,buzdin_kupr2}. Experimentally, this phenomenon
manifests in a non-monotonic dependence of the critical temperature on the
thickness of the junction observed in many experiments and discussed in
Section \ref{PRO-SF-TC}. Another manifestation of the transition from the $0$%
- state to the $\pi $- state is the sign reversal of the critical current
observed in the experiment by \citet{ryazanov} on $Nb/Cu_{x}Ni_{1-x}/Nb$
Josephson junctions (see FIG. \ref{Fig.5}). The {proper} choice of an alloy
with a weak ferromagnetic coupling was crucial for the observation of the
effect.

Subsequent experiments, \citet{kontos02}, \citet{palevski} and %
\citet{kontos03}, corroborated the observed change of the sign of
the Josephson coupling varying the thickness of the intermediate
$F$ layer. Qualitatively, the experimental data on the Josephson
effect in the $S/F/S$ structures are in agreement with the
theoretical works above mentioned. However, a more accurate
control and understanding of the $0$-$\pi $ transition demands
knowledge of the magnetic structure of the ferromagnetic
materials.

Almost in all theoretical papers very simplified models of the $S/F/S$
junction were analyzed. For example, \citet{blanter} assumed that the $F$
layer consisted either of one domain or two domains with the collinear
orientation of the magnetization. In this case and according to the
discussion of section \ref{EXO-INH} the LRTC is absent in the system.

If the $F$ layer is a single domain layer, the critical current $I_{c}$\ is
maximal at a non-zero external magnetic field $H_{ext}$\ equal to $-4\pi
M_{F}$, where $M_{F}$\ is the magnetization of the $F$ layer. At the same
time, in experiments {\cite
{ryazanov,kontos01,kontos02,palevski,strunk,sellier} } a decrease of the
current $I_{c}$\ with increasing field $H_{ext}$\ was observed and it was{\ }%
maximal at $H_{ext}=0$. This means, as it was assumed in these experimental
works, that the F layer in real junctions contains many magnetic domains. In
this case the Josephson critical current $I_{c}$ may change sign in the $%
S/F/S$ junctions with a multidomain magnetic structure even if the local
Josephson current density $j_{c}$ is always positive. The reason for the
sign reversal of $I_{c}$ in this case is a spatial modulation of the phase
difference $\phi (x)$ due to an alternating magnetization $M\left( x\right) $
in the domains \cite{volkov04}. In order to determine the mechanism that
leads to the sign reversal of the critical current further experiments are
needed.

In this chapter we discuss a new phenomenon, namely, {\ }how the Josephson
coupling between the $F$ layers in the $S/F$ structures is affected by the
LRTC.

First, we consider a planar $S/F/S$ Josephson junction with a ferromagnet
magnetization $\mathbf{M_{F}}$\ rotating in the direction normal to the
junction plane. This model is an idealization of a real multidomain
structure with different magnetization orientations. In this case, as we
discussed in Section \ref{EXO-LRPE}, the LRTC arising in the structure
affects strongly the critical current $I_{c}.$\

Next, we will analyze a multilayered $S/F/S/$... structure in which the
vector $\mathbf{M}_{F}$\ has a different direction in different $F$ layers.
Again, in this case the LRTC arises in the system. Interestingly, if the
thickness of the $F$ layers $d_{F}$\ is much larger than the penetration
length $\xi _{F}$ of the singlet component but less or of the order of $\xi
_{N}$, then the Josephson coupling between the $F$ layers is realized due to
the LRTC (odd triplet superconductivity in the transverse direction). At the
same time, the {\ }in-plane superconductivity is due to the conventional
singlet superconducting pairing.

Finally we will discuss the dc Josephson effect in a $SF/I/FS$ junction
(here $SF$ is a superconductor-ferromagnet bilayer and $I$ is a thin
insulating layer). In this structure, the exchange field may lead not only
to a suppression of the Josephson coupling as one could naively expect but,
under a certain condition, to its enhancement.

{Let us consider first }a planar $S/F/S$ Josephson junction. We assume the
following spacial dependence of the magnetization vector in the F layer: $%
\mathbf{M}_{F}=M_{F}(0,\sin (Qx),\cos (Qx))$, where the $x$-axis is normal
to the junction plane.

In this case, as we have seen in Section \ref{EXO-INH-DW}, the LRTC arises.
Due to the long range penetration into the ferromagnet the triplet component
can give a very important contribution to the Josephson current. A general
expression for the Josephson current can be written in the form
\begin{equation}
I_{J}=(L_{y}L_{z}/4e)\sigma _{F}(\pi T)\mathrm{Tr(}\hat{\sigma}_{0}\otimes
\hat{\tau}_{3}.\sum_{\omega }\check{f}_{\omega }\partial _{x}\check{f}%
_{\omega })\;.  \label{X.1}
\end{equation}
We assume that the impurity concentration is sufficiently high and therefore
the condensate function $\check{f}_{\omega }$ should be found from the
Usadel equation. In the limit of a weak proximity effect (the $S/F$
interface transparency is not too high) this equation can be linearized and
solved exactly. The solution for the $\check{f}_{\omega }$ matrix in the $F$
region can be found in a similar way as it was done in Section \ref
{EXO-INH-DW}. Due to the rotation of the magnetization the condensate
function contains the LRTC. We obtain for the Josephson current \cite
{BVE_josephson} the following expression
\begin{equation}
I_{J}=I_{c}\sin \phi  \label{X.2}
\end{equation}
where the critical current $I_{c}$ is equal to
\begin{equation}
I_{c}=(L_{y}L_{z}\sigma _{F}/l)\tilde{\gamma}_{F}^{2}\mathrm{Re}\sum_{\omega
>0}f_{s}^{2}\left[ \frac{e^{-\kappa _{+}d_{F}}}{\kappa _{+}l}+\frac{%
(Ql)^{2}e^{ -\kappa _{l}d_{F}}}{2(3h\tau )^{3/2}}\right] \;,
\label{ij_q}
\end{equation}
and $\kappa _{l}^{2}=2|\omega _{n}|/\Delta +Q^{2}$. {The parameter }$%
\tilde{\gamma}_{F}=(3/4)<\mu T(\mu )>${\ is an effective, averaged
over angles, transmittance coefficient which characterizes the S/F
interface transparency }and $\kappa _{+}$ is defined in Eq.
(\ref{3.2.10})

The first term in the brackets containing the parameter {\ }$\kappa _{+}$%
corresponds to Eq. (\ref{I_S1}). {\ }It decays by increasing the thickness $%
d_{F}$ over the short characteristic length $\xi _{F}=\sqrt{D_{F}/h}$ and
can change the sign. The second term in Eq. (\ref{ij_q}) {\ }originates from
{\ }the rotation of $h$ along the $x$-axis. It decays with the thickness $%
d_{F}$ over another characteristic length $\kappa _{l}^{-1}$ that can be
much larger than $\xi _{F}$. Therefore this term results in a drastic change
of the critical current.

The presence of the second term in Eq. (\ref{ij_q}) is especially
interesting in the case when the thickness $d_{F}$ of the ferromagnetic
spacer between the superconductors obeys $\xi _{F}<d_F<\kappa _{l}^{-1}$.
Then the main contribution to the Josephson coupling comes from the
long-range triplet component of the condensate. Another important feature of
this limit is that for sufficiently large values of $Ql$ the critical
current is always positive (no possibility for the $\pi $-contact). This can
be seen from FIG. \ref{Fig.7}.

The fact that the superconductivity looses its ``exotic properties'' at
large $Q$ is quite understandable. The superconductivity is sensitive not to
the local values of the exchange field but to its average on the scales of
the order of the superconducting coherence length. If the exchange field
oscillates very fast such that the period of the oscillations is much
smaller than the superconducting coherence length, its average on this scale
vanishes and therefore all new properties of the superconductivity
originating from the presence of the exchange field become negligible.

To conclude this introduction we summarize the results known for $S/F/S$
Josephson junctions.

When the magnetization in the ferromagnetic {\ }$F$ is homogeneous, we have
to distinguish between two different {\ }cases.

In the dirty limit ($h\tau \ll 1$) the change of the sign of the critical
current occurs if the thickness of the $F$ layer $d_{F}$ is of the order of $%
\xi _{F}$. The condensate function in the $F$ layer decays exponentially
over this $\xi _{h}$ and oscillates with the same period.

In the opposite clean limit, $h\tau \gg 1$, the condensate function
oscillates in space with the period $v_{F}/h$ and decays exponentially over
the mean free path $l$.

Finally, if the ferromagnetic region contains a domain wall described by a
vector $Q$, a long-range component of the condensate appears. It decays in
the $F$ film over a considerably larger length of the order $\xi _{N}=\sqrt{%
D/2\pi T}$ that can greater exceed the characteristic length ($\sim \sqrt{D/h%
}$) in a homogeneous $F$ layer ($Q=0$). In this case the coupling between
the superconductors survives even if the thickness of $F$ is larger than $%
\xi_F$.

It is clear that the presence of a domain wall between the superconductors
is something that cannot be controlled very well experimentally. Therefore
in the next section we discuss a possible experiment on $S/F$ multilayered
structures that may help in detecting {\ }the LRTC by measuring the
Josephson critical current.

\subsection{Josephson coupling between S layers via the triplet component}

\label{JOS-TRI} 
In this subsection we analyze another type of multilayered $S/F$ structure
in which the LRTC arises. This is a multilayered periodic $%
...S/F_{n-1}/S/F_{n}/S/F_{n+1}/S...$ structure with alternating
magnetization vector $\mathbf{M}_{F,n}$ in different $F$ layers. We assume
that the vector $\mathbf{M}_{F,n}$ is turned {\ }with respect to the vector $%
\mathbf{M}_{F,n-1}$ by an angle $2\alpha $, such that the angle increases
monotonously with increasing $n$. We call this arrangement of the
magnetization the one with a positive chirality.

In an infinite system the magnetization vector $\mathbf{M}_{F}$ averaged
over $n$ is equal to zero (it rotates when one moves from the $n-th$ to the (%
$n+1)-th,$ layer etc.). Another type of chirality (negative chirality) is
the arrangement {\ }when the angle between vectors $\mathbf{M}_{F,n} $ and $%
\mathbf{M}_{F,n-1}$ is equal to $2\alpha (-1)^{n}$. In this case the
averaged vector $\mathbf{M}_{F}$ is not zero.

In Section \ref{EXO-INH-FSF} we have seen that in a $F/S/F$ structure with a
non-collinear orientation of the magnetization vectors in the $F$ layers the
LRTC arises. If one assumes that the thickness of the $F$ layers $d_{F}$ is
larger than the coherence length in the normal metal {\ } $\xi _{N},$ the
overlap of the condensate functions created in a $F$ layer by neighboring $S$
layers is weak, and the solutions given by Eqs.(\ref{3.18}-\ref{3.27})
remain valid for the multilayered $S/F$ structure.

Using these solutions one can calculate the Josephson current between
neighboring $S$ layers. As the thickness $d_{F}$ is assumed to be much
larger than $\xi _{F}$ (as usual, we assume that $\xi _{F}<<\xi _{N}$), the
Josephson coupling between the $S$ layers is solely due to the LRTC. So, in
such systems we come to a new type of the superconductivity: an odd triplet
out-of-plane superconductivity and the conventional singlet in-plane
superconductivity (the triplet component gives {\ }only a small {\ }
contribution to the in-plane superconductivity).

Using the general Eqs. (\ref{3.18}-\ref{3.27}) one can perform explicit
calculations for this case without considerable difficulties. As a result,
the Josephson critical current $I_{c}$ can be written as follows \cite
{BVE_manifestation}
\begin{widetext}
\begin{equation}
eR_{F}I_{c}=\pm 2\pi T\sum_{\omega }\kappa _{\omega }d_{F}b_{1}^{2}(\alpha
)\left( 1+\tan ^{2}\alpha \right) e^{-d_{F}\kappa _{\omega }}\;,
\label{jt_current}
\end{equation}
where

\[
b_{1}(\alpha )=-f_{BCS}\sin \alpha \frac{\mbox{$\tilde{\kappa}$}_{S}^{2}(%
\mbox{$\tilde{\kappa}$}_{+}-\mbox{$\tilde{\kappa}$}_{-})\mbox{${\rm sgn}$}%
\omega }{\cosh ^{2}\Theta _{S}\left( M_{+}T_{-}+M_{-}T_{+}\right)
(g_{BCS}+\gamma _{F}\kappa _{\omega }\tanh \Theta _{F})}\;,
\]
$\Theta _{S}=\kappa _{s}d_{S}$, $\Theta _{F}=\kappa _{\omega }d_{F}$, $%
\mbox{$\tilde{\kappa}$}_{\pm }=\kappa _{\pm }/(g_{BCS}+\gamma _{F}\kappa
_{\pm })$, $\mbox{$\tilde{\kappa}$}=\kappa _{\omega }/(g_{BCS}+\gamma
_{F}\kappa _{\omega }\tanh \Theta _{F})$, $\mbox{$\tilde{\kappa}$}%
_{S}=\kappa _{S}/(g_{BCS}\gamma )$ and
\begin{eqnarray*}
M_{\pm } &=&T_{\pm }(\mbox{$\tilde{\kappa}$}_{S}\coth \Theta _{S}+%
\mbox{$\tilde{\kappa}$}\tanh \Theta _{F})+\tan ^{2}\alpha \,C_{\pm }(%
\mbox{$\tilde{\kappa}$}_{S}\tanh \Theta _{S}+\mbox{$\tilde{\kappa}$}\tanh
\Theta _{F}) \\
T_{\pm } &=&\mbox{$\tilde{\kappa}$}_{S}\tanh \Theta _{S}+\mbox{$\tilde{%
\kappa}$}_{\pm } \\
C_{\pm } &=&\mbox{$\tilde{\kappa}$}_{S}\coth \Theta _{S}+\mbox{$\tilde{%
\kappa}$}_{\pm }\;.
\end{eqnarray*}
\end{widetext}
$R_F$ is defined as $R_F=2d_F/(L_yL_z\sigma_F)$. Eq.
(\ref{jt_current}) describes the layered systems with both the
positive (``+'' sign) and negative (``-'' sign) chirality.

One can see from Eq. (\ref{jt_current}) that in the case of positive
chirality the critical current is positive, while if the chirality is
negative the system is in the $\pi $-state (negative current). This means
that changing the configuration of the magnetization, one can switch between
the $0$ and $\pi $ state.

It is important to emphasize that the nature of the $\pi $-contact obtained
here differs from that predicted by \citet{bula_kuzi} and observed by %
\citet{ryazanov}. In the latter cases the transition is due to the change of
the values of either the exchange field, the temperature or the thickness of
the $F$ film. In the case considered in this section, the negative Josephson
coupling originates from the presence of the triplet component and can be
realized in $S/F$ structures with negative chirality. Since for the{\ }%
positive chirality the Josephson current is positive, the result obtained
gives an unique opportunity to switch experimentally between the $0$ and $%
\pi $-states by changing the angles of the mutual magnetization of the
layers.

{A }similar dependence of the Josephson current $I_{c}$ on the chirality was
predicted in a Josephson junction $S_{m}IS_{m}$ ($I$ is an insulator)
between two magnetic superconductors $S_{m}$ by \citet{kulic_kulic}. For the
magnetic superconductors considered in that work, the magnetization vector $%
\mathbf{M}$ rotated with the angle of rotation equal to $\alpha =x\mathbf{%
Q\cdot n}_{x}\mathbf{,}$where $\mathbf{Q}$ is the wave vector of the x-
dependence of the angle $\alpha $, $\mathbf{n}_{x}$ is the unit vector
normal to the insulating layer $I.$ Therefore the chirality (or spiral
helicity, in terms of \citeauthor{kulic_kulic}) in this case is determined
by the sign of the product $\mathbf{Q}_{R}\mathbf{\cdot Q}_{L},$ where $%
\mathbf{Q}_{L,R}$ is the wave vector in the left (right) magnetic
superconductor.

However, there is an essential difference between the multilayered $S/F$
structure discussed here and the magnetic superconductors. In the magnetic
superconductors with a spiral magnetization the triplet component also
exists but, in contrast to the $S/F$ structures, the singlet and triplet
components cannot be separated. In particular, in the case of a collinear
alignment of $\mathbf{M,}$ the Josephson coupling in the $S/F$ structures
with thick $F$ layers disappears, whereas it remains finite {\ }in the $%
S_{m}IS_{m}$ system.
\begin{figure}[h]
\includegraphics[scale=0.35]{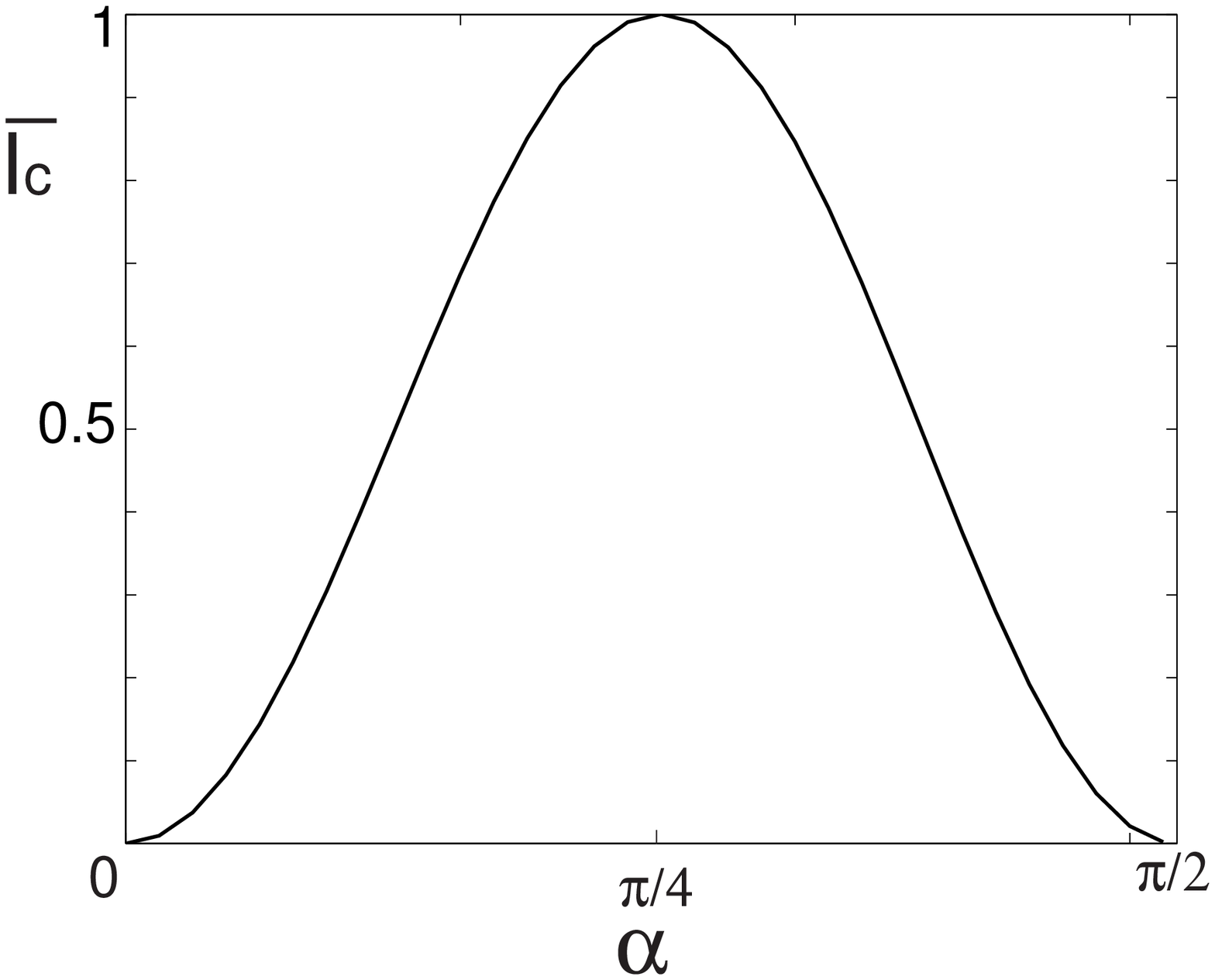}
\caption{ FIG. 16: Dependence of the critical current on the angle $\protect%
\alpha $. The value of the current is given in arbitrary units (from \citet*{%
BVE_manifestation}). }
\label{Fig.17}
\end{figure}

FIG. \ref{Fig.17} shows the dependence of the Josephson current $I_{c}$ on
the angle $\alpha $ given by Eq. (\ref{jt_current}). If the mutual {\ }%
orientation of $\mathbf{M}$ is parallel ($\alpha =0$) or antiparallel ($%
2\alpha =\pi $) the amplitude of the triplet component is zero and therefore
there is no coupling between the neighboring $S$ layers, i.e. $I_{c}=0$. For
any other angles between the magnetizations the amplitude of the triplet
component is finite and this leads to a non-zero critical current. At $%
2\alpha =\pi /2$ ( perpendicular orientation of $\mathbf{M}$) the Josephson
current $I_{c}$ reaches its maximum value.

\begin{figure}[h]
\includegraphics[scale=0.3]{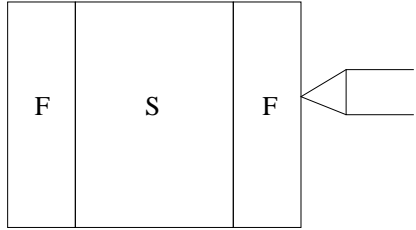}
\caption{Schematic: Measurement of the change of the density of states at
the outer F interface by tunnelling spectroscopy. \citet{kontos01} performed
such experiments on S/F structures. }
\label{Fig.18}
\end{figure}

Another possible experiment for detecting the long range triplet component
is the measurement of the density of states in the $F/S/F$ system as it is
shown in FIG. \ref{Fig.18}. \citet{kontos01} determined the spatial changes
of the DOS in a $PdNi/Al$ structure with the help of planar tunnelling
spectroscopy. This method could also {\ }be used in order to detect the
LRTC. It is clear that if the thickness of the $F$ layer in FIG. \ref{Fig.18}
is larger than the penetration of the short-range components, then any
change of the DOS at the outer boundary of the $F$ layer may occur only due
to the long range penetration of the {\ }triplet component. If the
magnetizations of both $F$ layer are collinear no effect is expected to {\ }%
be observed, while for a non-collinear magnetization a change of the DOS
should be seen.

\subsection{Enhancement of the critical Josephson current}

\label{JOS-ENH} 
Another interesting effect in the $S/F$ structures that we would like to
discuss is the enhancement of the Josephson critical current by the exchange
field. The common wisdom is that any exchange field should reduce or destroy
the singlet superconductivity. In the previous sections we have seen that
this is not always so and the superconductivity can survive in the presence
of a strong exchange field. But still, it is not so simple to imagine that
the superconducting properties can be enhanced by the exchange field.

\begin{figure}[h]
\includegraphics[scale=0.4]{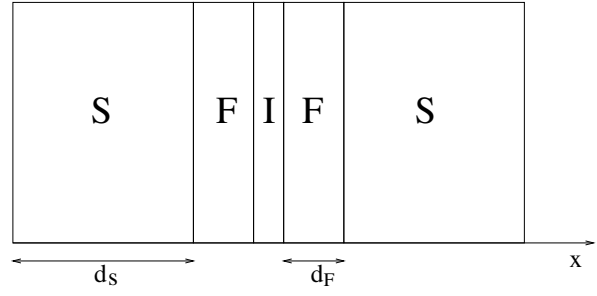}
\caption{The S/F/I/F/S system. I is an insulating thin layer.The relative
magnetization of the F layers can be switched. }
\label{Fig.19}
\end{figure}

Surprisingly, this possibility exists and we will demonstrate now how this
unusual phenomenon occurs. Although the LRTC is not essential to get the
critical current enhancement, the short-range triplet component arises in
this case and it plays a certain role in this effect. The enhancement of the
Josephson current in the $S/F/I/F/S$\ tunnel structures ($I$ stands for an
insulating layer, see FIG. \ref{Fig.19}) was predicted by \citet{BVE1} and
further considered in a subsequent work by \citet{golubov_sfifs}. As we will
see below, if the temperature is low enough and the $S/F$ interface
transparency is good, one can expect an enhancement of the critical current
with increasing the exchange field provided the magnetizations of the $F$
layers are antiparallel to each other.
This surprising result can be obtained in a quite simple way in the limit
when {\ }the thicknesses $d_{S}$ and $d_{F}$ of the $S$ and $F$ layers are
smaller than the superconducting coherence length $\xi _{S}\sim \sqrt{D/2\pi
T_{c}}$ and the penetration length of the condensate into the ferromagnet $%
\xi _{F}\sim \sqrt{D/h}$, respectively. In this case one can assume that the
quasiclassical Green's functions does not depend on the space coordinates
and, in particular, the superconducting order parameter $\Delta $ is a
constant in space. Moreover, instead of considering the dependence of the
exchange field $h$ on the coordinates one can replace it by a homogeneous
effective exchange field $h_{eff}$ with a reduced value. Therefore we use in
our calculations effective fields $\Delta _{eff}$ and $h_{eff}$ defined as
\begin{eqnarray}
\Delta _{eff}/\Delta &=&\nu _{S}d_{S}\left( \nu _{S}d_{S}+\nu
_{F}d_{F}\right) ^{-1}\;, \\
h_{eff}/h &=&\nu _{F}d_{F}\left( \nu _{S}d_{S}+\nu _{F}d_{F}\right) ^{-1}\;,
\end{eqnarray}
where $\nu _{S}$ and $\nu _{F}$ are the densities of states in the
superconductor and ferromagnet, respectively.

With this simplification, {\ }we can write the Gor'kov equations for the
normal and anomalous Green's functions in the spin-space as
\begin{eqnarray}
\left( i\omega +\xi -\mathbf{\sigma h}\right) \hat{G}_{\omega }+\hat{\Delta}%
\hat{F}_{\omega }^{+} &=&1  \label{4.B.3} \\
\left( -i\omega +\xi -\mathbf{\sigma h}\right) \hat{F}_{\omega }+\hat{\Delta}%
\hat{G}_{\omega } &=&0\;,  \label{4.B.4}
\end{eqnarray}
where $\mathbf{\sigma }=(\mbox{$\hat{\sigma}_1$},\mbox{$\hat{\sigma}_2$},%
\mbox{$\hat{\sigma}_3$})$ are the Pauli matrices and $\xi =\epsilon \left(
\mathbf{p}\right) -\epsilon _{F},$ $\varepsilon _{F}$ is the Fermi energy, $%
\varepsilon \left( \mathbf{p}\right) $ is the spectrum, and $\omega =\left(
2n+1\right) \pi T$ are Matsubara frequencies. (We omit the subscript $eff$ \
in Eqs. (\ref{4.B.3}-\ref{4.B.4}) and below).

In order to calculate the Josephson current $I_{J}$\ through the tunnel
junction represented in FIG. \ref{Fig.19} we use the well known standard
formula
\begin{equation}
I_{J}=\left( 2\pi T/eR\right) \mathrm{Tr}\sum_{n}\hat{f}_{\omega }(h_{1})%
\hat{f}_{\omega }(h_{2})\sin \varphi \;,  \label{je_a5}
\end{equation}
where
\begin{equation}
\hat{f}_{\omega }=\frac{i}{\pi }\int {\hat{F}_{\omega }\mathrm{d}\xi }
\label{je_a6}
\end{equation}
is the quasiclassical anomalous Green's function, $\varphi $ is the phase
difference between both the superconductors, $R$ is the barrier resistance
and $h_{1,2}$ are the exchange fields of the left and the right $F$-layers.

The only difference between Eqs. (\ref{je_a5}, \ref{je_a6}) and the
corresponding equations in the absence of the exchange field is the
dependence of the condensate function $\hat{f}_{\omega }$\ on $h$. This
dependence can be found immediately from Eqs. (\ref{4.B.3}, \ref{4.B.4}).
\begin{equation}
\hat{f}_{\omega }=\hat{\Delta}\left( \left( \omega +i\mathbf{\sigma h}%
\right) ^{2}+\Delta ^{2}\right) ^{-1/2}  \label{je_a61}
\end{equation}
What remains to be done is to insert the condensate function $\hat{f}$\ into
Eq. (\ref{je_a5}) for certain exchange fields $h_{1}$\ and $h_{2}$\ and to
calculate the sum over the Matsubara frequencies $\omega $. Although it is
possible to carry out these calculations for arbitrary vectors $h_{1}$ and $%
h_{2},$\ we restrict our consideration by the cases when the absolute values
the magnetizations $h_{1}$\ and $h_{2}$\ are equal but the magnetizations
are either parallel or antiparallel to each other. This simplifies the
computation of the Josephson current but, at the same time, captures the
essential physics of the phenomenon.

Using Eqs. (\ref{je_a5}-\ref{je_a61}) and assuming first that $h_{1}$\ and $%
h_{2}$\ are parallel to each other we write the expression for the critical
current as \cite{BVE1}
\begin{equation}
I_{J}=I_{c}\sin \varphi  \label{e10}
\end{equation}
\begin{equation}
I_{c}^{\left( p\right) }\!\!=\frac{\Delta ^{2}\left( T\right) 4\pi T}{eR}%
\sum_{\omega }\frac{\omega ^{2}+\Delta ^{2}\left( T,h\right) -h^{2}}{\left(
\omega ^{2}+\Delta ^{2}\left( T,h\right) -h^{2}\right) ^{2}+4\omega ^{2}h^{2}%
},  \label{je_a7}
\end{equation}
The corresponding equation for the antiparallel configuration is different
from Eq. (\ref{je_a7}) and can be written as
\begin{equation}
I_{c}^{\left( a\right) }\!\!=\frac{\Delta ^{2}\left( T\right) 4\pi T}{eR}%
\sum_{\omega }\frac{1}{\sqrt{\left( \omega ^{2}+\Delta ^{2}\left( T,h\right)
-h^{2}\right) ^{2}+4\omega ^{2}h^{2}}}.  \label{je_a8}
\end{equation}
One can easily check that the critical current $I_{c}^{\left( p\right) }$\
for the parallel configuration, Eq. (\ref{je_a7}), is always smaller than
the current $I_{c}^{\left( a\right) }$ for the antiparallel case. These two
expressions are equal to each other only in the absence of any magnetization.

\begin{figure}[h]
\includegraphics[scale=0.4]{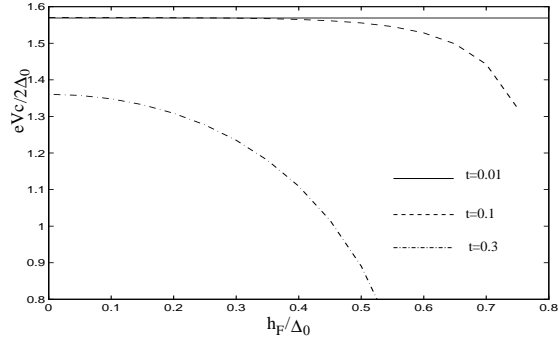}
\caption{Dependence of the normalized critical current on $h$ for different
temperatures in the case of a parallel orientation. Here $eV_{c}=eRI_{c}$, $%
h_{F}$ is the effective exchange field, $t=T/\Delta_{0}$ and $\Delta_{0}$ is
the superconducting order parameter at $T=0$ and $h=0$ }
\label{Fig.20}
\end{figure}

\begin{figure}[h]
\includegraphics[scale=0.35]{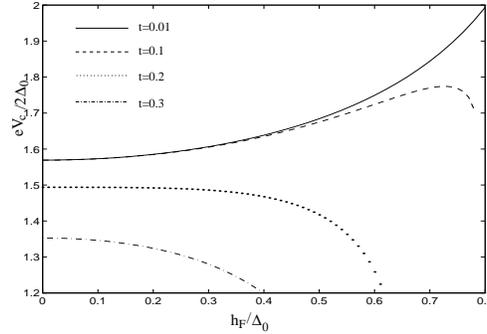}
\caption{The same dependence as in FIG. \ref{Fig.20} in the case of an
antiparallel orientation. }
\label{Fig.21}
\end{figure}

In FIGS. \ref{Fig.20} and \ref{Fig.21} we represent the dependence of the
critical current on the strength of the exchange field. {\ }We see from FIG.
\ref{Fig.20} that for the parallel configuration the exchange field reduces
the value of the Josephson current and this is exactly what one could
expect. At the same time, the critical current grows with the exchange field
for the antiparallel configuration at low temperatures, which is a new
intriguing effect (see FIG. \ref{Fig.21}).

This unexpected result can be understood from Eq. (\ref{je_a8}) rather
easily without making numerics. In the limit $T\rightarrow 0$ the sum over
the {\ }Matsubara frequencies can be replaced by an integral and one can
take for the superconducting order parameter $\Delta $ the values $\Delta
=\Delta _{0}$ if $h<\Delta _{0},$ and $\Delta =0$ if $h>\Delta _{0}$, where $%
\Delta _{0}$ is the BCS order parameter in the absence of an exchange field
(see, e.g. \citet{lo_state}).

Inserting this solution into Eq. (\ref{je_a8}) one can see that the
Josephson critical current $I_{c}^{\left( a\right) }$ grows with increasing
exchange field. Moreover, formally it diverges logarithmical when $%
h\rightarrow \Delta _{0}$%
\begin{equation}
I_{c}^{\left( a\right) }\left( h\rightarrow \Delta _{0}\right) \simeq \frac{%
I_{c}\left( 0\right) }{\pi }\ln \left( \Delta _{0}/\omega _{0}\right) \,,
\label{je_a12}
\end{equation}
where $I_{c}\left( 0\right) $ is the critical current in the absence of the
magnetic moment at $T=0$, and $\omega _{0}$ is a parameter needed to cut the
logarithm at low energies.

When deriving Eqs. (\ref{je_a7}, \ref{je_a8}) the conventional singlet
superconducting pairing was assumed. The electrons of a Cooper pair have the
opposite spins. This picture of a superconducting pairs with the opposite
spins of the electrons helps in the understanding of the effect.

If the magnetic moments in both the magnetic layers are parallel to each
other, they serve as an obstacle for the Cooper pair because the pairs
located in the region of the ferromagnet demand more energy. This leads to a
reduction of the Josephson current. However, if the exchange fields of the
different layers are antiparallel, they may favor the location of the Cooper
pairs in the vicinity of the Josephson junction. A certain probability
exists that one of the electrons of the pair is located in one layer,
whereas the other is in the second layer. Such a possibility is
energetically favorable because the spins of the electrons of the pair can
now have the same direction as the exchange fields of the layers. Then it is
more probable for the pairs to be near the junction even in comparison with
a junction without exchange fields and, as a result, the critical current
may increase.

The results presented above have been obtained for the $SF/I/FS$ structure
by \citet{BVE1}. Earlier a formula for the Josephson critical current in the
$S_{m}IS_{m}$ {(}$S_{m}$ {\ }is the magnetic superconductor) junction was
presented by \citet{kulic_kulic}. From that formulae one could, in
principle, derive an enhancement of the critical current for the
antiparallel $M$\ orientation in magnetic superconductors $S_{m}$.
Unfortunately, the authors seem to have missed this interesting effect.

Some remarks should be made at this point:

1) The results presented above are valid in the tunnelling regime, i.e. when
the transparency of the insulating barrier $I$ is low enough. %
\citet{golubov_sfifs} have shown that a smearing of the singularity of $%
I_{c}^{(a)}$ is provided by a finite temperature or a not very low barrier
transparency. The maximum of the critical current for the antiparallel
configuration $I_{c}^{(a)}$ decreases with decreasing resistance of the $I$
layer. The effect becomes weaker as the thickness\ of the $F$\ layer grows.

2) We assumed that the $S/F$ interface was perfect. In a structure with a
large $S/F$ interface resistance $R_{S/F}$ the bulk properties of the $S$
film are not considerably influenced by the proximity of the $F$ film (to be
more precise, the condition $R_{S/F}>(\nu _{F}d_{F}/\nu _{S}d_{S})\rho
_{F}\xi _{F}$ must be satisfied, where $\rho _{F}$ is the specific
resistance of the $F$ film). Then, as one can readily show (see section \ref
{PRO-SF}), a minigap $\epsilon _{bF}=\left( D\rho \right) _{F}/\left(
2R_{S/F}d_{F}\right) $ arises in the $F$ layer. The Green's functions in the
$F$ layers have the same form as before with $\Delta $ replaced by $\epsilon
_{bF}$. The singularity in $I_{c}(h)$ first occurs at $h$ equal to $\epsilon
_{bF}$.

A physical explanation for the singular behavior of the critical current $%
I_{c}^{(a)}$ was given by \citet{golubov_sfifs} These authors noticed that
the density of states in the $F$ layer has a singularity when $h=\epsilon
_{bF}$. At this value of $h$ the maximum of $I_{c}^{(a)}$ is achieved due to
an overlap of two $\epsilon ^{-1/2}$ singularities. This leads to the
logarithmic divergency of the critical current in the limit $T\rightarrow 0$
in analogy with the well known Riedel peak in $SIS$ tunnel junctions for the
voltage difference $2\Delta $. In the latter case the shift of the energy is
due to the electric potential.

\citet{golubov_sfifs} have also {\ }shown that for the parallel
configuration, at $h=\epsilon _{bF}$ the critical current changes its signs,
i.e. there is a transition from $0$ to a $\pi $ junction. Similar results
were obtained by \citet*{krivoruchko1, krivoruchko2}. The case of an
arbitrary $S/F$ transparency was also studied by %
\citet{li,chtchelkatchev,barash}. In the paper by \citet{barash} the authors
calculated the Josephson current as a function of the angle between the
magnetizations in the $F$ film.

\section{Reduction of the Magnetization due to Superconductivity: Inverse
Proximity Effect}

\label{RED} 
Until now we have been studying the superconducting properties of different $%
S/F$ structures for a fixed magnetization. This means that we assumed a
certain value for this quantity and its dependence on coordinates. The
implied justification of this assumption was that the ferromagnetism is a
stronger phenomenon than the superconductivity and the magnetic moment of
conventional ferromagnets can hardly be affected by the superconductivity.

This assumption is certainly correct in many cases but not always. Often the
presence of the superconductivity can drastically change magnetic properties
of the ferromagnets even if they are strong.

Experiments performed by \citet{muehge} and \citet{garifullin} showed that
the total magnetization of certain $S/F$ bilayers with strong ferromagnets
decreased with lowering the temperature below the critical superconducting
transition temperature $T_{c}$. As an explanation, it was suggested that due
to the proximity effect domains with different magnetization appeared in the
magnetic materials and this could reduce the total magnetization. At the
same time, quantitative estimates based on an existing theory \cite
{buzdin_crypto} led to a conclusion that this mechanism was not very
probable.

In this Chapter we address the problem of the reduction of the magnetic
moment by the presence of a superconductor assuming again that, in the
absence of the ferromagnet, we would have the conventional singlet
superconducting pairing. It turns out that two different and independent
mechanisms that lead to a decrease of the magnetization in $S/F$
heterostructures due to the proximity effect exist and we give a detailed
account of them.

In order to study the magnetic properties we have to choose a model. One can
distinguish two different types of the ferromagnetism: a) itinerant
ferromagnetism due to the spin ordering of free electrons and b)
ferromagnetism caused by localized spins. Most of ferromagnetic metals show
both of the types of ferromagnetism simultaneously, \textit{i.e.} their
magnetization consists of both contributions.

We consider a model in which the conducting electrons interact
with the localized moments via an effective exchange interaction.
The corresponding term in the Hamiltonian is taken in {\ }the form
(see Appendix \ref{ApA}):
\begin{equation}
-\int d^{3}r\psi ^{\dagger }(\mathbf{r})_{\alpha }\left( J\mathbf{S(r)}%
\mathbf{\sigma })\right) _{\alpha \beta }\psi (\mathbf{r})_{\beta }\;.
\label{inv_exchange}
\end{equation}
This term is suitable to describe $s-d$ or $s-f$ interaction between the $s$
and localized $d$ and $f$ electrons. We also consider the ferromagnetic
interaction between the localized moments. This interaction can be very
complicated and to determine it, one should know the detailed band structure
of the metal as well as different parameters. {\ }However, all these details
are not important for us and we write the interaction between the localized
spins {\ }phenomenologically as
\begin{equation}
-\sum_{ij}\mathcal{J}_{ij}\mathbf{S}_{i}\mathbf{S}_{j}\;.
\label{inv_loc_mom}
\end{equation}
It is assumed that $\mathcal{J}$ is positive. This interaction, Eq. (\ref
{inv_loc_mom}) {, }is responsible for the ferromagnetic alignment of the
localized moments and is known as the Heisenberg Hamiltonian.

So {,} we consider a metallic ferromagnet in which the conduction electrons
interact with localized magnetic moments. The ferromagnetic interaction (\ref
{inv_loc_mom}) assures a finite magnetic moment of the background. The total
magnetization is the sum of the background magnetization (localized moments)
and the magnetization of the polarized free electrons.

In the next two sections we discuss the {\ }two different mechanisms that
lead to a decrease of the magnetization at low temperatures. In Section \ref
{RED-CRY} we consider a possibility of changing the magnetic order of the
localized magnetic spins in a $F$ film deposited on top of a bulk
superconductor. The contribution from free electrons to the magnetization is
first assumed to be small. We will see that for not too strong ferromagnetic
coupling $\mathcal{J}$ the proximity effect may lead to an inhomogeneous
magnetic state. Contrary to this case, we consider in Section \ref{RED-INV}
an itinerant ferromagnet in which the main contribution to the magnetization
is due to free electrons. We will show that the magnetization of free
electrons may decrease at low temperatures {\ }due to a some kind of spin
screening. Thus, both effects may lead to the decrease in the magnetization
observed in experiments \cite{muehge,garifullin}.

\subsection{Cryptoferromagnetic state}

\label{RED-CRY} 
In \citeyear{anderson} \citeauthor*{anderson} suggested that
superconductivity could coexist with a nonhomogeneous magnetic order in some
type of materials. \citeauthor*{anderson} called this state \textit{%
cryptoferromagnetic } state.

The reason for this coexistence is that, if the magnetization direction
varies over a scale smaller than the superconducting coherence length, the
superconductivity may survive despite the ferromagnetic background. This is
due to the fact that the superconductivity is sensitive to the ferromagnetic
moment averaged on the scale of the size of Cooper pairs rather than to its
local values.

In 1988 \citeauthor{buzdin_crypto} discussed properties of a bilayer system
consisting of a conventional superconductor in contact with a ferromagnet.
They have shown that the magnetic ordering in the magnet might take the form
of a structure consisting of small size domains, such that the
superconductivity is not destroyed. Of course, as follows from Eq. (\ref
{inv_loc_mom}), the formation of a domain-like structure costs a magnetic
energy but this is compensated by the energy of the superconductor that
would have been lost if the magnetic order remained ferromagnetic.

This is only possible if the stiffness of the magnetic order parameter ($%
\mathcal{J}$) is not too large. For instance this nonhomogeneous
magnetization occurs in magnetic superconductors as those studied by %
\citet{bula_adv}. But can one see it in the heterostructures containing
strong ferromagnets like $Fe$\ or $Ni$\ in contact with conventional
superconductors?

At first glance, it seems impossible, since the Curie temperature of, for
example, iron is hundred times or more larger than the critical temperature
of a conventional superconductor. Therefore any change of the ferromagnetic
order look much less favorable energetically than the destruction of the
superconductivity in the vicinity of the $S/F$ interface.

This simple argument was however questioned in the experiments performed by %
\citet{muehge} on $Fe/Nb$ bilayers and by \citet{garifullin} on $%
V/Pd_{1-x}Fe_{x}$ structures. Direct measurements of the ferromagnetic
resonance has shown that in several samples with thin ferromagnetic layers
the average magnetic moment started to decrease below the superconducting
transition temperature $T_{c}$.

Of course, one can reduce the influence of the ferromagnet on the
superconductor by diminishing the thickness of the ferromagnet. Using the
formulae obtained by \citet{buzdin_crypto}, \citet{muehge} estimated the
thickness of the ferromagnet for which the superconductivity was still
possible and got a value of the order of $1$\AA , which created a doubt on
the explanation of the experiment in this way.

At the same time, the use of the formulae derived by %
\citeauthor{buzdin_crypto} was not really justified because the calculations
were done for thick but weak ferromagnets assuming a strong anisotropy of
the ferromagnet that was necessary for a formation of the domain walls with
the magnetization vector changing its sign but not its axis.

\citet*{BEL} investigated theoretically the possibility of a
cryptoferromagnetic-like (CF) state in $S/F$ bilayers with parameters
corresponding to the experiments by \citeauthor{muehge} and %
\citeauthor{garifullin}. In that work a CF state with a magnetic moment that
rotates in space was considered. This corresponds to a weak anisotropy of
the ferromagnet, which was the case in the samples studied in \citet{muehge}%
. In particular, \citet{BEL} studied a phase transition between the CF and
the ferromagnetic (FM) phases. The calculations were carried out in the
limit $d_F\ll \xi _{h}=v_{0}/h,$ $T_{c}\ll h\ll \epsilon _{0}$, $v_{0}$ and $%
\epsilon _{0}$ are the Fermi velocity and Fermi energy, respectively. This
limit is consistent with the parameters of the experiment of \citet{muehge}, %
\citet{garifullin}. We present here the main ideas of this work.

\begin{figure}[h]
\includegraphics[scale=0.4]{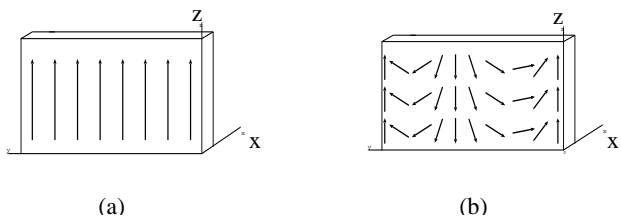}
\caption{A S/F bilayer consisting of a thin ferromagnet attached to a bulk
superconductor. The ferromagnet may be either in the (a) ferromagnetic or
the (b) cryptoferromagnetic phase. }
\label{Fig.22}
\end{figure}

The Hamiltonian describing the bilayer structure in FIG. \ref{Fig.22} can be
written as
\begin{equation}
H\left( \gamma \right) =H_{0}+H_{BCS}-\gamma \int d\mathbf{r}\Psi
_{\alpha }^{+}(\mathbf{r})\left[ \mathbf{h}(\mathbf{r})\mathbf{\sigma }%
\right] _{\alpha \beta }\Psi _{\beta }(\mathbf{r})+H_{M}\;,
\label{crypto_hamiltonian}
\end{equation}
where the integration must be taken in the region $-d<x<0$. Here
$H_{0}$ is the one-particle electron energy (including an
interaction with impurities), $H_{BCS}$ is the usual term
describing the conventional BCS superconductivity in the
superconductor $S$ and the third term describes
the interaction between localized moments and conduction electrons, where $%
\gamma $ is a constant that will be put to $1$ at the end (see
Appendix \ref{ApA}).

The term $H_{M}$ describes the interaction between the localized moments in
the ferromagnet (\textit{cf.} Eq.(\ref{inv_loc_mom})). We assume that the\
magnetization of the localized spins is described by classical vectors and
take into account the interaction between neighboring spins only. In the
limit of slow variations of the magnetic moment in space with account of Eq.
(\ref{inv_loc_mom}), the Hamiltonian $H_{M}$ can be written in the form
\begin{equation}
H_{M}=\int \mathcal{J}\left[ \left( \mathbf{\nabla }S_{x}\right) ^{2}+\left(
\mathbf{\nabla }S_{y}\right) ^{2}+\left( \mathbf{\nabla }S_{z}\right) ^{2}%
\right] dV,  \label{cryp_exchange-energy}
\end{equation}
where the magnetic stiffness $\mathcal{J}$ characterizes the strength of the
coupling between the localized moments in the $F$ layer and the $S_{i}$ are
the components of a unit vector that are parallel to the local direction of
the magnetization.

We assume that the magnetic moments are directed parallel to the $S/F$
interface and write the spin vector $\mathbf{S}$ as $\mathbf{S}=\left(
0,-\sin \theta ,\cos \theta \right) $. A perpendicular component of the
magnetization would induce strong Meissner currents in the superconductor,
which would require a greater additional energy.

The condition for an extremum of the energy $H_{M}$, Eq. (\ref
{cryp_exchange-energy}) can be written as
\begin{equation}
\Delta \theta =0  \label{e12}
\end{equation}
Solutions of Eq. (\ref{e12}) can be written in the form $\theta =Qy$, where $%
Q$\ is the wave vector characterizing the rotation in space (see FIG. \ref
{Fig.22}). The value $Q=0$\ corresponds to the ferromagnetic state.

What we want to do now is to compare the energies of the ferromagnetic and
cryptoferromagnetic states. The latter will be considered for the case with
a rotating in space magnetic moment $\theta =Qy$. This should be
energetically more favorable than the domain-like structure one provided the
magnetic anisotropy of $F$\ is low. Such a CF state corresponds to a so
called Neel wall (see for example \citet{aharoni}).

Strictly speaking, one has to take into account also a magnetostatic energy\
due to a purely magnetic interaction of the magnetic moments. However, if
the condition{\
\begin{equation}
\frac{\mathcal{J}}{M_{s}^{2}}\gg d^{2}  \label{condition1}
\end{equation}
where $M_{s}$ is the magnetic moment per volume, is fulfilled one can
neglect its contribution with respect to the one of the exchange energy \cite
{aharoni}. }

{\ }Taking typical values of the parameters for $Fe$: $M_{s}=800$emu/cm$^{3}$
and $J=2.10^{-6}$erg/cm one can see that the condition (\ref{condition1})
requires that the thickness $d$ of the ferromagnet is smaller than $10nm$,
which corresponds to comparatively thick layers. Throughout this section
this condition is assumed to be fulfilled.

In this case the magnetic energy $\Omega _{M}$ (per unit surface area) is
given by the simple expression
\begin{equation}
\Omega _{M}=JdQ^{2}\;.  \label{crypto_m}
\end{equation}

In order to calculate the superconducting energy $\Omega _{S}$ one has to
take into account the fact that the order parameter should be destroyed, at
least partially, near the contact with the ferromagnet. This means that the
order parameter $\Delta $\ is a function of the coordinate $x$\
perpendicular to the interface. As we want to minimize the energy we should
look for a non-homogeneous solution for $\Delta \left( x\right) $\ of
non-linear equations describing the superconductivity. Near the critical
temperature $T_{c}$\ one can use Ginzburg-Landau equations. The proper
solution of these equations can be written in the form
\begin{equation}
\Delta \left( x\right) =\Delta _{0}\tanh \left( \frac{x}{\sqrt{2}\xi
_{GL}\left( T\right) }+C\right)  \label{e13}
\end{equation}
where $\Delta _{0}$\ the value of the order parameter in the bulk, and $\xi
_{GL}$\ is the correlation length of the superconductor defined in Eq. (\ref
{e1a}). Near $T_c$ this length can be much larger than the length $\xi _{S}$%
. The parameter $C$\ in Eq. (\ref{e13}) is a number that has to be found
from boundary conditions.

The solution for $\Delta \left( x\right) $, Eq. (\ref{e13}) is applicable at
distances exceeding the length $\xi _{S}$\ and therefore we cannot use it
near the interface.

Having fixed the constant $C$\ one can compute the decrease of the
superconducting energy due to the suppression of superconductivity in the $S$
layer using the Ginzburg-Landau free energy functional %
\citep[e.g.][]{degennes_book}. The decrease of the superconducting energy $%
\Omega _{S}$ per unit area at the $F/S$ interface is a function of $C$ and
can be written as
\begin{equation}
\Omega _{S}=\frac{\sqrt{\pi }}{6\sqrt{2}}|\tau |^{3/2}\left( 2+K\right)
(1-K)^{2}\;,  \label{crypto_s}
\end{equation}
where $K=\tanh C$, and $\tau =(T-T_{c})/T_{c}.$

It remains only to determine the contribution from the third term of the
Hamiltonian (\ref{crypto_hamiltonian}). The corresponding free energy $%
\Omega _{M/S}$ is given by the expression:
\begin{equation}
\Omega _{M/S}=-i\pi T\nu _{0}\frac{\mathrm{Tr}}{2}\sum_{\omega
}\int_{0}^{1}d\gamma \int d^{3}\mathbf{r}(\mathbf{h\sigma })\left\langle
\hat{g}\right\rangle _{0}\;,  \label{crypto_ms}
\end{equation}
where $\nu _{0}$ is the density of states and $\left\langle \hat{g}%
\right\rangle _{0}$ is the quasiclassical Green's function
averaged over all directions of the Fermi velocity.

Since the exchange field $h$ in a strong ferromagnet may be much higher than
the value of {\ }$\tau ^{-1}$ (here $\tau $ is the momentum relaxation
time), one has to solve the Eilenberger equation in the $F$ region and the
Usadel equation in the $S$ region. Solutions for these equations in both the
superconductor and ferromagnet were obtained by \citet{BEL}.

Thus, the total energy is given by $\Omega =\Omega _{M}+\Omega _{S}+\Omega
_{M/S}$, Eqs. (\ref{crypto_m}, \ref{crypto_s}, \ref{crypto_ms}). As a
result, one can express the free energy as a function of two unknown
parameters, $K$ and $Q$. One can find these parameters from the condition
that the free energy must be minimal, which leads to the equations
\begin{equation}
\partial {\ \Omega }/\partial K=\partial {\Omega }/\partial Q=0  \label{e15}
\end{equation}
\begin{figure}[h]
\includegraphics[scale=0.3]{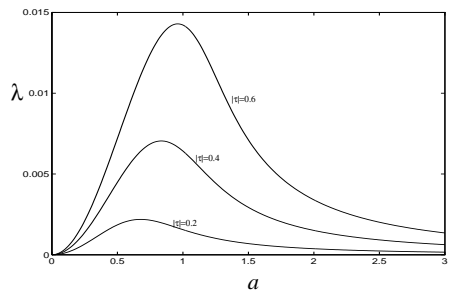}
\caption{Phase diagrams ($\protect\lambda ,a$) for different values of $|%
\protect\tau|=\frac{T_c-T}{T_c}$. The area above (below) the curves
corresponds to the F (CF) state. }
\label{Fig.23}
\end{figure}

One can show that the CF-F transition is of second order, which means that
near the transition the parameter $Q$\ is small. At the transition it
vanishes and this gives an equation binding the parameters. Solving the
equation numerically we come to the phase diagram of FIG. \ref{Fig.23}
determining the boundary between the ferromagnetic and cryptoferromagnetic
states. The parameters $a$ and $\lambda $ used in FIG. \ref{Fig.23} are
defined as
\begin{equation}
a^{2}\equiv \frac{2h^{2}d_{f}^{2}}{DT_{c}\eta ^{2}},\;\;\;\lambda \equiv
\frac{\mathcal{J}d_{F}}{\nu _{F}\sqrt{2T_{c}D^{3}}}\frac{7\zeta (3)}{2\pi
^{2}},  \label{e16}
\end{equation}
where $\eta $ is the ratio between the Fermi velocities $v_{0}^{F}/v_{0}^{S}$%
. It is clear from Eqs. {\ }(\ref{e16}) that the parameter $a$ is related to
the exchange energy $h$, while $\lambda $ is the related to the magnetic
stiffness $\mathcal{J}$.

The conclusion that the phase transition between F and CF states
should be of the second order was drawn neglecting the
magnetostatic interaction. {The direct magnetic interaction can
change this transition to a first order one \cite{buzdinPC}.
However,} in the limit of Eq. (\ref {condition1}), this first
order transition will be inevitably close to the second order one.
Such a modification of the type of the phase transition is out of
the focus of this review.

Let us make estimates for the materials used in the experiments. Performing
ferromagnetic resonance measurements, \citet{muehge} have observed a
decrease of the effective magnetization of a $Nb/Fe$ bilayer. The stiffness $%
\mathcal{J}$ for materials like $Fe$ and $Ni$ is $\approx 60K/$\AA . The
parameters characterizing $Nb$ can be estimated as follows: $T_{c}\!=\!10K$
, $v_{F}\!\cong \!10^{8}$cm/s, and $l\!\cong \!100$\AA . The thickness of
the magnetic layer is of order $d\!=\!10$\AA , and the exchange field $%
h\!\cong \!10^{4}K$ which is proper for iron.

Assuming that the Fermi velocities and energies of the ferromagnet and
superconductor are close to each other, we obtain $a\approx 25$ and $\lambda
\sim 6.10^{-3}$. It is clear from FIG. \ref{Fig.23} that the
cryptoferromagnetic state is hardly possible in the $Fe/Nb$ samples used in
the experiment \citet{muehge}.

However, one can in principle explain the observed, decrease of the
magnetization taking a closer look at the structure of the $S/F$ interface.
In the samples analyzed by \citeauthor{muehge} the interface between the $Nb
$ and $Fe$ layers is rather rough. So, one can expect that in the magnetic
layers there were ``islands'' with smaller values of $\mathcal{J}$ and/or $h$%
. A reduction of these parameters in the $Fe/Nb$ bilayers is not unrealistic
because of the formation of non-magnetic ``dead'' layers that can also
affect the parameters of the ferromagnetic layers. If the
cryptoferromagnetic state were realized only on the islands, the average
magnetic moment would be reduced but would remain finite. Such a conclusion
correlates with what one observes experimentally. One can also imagine
islands very weakly connected to the rest of the layer, which would lead to
smaller energies of a non-homogeneous state.

Let us now consider the experiment by \citet{garifullin} on $%
Pd_{0.97}Fe_{0.03}/V$. Due to the low concentration of iron, the magnetic
stiffness and the exchange field of the $F$-layers is much lower than the
one in the case of a pure iron. For this system, one estimates the
parameters as(see \citet{garifullin}) $J\sim 60K/nm$, $h\sim 100K$. Assuming
again that the Fermi velocities of $V$ and $Pd_{1-x}Fe_{x}$ are close to
each other, \citet{garifullin} obtained for the sample with $d_{F}=1.2$nm
the following values of the parameters $a\sim 1.2$ and $\lambda \sim
1.3.10^{-3}$.

Using these values for $a$ and $\lambda $ one can see from the phase diagram
in FIG. \ref{Fig.23} that {\ }there can be a transition from the F to the CF
state at $|\tau |\sim 0.2$, which corresponds to $T\sim 2.4K$. The decrease
of the effective magnetization $M_{eff}$ with decreasing temperature was not
observed in samples with larger $F$ thickness $d_{F}$: $M_{eff}$ was a
temperature-independent constant for the sample with $d_{F}=4.4nm$ and $%
d_{S}=37.2nm$. In the sample with $d_{F}=1.2nm$ and $d_{S}=40nm$ the
effective magnetization $M_{eff}$ decreased by $\approx 50\%$ with cooling
from $T\approx 4K$ to $T\approx 1.5K$. This fact is again in accordance with
the predictions of \citet{BEL}.

The results of this section demonstrate that not only ferromagnets change
superconducting properties but also superconductivity can affect
ferromagnetism. This result is valid, in particular, for strong
ferromagnets, although the thickness of the ferromagnetic layers must be
small in this case.

The exchange interaction between the superconducting condensate and the
magnetic order parameter reduces the energy of the system if the direction
of the magnetization\ vector $M_{F}$\ is not constant in space but
oscillates. Provided the energy of the anisotropy is small, this interaction
leads to the formation of a spiral magnetic structure in the $F$\ film.

As we will see in the next section the appearance of the CF-state is not the
only effect that leads to a reduction of the effective magnetization in S/F
structures. We will show that the proximity effect may also lead to a change
of the absolute value of the magnetic moment $M_{F}$\ in the ferromagnet and
to an induced magnetization $M_{S}$ in the superconductor.

\subsection{Ferromagnetism induced in a superconductor}

\label{RED-INV} 
In the previous section we have seen that the superconductivity can affect
the magnetic ordering changing the orientation of magnetic moments in the
ferromagnetic film. In this section we want to demonstrate that another
mechanism for a change of the total magnetization of a $S/F$\ system exists.
In contrast to the phenomenon discussed in the previous Section, the
orientation of the magnetic moments in the $F$\ film does not change but the
magnitude of the magnetization both in the $F$\ and $S$\ films does.

This change is related to the contribution of free electrons both in the
ferromagnet ($\delta M_{F}$) and in the superconductor\ ($M_{S}$) to the
total magnetization. On one hand, the DOS in the $F$\ film is reduced due to
the proximity effect and therefore $\delta M_{F}$\ is reduced. On the other
hand, the Cooper pairs in the $S$\ film are polarized in the direction
opposite to $M_{F}$, where $M_{F}$\ is the magnetization of free electrons
in the ferromagnet.

Let us consider first a bulk ferromagnet and derive a relation between the
exchange field and the magnetization of the free electrons. {The exchange
field }$h=JS$ in the ferromagnet can be due to the {\ }localized moments
(see Eq. (\ref{inv_exchange})) or due to the free electrons in the case of
an itinerant ferromagnet\footnote{%
In many papers the exchange ''field'' $h${\ is defined in another
way (}$h=-JS${) so that the energy minimum corresponds to
orientation of the vector }$<\sigma >${\ antiparallel to the
vector }$h.${\ In this case the magnetic moment }$m=-\mu
_{B}<\sigma >${\ is parallel to }$h${. Both definitions lead to
the same results.}} In some ferromagnets both the {\ }localized
and itinerant moments contribute to the magnetization.

The magnetization of the free electrons is given by
\begin{equation}
M=\frac{i}{4}\mu _{B}\int \frac{d\omega }{2\pi }\int \frac{d^{3}p}{(2\pi
)^{3}}\mathrm{Tr}\mbox{$\hat{\tau}_3$}\mbox{$\hat{\sigma}_3$}\left( \check{G}%
^{R}-\check{G}^{A}\right) n_{p}\;,  \label{5.1}
\end{equation}
where $\mu _{B}$ is an effective Bohr magneton and $n_{P}$ {is }the Fermi
distribution function of the free electrons. The {expression} in front of $%
n_{P}$ in Eq. (\ref{5.1}) determines the DOS {that} depends on the exchange
field $\mathbf{h}$. We assume that the magnetization is oriented along the $%
z $-axis.

Using Eq. (\ref{5.1}) one can easily compute the contribution of the free
electrons to the magnetization in a bulk ferromagnet. In the simplest case
of a normal metal with a quadratic energy spectrum we have
\begin{equation}
M_{F}=\frac{\mu _{B}}{(2\pi )^{2}}\int p^{2}dp\left[ n(\xi _{p}-h)-n(\xi
_{p}+h)\right] \;,  \label{5.2}
\end{equation}
where $\xi _{p}=p^{2}/2m-\epsilon _{F}$. At $T=0$ the {\ }magnetization is
given by:
\begin{equation}
M_{F0}=\frac{\mu _{B}}{2(3\pi ^{2})}\left( p_{+}^{3}-p_{-}^{3}\right) \;
\label{5.3}
\end{equation}
where $p_{\pm }=\sqrt{2m(\epsilon _{F}\pm h)}$ are the Fermi momenta for
spin up and spin down electrons. In the quasiclassical limit it is assumed
that $h\ll \epsilon _{F}$, and therefore
\begin{equation}
M_{F0}\cong \mu _{B}\nu h\,,  \label{5.4}
\end{equation}
where $\nu =p_{F0}m/\pi ^{2}$ is the density of states at the Fermi level,
and $p_{F0}=\sqrt{2m\epsilon _{F}}$ is the Fermi momentum in the absence of
the exchange field\footnote{%
Actually Eq.(\ref{5.4}) is valid not only in the case of a quadratic
spectrum but also in a more general case.}. For the temperature range $T\ll
h $ we are interested in, one can assume that the magnetization of the
ferromagnet does not depend on $T$ and is given by Eq. (\ref{5.4}).

\begin{figure}[h]
\includegraphics[scale=0.3]{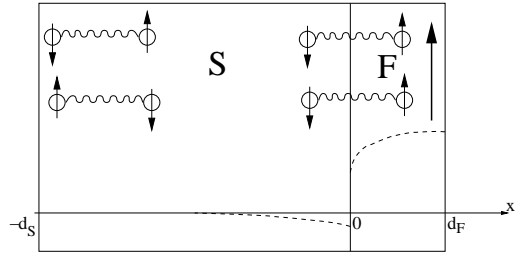}
\caption{S/F structure and schematic representation of the inverse proximity
effect. The dashed curves show the local magnetization. }
\label{Fig.24}
\end{figure}

Now let us consider a $S/F$ system with a thin F layer (see FIG. \ref{Fig.24}%
) and ask a question: Is the magnetization of the itinerant electrons
modified by the proximity effect? We assume that the exchange field of the
ferromagnet $F$ is homogeneous and aligned in the $z$- direction, which is
the simplest situation.

At first glance, it is difficult to expect anything interesting in this
situation and, to the best of our knowledge, such a system has not been
discussed until recently.

However, physics of this heterostructure is actually very interesting and is
general for any shape of the $S$\ and $F$\ regions. It turns out that the
proximity effect reduces the total magnetization of the system and this
effect can be seen as a certain kind of ``spin screening''.

Before doing explicit calculations we would like to explain the phenomenon
in simple words. If the temperature is above $T_{c}$, the total
magnetization of the system $M_{tot}$ equals $M_{0F}d_{F}$, where $d_{F}$ is
the thickness of the $F$-layer. When the temperature is lowered below $%
T_{c}, $ the $S$ layer becomes superconducting and the Cooper pairs with the
size of the order of $\xi _{S}\cong \sqrt{D_{S}/2\pi T_{c}}$ arise in the
superconductor. Due to the proximity effect the Cooper pairs penetrate the
ferromagnet. In the case of a homogeneous magnetization the Cooper pairs
consist, as usual, of electrons with the opposite spins, such that the total
magnetic moment of a pair is equal to zero. The exchange field is assumed to
be {\ }not too strong, otherwise the pairs would break down.

It is clear from this simple picture that pairs located entirely in the
superconductor cannot contribute to the magnetic moment of the
superconductor because their magnetic moment is simply zero, which is what
one could expect. Nevertheless, some pairs are located in space in a more
complicated manner: one of the electrons of the pair is in the
superconductor, while the other moves in the ferromagnet. These are the
pairs that create the magnetic moment in the superconductor. This follows
from the simple fact that the direction along the magnetic moment $\mathbf{M}
$ in the ferromagnet is preferable for the electron located in the
ferromagnet (we assume a ferromagnetic type of exchange field) and this
makes the spin of the other electron of the pair be antiparallel to $\mathbf{%
M}$. So, all such pairs with one electron in the ferromagnet and one in the
superconductor equally contribute to the magnetic moment in the bulk of the
superconductor. As a result, a ferromagnetic order is created in the
superconductor, the direction of the magnetic moment in this region being
opposite to the direction of the magnetic moment $\mathbf{M}$ in the
ferromagnet. Moreover, the induced magnetic moment penetrates the
superconductor over the size of the Cooper pairs $\xi _{S}$ that can be much
larger than {\ }$d_{F}$.

{This means that although the magnetization }$M_{S}${\ induced
in the superconductor is less than the magnetization in the ferromagnet }$%
M_{F0}${, the total magnetic moment in the superconductor }$\bar{M}%
_{S}=\int_{S}d^{3}rM_{S}(r)${\ may be comparable with the magnetic
moment of the ferromagnet in the normal state }$\bar{M}_{F0}=M_{F0}V_{F}$%
{, where }$V_{F}=d_{F}${\ in the case of a flat geometry (}$%
\bar{M}_{F0}${\ is the magnetic moment per unit square) and }$%
V_{F}=4\pi a_{F}^{3}/3${\ is the volume of the spherical
ferromagnetic grain. It turns out that the total magnetic moment
of the
ferromagnetic region ( film or grain) }$\bar{M}_{F0}=\mu _{B}\nu _{F}hV_{F}$%
{\ due to free electrons is compensated at zero temperature by the
total magnetic moment }$\bar{M}_{S}${\ induced in the
superconductor. This statement is valid if the condition}
\begin{equation}
\Delta <<h<<E_{Th}=D_{F}/d_{F}^{2}  \label{46a}
\end{equation}
{is fulfilled.} {If the thickness of the F film (or radius of the
F grain) is not small in comparison with the correlation length
}$\xi
_{S}${, the situation changes: the induced magnetic moment }$\bar{M}%
_{S}${\ is much smaller than }$\bar{M}_{F0}${\ but a variation
of the magnetic moment of the ferromagnetic film (or grain) }$\delta M_{F}$%
{\ becomes comparable with }$\bar{M}_{F0}${. The latter is caused
by a change in the density of states of the ferromagnet due to the
proximity effect. However, the case of a large ferromagnet size is
less
interesting because the exchange field }$h${\ should be smaller than }%
$\Delta ${\ (the full screening of }$\bar{M}_{F0}${\ occurs only
if the second condition in Eq.(\ref{46a}) is fulfilled).}

Using similar arguments we can come to a related effect: the magnetic moment
in the ferromagnet should be reduced in the presence of the
superconductivity because some of the electrons located in the ferromagnet
condensate into Cooper pairs and do not contribute to the magnetization.

>From this qualitative and somewhat oversimplified picture one can expect
that the total magnetization of the $S/F$ system will be reduced for
temperatures below $T_{c}$. Both the mechanism studied here and that of the
last section lead to a negative change of the total magnetization. Thus,
independently of the origin of ferromagnetism, they can explain, at least
qualitatively, the experimental data of \citet{muehge} and \citet{garifullin}%
.

The ideas presented above can be confirmed by calculations based on the
Usadel equation. In order to determine the change of the magnetization it is
enough to compute the quasiclassical Green's functions $\mbox{$\check{g}$}%
^{R(A)}=(i/\pi )\int d\xi \check{G}^{R(A)}$ and, in particular, the
component proportional to $\mbox{$\hat{\tau}_3$}\mbox{$\hat{\sigma}_3$}$.

The matrix Green's function has the form (we write
$\mbox{$\check{g}$}$ in Matsubara representation:
$\mbox{$\check{g}$}(\omega )=\check{g}^{R}(i\omega )$ for positive
$\omega $)
\begin{equation}
\mbox{$\check{g}$}=\mbox{$\hat{\tau}_3$}\mbox{$\hat{g}$}+i%
\mbox{$\hat{\tau}_2$}\mbox{$\hat{f}$}\;.  \label{5.5}
\end{equation}
In the ferromagnet we represent, for convenience, the matrix $%
\mbox{$\hat{f}$}$ in the spin-space as
\begin{equation}
\left(
\begin{array}{cc}
f_{+} & 0 \\
0 & f_{-}
\end{array}
\right)  \label{5.6}
\end{equation}
The diagonal form of the matrix is a consequence of the uniformity of the
exchange field $h$. The matrix $\mbox{$\hat{g}$}$ has the same form.

In order to find the function $g_{3}$\ that determines the magnetization, we
have to solve the Usadel equation (\ref{usadel}) in the $F$\ and $S$\ region
and to match the corresponding solutions with the help of the boundary
conditions (\ref{APboundary-condition}).

\begin{figure}[h]
\includegraphics[scale=0.4]{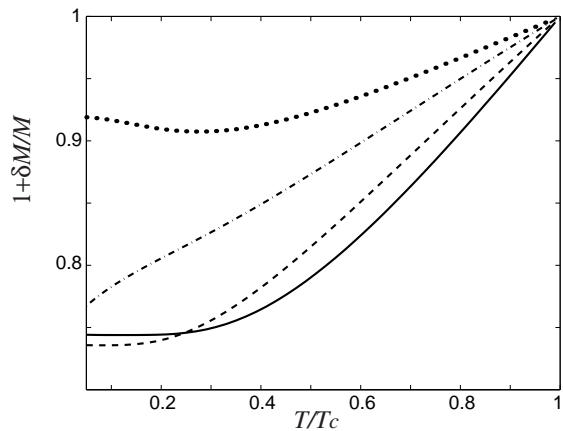}
\caption{Change of the magnetization of a F/S bilayer as a function of the
temperature.}
\label{Fig.25}
\end{figure}

The simplest case when the Usadel equation can be solved analytically {\ }is
the case of a thin $F$ layer. We suppose that the thickness $d_{F}$ of the $%
F $ layer is small compared with the characteristic length $\xi
_{F}$ of the condensate penetration into the ferromagnet (this
condition is fulfilled in the experiments by \citet{garifullin}).
In this case we can average the exact Usadel equation
(\ref{usadel}) over $x$ in the $F$ layer assuming that the Green's
functions are almost constant in space. In addition, provided the
ratio $\sigma _{F}/\sigma _{S}$ is small enough, the Green's
functions in the superconductor are close to the bulk values
$f_{BCS}$ and $g_{BCS}$. This allows us to linearize the Usadel
equation in the superconductor. The component of the Green's
function in $S$ that enters the expression for the magnetization
can be obtained from the boundary condition (\ref
{APboundary-condition}) and is given by
\begin{equation}
g_{S3}(x)=-\frac{1}{\gamma _{S}\kappa _{S}}\left(
-g_{BCS}f_{F0}+f_{BCS}g_{F3}\right) e^{\kappa _{s}x}\;,  \label{magn_gs3}
\end{equation}
where $\kappa _{S}^{2}=2\sqrt{\omega ^{2}+\Delta ^{2}}/D_{S}$, $%
f_{F0}=(f_{+}+f_{-})/2$, $g_{F3}=(g_{+}-g_{-})/2$ and $g_{\pm }$ and $f_{\pm
}$ are the components of the matrices $\hat{g}$ and $\hat{f}$. They are
defined as
\begin{equation}
g_{F\pm }=\tilde{\omega}_{\pm }/\zeta _{\omega \pm },\text{ \ }f_{F\pm }=\pm
\epsilon _{bF}f_{BCS}/\zeta _{\omega \pm }\;,  \label{gfPM}
\end{equation}
where $\tilde{\omega}_{\pm }=\omega +\epsilon _{bF}g_{BCS}\mp ih$, $\zeta
_{\omega \pm }=\sqrt{\tilde{\omega}_{\pm }^{2}-(\epsilon _{bF}f_{BCS})^{2}}$%
, $\epsilon _{bF}=D_{F}/(2\gamma _{F}d_{F})$. The magnetization variation is
determined by the expression
\begin{equation}
\delta M=-i\pi \nu T\sum_{\omega =-\infty }^{\infty }\mathrm{Tr}\text{ }(%
\hat{g}\cdot \hat{\sigma}_{3})\;,  \label{5.7}
\end{equation}
Using Eqs. (\ref{magn_gs3}-\ref{5.7}) for $\mathrm{Tr}(\hat{g}\cdot \hat{%
\sigma}_{3})/2\equiv g_{3}=(g_{+}-g_{-})/2,$ one can easily calculate $%
\delta M$. In FIG. \ref{Fig.25} we show the change of the magnetization $%
\delta M$ induced in the superconductor as a function of the temperature. We
see that for low enough temperatures the decrease of the magnetization can
be very large. At the same time, the change of the magnetization in the
ferromagnet is small \cite{BVE_inverse}.

It is interesting to calculate the total magnetic moment $\delta \bar{M}_{S}$
induced in the superconducting film and compare it with the total
magnetization of the ferromagnet $M_{F0}d_{F}$ (as we have mentioned, the
magnetization variation $\delta M_{F}$ in the ferromagnet is small and can
be neglected).

The total magnetization of the superconductor is given by
\[
\delta \bar{M}_{S}=\int_{-d_{s}}^{0}dx\delta M_{S}(x)\;.
\]
Assuming that $h\ll \epsilon _{bF}=D_{F}/(2\gamma _{F}d_{F})$ {\ or }$%
h\ll \lbrack D_{F}/(2d_{F}^{2})](\rho _{F}d_{F}/R_{b})$, we can easily
compute the ratio
\begin{equation}
\frac{\delta \bar{M}_{S}}{M_{F0}d_{F}}\approx -\pi \frac{D_{S}\nu _{S}\Delta
^{2}T}{d_{F}\gamma _{S}\nu _{F}\epsilon _{bF}}\sum_{\omega }\frac{1}{(\omega
^{2}+\Delta ^{2})^{3/2}}=-1\;,  \label{5.8}
\end{equation}
{where }$\rho _{F}${\ is the resistivity of the F region.}

We see that in the case of a thin ferromagnet at low temperatures and a not
too strong exchange field the magnetization induced in the superconductor
compensates completely the magnetization in the ferromagnet. This result
follows from the fact that the magnetization induced in the superconductor
(it is proportional to $g_{S3}$) spreads over distances of the order of $\xi
_{S}$. In view of this result one can expect that the magnetic moment of a
small ferromagnetic particle embedded in a superconductor should be
completely screened by the Cooper pairs. We discuss the screening of a
ferromagnet particle by the Cooper pairs in the next subsection.

It is worth mentioning that the problem of finding the magnetization in a $%
S/F$ structure consisting of thin $S$ ($d_{S}<\xi _{S}$) and $F$ ($d_{F}<\xi
_{F}$) layers is equivalent to the problem of magnetic superconductors where
ferromagnetic (exchange) interaction and superconducting correlations
coexist. If we assume a strong coupling between the thin $S$ and $F$ layers,
we can again average the equations over the thickness of the structure and
arrive at the Usadel equation for the averaged Green's function with an
effective exchange field $\tilde{h}=hd_{F}/d$ and an effective order
parameter $\tilde{\Delta}=\Delta d_{S}/d,$ where $d=d_{S}+d_{F}$. In this
case the magnetization is given by $M=g\mu _{B}\nu \sqrt{\tilde{h}^{2}-%
\tilde{\Delta}^{2}}\Theta (\tilde{h}-\tilde{\Delta})$, where {\ }$\Theta
\left( x\right) $ {\ }is the step function. This means that the total
magnetization $M$ is zero for $\tilde{h}<\tilde{\Delta}$. This result agrees
with those obtained by \citet{littlewood,shen} who studied the problem of
the coexistence of superconductivity and itinerant ferromagnetism in
magnetic superconductors.

One of the assumptions made for obtaining the previous results is the
quasiclassical condition $h/\epsilon _{F}\ll 1$. For some materials the
latter is not fulfilled and one has to go beyond the quasiclassical
approach. \citet{halterman_02} studied the imbalance of spin up and spin
down electrons in pure $S/F$ structures (\textit{i.e.} without impurities)
in the case of strong exchange fields ($h/\epsilon _{F}\leq 1$). In that
case superconductivity is strongly suppressed at the $S/F$ interface.
Solving the Bogoliubov-de Gennes equations numerically the authors showed
that there was a magnetic ``leakage'' from the ferromagnet into the
superconductor, which lead to a polarization of the electrons in $S$ over
the short length scale $\lambda _{F}$. The direction of the induced magnetic
moment in the superconductor was parallel to that in the ferromagnet, which
contrasts our finding.

At the same time, the limit of a very strong exchange field considered by %
\citet{halterman_02} differs completely from ours. It is clear that due to
the strong suppression of the superconductivity at the $S/F$ interface, the
magnetic moment cannot be influenced by the superconductivity and therefore
thick ferromagnetic layers with exchange energies of the order of the Fermi
energy are not suitable for observing the reduction of the magnetization
described above.

The DOS for states with spin-up and spin-down electrons in a $S/F$\
structure has been calculated on the basis of the Usadel equation by %
\citet{fazio}. The authors have found that the DOS of these states was
different in the superconductor over the length of the order $\xi _{S}$.
However, the change of the magnetization has not been calculated in this
work.

This has been done later by \citet{krivoruchko3} for a $S/F$\ structure.
Using the Usadel equation, the authors numerically calculated the
magnetization induced in the superconductor. They found that the magnetic
moment leaked from the $F$\ layer into the $S$\ layer and changed the sign
at some distance of the order of $\xi _{S}$, thus becoming negative at
sufficiently large distances only. In our opinion, the ``leakage'' of the
magnetic moment $M_{S}$\ obtained in that paper is a consequence of the use
by the authors of a wrong expression for the magnetic moment. They did not
add to the formula obtained in the quasiclassical approximation a
contribution from the energies levels located far from the Fermi energy. The
latter contribution is not captured by the quasiclassical approach and
should be written additionally.

{We have seen that under certain conditions a finite magnetic
moment is induced inside the superconductor. Does this magnetic
moment affect the superconductivity? The magnetic field $B_{S}$ in
the superconductor equals the magnetization $4\pi M_{s}$. The
induced magnetization in the superconductor $M_{S}$\ is smaller
than the magnetization in the ferromagnet: $M_{S}=M_{F}\max
(d_{F}/\{\xi _{S},d_{S}\})$. The critical field for
superconducting thin films is given by the expression $H_{c}\sim
(\lambda _{L}/d_{S})H_{bulk}$, where $\lambda _{L}$\ is the London
penetration depth, and\ $H_{bulk}$\ is the critical field of the
bulk
material. The superconductivity is not affected by the induced field $B_{S}$%
\ if the field $B_{S}\approx 4\pi M_{F}(d_{F}/\xi _{S})$\ (we set $%
d_{S}\approx \xi _{S}$) is smaller than $H_{c}$. Therefore the condition $%
4\pi M_{F}<(\lambda _{L}/d_{F})H_{bulk}$\ should be satisfied. If we take $%
\lambda _{L}\approx 1\mu m$\ and $d_{F}\approx 50$\AA , we arrive at the
condition $4\pi M_{F}<200H_{bulk}$. This condition is fulfilled easily for
the case of not too strong ferromagnets. Due to the presence of the
magnetization in the ferromagnet and superconductor spontaneous currents
arise in the system. The spontaneous Meissner currents induced by the
magnetization in S/F structures were studied by %
\citet{BVE_josephson,annett04}}.

The phenomenon discussed in this section can be considered as an alternative
mechanism of the decrease of the total magnetic moment observed by
\citet
{garifullin}. In order to clarify which of these two effects is more
important for the experimental observations one needs more information.

The most direct check for the cryptoferromagnetic phase would be
measurements with polarized neutrons. In a recent work by \cite{bernhard},
in which a multilayered $S/F/S/F...$ structure was studied. This structure
consists of the high $T_{c}$ superconductor $YBa_{2}Cu_{3}O_{7}$\ ($S$\
layer) and of the ferromagnet $La_{2/3}Ca_{1/3}MnO_{3}$ ($F$\ layer). Two
samples with the $S$\ and $F$\ layers of the same thickness were used.
Layers of sample 1(2) are $98$\AA ($160$\AA ) thick. The Curie temperature
of the ferromagnet and the temperature of the superconducting transition are
equal to $165K$ and $75K$\ respectively. By using neutron reflectometry the
authors obtained an information about the spatial distribution of the
magnetic moment in the structure. Analyzing the temperature dependence of
the Bragg peaks intensity they came to the conclusion that the most probable
scenario to explain important features of this dependence observed was the
assumption that an induced magnetization arises in the $S$\ layers. If this
explanation was correct, the sign of the induced magnetization had to be
opposite to the sign of the magnetization in the $F$\ layers. It is quite
reasonable to think that the mechanism discussed above for conventional
superconductors should be present also in high $T_{c}$\ superconductors and
then the theoretic scenario analyzed in this section can serve as an
explanation of the experiment.

\subsection{Spin screening of the magnetic moment of a ferromagnetic
particle in a superconductor}

\label{RED-SCR} 
\begin{figure}[h]
\includegraphics[scale=0.5]{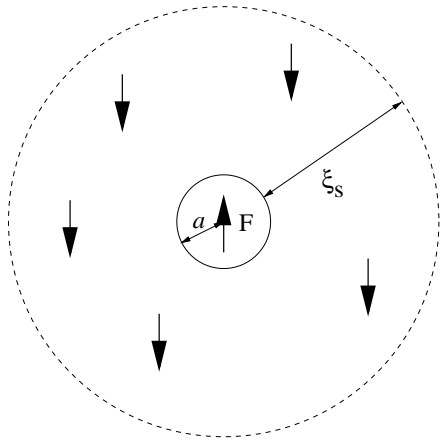}
\caption{ Ferromagnetic grain embedded into a superconductor. Due to the
inverse proximity effect the magnetic moment of the grain is screened by the
electrons of the superconductor. }
\label{Fig.26}
\end{figure}

Let us {\ }consider now a ferromagnetic particle (grain) embedded into a
superconductor (see FIG. \ref{Fig.26}). As in the previous subsection, we
analyze the magnetic moment induced in the superconductor around the
particle and compare it with the magnetic moment of the $F$ particle $(4\pi
a^{3}/3)M_{F0} $ (we assume that the particle has a spherical form and
radius $a$).

It is well known that the superconducting currents (Meissner currents) in a
superconductor screen a magnetic field that decays from the surface over{\ }%
the London penetration length $\lambda _{L}$ and vanishes in the bulk of the
superconductor. The same length characterizes the decay of the magnetic
field created by a ferromagnetic $(F)$ grain embedded in a superconductor if
the radius {\ }of the grain $a$\ is larger than $\lambda _{L}$. However, if
the radius $a$\ is small, the Meissner effect can be neglected and a stray
magnetic field around the grain should decay, as in a normal metal, over a
length of the order $a$. We consider now just this case.

Above the critical temperature $T_{c}$\ the stray magnetic field polarizes
the spins of free electrons and induces a magnetic moment. This magnetic
moment is very small because the Pauli paramagnetism is weak ($\mu
_{B}^{2}\nu \sim 10^{-6}$). In addition, the total magnetic moment induced
by the stray magnetic field is zero. The penetration depth $\lambda _{L}$
can be of the order of hundreds of interatomic distances or larger,\ so that
if $a$\ is smaller or of the order of $10nm$, the Meissner effect can be
neglected.

{The screening of the magnetic moment is a phenomenon specific for
superconductors.\ It is usually believed that in a situation, when the
screening due to the orbital }electron{\ motion can be neglected (}small
grains and thin films){, the total magnetic moment is just the magnetic
moment of the ferromagnetic particle and no additional magnetization is
induced by the electrons of the superconductor. }

{This common wisdom is quite natural because in conventional
superconductors the total spin of a Cooper pair is equal to zero
and the polarization of the conduction electrons is even smaller
than in the normal metal. Spin-orbit interactions may lead to a
finite magnetic susceptibility of the superconductor but it is
positive and smaller anyway than the one in the normal state
\cite{AG_knight,abrikosov_book}. }

Let us now take a closer look at the results of the last subsection. We have
seen that the proximity effect induces in the superconductor a magnetic
moment with the sign opposite to the one in the ferromagnet. In view of this
result it is quite natural to expect that the magnetic moment of a small
ferromagnetic particle embedded in a superconductor may be screened by the
Cooper pairs as it is sketched in FIG. \ref{Fig.26}. So, let us consider
this situation in more detail.

We consider a ferromagnetic grain of radius $a$ embedded in a bulk
superconductor. If the size of the particle is smaller than the
length $\xi _{F}$ we can again assume that the quasiclassical {\
}Green's functions in the
$F$ region are almost constant and given by Eq. (\ref{gfPM}), where now $%
\epsilon _{bF}=3D_{F}/(2\gamma _{F}a)$. In the superconductor we have to
solve the linearized Usadel equation for the component $g_{S3}$ determining
the magnetization
\begin{equation}
\nabla ^{2}g_{S3}-\kappa _{S}^{2}g_{S3}=0\;,  \label{EqGrain}
\end{equation}
where $\nabla ^{2}=\partial _{rr}+(2/r)\partial _{r}$ is the Laplace
operator in spherical coordinates.

Using the boundary conditions Eq. (\ref{APboundary-condition}) we write the
solution of this equation as
\begin{equation}
g_{S3}=\frac{f_{BCS}}{\gamma _{S}}\left( g_{BCS}f_{F0}-f_{BCS}g_{F3}\right)
\frac{a^{2}}{1+\kappa _{S}a}\frac{e^{-\kappa _{S}(r-a)}}{r}\;,  \label{gS3}
\end{equation}
where $f_{F0}=(f_{F+}+f_{F-})/2$ and $g_{F3}=(g_{F+}-g_{F-})/2$.

We assume again that the transmission coefficient through the $S/F$
interface is not small and the condition $\Delta <<h\leq (D_{F}/a^{2})$ is
fulfilled. In this case the expression for $g_{S3}$ drastically simplifies.
{Indeed, in this limit }$g_{F3}=f_{F0}f_{BCS}/g_{BCS}${\ and }$%
f_{F0}\cong ihf_{BCS}g_{BCS}/\epsilon _{bF}${. Therefore Eq.(\ref{gS3}%
) acquires the form}
\begin{equation}
\mathbf{g}_{S3}\mathbf{=}\frac{f_{BCS}^{2}}{\gamma _{S}}\frac{a^{2}}{r}\frac{%
ih}{\epsilon _{bF}}\mathbf{e}^{-\kappa _{S}(r-a)}\mathbf{\;,}
\end{equation}
{This solution can be obtained from Eq.(\ref{EqGrain}) if one
writes down the term }$4\pi A\delta (r)${\ on the right-hand side
of this equation with }$A=f_{BCS}^{2}a^{2}ih/(\gamma _{S}\epsilon
_{bF}).${\ This means that the ferromagnetic grain acts on Cooper
pairs as a magnetic impurity embedded into a dirty superconductor.
It induces a ferromagnetic
cloud of the size of the order }$\xi _{S}${\ with a magnetic moment }$%
\sim -\mu _{B}\nu hV_{F}.$

{In order to justify the assumptions made above} we estimate the
energy $D_{F}/a^{2}$ assuming that the mean free path is of the order of $a$%
. For $a=30$\AA\ and $v_{F}=10^{8}cm/\sec $ we get $D_{F}/a^{2}\sim 1000K;$.
This condition is fulfilled for ferromagnets with the exchange energy of the
order of several hundreds $K$.

In the limit of {\ }low temperatures the calculation of the magnetic moment
becomes very easy and we obtain for the magnetic moment $\bar{M}_{S}$
induced in the superconductor the following expression
\begin{equation}
\frac{\bar{M}_{S}}{M_{F0}(4\pi a^{3}/3)}=-1  \label{e17}
\end{equation}

This is a remarkable result which shows that the induced magnetic moment is
opposite in sign to the moment of the ferromagnetic particle and their
absolute values are equal to each other. In other words, the magnetic moment
of the ferromagnet is completely screened by the superconductor \cite
{BVE_screening}. The characteristic radius of the screening is the coherence
length $\xi _{S}$, which contrasts the orbital screening due to the Meissner
effect characterized by the London penetration depth $\lambda _{L}$.

To avoid misunderstanding we emphasize once again that the full
screening occurs only if the magnetization (per unit volume) of
the ferromagnetic grain $M_{F0}$\ is given by Eq.(\ref{5.4}),
which means that the ferromagnetic grain is an itinerant
ferromagnet. If the magnetization of the ferromagnet is caused by
both localized moments ($M_{loc})$\ and itinerant electrons
($M_{itin})$, the full screening is not achieved. Moreover, the
magnetization $M_{loc}$\ may be larger than $M_{itin}$\ and have
opposite direction. In this case we would have an anti-screening
\cite{garcia}.

Actually, we have discussed the diffusive case only. {However, it
turns out that the spin screening occurs also in the clean case
provided the
exchange field is not too high: }$h<<v_{F}/d_{F}${, where }$v_{F}$%
{\ and }$d_{F}${\ are the Fermi velocity and the thickness
(radius) of the ferromagnetic film or grain
\cite{kharitonov05,bergeret05}. }

The energy spectrum of a superconductor with a point-like classical magnetic
moment was studied many years ago by \citet{shiba}, \citet{sakurai} and %
\citet{rusinov}, and more recently by \citet{schrieffer}. The magnetic
impurity leads to a bound state $\beta _{0}$ inside the superconducting
energy gap. There is some critical strength $h_{c}\sim \epsilon _{F}$ of the
exchange coupling $h$ that separates two different ground states of the
system denoted by $\psi $ if $h<h_{c}$ and $\psi ^{\prime }$ if $h>h_{c}$.
The bound state $\beta _{0}$ corresponds to a localized quasiparticle with
spin ``up''\footnote{%
One assumes that the magnetic impurity has spin up.}. Since the total
electronic spin in the state $\psi $ is zero one says that the continuum
localizes a spin ``up''. The energy needed to create a quasiparticle
excitation decreases when increasing $h$. At $h=h_{c}$ the state $\psi $
becomes unstable against a spontaneous creation of an excitation with spin
``up'' and the transition to the state $\psi ^{\prime }$ occurs. In this
state the electronic spin at the impurity site is now equal to $-1/2$. All
the works considering this problem focused the attention on the subgap
structure of the spectrum and did not addressed the problem of the screening
of the magnetic moment by the continuum spectrum. This is of no surprise
because a sufficiently large magnetic moment of the impurity ($S\gg 1$)
cannot be screened by the quasiparticles.

\subsection{Spin-orbit interaction and its effect on the proximity effect}

\label{RED-SO} 
In this section we discuss the influence of the spin-orbit (SO) interaction
on the proximity effect. Although in general its characteristic energy scale
is much smaller than the exchange energy $h$, it can be comparable with the
superconducting gap $\Delta $ and therefore this effect {can be very
important}. Since the SO scattering leads to a mixing of the spin channels,
we expect that it will affect not only the singlet component of the
condensate but also the triplet one in the ferromagnet.

In conventional superconductors the SO interaction does not affect
thermodynamic properties. However, a non-vanishing magnetic susceptibility
at zero temperature (Knight shift) observed in small superconducting samples
and films can be explained only if the SO interaction is taken into account
\cite{AG_knight}. In the $F/S$ structures considered here the exchange field
$h$ breaks the time-reversal symmetry in analogy to the external magnetic
field in the Knight shift problem. Therefore the SO interaction in the
superconductor is expected to influence the inverse proximity effect studied
in this Chapter.

In this Section we will generalize the analysis of the long-range proximity
effect and the inverse proximity effect presented above taking the SO
interaction into account. The quasiclassical equations in the presence of
the SO interaction were derived by \citet{alexander} and used for the first
time for the $F/S$ systems by \citet{demler}.

The derivation of these equations is presented in the Appendix
\ref{ApA}. The resulting Usadel equation takes the form
\begin{eqnarray}
-iD\partial _{\mathbf{r}}(\mbox{$\check{g}$}\partial _{\mathbf{r}}%
\mbox{$\check{g}$})&+&i\left( \mbox{$\hat{\tau}_3$}\partial _{t}%
\mbox{$\check{g}$}+\partial _{t^{\prime }}\mbox{$\check{g}$}%
\mbox{$\hat{\tau}_3$}\right) +\left[ \check{\Delta},\mbox{$\check{g}$}\right]
+\left[ \mathbf{h}\mathbf{\check{S}},\mbox{$\check{g}$}\right]  \nonumber \\
&+&\frac{i}{\tau _{s.o.}}\left[ \mathbf{\check{S}}\mbox{$\hat{\tau}_3$}%
\mbox{$\check{g}$}\mbox{$\hat{\tau}_3$}\mathbf{\check{S}},\mbox{$\check{g}$}%
\right] =0\;.  \label{so_usadel}
\end{eqnarray}
All symbols are defined in the Appendix \ref{ApA}. The spin-orbit relaxation time $%
\tau _{s.o.}$ takes very different values depending on the material used in
the experiments. Some estimates for the values of $1/h\tau _{s.o.}$can be
found in \citet{beasley}. For example, for transition metals like $Fe$ one
obtains $1/h\tau _{s.o.}\sim 10^{-2}$, while for a typical magnetic rare
earth the value {\ }$1/h\tau _{s.o.}\sim 0.3$ is more typical. In the latter
case the SO interaction should clearly affect the penetration of the
condensate into the ferromagnet.

In order to study the influence of the SO interaction on both the long-range
and the inverse proximity effect we will use Eq.(\ref{so_usadel}). We
consider first the well known problem of the Knight shift. This example will
show the convenience of using the quasiclassical approach.

\subsubsection*{The Knight shift in superconductors}

Since the pioneering work of \citet{AG_knight} it is well established that
the magnetic susceptibility of small superconducting samples is not zero due
to the spin-orbit interaction. This explains the experiments performed for
the first time many years ago by \citet{knight} who used the nuclear
magnetic resonance technique.

Let us consider a superconductor in an external magnetic field $H$. In the
Usadel equation, Eq. (\ref{so_usadel}), the field $H$ plays the role of the
exchange energy $h$. We are interested in the linear response to this filed,
\textit{i.e.} in the magnetic susceptibility $\chi _{S}$ of the
superconductor. We assume that the superconductor is homogeneous and
therefore we drop the gradient term in Eq. (\ref{so_usadel}):
\begin{widetext}
\begin{eqnarray}
-\omega \left[ \hat{\tau}_{3},\check{g}\right] +i\left[ \check{\Delta},%
\mbox{$\check{g}$}\right] +iH\left[
\check{n},\mbox{$\check{g}$}\right] -({
1}/\tau _{s.o.})\left[ \mathbf{\check{S}}\mbox{$\hat{\tau}_3$}%
\mbox{$\check{g}$}\mbox{$\hat{\tau}_3$}\mathbf{\check{S}},\mbox{$\check{g}$}%
\right] &=&0\;,  \label{usadel_knight} \\
\check{g}^{2}&=&1\;.  \label{norm_k}
\end{eqnarray}
\end{widetext}
The solution of Eq. (\ref{usadel_knight}) has the form
\begin{equation}
\check{g}=\left( g_{BCS}+g_{3}\hat{\sigma}_{3}\right) \hat{\tau}_{3}+\left(
f_{BCS}\hat{\sigma}_{3}+f_{0}\right) i\hat{\tau}_{2}\;,  \label{eq.f}
\end{equation}
where the functions $g_{3}$ and $f_{0}$ are corrections to the normal $%
g_{BCS}$ and anomalous $f_{BCS}$ Green's functions. In the
particle-hole space
the matrix $\check{g}$ has the usual form, i.e. it is expanded in matrices $%
\mbox{$\hat{\tau}_3$}$ and $i\mbox{$\hat{\tau}_2$}$. In the spin space the
triplet component (the $g_{3}$ and $f_{0}$ terms) appears due to the
magnetic field acting on the spins. Using Eqs. (\ref{usadel_knight}-\ref
{eq.f}) one can readily obtain
\begin{equation}
g_{3}=-i\frac{\Delta ^{2}H}{E_{\omega }^{2}(E_{\omega }+4/\tau _{s.o.})}\;.
\label{e18}
\end{equation}
where $E_{\omega }=\sqrt{\Delta ^{2}+\omega ^{2}}$.

Substituting Eq. (\ref{e18}) into Eq. (\ref{5.7}) we can write the
magnetization $M$ as follows
\begin{equation}
M=M_{0}-\mu _{B}\nu \left( 2\pi T\Delta ^{2}\sum_{\omega }\frac{1}{E_{\omega
}^{2}(E_{\omega }+4/\tau _{s.o.})}\right) H  \label{knight_mag}
\end{equation}
The first term in Eq. (\ref{knight_mag}) cannot be calculated in the
framework of the quasiclassical theory and one should use exact Green's
functions. It corresponds to the Pauli paramagnetic term given by $M_{0}=\mu
_{B}\nu H$. In the quasiclassical approach this term is absent. This term
does not depend on temperature on the energy scale of the order of $T_{c}$
and originates from a contribution of short distances where the
quasiclassical approximation fails.

This situation is rather typical for the quasiclassical approach and one
usually adds to formulae obtained within this approach contributions coming
from short distances or times by hand (see, for example, %
\citet{artemenko_volkov,kopnin,rammer}). Eq. (\ref{knight_mag}) was first
obtained by \citet{AG_knight}.

In the absence of the spin orbit interaction the magnetization at $T=0$ is,
as expected, equal to zero. However, if the SO interaction is finite the
spin susceptibility $\chi _{S}$ does not vanish at $T=0$. It is interesting
that, as follows from Eq. (\ref{usadel_knight}), the singlet component of
the condensate is not affected by the SO interaction. The origin of the
finite susceptibility is the existence of the triplet component $f_{0}$ of
the condensate.

In the $S/F$ structures there is no exchange field in the superconductor and
therefore the situation is in principle different. However, we have seen
that due to the proximity effect the triplet component $f_{0}$ is induced in
the superconductor.

From the above analysis one expects that the SO interaction may
affect the penetration length of such component in the
superconductor. In the next
sections we consider the influence of the SO the superconducting condensate{%
\ }in both the ferromagnet and the superconductor.

\subsubsection{Influence of the Spin-Orbit interaction on the long-range
Proximity Effect}

Now we consider again the $S/F/S/F/S$ structure of Section \ref{JOS-TRI} and
assume that the long-range triplet component is created, which is possible\
provided the angle $\alpha $ between the magnetizations differs from $0$ and
$\pi $. In order to understand how the SO interaction affects the triplet
component it is convenient to linearize Eq. (\ref{so_usadel}) in the $F$%
-layer assuming, for example, that the proximity effect is weak. One can
easily obtain a linearized equation similar to Eq. (\ref{3.14}) for the
condensate function $\mbox{$\check{f}$}$. The solution of this equation is
represented again in the form
\begin{equation}
\mbox{$\check{f}$}(x)=i\mbox{$\hat{\tau}_2$}\otimes (f_{0}(x)%
\mbox{$\hat{\sigma}_0$}+f_{3}(x)\mbox{$\hat{\sigma}_3$})+i%
\mbox{$\hat{\tau}_1$}\otimes f_{1}(x)\mbox{$\hat{\sigma}_1$}\;.
\end{equation}
The functions $f_{i}(x)$ are given as before {\ }by $f_{i}(x)=\sum_{j}b_{j}%
\exp [\kappa _{j}x]$ but now the new eigenvalues $\kappa _{j}$ are written
as
\begin{eqnarray}
\kappa _{\pm }^{2} &=&\pm \frac{2i}{D_{F}}\sqrt{h^{2}-\left( \frac{4}{\tau
_{so}}\right) ^{2}}+\frac{4}{\tau _{so}D_{F}} \\
\kappa _{0}^{2} &=&\kappa _{\omega }^{2}+2\left( \frac{4}{\tau _{s.o.}D_{F}}%
\right) \;.
\end{eqnarray}
We see from these equations that both the singlet and triplet components are
affected by the spin-orbit interaction making the decay of the condensate in
the ferromagnet faster. In the limiting case, when {\ }$4/\tau _{so}>h,T_{c}$%
, both the components penetrate over the same distance $\xi _{s.o.}=\sqrt{%
\tau _{so}D_{F}}$ and therefore the long-range effect is suppressed. In this
case the characteristic oscillations of the singlet component are destroyed
\cite{demler}. In the more interesting case $4/\tau _{so}\sim T_{c}<h$, the
singlet component does not change and penetrates over the short distance $%
\xi _{F}$ . At the same time, the triplet component is more sensitive to the
spin-orbit interaction and the penetration length equals $\min $($\xi
_{so},\xi _{T}$)$>\xi _{F}$.

Therefore, if the spin-orbit interaction is not very strong, the penetration
of triplet condensate over the long distances discussed in the preceding
sections is still possible, although the penetration length is reduced.

\subsubsection{Spin-Orbit Interaction and the Inverse Proximity Effect}

Studying a $S/F$ bilayer we have seen that the induced magnetic moment in
the superconductor $S$ is related to the appearance of the triplet component
$f_{0}$. Moreover, we have shown that this component is affected by the SO
interaction, while the singlet one $f_{3}$ is not. So, one should expect
that the SO interaction may change the scale over which the magnetic moment
is induced in the superconductor and one can estimate easily this length.

Assuming that the Green's functions in the superconductor take
values close to the bulk values we linearize the Usadel equation
(\ref{so_usadel}) in the
superconductor. The solution has the same form as before, Eq. (\ref{magn_gs3}%
), but $\kappa _{S}$ should be replaced by
\begin{equation}
\kappa _{S}^{2}\rightarrow \kappa _{S}^{2}+\kappa _{so}^{2},  \label{e4}
\end{equation}
where $\kappa _{so}^{2}=8D_{S}/\tau _{so}$. Therefore, the length of the
penetration of $g_{S3}$ and, in its turn, of {\ }$M_{S}$ into the $S$ region
decreases if $\kappa _{S}^{2}\sim \xi _{S}^{-2}<\kappa _{so}^{2}$.

In principle, one can measure the spatial distribution of the magnetic
moment in the $S$ region as it was done by \citet{muon} by means of muon
spin rotation and get an information about the SO interaction in
superconductors. As Eq. (\ref{e4}) shows, this would be an alternative
method to measure the strength of the SO interaction in superconductors,
complementary to the measurement of the Knight shift \cite{knight}.

\section{Discussion of the results and outlook}

In this review we have discussed new unusual properties of structures
consisting of conventional superconductors in a contact with ferromagnets.
It has been known that such systems might exhibit very interesting
properties like a non-monotonous reduction of the superconducting
temperature as a function of the thickness of the superconductor,
possibility of a $\pi $-contact in Josephson junctions with ferromagnetic
layers, etc.

However, as we have seen, everything is even more interesting and
some spectacular phenomena are possible that even might look at
first glance as a paradox. The common feature of the effects
discussed in this review is that almost all of them originate in
situations when the exchange field is not homogeneous. As a
consequence of the inhomogeneity, the spin structure of the
superconducting condensate function becomes very non-trivial and,
in particular, the triplet components are generated. In the
presence of the inhomogeneous exchange field, the total spin of a
Cooper pair is not necessarily equal to zero and the total spin
equal to unity with all projections onto the direction of the
exchange field is possible.

We have discussed the main properties of the odd triplet superconductivity
in the $S/F$ structures. This superconductivity differs from the well known
types of superconductivity: a) singlet superconductivity with the s-wave
(conventional $T_{c}$ superconductors) and d-wave (high $T_{c}$
superconductors) types of pairing; b) odd in momentum $p$ and even in
frequency $\omega $ triplet superconductivity observed, e.g., in $%
Sr_{2}RuO_{4}$.

The odd triplet superconductivity discussed in this Review has a condensate
(Gor'kov) function that is an odd function of the Matsubara frequency $%
\omega $ and an even function (in the main approximation) of the momentum $p$
in the diffusive limit. It is insensitive to the scattering on nonmagnetic
impurities and therefore may be realized in thin film $S/F$ structures where
the mean free path is very short.

For the first time, the condensate function of this type has been suggested
by \citet{berezin} many years ago as a possible candidate to describe
superfluidity in $He^{3}$. Later, it has been established {\ }that the
superfluid condensate in $He^{3}$ had a different structure - it was odd in $%
p$ and even in $\omega $. In principle, there is an important difference
between the triplet superconductivity discussed here and that predicted by
Berezinskii who assumed that the order parameter $\Delta $ was also an odd
function of $\omega $. In our case the order parameter $\Delta $ is
determined by the singlet, s-wave condensate function and has the ordinary
BCS structure (i.e., it does not depend on the momentum $p$ and frequency $%
\omega $). On the other hand the structure of the triplet condensate
function $\;\check{f}$ in the diffusive case considered here {\ }coincides
with that suggested by Berezinskii: it is an odd function of the Matsubara
frequency $\omega $ and, in the main approximation, is constant in the
momentum space. The antisymmetric part of $\;\check{f}$ \ is small compared
with the symmetric part, {\ }being odd in $p$\ and even in $\omega .$

The triplet component with the projection of the total spin $S_{z}=\pm 1$
penetrates the ferromagnet over a long distance of the order of $\xi
_{N}\approx \sqrt{D_{F}/2\pi T},$ which shows that the exchange field does
not affect the triplet part of the condensate. {\ }At the same time, the
exchange field suppresses the amplitude of the singlet component at the $S/F$
interface that determines the amplitude of the triplet component. The
long-range triplet component arises only in the case of a nonhomogeneous
magnetization. The triplet component appears also in a system with a
homogeneous magnetization but in this case it corresponds to the projection $%
S_{z}=0$ and penetrates the ferromagnet over a short length $\xi _{F}=\sqrt{%
D_{F}/h}<<\xi _{N}$.

The triplet component exists also in magnetic superconductors \cite
{bula_adv,kulic_kulic} with a spiral magnetic structure. However, it always
coexists with the singlet component and cannot be separated from it. In
contrast, in the {\ }multilayered $S/F$ structures with a nonhomogeneous
magnetization and with the thickness of the $F$ layers $d_{F}$ exceeding $%
\xi _{F}$, the Josephson coupling between $S$ layers is realized only
through the long-range triplet component and this separates the singlet and
triplet components from each other. As a result, the ``real'' odd triplet
superconductivity may be realized in the transverse direction in such
structures.

Another interesting peculiarity of the $S/F$ structures is the inverse
proximity effect, namely, the penetration of the magnetic order parameter
(spontaneous magnetic moment $M$) into the superconductor and a spatial
variation of the magnetization direction in the ferromagnet under the
influence of the superconductivity. It turns out that both effects are
possible. A homogeneous distribution of the magnetization $M_{F}$ in the $%
S/F $ bilayer structures may be energetically unfavorable in $F$ even in a
one-domain case resulting in a nonhomogeneous distribution of $\mathbf{M}%
_{F} $ in the ferromagnet.

Moreover, the magnetic moment penetrates the superconductor
(induced ferromagnetism) changing sign at the $S/F$ interface.
Therefore the total magnetic moment of the system is reduced.
Under some condition the full spin screening of $M_{F}$ occurs.
For example, at zero temperature the itinerant magnetic moment of
a ferromagnetic grain embedded into a superconductor is completely
screened by spins of the Cooper pairs in $S$. The radius of the
screening cloud is of the order of the superconducting coherence
length $\xi _{S}$. If the magnetization vector $\mathbf{M}_{F}$ is
oriented in the opposite direction to the ferromagnetic exchange
field\ $\mathbf{h}$, the anti-screening is possible.

As concerns the experimental situation, certainly there are indications in
favor of the long-range triplet component, although an unambiguous evidence
does not exist so far. For example, the resistance of ferromagnetic films or
wires in the $S/F$ structures changes on distances that exceed the length of
the decay of the singlet component $\xi _{h}$ \cite
{chandrasekhar,petrashov,pannetier}. A possible reason for this long-range
proximity effect in the $S/F$ systems is the long-range penetration of the
triplet component. However a simpler effect might also be the reason for
this long-range proximity effect. It is related to a rearrangement of a
domain structure in the ferromagnet when the temperature lowers below $T_{c}$%
. The Meissner currents induced in the superconductor by a stray magnetic
field affect the domain structure, and the resistance of the ferromagnet may
change \cite{Geim}. At the same time, the Meissner currents should be
considerably reduced in an one dimensional geometry for the ferromagnet like
that used in \cite{pannetier} and the explanation in terms of the long range
penetration of the triplet component are more probable here.

\citet{pena} also obtained some indications on the existence of a triplet
component in a multilayered $S/F/S/F$... structure. The samples used by %
\citeauthor{pena} contained the high $T_{c}$\ material $YBa_{2}Cu_{3}O_{7}$\
(as a superconductor) and the half-metallic ferromagnet $%
La_{0.7}Ca_{0.3}MnO_{3}$\ (as a ferromagnet). They found that
superconductivity persisted even in the case when the thickness of the $F$\
layers $d_{F}$ essentially exceeded $\xi _{F}$ ($d_{F}\gtrsim 10nm$ and $\xi
_{F}\approx 5nm$). In a half-metal ferromagnet with spins of free electrons
aligned in one direction the singlet\ Cooper pairs cannot exist. Therefore
it is reasonable to assume that the superconducting coupling between
neighboring $S$\ layers is realized via the triplet component \cite
{BVE3,schoen_half}.

A reduction of the magnetic moment of the $S/F$ structures due to
superconducting correlations {\ }has been observed already \cite{garifullin}%
. This reduction may be caused both by the spin screening of the magnetic
moment $M_{F}$ and by the rotation of $M_{F}$ in space \cite{BEL,BVE_inverse}%
. Perhaps, the spin screening can be observed directly by probing the
spatial distribution of the magnetic field (or magnetic moment $M $) with
the aid of the muon spin rotation {\ }technique \cite{muon}. The variation
of the magnetic moment $M$ occurs on a macroscopic length $\xi _{S}$ and
therefore can be detected.

An evidence in favor of the inverse proximity effect has also been obtained
in another experimental work \cite{bernhard}. Analyzing data of neutron
reflectometry on a multilayered $YBa_{2}Cu_{3}O_{7}/La_{2/3}Ca_{1/3}MnO_{3}$
structure, the authors concluded that a magnetic moment was induced in the
superconducting $YBa_{2}Cu_{3}O_{7} $ layers. The sign of this induced
moment was opposite to the sign of the magnetic moment in the ferromagnetic $%
La_{2/3}Ca_{1/3}MnO_{3}$ layers, which correlates with our prediction.

In spite of these experimental results that may be considered as, at least
preliminary, confirmation of the existence of the triplet component in the $%
S/F$ structures, there is a need in additional experimental studies of the
unconventional superconductivity discussed in this review. One of the
important issues would be to understand whether the long range proximity
effects already observed experimentally are due to the triplet pairing or to
a simple redistribution of the domain walls by the Meissner currents. We
believe that measurements on thin ferromagnetic wires where the Meissner
currents are reduced may clarify the situation.

It is very interesting to distinguish between the two possible inverse
proximity effects experimentally. Although both the formation of the
cryptoferromagnetic state and the induction of the magnetic moments in the
superconductors are very interesting effects, it is not clear yet which of
these effects causes the magnetization reduction observed by %
\citet{muehge,garifullin}.

The enhancement of the Josephson current by the presence of the
ferromagnet near the junction is one more theoretical prediction
that has not been observed yet but, certainly, this effect
deserves an attention. {An overview for experimentalists
interested in all these subjects is presented in Appendix
\ref{ApB}, where we discuss briefly different experiments on S/F
structures, focusing our attention on the materials for which, we
expect, the main effects discussed in this review may be
observed.}

In addition, further theoretical investigations are needed. The odd triplet
component has been studied mainly in the diffusive limit ($h\tau <<1$). It
would be interesting to investigate the properties of the triplet component
for an arbitrary impurity concentration ($h\tau \gtrless 1$). No theoretical
work on dynamics of magnetic moments in the $S/F$\ structures has been
performed yet, although the triplet component may play a very important role
in the dynamics of the $S/F$\ structures. Transport properties of the $S/F$\
structures require also further theoretical considerations. It would be
useful to study the influence of domain structures on properties of the $S/F$%
\ structures, etc. In other words, physics of the proximity effects in the
superconductor-ferromagnet structures is evolving into a very popular field
of research, both experimentally and theoretically.

The study of the proximity effect in $S/F$ structures may be extended to
include ferromagnets in contact with high temperature superconductors.{\ }%
Although some experiments have been done already \cite{pena,bernhard}, one
can expect much more broad experimental investigations in the future. The
modern technique allows the preparation of multilayered $S/F/S/F$..
structures consisting of thin ferromagnetic layers (as $%
La_{2/3}Ca_{1/3}MnO_{3}$) and thin layers of high $T_{c}$ superconductor (as
$YBa_{2}Cu_{3}O_{7}$) with variable thicknesses. It would be very
interesting to study, both experimentally and theoretically, such a system
with non-collinear magnetization orientations. In this case d-wave singlet
and odd triplet superconductivity should coexist in the system. It is known
that many properties of the ordinary BCS superconductivity remain unchanged
in the high $T_{c}$ superconductors. This means that many effects considered
in this review can also occur in S/F structures containing high $T_{c}$\
materials, but there will certainly be differences with respect to the
conventional superconductors with the s-pairing.

We hope that this review will encourage experimentalists and theoreticians
to make further investigations in this fascinating field of research.

\section*{Acknowledgements}

We appreciate fruitful discussions with A. I. Buzdin, Y. V.
Fominov, A. A. Golubov, I. A. Garifullin, A. Gerber, A. Palevski,
L. R. Tagirov, K. Westerholt and H. Zabel. We would like to thank
SFB 491 for financial support. F.S.B would like to thank the E.U.
network DIENOW for financial support.

\appendix

\section{Basic equations}\label{ApA}

\label{APP}

Throughout this review we use mainly the well established method
of quasiclassical Green's functions. Within this method the
Gor'kov equations can be drastically simplified by integrating the
Green's function over the
momentum. This method was first introduced by \citet{larkin_ovch} and %
\citet{eilenberger}, and then extended by \citet*{usadeleq} for a dirty case
and by \citet{eliashberg} for a non-equilibrium case . The method of the
quasiclassical Green's functions is discussed in many reviews %
\citet{serene,rammer,lo_book,belzig_rev} and in the book by
\citet{kopnin}. In this Appendix we present a brief derivation of
equations for the quasiclassical Green's functions and write
formulae for the main observable quantities in terms of these
functions. A special attention will be paid to the dependence of
these functions on the spin variables that play a crucial role in
the S/F structures. In particular, we take into account the
spin-orbit interaction alongside with the exchange interaction in
the ferromagnet.

We start with a general Hamiltonian describing a conventional
BCS-superconductor/ferromagnet structure:
\begin{eqnarray}
\hat{H}&=&\sum_{\{p,s\}}\left\{ a_{sp}^{+}\left[ \left( \left( \xi
_{p}\delta _{pp^{\prime }}+eV\right) +U_{imp}\right) \delta _{ss^{\prime
}}+U_{s.o.}\right.\right.  \nonumber \\
& &\left.\left. -\left(\mathbf{h}.\sigma \right) \right] a_{s^{\prime
}p^{\prime }}-\left( \Delta a_{\overline{s}\overline{p}}^{\dagger
}a_{s^{\prime }p^{\prime }}^{\dagger }+c.c.\right) \right\} \;.
\label{hamiltonian}
\end{eqnarray}
The summation is carried out over all momenta $(p,p^{\prime })$ and spins $%
(s,s^{\prime })$ (the notation $\overline{s}$, $\overline{p}$ means
inversion of both spin and momentum), $\xi _{p}=p^{2}/2m-\epsilon _{F}$ is
the kinetic energy counted from the Fermi energy $\epsilon _{F}$, $V$ is a
smoothly varying electric potential. The superconducting order parameter $%
\Delta $ must be determined self-consistently. It vanishes in the
ferromagnetic regions. The potential $U_{imp}=U(p-p^{\prime })$ describes
the interaction of the electrons with nonmagnetic impurities, and $U_{s.o.}$
describes a possible spin-orbit interaction \cite{AG_knight}:
\[
U_{s.o.}=\sum_{i}\frac{u_{s.o.}^{(i)}}{p_{F}^{2}}\left( \mathbf{p\times
p^{\prime }}\right) \mathbf{\sigma }\;.
\]
Here the summation is performed over all impurities.

The representation of the Hamiltonian in the form (\ref{hamiltonian})
implies that we use the mean-field approximation for the superconducting ($%
\Delta $) and magnetic ($\mathbf{h}$) order parameter. The exchange field $%
\mathbf{h}$ is parallel to the magnetization $\mathbf{M}_{F}$ in $F$%
\footnote{%
(we remind that the exchange field $h$ is measured in energy units, see also
the Footnote on page 78)}. In strong ferromagnets the magnitude of $\mathbf{h%
}$ is much higher than $\Delta $ and corresponds to an effective magnetic
field $H_{exc}=h/\mu _{B}$ of the order $10^{6}Oe$ (where $\mu _{B}=g\mu
_{Bohr},$ $g$ is the $g$-factor and $\mu _{Bohr}$ is the Bohr magneton).

In order to describe the ferromagnetic region we use a simplified model that
catches all physics we are interested in. Ferromagnetism in metals is caused
by the electron-electron interaction between electrons belonging to
different bands that can correspond to localized and conducting states. Only
the latter participate in the proximity effect. If the contribution of free
electrons strongly dominates (an itinerant ferromagnet), one has {\ }$%
M_{F}\cong M_{e}$ and the exchange energy is caused mainly by free electrons.

If the polarization of the conduction electrons is due to the interaction
with localized magnetic moments, the Hamiltonian $\hat{H}_{F}$ can be
written in the form
\begin{equation}
\ \hat{H}_{F}=-h_{1}\sum_{\{p,s\}}\left\{ a_{sp}^{+}\mathbf{S}\ast \mathbf{%
\sigma }_{ss^{\prime }}a_{s^{\prime }p^{\prime }}\right\}  \label{Ham_F1}
\end{equation}
where $\mathbf{S}=\sum_{a}\mathbf{S}_{a}\delta (r-r_{a})$, $\mathbf{S}_{a}$\
is the spin of a particular ion. A constant $h_{1}$\ is related to $h$\ via
the equation: $h=h_{1}n_{M}S_{0}$\ , where $n_{M}$ is the concentration of
magnetic ions and $S_{0}$\ is a maximum value of $S_{a}$\ (we consider these
spins as classical vectors; see Ref. \cite{GR}). In this case the
magnetization is a sum: $M=M_{loc}+M_{e}$, and the magnetization $M_{e}$\
can be aligned parallel ($h_{1}>0$, the ferromagnetic type of the exchange
field) \ to $M$\ or antiparallel ($h_{1}<0$, the antiferromagnetic type of
the exchange field). In the following we will assume a ferromagnetic
exchange interaction ($M_{e}$ and $M$\ are oriented in the same direction$)$. In principle, \ one can add to Eq. (\ref{Ham_F1}) the term $%
\sum_{\{a,b\}}\left\{ \mathbf{S}_{a}\ast \mathbf{S}_{b}\right\} $\
describing a direct interaction between localized magnetic moments but in
the most part of the review this term is not important except Section V.A.
where the cryptoferromagnetic state is discussed.

Starting from the Hamiltonian (\ref{hamiltonian}) and using a standard
approach \cite{lo_book}, one can derive the Eilenberger and Usadel
equations. Initially these equations have been derived for $2\times 2$
matrix Green's functions $g_{n,n^{\prime }}$, where indices $n,n^{\prime }$
relate to normal $(g_{11},g_{22})$ and anomalous or condensate $%
(f_{12},f_{21})$ Green's functions. These functions describe the singlet
component. In the case of a non-homogenous magnetization considered in this
review one has to introduce additional Green's functions depending on spins
and describe not only the singlet but also the triplet component. These
matrices depend not only on $n,n^{\prime }$ indices but also on the spin
indices $s,s^{\prime },$ and are $4\times 4$ matrices in the spin and
Gor'kov space (sometimes the $n,n^{\prime }$ space is called the Nambu or
Nambu-Gor'kov space).

In order to define the Green's functions in a customary way it is convenient
to write the Hamiltonian (\ref{hamiltonian}) in terms of new operators $%
c_{nsp}^{\dagger }$ and $c_{nsp}$ that are related to the creation and
anhilation operators $a_{s}^{+}$ and $a_{s}$ by the relation (we drop the
index $p$ related to the momentum)
\begin{equation}
c_{ns}=\left\{
\begin{array}{c}
a_{s},n=1 \\
a_{\overline{s}}^{\dagger },n=2\;.\label{c_operators}
\end{array}
\right.
\end{equation}
These operators (for $s=1$) were introduced by Nambu \cite{Nambu} . The new
operators allow one to express the anomalous averages $<a_{\uparrow }\cdot
a_{\uparrow }>$ introduced by Gor'kov as the conventional averages $%
<c_{1}\cdot c_{2}^{+}>$ and therefore the theory of superconductivity can be
constructed by analogy with a theory of normal systems. Thus, the index $n$
operates in the particle-hole (Numbs-Gor'kov) space, while the index $s$
operates in the spin space. In terms of the $c_{ns}$ operators the
Hamiltonian can be written in the form
\begin{equation}
H=\sum_{\{p,n,s\}}c_{ns}^{+}\mathcal{H}_{(nn^{\prime })(ss^{\prime
})}c_{n^{\prime }s^{\prime }}\;,
\end{equation}
where the summation is performed over all momenta, particle-hole and spin
indices. The matrix $\check{\mathcal{H}}$ is given by
\begin{eqnarray}
\check{\mathcal{H}}&=&\frac{1}{2}\left\{ \left[ \left( \xi _{p}\delta
_{pp^{\prime }}+eV\right) +U_{imp}\right] \hat{\tau}_{3}\otimes \hat{\sigma}%
_{0}+\widetilde{\hat{\Delta}}\otimes \hat{\sigma}_{3}-\mathbf{h}\hat{\tau}%
_{3}\mathbf{\check{S}}\right.  \nonumber \\
& &\left.+\sum_{i}\frac{u_{s.o.}^{(i)}}{p_{F}^{2}}\left( \mathbf{p\times
p^{\prime }}\right) \mathbf{\check{S}}\right\} \;.  \label{H_2}
\end{eqnarray}
The matrices $\hat{\tau}_{i}$ and $\hat{\sigma}_{i}$ are the Pauli matrices
in the particle-hole and spin space respectively; $i=0,1,2,3$, where $\hat{%
\tau}_{0}$ and $\sigma _{0}$ are the corresponding unit matrices. The matrix
vector $\mathbf{\check{S}}$ is defined as
\[
\mathbf{\check{S}}=(\hat{\sigma}_{1},\hat{\sigma}_{2},\hat{\tau}_{3}\hat{%
\sigma}_{3})\;,
\]
and the matrix order parameter equals $\widetilde{\hat{\Delta}}=\hat{\tau}%
_{1}Re\Delta -\hat{\tau}_{2}Im\Delta $. Now we can define the
matrix Green's functions (in the particle-hole$\otimes $spin
space) in the Keldysh representation in a standard way
\begin{equation}
\check{G}(t_{i},t_{k}^{\prime })=\frac{1}{i}\left\langle T_{C}\left(
c_{ns}(t_{i})c_{n^{\prime }s^{\prime }}^{\dagger }(t_{k}^{\prime })\right)
\right\rangle ,
\end{equation}
where the temporal indices take the values $1$ and $2$, which correspond to
the upper and lower branch of the contour $C$, running from $-\infty $ to $%
+\infty $ and back to $-\infty $.

One can introduce a matrix in the Keldysh space of the form
\begin{equation}
\mathbf{\check{G}}(t,t^{\prime })=\left(
\begin{array}{cc}
\check{G}(t,t^{\prime })^{R} & \check{G}(t,t^{\prime })^{K} \\
0 & \check{G}(t,t^{\prime })^{A}
\end{array}
\right)  \label{MatrixG}
\end{equation}
where the retarded (advanced) Green's functions
$\check{G}(t,t^{\prime
})^{R(A)}$ are related to the matrices $\check{G}(t_{i},t_{k}^{\prime }):$ $%
\check{G}(t,t^{\prime })^{R(A)}=\check{G}(t_{1},t_{1}^{\prime })-\check{G}%
(t_{1(2)},t_{2(1)}^{\prime })$. All these elements are $4\times 4$ matrices.
These functions determine thermodynamic properties of the system (density of
states, the Josephson current etc). The matrix $\check{G}(t,t^{\prime
})^{K}= $ $\check{G}(t_{1},t_{2}^{\prime })+\check{G}(t_{2},t_{1}^{\prime })$
is related to the distribution function and has a nontrivial structure only
in a nonequilibrium case. In the equilibrium case it is equal to: $\check{G}%
(\epsilon )^{K}=\int d(t-t^{\prime })\check{G}(t-t^{\prime })^{K}\exp
(i\epsilon (t-t^{\prime }))=[\check{G}(\epsilon )^{R}-\check{G}(\epsilon
)^{A}]\tanh (\epsilon /2T).$

In order to obtain the equations for the quasiclassical Green's
functions, we follow the procedure introduced by \citet*{lo_book}.
The equation of motion for the Green's functions is
\begin{equation}
\left( i\partial _{t}-\check{H}-\check{\Sigma}_{imp}-\check{\Sigma}%
_{s.o.}\right) \mathbf{\check{G}}=\check{1}\;,  \label{dyson}
\end{equation}
where
\[
\check{H}=-\mbox{$\hat{\tau}_3$}\frac{\partial _{\mathbf{r}}^{2}}{2m}%
-\epsilon _{F}-\mathbf{h}\hat{\tau}_{3}\mathbf{\check{S}}+\widetilde{\hat{%
\Delta}}\otimes \hat{\sigma}_{3}
\]
and $\check{\Sigma}_{imp}$ and $\check{\Sigma}_{s.o.}$ are the self-energies
given in the Born approximation by
\begin{eqnarray}
\check{\Sigma}_{imp} &=&N_{imp}u_{imp}^{2}\hat{\tau}_{3}\langle \mathbf{%
\check{G}}\rangle \hat{\tau}_{3},\;\;\langle \mathbf{\check{G}}\rangle =\nu
\int d\xi _{p}\int \frac{d\Omega }{4\pi }\mathbf{\check{G}}  \nonumber \\
\check{\Sigma}_{s.o.} &=&N_{imp}u_{s.o.}^{2}\langle \mathbf{\check{G}}%
\rangle _{s.o.},  \nonumber \\
\langle \mathbf{\check{G}}\rangle _{s.o.} &=&\nu \int d\xi _{p}\int \frac{%
d\Omega }{4\pi }(\mathbf{n\times n^{\prime }})\mathbf{\check{S}\check{G}%
\check{S}}(\mathbf{n\times n^{\prime }})\;.
\end{eqnarray}
Here $N_{imp}$ is the impurity concentration, $\nu $ is the density of
states at the Fermi level and $\mathbf{n}$ is a unit vector parallel to the
momentum.

Next step is to subtract from Eq.(\ref{dyson}), multiplied by $\hat{\tau}%
_{3} $ from the left, its conjugate equation multiplied by
$\hat{\tau}_{3}$ from the right. Then one has to go from the
variables ({$\mathbf{r,r^{\prime }}$}) to ($\mathbf{(r+r^{\prime
})/2,r-r^{\prime }}$) and to perform a Fourier transformation with
respect to the relative coordinate. By making use of the fact that
the Green's functions are peaked at the Fermi surface, one can
integrate the resulting equation over $\xi _{p},$ and finally one
obtains
\begin{eqnarray}
\hat{\tau}_{3}\partial_t \mathit{\check{g}} +\partial_{t^{\prime}} \mathit{%
\check{g}}\hat{\tau}_{3}&+&\mathbf{v_{F}}\nabla\mathit{
\check{g}}-i\left[ \mathbf{h}\mathbf{\check{S}},\mathit{%
\check{g}}\right] -i\left[ \check{\Delta},\mathit{\check{g}}\right]
\nonumber \\
+\frac{1}{2\tau }\left[ \langle \mathit{\check{g}}\rangle ,\mathit{\check{g%
}}\right] &+&\frac{1}{2\tau _{s.o.}}\left[ \hat{\tau}_{3}\langle \mathit{%
\check{g}}\rangle _{s.o.}\hat{\tau}_{3},\mathit{\check{g}}\right]
=0 \label{eilenberger}
\end{eqnarray}
where $\check{\Delta}=\hat{\tau}_{3}\widetilde{\hat{\Delta}}$ and the
quasiclassical Green's functions $\mathit{\check{g}}(t_{i},t_{k}^{\prime })$
are defined as
\begin{equation}
\mathit{\check{g}}(\mathbf{p}_{\mathbf{F}},\mathbf{r})=\frac{i}{\pi }\left(
\hat{\tau}_{3}\otimes \hat{\sigma}_{0}\right) \int d\xi _{p}\mathbf{\check{G}%
}(t_{i},t_{k}^{\prime };\mathbf{p},\mathbf{r})\;,
\end{equation}
and $\mathbf{v_{F}}$ is the Fermi velocity. The scattering times appearing
in Eq. (\ref{eilenberger}) are defined as
\begin{eqnarray}
\tau ^{-1} &=&2\pi \nu N_{imp}u_{imp}^{2} \\
\tau _{s.o.}^{-1} &=&\frac{1}{3}\pi \nu N_{imp}\int \frac{d\Omega }{4\pi }%
u_{s.o.}^{2}\sin ^{2}\theta
\end{eqnarray}

Eq. (\ref{eilenberger}) is a generalization of an equation derived by \citet*%
{larkin_ovch}, and \citet*{eilenberger} for a general
nonequilibrium case. This generalization (in the absence of
spin-dependent interactions) has been done by \citet{eliashberg}
and Larkin and Ovchinnikov \cite{lo_book}. A solution for
Eq.(\ref{eilenberger}) is not unique. The proper solutions must
obey the so called normalization condition
\begin{equation}
\int (d\epsilon _{1}/2\pi )\mathit{\check{g}}(\mathbf{p}_{\mathbf{F}},%
\mathbf{r;}\epsilon ,\epsilon _{1}).\mathit{\check{g}}(\mathbf{p}_{\mathbf{F}%
},\mathbf{r;}\epsilon _{1},\epsilon ^{\prime })=1  \label{NormCond}
\end{equation}
Generalization for the case of exchange and spin-orbit interaction was
presented in \cite{BEL} and \cite{BVE_josephson}. The solution for Eq.(\ref
{eilenberger}) can be obtained in some limiting cases, for example, in a
homogeneous case. However finding its solution for nonhomogeneous structures
with an arbitrary impurity concentration may be a quite difficult task.
Further simplifications can be made in the case of a dirty superconductor
when the energy $\tau ^{-1}$ related to the elastic scattering by
nonmagnetic impurities is larger than all other energies involved in the
problem, and the mean free path $l$ is smaller than all characteristic
lengths (except the Fermi wave length that is set in the quasiclassical
theory to zero). In this case one can expand the solution of Eq. (\ref
{eilenberger}) in terms of spherical harmonics and retain only the first two
of them, i.e.
\begin{equation}
\mathit{\check{g}}(\mathbf{p}_{\mathbf{F}},\mathbf{r;})=\mathit{\check{g}}%
_{s}(\mathbf{r})+(\mathbf{p}_{F}/p_{F})\mathbf{\check{g}}_{a}(\mathbf{r}%
)\;,  \label{gUs}
\end{equation}
where $\mathit{\check{g}}_{s}(\mathbf{r})$ is a matrix that
depends only on coordinates. The second term is the antisymmetric
part (the first Legendre polynomial) that determines the current.
It is assumed that the second term is smaller than the first one.
The parameter $l/x_{0}$ determines it's smallness, where $l$ is
the mean free path and $x_{0}$ is a characteristic length of the
problem. In $S/F$ structures $x_{0}\approx \sqrt{D_{F}/h\text{
}}$ is the shortest length because usually $h>\Delta .$ In the limit $%
l/x_{0}<<1,$ that is, if the product $h\tau $ is small, one can express $%
\mathbf{\check{g}}_{a}(\mathbf{r})$ from Eq.(\ref{eilenberger}) in
terms of $\check{g}_{s}(\mathbf{r})$
\begin{equation}
\mathbf{\check{g}}_{a}(\mathbf{r;}\epsilon ,\epsilon ^{\prime })=-l\mathit{%
\check{g}}_{s}(\mathbf{r;}\epsilon ,\epsilon _{1})\nabla \mathit{\check{g}}%
_{s}(\mathbf{r;}\epsilon _{1},\epsilon ^{\prime }),  \label{g_as}
\end{equation}
When obtaining Eq.(\ref{g_as}), we used the relations
\begin{eqnarray}
&&\mathit{\check{g}}_{s}(\mathbf{r;}\epsilon ,\epsilon _{1})\circ \mathit{%
\check{g}}_{s}(\mathbf{r;}\epsilon _{1},\epsilon ^{\prime })=1\;,
\label{NormCond1} \\
&&\mathbf{\check{g}}_{as}(\mathbf{r;}\epsilon ,\epsilon _{1})\circ \mathit{%
\check{g}}_{s}(\mathbf{r;}\epsilon _{1},\epsilon ^{\prime })+\mathit{%
\check{g}}_{s}(\mathbf{r;}\epsilon ,\epsilon _{1})\circ \mathbf{\check{g}}%
_{a}(\mathbf{r;}\epsilon _{1},\epsilon ^{\prime })=0  \nonumber
\end{eqnarray}
The symbolically written products in
Eqs.(\ref{g_as}),(\ref{NormCond1}) imply an integration over the
internal energy $\epsilon _{1}$ as it is shown in
Eq.(\ref{NormCond}).

The equation for the isotropic component of the Green's function
after averaging over the direction of the Fermi velocity
$\mathbf{v_{F}}$ reads
\begin{eqnarray}
-iD\nabla(\mbox{$\check{g}$}\nabla\mbox{$\check{g}$}) &+&i\left( \mbox{$\hat{\tau}_3$}\partial _{t}%
\mbox{$\check{g}$}+\partial _{t^{\prime }}\mbox{$\check{g}$}%
\mbox{$\hat{\tau}_3$}\right) +\left[ \check{\Delta},\mbox{$\check{g}$}\right]
+\left[ \mathbf{h}\mathbf{\check{S}},\mbox{$\check{g}$}\right]  \nonumber \\
&+&\frac{i}{\tau _{s.o.}}\left[ \mathbf{\check{S}}\mbox{$\hat{\tau}_3$}%
\mbox{$\check{g}$}\mbox{$\hat{\tau}_3$}\mathbf{\check{S}},\mbox{$\check{g}$}%
\right] =0\;,  \label{usadel}
\end{eqnarray}
where $D$ is the diffusion coefficient.

If we take the elements $(11)$ or $(22)$ of the supermatrix $\mathit{\check{g%
}}$, we obtain the Usadel equation for the retarded and advanced
Green's functions $\check{g}^{R(A)}(t,t^{\prime })$ generalized
for the case of the exchange field acting on the spins of
electrons. In this review we are
interested mainly in stationary processes, when the matrices $\check{g}%
^{R(A)}(t,t^{\prime })$ depend only on the time difference $(t-t^{\prime }).$
Performing the Fourier transformation $\check{g}^{R(A)}(\epsilon )=\int
d(t-t^{\prime })\check{g}^{R(A)}(t-t^{\prime })\exp (i\epsilon (t-t^{\prime
}))$, we obtain for $\check{g}^{R(A)}(\epsilon )$ the following equation (we
drop the indices $R(A)$)
\begin{widetext}
\begin{equation}
D\partial_x \left( \mbox{$\check{g}$}\partial_x \check{g}\right)
+i\epsilon \left[ \mbox{$\hat{\tau}_3$}\mbox{$\hat{\sigma}_0$},%
\mbox{$\check{g}$}\right] +ih\left\{ \left[ \mbox{$\hat{\tau}_3$}%
\mbox{$\hat{\sigma}_3$},\mbox{$\check{g}$}\right] \cos \alpha (x)+ \left[ \mbox{$\hat{\tau}_0$}\mbox{$\hat{\sigma}_2$},%
\mbox{$\check{g}$}\right] \sin \alpha (x)\right\} +i\left[ \check{\Delta},%
\mbox{$\check{g}$}\right] +\frac{i}{\tau _{s.o.}}\left[ \mathbf{\check{S}}%
\hat{\tau}_{3}\check{g}\hat{\tau}_{3}\mathbf{\check{S}},\check{g}\right]
=0\;.  \label{Usad}
\end{equation}
\end{widetext}
It is assumed here that $\mathbf{h}$ has the components $h(0,\sin \alpha
,\cos \alpha )$. This equation was first obtained by \citet{usadeleq} and it
is known as the Usadel equation. An inclusion of the exchange and spin-orbit
interaction was made in \cite{alexander,demler}.

Eq. (\ref{usadel}) can be solved analytically in many cases and it is used
in most of previous sections in order to describe different $S/F$
structures. Solutions for the Usadel equation must obey the normalization
condition
\begin{equation}
\check{g}(\mathbf{p}_{\mathbf{F}},\mathbf{r;}\epsilon ).\check{g}(\mathbf{p}%
_{\mathbf{F}},\mathbf{r;}\epsilon )=1  \label{NrmCondUs}
\end{equation}
The Usadel equation is complemented by the boundary conditions presented by %
\citet{kuprianov} on the basis of the Zaitsev's boundary conditions \cite
{zaitsev}. Various aspects of the boundary conditions have been discussed by %
\citet{lambert_bc,nazarov_bc,kopu,bauer02}. In the absence of spin-flip
processes at the interface they take the form:
\begin{equation}
\check{g}_{1}\partial _{x}\check{g}_{1}=\frac{1}{2\gamma _{a}}\left[ \check{g%
}_{1},\check{g}_{2}\right] \;,  \label{APboundary-condition}
\end{equation}
where $\gamma _{1}=R_{b}\sigma _{1}$, $\sigma _{1}$ is the conductivity of
the conductor $1$ and $R_{b}$ is the interface resistance per unit area, the
$x$-coordinate \ is assumed to be normal to the plane of the interface.

{The boundary condition (\ref{APboundary-condition}) implies that
we accept the simplest model of the S/F interface which is used in
most papers on S/F structures. We assume that the interface
separates two dirty regions: a singlet superconductor and a
ferromagnet. The superconductor and the ferromagnet are described
in the mean field approximation with different order parameters:
the off-diagonal order parameter }$\Delta ${\ in
the superconductor (in the weak coupling limit) and the exchange field }$\ h$%
{\ in the ferromagnet acting on the spins of free electrons. No
spin-flip scattering processes are assumed at the S/F interface. A
generalization of the boundary conditions we use to the case of a
spin-active S/F interface was carried out in the papers by
\citet{millis,kopu,fogel00,eschrig00}}.

Eqs. (\ref{usadel}) and (\ref{APboundary-condition}) together with
the self-consistency equation that determines the superconducting
order parameter $\Delta $, constitute a complete set of equations
from which one can obtain the Green's functions.

The Usadel equation can be solved in some particular cases. We
often use the linearized Usadel equation. In order to obtain the
linearized Usadel equation we represent the Green's functions
$\check{g}$ in the superconductor in the form
\begin{equation}
\check{g}(\mathbf{p}_{\mathbf{F}},\mathbf{r;}\omega )=\check{g}_{BCS}(\omega
)+\delta \check{g}_{S}+\mbox{$\delta
\check f_S$}\;,  \label{Linear}
\end{equation}
where $\check{g}_{BCS}(\omega )=\hat{\tau}_{3}g_{BCS}(\omega )+i\hat{\tau}%
_{2}f_{BCS}$, $g_{BCS}(\omega )=(i\omega /\Delta )f_{BCS},$ $f_{BCS}=\Delta
/i\sqrt{\omega ^{2}+\Delta ^{2}}$. We have written the matrix $\check{g}$ in
the so called Matsubara representation. This means that a substitution $%
\epsilon \Longrightarrow i\omega $ ($\omega =\pi T(2n+1),n=0,\pm 1,\pm 2,...$%
..) is done and $\check{g}(\omega )$ coincides with $\check{g}^{R}(\epsilon
) $ for positive $\omega $ and with $\check{g}^{A}(\epsilon )$ for negative $%
\omega $. The linearized Usadel equation has the form
\begin{equation}
\partial ^{2}_{xx}\mbox{$\delta \check f_S$}-\kappa _{S}^{2}%
\mbox{$\delta \check f_S$}=2i(\delta \breve{\Delta}/D_{S})g_{BCS}^{2}
\label{UsadS}
\end{equation}
in the $S$ region and
\begin{widetext}
\begin{equation}
\partial ^{2}_{xx}\mbox{$\delta \check f$}-\kappa _{\omega }^{2}%
\mbox{$\delta \check f$} +i\kappa
_{h}^{2}\left\{[\hat{\sigma}_3,\mbox{$\delta \check f$}]_{+} \cos
\alpha \pm \hat{\tau}_3[\hat{\sigma}_2,\mbox{$\delta \check f$}
]_{-}\sin \alpha \right\}=0  \label{UsadF}
\end{equation}
\end{widetext}
in the $F$ region. Here $\kappa _{S}^{2}=2E_{\omega }/D_{S}$, $\kappa
_{\omega }^{2}=2|\omega |/D_{F}$, $\kappa _{h}^{2}=h$ $\mbox{${\rm sgn}$}%
\omega /D_{F}$ and $[A,B]_{\pm }=AB\pm BA,\delta \breve{\Delta}=i\hat{\tau}%
_{2}\hat{\sigma}_{3}\delta \Delta $. The signs $\pm $ in Eq. (\ref{UsadF})
correspond to the right and left layer respectively.

\bigskip The boundary conditions for $\mbox{$\delta \check f_S$}$ and $%
\mbox{$\delta \check f$}_{F}\equiv f$ (in zero-order approximation $%
\mbox{$\delta \check f$}_{F}=0$) are obtained from Eq.(\ref
{APboundary-condition}). They have the form
\begin{equation}
\partial_x \mbox{$\delta \check f_S$}=(1/\gamma _{S})[g_{BCS}^{2}%
\mbox{$\delta \check f$}-g_{BCS}\hat{f}_{BCS}\hat{\sigma}_{3}g_{F3}-g_{BCS}%
\mbox{$\check f_S$}]  \label{BCs}
\end{equation}
\begin{equation}
\partial_x \mbox{$ \check f_F$}=(1/\gamma _{F})[g_{BCS}%
\mbox{$\delta \check f$}-\mbox{$\check f_S$}]  \label{BCf}
\end{equation}

where $\gamma _{F,S}=R_{b}\sigma _{F,S}$.

If the Green's functions are known, one can calculate macroscopic
quantities such as the current, magnetic moment etc. For example,
the current is given by \citet{lo_book}
\begin{equation}
I_{S}=(L_{y}L_{z}/16)\sigma _{F}\mathrm{Tr}\left( \mbox{$\hat{\tau}_3$}%
\mbox{$\hat{\sigma}_0$}\right) \int d\epsilon (\mathit{\check{g}}%
_{s}\partial \mathit{\check{g}}_{s}/\partial x)_{12}
\label{gen_current}
\end{equation}
where $L_{y,z}$ are the widths of the films in $y$ and $z$ direction (the
current flows in the transverse $x$-direction) and subscript $(12)$ shows
that one has to take the Keldysh component of the supermatrix $\mathit{%
\check{g}}_{s}\partial \mathit{\check{g}}_{s}/\partial x$. A
variation of the magnetic moment due to proximity effect is
determined by formulae
\begin{equation}
\delta M_{z}=\mu _{B}\nu (1/2)i\pi T\sum_{\omega
}Tr(\hat{\tau}_{3}\otimes \hat{\sigma}_{3}\delta \check{g})
\label{M_z}
\end{equation}
\begin{equation}
\delta M_{x,y}=\mu _{B}\nu (1/2)i\pi T\sum_{\omega }Tr(\hat{\tau}%
_{0}\otimes \hat{\sigma}_{1,2}\delta \check{g})  \label{M_x,y}
\end{equation}
where $\nu $ is the density-of-states at the Fermi level in the normal state
and $\mu _{B}=g\mu _{Bohr}$ is an effective Bohr magneton.

{Finally, it is important to make remarks concerning the notations used in
this review. In most works where the }${S/F}${\ structures with homogeneous
magnetization are studied, the Green's function $\check{g}$ is a }${2\times 2}$%
{\ matrix with the usual normal and Gor'kov's components. Of
course, this simplification can be made provided the
magnetizations of the }${F}${\ layers involved in the problem are
aligned in one direction. } {However, this simple form leads to
erroneous results if the magnetizations are arbitrarily oriented
with respect to each other. The }${4\times 4}${\ form of the
Green's function is unavoidable if one studies structures with a
non-homogeneous magnetization. Of course, the $c$- operators in
Eq. (\ref
{c_operators}) can be defined in different ways. For example, %
\citet{maki_book} introduced a spinor representation of the field operators,
which is equivalent to letting in Eq. (\ref{c_operators}) the spin index of
the operator $a$ unchanged when $n=2$. This notation was used in later works
\cite[e.g]{demler,alexander} in which the Green's functions have a }${2\times 2%
}${\ block matrix form. The diagonal blocks represent the normal
Green's functions, while the off-diagonal blocks represent the
anomalous one. With this notation the matrix, Eq. (\ref{H_2}),
changes its form. For example, the term containing $\Delta $ is
proportional to $i\mbox{$\hat{\sigma}_2$}$ and not to
$\mbox{$\hat{\sigma}_3$}$. The choice of the notation depends on
the problem to solve. In order to study the triplet
superconductivity
induced in }${S/F}${\ systems and to see explicitly the three projections ($%
S_{z}=0,\pm 1$) of the condensate function it is more convenient to use the
operators defined in Eq.(\ref{c_operators}) (see for example %
\citet{golubov_triplet} and \citet{BVE_josephson}). }

\section{Future direction of the experimental
research}\label{ApB}

As we have seen throughout the paper there is a great number of
experiments on S/F structures. The variety of superconducting and
ferromagnetic materials is very large. In this section we review
briefly some of these experiments. We will not dwell on specific
fabrication techniques but rather focus on the discussion: which
pairs of material (S and F) are more appropriate for the
observation of the effects studied in this review.

 First
experiments on S/F structures used strong ferromagnets (large
exchange fields) as Fe, Ni, Co or Gd and conventional
superconductors like Nb, Pb, V, etc. \cite{werthamer_sf}. In these
experiments the dependence of the superconducting transition
temperature on the thicknesses of the S and F layers has been
measured. In other words the suppression of the superconductivity
due to the strong exchange field of the ferromagnet was analyzed.
It is clear that for such strong ferromagnets the spin splitting
is large and therefore a mismatch in electronic parameters of the
S and F regions is large. This leads to a low interface
transparency and a weak
proximity effect. This was confirmed by \citet{aarts} in experiments on V/V$%
_{1-x}$Fe$_{x}$ multilayers. Varying the concentration of Fe in
the VFe alloys they could change the values of the exchange field
and indirectly the transparency of the interface. Such systems
consisting of a conventional superconductor and a ferromagnetic
alloy, both with similar band structure (in the above experiment
the mismatch was $<5\%$) , are good candidates for observing the
effects discussed in sections \ref{JOS-TRI}., \ref{RED-INV}. and
\ref{RED-SCR}.

Weak ferromagnets have been used in the last years in many
experiments on S/F structures. Before we turn our attention to
ferromagnets
with small exchange fields it is worth mentioning the experiment by %
\citet{rusanov}. They analyzed the so-called spin switch effect.
In particular they studied the transport properties of
Permalloy(Py)/Nb bilayers. They observed an enhancement of
superconductivity in the resistive transition in the field range
where the magnetization of the Py switches and many domains are
present. Interesting for us is that Py shows a well defined
magnetization switching at low fields and therefore it could be
used in order to detect the long-range triplet component that
appears when the magnetization of the ferromagnet is not
homogeneous (see section \ref {EXO-INH}). Finally, a magnetic
configuration analysis of the strong-ferromagnetic structures used
in transport experiments as those performed by
\citet{petrashov,pannetier} may also serve to confirm the
predictions of section \ref{EXO-INH}. As it was discussed before
the increase in the conductance of the ferromagnet for
temperatures below the superconducting $T_{c}$ may be explained
assuming a long-range proximity effect.

The proximity effect in S/F is stronger if one uses dilute
ferromagnetic alloys. Thus, such materials are the best candidates
in order to observed most of the effects discussed in this review.
The idea of using ferromagnetic alloys with small exchanges field
was used by \citet{ryazanov}. They were the first in observing the
sign reversal of the critical current
in a S/F/S Josephson junction. Nb was used as superconductor while Cu$%
_{0.48} $Ni$_{0.52}$ alloy as a ferromagnet (exchange field $\sim 25K$).
(Later on similar results were obtained by \citet{kontos02} on Nb/Al/Al$_{2}$%
O$_{3}$/PdNi/Nb structures). The CuNi alloy was also used in the
experiment by \citet{argonne} on F/S/F structures. In this
experiment the authors determined the dependence of the
superconducting transition temperature on the relative
magnetization-orientation of the two F layers. In order to get
different alignments between the two CuNi layers an
exchange-biased spin-valve stack of
CuNi/Nb/CuNi/Fe$_{50}$Mn$_{50}$ was employed. With a small
magnetic field the authors could switch the magnetization
direction of the free NiCu layer. This technique could be very
useful in order to observe the Josephson coupling via the triplet
component as described in section \ref {JOS-TRI}.

Finally, it is worth mentioning the experiment by \citet{bernhard}
on YBa$_{2}$Cu$_{3}$O$_{7}$/La$_{2/3}$Ca$_{1/3}$MnO$_{3}$. Using
the neutron reflectometry technique they observed a induced
magnetic moment in the superconductor. Although the materials
employed in this experiment cannot be quantitatively described
with the methods presented in this review (the ferromagnet used by
\citet{bernhard} is a half-metal with a exchange field comparable
to the Fermi-energy and the superconductor is unconventional), the
experimental technique may be used in other experiments in order
to detected the induced magnetization predicted in sections
\ref{RED-INV}. and \ref{RED-SCR}.

\section*{List of Symbols and Abbreviations}

\begin{tabular}[t]{lp{20mm}p{11cm}}
S &  & superconductor \\
N &  & nonmagnetic normal metal \\
F &  & ferromagnetic metal \\
I &  & insulator \\
LRTC &  & long-range triplet component \\
$\hat{\tau}_i$, $i=1,2,3$ &  & pauli matrices in particle-hole space \\
$\hat{\sigma}_i$, $i=1,2,3$ &  & Pauli matrices in spin space \\
$\hat{\tau}_0$, $\hat{\sigma}_0$ &  & unit matrices. \\
$D$ &  & diffusion coefficient \\
$\nu$ &  & density of states \\
$\omega=\pi T(2n+1)$ &  & Matsubara Frequency \\
$\epsilon$ &  & real frequency (energy) \\
$g_{BCS}$ &  & quasiclassical normal Green's function for a bulk
superconductor \\
$f_{BCS}$ &  & quasiclassical anomalous Green's function for a bulk
superconductor \\
$T_c$ &  & superconducting critical temperature \\
$I_c$ &  & Josephson critical current \\
$R_b$ &  & interface resistance per unit area \\
$\epsilon_{bN}=D_N/2R_b\sigma_Nd_N$ &  & minigap induced in a normal metal
\\
$\sigma_{S,F}$ &  & conductivity in the normal state \\
$\gamma_{S,F}$ &  & $R_b\sigma_{S,F}$ \\
$\gamma$ &  & ratio $\sigma_F/\sigma_S$ \\
$\mathcal{J}$ &  & magnetic coupling between localized magnetic moments. \\
$h$ &  & exchange field acting on the spin of conducting electrons \\
$\xi_N=\sqrt{\frac{D_N}{2\pi T}}$ &  & characteristic penetration length of
the condensate into a dirty normal metal \\
$\xi_F=\sqrt{\frac{D_F}{h}}$ &  & characteristic penetration length of the
condensate into a dirty ferromagnet \\
$\xi_S=\sqrt{\frac{D_S}{2\pi T_c}}$ &  & superconducting coherence length
for a dirty superconductor
\end{tabular}

\clearpage


\begin{thebibliography}{233}
\expandafter\ifx\csname
natexlab\endcsname\relax\def\natexlab#1{#1}\fi
\expandafter\ifx\csname bibnamefont\endcsname\relax
  \def\bibnamefont#1{#1}\fi
\expandafter\ifx\csname bibfnamefont\endcsname\relax
  \def\bibfnamefont#1{#1}\fi
\expandafter\ifx\csname citenamefont\endcsname\relax
  \def\citenamefont#1{#1}\fi
\expandafter\ifx\csname url\endcsname\relax
  \def\url#1{\texttt{#1}}\fi
\expandafter\ifx\csname
urlprefix\endcsname\relax\def\urlprefix{URL }\fi
\providecommand{\bibinfo}[2]{#2}
\providecommand{\eprint}[2][]{\url{#2}}

\bibitem[{\citenamefont{Aarts} \emph{et~al.}(1997)\citenamefont{Aarts, Geers,
  Br\"uck, Golubov, and Coehoorn}}]{aarts}
\bibinfo{author}{\bibnamefont{Aarts}, \bibfnamefont{J.}},
  \bibinfo{author}{\bibfnamefont{J.~M.~E.} \bibnamefont{Geers}},
  \bibinfo{author}{\bibfnamefont{E.}~\bibnamefont{Br\"uck}},
  \bibinfo{author}{\bibfnamefont{A.~A.} \bibnamefont{Golubov}}, and
  \bibinfo{author}{\bibfnamefont{R.}~\bibnamefont{Coehoorn}},
  \bibinfo{year}{1997}, \bibinfo{journal}{Phys. Rev. B}
  \textbf{\bibinfo{volume}{56}}, \bibinfo{pages}{2779}.

\bibitem[{\citenamefont{Abrahams} \emph{et~al.}(1993)\citenamefont{Abrahams,
  Balatsky, Schrieffer, and Allen}}]{abrahams}
\bibinfo{author}{\bibnamefont{Abrahams}, \bibfnamefont{E.}},
  \bibinfo{author}{\bibfnamefont{A.~V.} \bibnamefont{Balatsky}},
  \bibinfo{author}{\bibfnamefont{J.~R.} \bibnamefont{Schrieffer}}, and
  \bibinfo{author}{\bibfnamefont{P.~B.} \bibnamefont{Allen}},
  \bibinfo{year}{1993}, \bibinfo{journal}{Phys. Rev. B}
  \textbf{\bibinfo{volume}{47}}, \bibinfo{pages}{513}.

\bibitem[{\citenamefont{Abrikosov}(1988)}]{abrikosov_book}
\bibinfo{author}{\bibnamefont{Abrikosov}, \bibfnamefont{A.~A.}},
  \bibinfo{year}{1988}, \emph{\bibinfo{title}{Fundamentals of the Theory of
  Metals}} (\bibinfo{publisher}{North-Holland}, \bibinfo{address}{Amsterdam}).

\bibitem[{\citenamefont{Abrikosov and Gor'kov}(1962)}]{AG_knight}
\bibinfo{author}{\bibnamefont{Abrikosov}, \bibfnamefont{A.~A.}}, and
  \bibinfo{author}{\bibfnamefont{L.~P.} \bibnamefont{Gor'kov}},
  \bibinfo{year}{1962}, \bibinfo{journal}{Zh. Eksp. Teor. Fiz.}
  \textbf{\bibinfo{volume}{42}}, \bibinfo{pages}{1088}, \bibinfo{note}{[Sov.
  Phys. JETP {\bf 15},752 (1962)]}.

\bibitem[{\citenamefont{Adkins and Kington}(1969)}]{adkins}
\bibinfo{author}{\bibnamefont{Adkins}, \bibfnamefont{C.~J.}}, and
  \bibinfo{author}{\bibfnamefont{B.~W.} \bibnamefont{Kington}},
  \bibinfo{year}{1969}, \bibinfo{journal}{Phys. Rev.}
  \textbf{\bibinfo{volume}{177}}, \bibinfo{pages}{777}.

\bibitem[{\citenamefont{Aharoni}(1996)}]{aharoni}
\bibinfo{author}{\bibnamefont{Aharoni}, \bibfnamefont{A.}},
  \bibinfo{year}{1996}, \emph{\bibinfo{title}{Introduction to the Theory of
  Ferromagnetism}} (\bibinfo{publisher}{Claredon Press (Oxford)}).

\bibitem[{\citenamefont{Alexander} \emph{et~al.}(1985)\citenamefont{Alexander,
  Orlando, Rainer, and Tedrow}}]{alexander}
\bibinfo{author}{\bibnamefont{Alexander}, \bibfnamefont{J.~A.~X.}},
  \bibinfo{author}{\bibfnamefont{T.~P.} \bibnamefont{Orlando}},
  \bibinfo{author}{\bibfnamefont{D.}~\bibnamefont{Rainer}}, and
  \bibinfo{author}{\bibfnamefont{P.~M.} \bibnamefont{Tedrow}},
  \bibinfo{year}{1985}, \bibinfo{journal}{Phys. Rev. B}
  \textbf{\bibinfo{volume}{31}}, \bibinfo{pages}{5811}.

\bibitem[{\citenamefont{Altland} \emph{et~al.}(2000)\citenamefont{Altland,
  Taras-Semchuk, and Simons}}]{altland}
\bibinfo{author}{\bibnamefont{Altland}, \bibfnamefont{A.}},
  \bibinfo{author}{\bibfnamefont{D.}~\bibnamefont{Taras-Semchuk}}, and
  \bibinfo{author}{\bibfnamefont{B.~D.} \bibnamefont{Simons}},
  \bibinfo{year}{2000}, \bibinfo{journal}{Adv. Phys.}
  \textbf{\bibinfo{volume}{49}}, \bibinfo{pages}{321}.

\bibitem[{\citenamefont{Anderson and Suhl}(1959)}]{anderson}
\bibinfo{author}{\bibnamefont{Anderson}, \bibfnamefont{P.}}, and
  \bibinfo{author}{\bibfnamefont{H.}~\bibnamefont{Suhl}}, \bibinfo{year}{1959},
  \bibinfo{journal}{Phys. Rev.} \textbf{\bibinfo{volume}{116}},
  \bibinfo{pages}{898}.

\bibitem[{\citenamefont{Andreev}(1964)}]{andreev}
\bibinfo{author}{\bibnamefont{Andreev}, \bibfnamefont{A.}},
  \bibinfo{year}{1964}, \bibinfo{journal}{Sov. Phys. JETP}
  \textbf{\bibinfo{volume}{19}}, \bibinfo{pages}{1228}.

\bibitem[{\citenamefont{Androes and Knight}(1961)}]{knight}
\bibinfo{author}{\bibnamefont{Androes}, \bibfnamefont{G.~M.}}, and
  \bibinfo{author}{\bibfnamefont{W.~D.} \bibnamefont{Knight}},
  \bibinfo{year}{1961}, \bibinfo{journal}{Phys. Rev.}
  \textbf{\bibinfo{volume}{121}}, \bibinfo{pages}{779}.

\bibitem[{\citenamefont{Anthore} \emph{et~al.}(2003)\citenamefont{Anthore,
  Pothier, and Esteve}}]{esteve03}
\bibinfo{author}{\bibnamefont{Anthore}, \bibfnamefont{A.}},
  \bibinfo{author}{\bibfnamefont{H.}~\bibnamefont{Pothier}}, and
  \bibinfo{author}{\bibfnamefont{D.}~\bibnamefont{Esteve}},
  \bibinfo{year}{2003}, \bibinfo{journal}{Phys. Rev. Lett.}
  \textbf{\bibinfo{volume}{90}}, \bibinfo{pages}{127001}.

\bibitem[{\citenamefont{Artemenko} \emph{et~al.}(1979)\citenamefont{Artemenko,
  Volkov, and Zaitsev}}]{artemenko}
\bibinfo{author}{\bibnamefont{Artemenko}, \bibfnamefont{S.~N.}},
  \bibinfo{author}{\bibfnamefont{A.}~\bibnamefont{Volkov}}, and
  \bibinfo{author}{\bibfnamefont{A.~V.} \bibnamefont{Zaitsev}},
  \bibinfo{year}{1979}, \bibinfo{journal}{Sol. St. Comm.}
  \textbf{\bibinfo{volume}{30}}, \bibinfo{pages}{771}.

\bibitem[{\citenamefont{Artemenko and Volkov}(1980)}]{artemenko_volkov}
\bibinfo{author}{\bibnamefont{Artemenko}, \bibfnamefont{S.~N.}}, and
  \bibinfo{author}{\bibfnamefont{A.~F.} \bibnamefont{Volkov}},
  \bibinfo{year}{1980}, \bibinfo{journal}{Sov. Phys. Uspejhi}
  \textbf{\bibinfo{volume}{22}}, \bibinfo{pages}{295}.

\bibitem[{\citenamefont{Aumentado and Chandrasekhar}(2001)}]{chandrasekhar}
\bibinfo{author}{\bibnamefont{Aumentado}, \bibfnamefont{J.}}, and
  \bibinfo{author}{\bibfnamefont{V.}~\bibnamefont{Chandrasekhar}},
  \bibinfo{year}{2001}, \bibinfo{journal}{Phys. Rev. B}
  \textbf{\bibinfo{volume}{64}}, \bibinfo{pages}{054505}.

\bibitem[{\citenamefont{Bagrets} \emph{et~al.}(2003)\citenamefont{Bagrets,
  Lacroix, and Vedyayev}}]{bagrets}
\bibinfo{author}{\bibnamefont{Bagrets}, \bibfnamefont{A.}},
  \bibinfo{author}{\bibfnamefont{C.}~\bibnamefont{Lacroix}}, and
  \bibinfo{author}{\bibfnamefont{A.}~\bibnamefont{Vedyayev}},
  \bibinfo{year}{2003}, \bibinfo{journal}{Phys. Rev. B}
  \textbf{\bibinfo{volume}{68}}, \bibinfo{pages}{054532}.

\bibitem[{\citenamefont{Baladie and Buzdin}(2001)}]{baladie}
\bibinfo{author}{\bibnamefont{Baladie}, \bibfnamefont{I.}}, and
  \bibinfo{author}{\bibfnamefont{A.}~\bibnamefont{Buzdin}},
  \bibinfo{year}{2001}, \bibinfo{journal}{Phys. Rev. B}
  \textbf{\bibinfo{volume}{64}}, \bibinfo{pages}{224514}.

\bibitem[{\citenamefont{Baladie and Buzdin}(2003)}]{baladie_tc}
\bibinfo{author}{\bibnamefont{Baladie}, \bibfnamefont{I.}}, and
  \bibinfo{author}{\bibfnamefont{A.}~\bibnamefont{Buzdin}},
  \bibinfo{year}{2003}, \bibinfo{journal}{Phys. Rev. B}
  \textbf{\bibinfo{volume}{67}}, \bibinfo{pages}{014523}.

\bibitem[{\citenamefont{Baladie} \emph{et~al.}(2001)\citenamefont{Baladie,
  Buzdin, Ryzhanova, and Vedyayev}}]{buzdin_wr}
\bibinfo{author}{\bibnamefont{Baladie}, \bibfnamefont{I.}},
  \bibinfo{author}{\bibfnamefont{A.}~\bibnamefont{Buzdin}},
  \bibinfo{author}{\bibfnamefont{N.}~\bibnamefont{Ryzhanova}}, and
  \bibinfo{author}{\bibfnamefont{A.}~\bibnamefont{Vedyayev}},
  \bibinfo{year}{2001}, \bibinfo{journal}{Phys. Rev. B}
  \textbf{\bibinfo{volume}{63}}, \bibinfo{pages}{054518}.

\bibitem[{\citenamefont{Balatsky and Abrahams}(1992)}]{balatsky92}
\bibinfo{author}{\bibnamefont{Balatsky}, \bibfnamefont{A.}}, and
  \bibinfo{author}{\bibfnamefont{E.}~\bibnamefont{Abrahams}},
  \bibinfo{year}{1992}, \bibinfo{journal}{Phys. Rev. B}
  \textbf{\bibinfo{volume}{45}}, \bibinfo{pages}{13125}.

\bibitem[{\citenamefont{Balatsky} \emph{et~al.}(1995)\citenamefont{Balatsky,
  Abrahams, Scalapino, and Schrieffer}}]{balatsky95}
\bibinfo{author}{\bibnamefont{Balatsky}, \bibfnamefont{A.}},
  \bibinfo{author}{\bibfnamefont{E.}~\bibnamefont{Abrahams}},
  \bibinfo{author}{\bibfnamefont{D.~J.} \bibnamefont{Scalapino}}, and
  \bibinfo{author}{\bibfnamefont{J.~R.} \bibnamefont{Schrieffer}},
  \bibinfo{year}{1995}, \bibinfo{journal}{Phys. Rev. B}
  \textbf{\bibinfo{volume}{52}}, \bibinfo{pages}{1271}.

\bibitem[{\citenamefont{Barash} \emph{et~al.}(2002)\citenamefont{Barash,
  Bobkova, and T.Kopp}}]{barash}
\bibinfo{author}{\bibnamefont{Barash}, \bibfnamefont{Y.~S.}},
  \bibinfo{author}{\bibfnamefont{I.~V.} \bibnamefont{Bobkova}}, and
  \bibinfo{author}{\bibnamefont{T.Kopp}}, \bibinfo{year}{2002},
  \bibinfo{journal}{Phys. Rev. B} \textbf{\bibinfo{volume}{66}},
  \bibinfo{pages}{140503}.

\bibitem[{\citenamefont{Bardeen} \emph{et~al.}(1957)\citenamefont{Bardeen,
  Cooper, and Schrieffer}}]{BCS}
\bibinfo{author}{\bibnamefont{Bardeen}, \bibfnamefont{J.}},
  \bibinfo{author}{\bibfnamefont{L.~N.} \bibnamefont{Cooper}}, and
  \bibinfo{author}{\bibfnamefont{J.~R.} \bibnamefont{Schrieffer}},
  \bibinfo{year}{1957}, \bibinfo{journal}{Phys. Rev.}
  \textbf{\bibinfo{volume}{106}}, \bibinfo{pages}{162}.

\bibitem[{\citenamefont{Barone and Paterno}(1982)}]{barone_book}
\bibinfo{author}{\bibnamefont{Barone}, \bibfnamefont{A.}}, and
  \bibinfo{author}{\bibfnamefont{G.}~\bibnamefont{Paterno}},
  \bibinfo{year}{1982}, \emph{\bibinfo{title}{Physics and Applications of the
  Josephson Effect}} (\bibinfo{publisher}{Wiley}, \bibinfo{address}{New York}).

\bibitem[{\citenamefont{Baselmans} \emph{et~al.}(1999)\citenamefont{Baselmans,
  Morpurgo, van Wees, and Klapwijk}}]{vanWees}
\bibinfo{author}{\bibnamefont{Baselmans}, \bibfnamefont{J.~J.~A.}},
  \bibinfo{author}{\bibfnamefont{A.}~\bibnamefont{Morpurgo}},
  \bibinfo{author}{\bibfnamefont{B.~J.} \bibnamefont{van Wees}}, and
  \bibinfo{author}{\bibfnamefont{T.~M.} \bibnamefont{Klapwijk}},
  \bibinfo{year}{1999}, \bibinfo{journal}{Nature}
  \textbf{\bibinfo{volume}{397}}, \bibinfo{pages}{43}.

\bibitem[{\citenamefont{Bauer} \emph{et~al.}(2004)\citenamefont{Bauer, Bentner,
  Aprili, Rocca, Reinwald, Wegscheider, and Strunk}}]{strunk04}
\bibinfo{author}{\bibnamefont{Bauer}, \bibfnamefont{A.}},
  \bibinfo{author}{\bibfnamefont{J.}~\bibnamefont{Bentner}},
  \bibinfo{author}{\bibfnamefont{M.}~\bibnamefont{Aprili}},
  \bibinfo{author}{\bibfnamefont{M.~L.~D.} \bibnamefont{Rocca}},
  \bibinfo{author}{\bibfnamefont{M.}~\bibnamefont{Reinwald}},
  \bibinfo{author}{\bibfnamefont{W.}~\bibnamefont{Wegscheider}}, and
  \bibinfo{author}{\bibfnamefont{C.}~\bibnamefont{Strunk}},
  \bibinfo{year}{2004}, \bibinfo{journal}{Physical Review Letters}
  \textbf{\bibinfo{volume}{92}}, \bibinfo{pages}{217001}.

\bibitem[{\citenamefont{Beckmann} \emph{et~al.}(2004)\citenamefont{Beckmann,
  Weber, and v.~L\"{o}hneysen}}]{Beck}
\bibinfo{author}{\bibnamefont{Beckmann}, \bibfnamefont{D.}},
  \bibinfo{author}{\bibfnamefont{H.~B.} \bibnamefont{Weber}}, and
  \bibinfo{author}{\bibfnamefont{H.}~\bibnamefont{v.~L\"{o}hneysen}},
  \bibinfo{year}{2004}, \bibinfo{journal}{cond-mat/0404360} .

\bibitem[{\citenamefont{Bednorz and M\"{u}ller}(1986)}]{mueller}
\bibinfo{author}{\bibnamefont{Bednorz}, \bibfnamefont{J.~G.}}, and
  \bibinfo{author}{\bibfnamefont{K.~A.} \bibnamefont{M\"{u}ller}},
  \bibinfo{year}{1986}, \bibinfo{journal}{Z. Phys. B}
  \textbf{\bibinfo{volume}{64}}, \bibinfo{pages}{189}.

\bibitem[{\citenamefont{Beenakker}(1997)}]{beenakker_rev}
\bibinfo{author}{\bibnamefont{Beenakker}, \bibfnamefont{C.~W.~J.}},
  \bibinfo{year}{1997}, \bibinfo{journal}{Rev. Mod. Phys.}
  \textbf{\bibinfo{volume}{69}}, \bibinfo{pages}{731}.

\bibitem[{\citenamefont{Belzig} \emph{et~al.}(2000)\citenamefont{Belzig,
  Brataas, Nazarov, and Bauer}}]{nazarov_belzig}
\bibinfo{author}{\bibnamefont{Belzig}, \bibfnamefont{W.}},
  \bibinfo{author}{\bibfnamefont{A.}~\bibnamefont{Brataas}},
  \bibinfo{author}{\bibfnamefont{Y.~V.} \bibnamefont{Nazarov}}, and
  \bibinfo{author}{\bibfnamefont{G.~E.~W.} \bibnamefont{Bauer}},
  \bibinfo{year}{2000}, \bibinfo{journal}{Phys. Rev. B}
  \textbf{\bibinfo{volume}{62}}, \bibinfo{pages}{9726}.

\bibitem[{\citenamefont{Belzig} \emph{et~al.}(1999)\citenamefont{Belzig,
  Wilhelm, Bruder, Schön, and Zaikin}}]{belzig_rev}
\bibinfo{author}{\bibnamefont{Belzig}, \bibfnamefont{W.}},
  \bibinfo{author}{\bibfnamefont{F.}~\bibnamefont{Wilhelm}},
  \bibinfo{author}{\bibfnamefont{C.}~\bibnamefont{Bruder}},
  \bibinfo{author}{\bibfnamefont{G.}~\bibnamefont{Schön}}, and
  \bibinfo{author}{\bibfnamefont{A.}~\bibnamefont{Zaikin}},
  \bibinfo{year}{1999}, \bibinfo{journal}{Superlattices and Microstructures}
  \textbf{\bibinfo{volume}{25}}, \bibinfo{pages}{1251}.

\bibitem[{\citenamefont{Berezinskii}(1975)}]{berezin}
\bibinfo{author}{\bibnamefont{Berezinskii}, \bibfnamefont{V.~L.}},
  \bibinfo{year}{1975}, \bibinfo{journal}{JETP Lett.}
  \textbf{\bibinfo{volume}{20}}, \bibinfo{pages}{287}.

\bibitem[{\citenamefont{Bergeret} \emph{et~al.}(2005)\citenamefont{Bergeret,
  Yeyati, and Mart\'{i}n-Rodero}}]{bergeret05}
\bibinfo{author}{\bibnamefont{Bergeret}, \bibfnamefont{F.}},
  \bibinfo{author}{\bibfnamefont{A.~L.} \bibnamefont{Yeyati}}, and
  \bibinfo{author}{\bibfnamefont{A.}~\bibnamefont{Mart\'{i}n-Rodero}},
  \bibinfo{year}{2005}, \bibinfo{journal}{to be published} .

\bibitem[{\citenamefont{Bergeret} \emph{et~al.}(2000)\citenamefont{Bergeret,
  Efetov, and Larkin}}]{BEL}
\bibinfo{author}{\bibnamefont{Bergeret}, \bibfnamefont{F.~S.}},
  \bibinfo{author}{\bibfnamefont{K.~B.} \bibnamefont{Efetov}}, and
  \bibinfo{author}{\bibfnamefont{A.~I.} \bibnamefont{Larkin}},
  \bibinfo{year}{2000}, \bibinfo{journal}{Phys. Rev. B}
  \textbf{\bibinfo{volume}{62}}, \bibinfo{pages}{11872}.

\bibitem[{\citenamefont{Bergeret and Garc\'ia}(2004)}]{garcia}
\bibinfo{author}{\bibnamefont{Bergeret}, \bibfnamefont{F.~S.}}, and
  \bibinfo{author}{\bibfnamefont{N.}~\bibnamefont{Garc\'ia}},
  \bibinfo{year}{2004}, \bibinfo{journal}{Phys. Rev. B}
  \textbf{\bibinfo{volume}{70}}, \bibinfo{pages}{052507}.

\bibitem[{\citenamefont{Bergeret}
  \emph{et~al.}(2002{\natexlab{a}})\citenamefont{Bergeret, Pavlovskii, Volkov,
  and Efetov}}]{BPVE}
\bibinfo{author}{\bibnamefont{Bergeret}, \bibfnamefont{F.~S.}},
  \bibinfo{author}{\bibfnamefont{V.~V.} \bibnamefont{Pavlovskii}},
  \bibinfo{author}{\bibfnamefont{A.~F.} \bibnamefont{Volkov}}, and
  \bibinfo{author}{\bibfnamefont{K.~B.} \bibnamefont{Efetov}},
  \bibinfo{year}{2002}{\natexlab{a}}, \bibinfo{journal}{Int. Jour. Mod. Phys.
  B} \textbf{\bibinfo{volume}{16}}, \bibinfo{pages}{1459}.

\bibitem[{\citenamefont{Bergeret}
  \emph{et~al.}(2001{\natexlab{a}})\citenamefont{Bergeret, Volkov, and
  Efetov}}]{BVE2}
\bibinfo{author}{\bibnamefont{Bergeret}, \bibfnamefont{F.~S.}},
  \bibinfo{author}{\bibfnamefont{A.~F.} \bibnamefont{Volkov}}, and
  \bibinfo{author}{\bibfnamefont{K.~B.} \bibnamefont{Efetov}},
  \bibinfo{year}{2001}{\natexlab{a}}, \bibinfo{journal}{Phys. Rev. Lett.}
  \textbf{\bibinfo{volume}{86}}, \bibinfo{pages}{4096}.

\bibitem[{\citenamefont{Bergeret}
  \emph{et~al.}(2001{\natexlab{b}})\citenamefont{Bergeret, Volkov, and
  Efetov}}]{BVE1}
\bibinfo{author}{\bibnamefont{Bergeret}, \bibfnamefont{F.~S.}},
  \bibinfo{author}{\bibfnamefont{A.~F.} \bibnamefont{Volkov}}, and
  \bibinfo{author}{\bibfnamefont{K.~B.} \bibnamefont{Efetov}},
  \bibinfo{year}{2001}{\natexlab{b}}, \bibinfo{journal}{Phys. Rev. Lett.}
  \textbf{\bibinfo{volume}{86}}, \bibinfo{pages}{3140}.

\bibitem[{\citenamefont{Bergeret}
  \emph{et~al.}(2001{\natexlab{c}})\citenamefont{Bergeret, Volkov, and
  Efetov}}]{BVE_josephson}
\bibinfo{author}{\bibnamefont{Bergeret}, \bibfnamefont{F.~S.}},
  \bibinfo{author}{\bibfnamefont{A.~F.} \bibnamefont{Volkov}}, and
  \bibinfo{author}{\bibfnamefont{K.~B.} \bibnamefont{Efetov}},
  \bibinfo{year}{2001}{\natexlab{c}}, \bibinfo{journal}{Phys. Rev. B}
  \textbf{\bibinfo{volume}{64}}, \bibinfo{pages}{134506}.

\bibitem[{\citenamefont{Bergeret}
  \emph{et~al.}(2002{\natexlab{b}})\citenamefont{Bergeret, Volkov, and
  Efetov}}]{BVE_dos}
\bibinfo{author}{\bibnamefont{Bergeret}, \bibfnamefont{F.~S.}},
  \bibinfo{author}{\bibfnamefont{A.~F.} \bibnamefont{Volkov}}, and
  \bibinfo{author}{\bibfnamefont{K.~B.} \bibnamefont{Efetov}},
  \bibinfo{year}{2002}{\natexlab{b}}, \bibinfo{journal}{Phys. Rev. B}
  \textbf{\bibinfo{volume}{65}}, \bibinfo{pages}{134505}.

\bibitem[{\citenamefont{Bergeret} \emph{et~al.}(2003)\citenamefont{Bergeret,
  Volkov, and Efetov}}]{BVE_manifestation}
\bibinfo{author}{\bibnamefont{Bergeret}, \bibfnamefont{F.~S.}},
  \bibinfo{author}{\bibfnamefont{A.~F.} \bibnamefont{Volkov}}, and
  \bibinfo{author}{\bibfnamefont{K.~B.} \bibnamefont{Efetov}},
  \bibinfo{year}{2003}, \bibinfo{journal}{Phys. Rev. B}
  \textbf{\bibinfo{volume}{68}}, \bibinfo{pages}{064513}.

\bibitem[{\citenamefont{Bergeret}
  \emph{et~al.}(2004{\natexlab{a}})\citenamefont{Bergeret, Volkov, and
  Efetov}}]{BVE_inverse}
\bibinfo{author}{\bibnamefont{Bergeret}, \bibfnamefont{F.~S.}},
  \bibinfo{author}{\bibfnamefont{A.~F.} \bibnamefont{Volkov}}, and
  \bibinfo{author}{\bibfnamefont{K.~B.} \bibnamefont{Efetov}},
  \bibinfo{year}{2004}{\natexlab{a}}, \bibinfo{journal}{Phys. Rev. B}
  \textbf{\bibinfo{volume}{69}}, \bibinfo{pages}{174504}.

\bibitem[{\citenamefont{Bergeret}
  \emph{et~al.}(2004{\natexlab{b}})\citenamefont{Bergeret, Volkov, and
  Efetov}}]{BVE_screening}
\bibinfo{author}{\bibnamefont{Bergeret}, \bibfnamefont{F.~S.}},
  \bibinfo{author}{\bibfnamefont{A.~F.} \bibnamefont{Volkov}}, and
  \bibinfo{author}{\bibfnamefont{K.~B.} \bibnamefont{Efetov}},
  \bibinfo{year}{2004}{\natexlab{b}}, \bibinfo{journal}{Europhys. Lett.} .

\bibitem[{\citenamefont{Blanter and Hekking}(2004)}]{blanter}
\bibinfo{author}{\bibnamefont{Blanter}, \bibfnamefont{Y.~M.}}, and
  \bibinfo{author}{\bibfnamefont{F.~W.~J.} \bibnamefont{Hekking}},
  \bibinfo{year}{2004}, \bibinfo{journal}{Phys. Rev. B}
  \textbf{\bibinfo{volume}{69}}, \bibinfo{pages}{024525}.

\bibitem[{\citenamefont{Blum} \emph{et~al.}(2002)\citenamefont{Blum,
  A.~Tsukernik, and Palevski}}]{palevski}
\bibinfo{author}{\bibnamefont{Blum}, \bibfnamefont{Y.}},
  \bibinfo{author}{\bibfnamefont{M.~K.} \bibnamefont{A.~Tsukernik}}, and
  \bibinfo{author}{\bibfnamefont{A.}~\bibnamefont{Palevski}},
  \bibinfo{year}{2002}, \bibinfo{journal}{Phys. Rev. Lett.}
  \textbf{\bibinfo{volume}{89}}, \bibinfo{pages}{187004}.

\bibitem[{\citenamefont{Bourgeois and Dynes}(2002)}]{dynes02}
\bibinfo{author}{\bibnamefont{Bourgeois}, \bibfnamefont{O.}}, and
  \bibinfo{author}{\bibfnamefont{R.~C.} \bibnamefont{Dynes}},
  \bibinfo{year}{2002}, \bibinfo{journal}{Phys. Rev. B}
  \textbf{\bibinfo{volume}{65}}, \bibinfo{pages}{144503}.

\bibitem[{\citenamefont{Bulaevskii}
  \emph{et~al.}(1985)\citenamefont{Bulaevskii, Buzdin, and
  Panyukov}}]{bula_adv}
\bibinfo{author}{\bibnamefont{Bulaevskii}, \bibfnamefont{L.~N.}},
  \bibinfo{author}{\bibfnamefont{M.~L.} \bibnamefont{Buzdin},
  \bibfnamefont{A.~I.and~Kulic}}, and \bibinfo{author}{\bibfnamefont{S.~V.}
  \bibnamefont{Panyukov}}, \bibinfo{year}{1985}, \bibinfo{journal}{Adv. Phys.}
  \textbf{\bibinfo{volume}{34}}, \bibinfo{pages}{175}.

\bibitem[{\citenamefont{Bulaevskii}
  \emph{et~al.}(1977)\citenamefont{Bulaevskii, Kuzii, and
  Sobyanin}}]{bula_kuzi}
\bibinfo{author}{\bibnamefont{Bulaevskii}, \bibfnamefont{L.~N.}},
  \bibinfo{author}{\bibfnamefont{V.~V.} \bibnamefont{Kuzii}}, and
  \bibinfo{author}{\bibfnamefont{A.~A.} \bibnamefont{Sobyanin}},
  \bibinfo{year}{1977}, \bibinfo{journal}{Pis'ma Zh. Eksp. Teor. Fiz.}
  \textbf{\bibinfo{volume}{25}}, \bibinfo{pages}{314}, \bibinfo{note}{[JETP
  Lett. {\bf 25}, 290 (1977)}.

\bibitem[{\citenamefont{Buzdin}(2000)}]{buzdin}
\bibinfo{author}{\bibnamefont{Buzdin}, \bibfnamefont{A.}},
  \bibinfo{year}{2000}, \bibinfo{journal}{Phys. Rev. B}
  \textbf{\bibinfo{volume}{62}}, \bibinfo{pages}{11377}.

\bibitem[{\citenamefont{Buzdin}(2005{\natexlab{a}})}]{buzdin_rev}
\bibinfo{author}{\bibnamefont{Buzdin}, \bibfnamefont{A.}},
  \bibinfo{year}{2005}{\natexlab{a}}, \bibinfo{note}{to be published in Rev.
  Mod .Phys.}

\bibitem[{\citenamefont{Buzdin}(2005{\natexlab{b}})}]{buzdinPC}
\bibinfo{author}{\bibnamefont{Buzdin}, \bibfnamefont{A.}},
  \bibinfo{year}{2005}{\natexlab{b}}, \bibinfo{journal}{Private communication}
  .

\bibitem[{\citenamefont{Buzdin and Baladie}(2003)}]{buzdin_pi}
\bibinfo{author}{\bibnamefont{Buzdin}, \bibfnamefont{A.}}, and
  \bibinfo{author}{\bibfnamefont{I.}~\bibnamefont{Baladie}},
  \bibinfo{year}{2003}, \bibinfo{journal}{Phys. Rev. B}
  \textbf{\bibinfo{volume}{67}}, \bibinfo{pages}{184519}.

\bibitem[{\citenamefont{Buzdin and Bulaevskii}(1988)}]{buzdin_crypto}
\bibinfo{author}{\bibnamefont{Buzdin}, \bibfnamefont{A.~I.}}, and
  \bibinfo{author}{\bibfnamefont{L.~N.} \bibnamefont{Bulaevskii}},
  \bibinfo{year}{1988}, \bibinfo{journal}{Sov. Phys.JETP}
  \textbf{\bibinfo{volume}{67}}, \bibinfo{pages}{576}.

\bibitem[{\citenamefont{Buzdin and Kupriyanov}(1990)}]{buzdin_kupr_tc}
\bibinfo{author}{\bibnamefont{Buzdin}, \bibfnamefont{A.~I.}}, and
  \bibinfo{author}{\bibfnamefont{M.~Y.} \bibnamefont{Kupriyanov}},
  \bibinfo{year}{1990}, \bibinfo{journal}{Pis'ma Zh. Eksp. Teor. Fiz.}
  \textbf{\bibinfo{volume}{52}}, \bibinfo{pages}{1089}, \bibinfo{note}{[JETP
  Lett. {\bf 52}, 487 (1990)]}.

\bibitem[{\citenamefont{Buzdin and Kupriyanov}(1991)}]{buzdin_kupr1}
\bibinfo{author}{\bibnamefont{Buzdin}, \bibfnamefont{A.~I.}}, and
  \bibinfo{author}{\bibfnamefont{M.~Y.} \bibnamefont{Kupriyanov}},
  \bibinfo{year}{1991}, \bibinfo{journal}{Pis'ma Zh. Eksp. Teor. Fiz.}
  \textbf{\bibinfo{volume}{53}}, \bibinfo{pages}{308}, \bibinfo{note}{[JETP
  Lett. {\bf 53}, 321 (1991)]}.

\bibitem[{\citenamefont{Buzdin} \emph{et~al.}(1982)\citenamefont{Buzdin,
  L.N.Bulaevskii, and Panyukov}}]{buzdin_sfs}
\bibinfo{author}{\bibnamefont{Buzdin}, \bibfnamefont{A.~I.}},
  \bibinfo{author}{\bibnamefont{L.N.Bulaevskii}}, and
  \bibinfo{author}{\bibfnamefont{S.~V.} \bibnamefont{Panyukov}},
  \bibinfo{year}{1982}, \bibinfo{journal}{Pis'ma Zh. Eksp. Teor. Fiz.}
  \textbf{\bibinfo{volume}{35}}, \bibinfo{pages}{147}, \bibinfo{note}{[JETP
  Lett. {\bf 35}, 178 (1982)]}.

\bibitem[{\citenamefont{Buzdin and Vujicic}(1992)}]{buzdin_kupr2}
\bibinfo{author}{\bibnamefont{Buzdin}, \bibfnamefont{A.~I.}}, and
  \bibinfo{author}{\bibfnamefont{M.~Y.} \bibnamefont{Vujicic},
  \bibfnamefont{B.and~Kupriyanov}}, \bibinfo{year}{1992}, \bibinfo{journal}{Zh.
  Eksp. Teor. Fiz.} \textbf{\bibinfo{volume}{101}}, \bibinfo{pages}{231},
  \bibinfo{note}{[JETP {\bf 74}, 124 (1992)]}.

\bibitem[{\citenamefont{Champel and Eschrig}(2005{\natexlab{a}})}]{eschrig_pp}
\bibinfo{author}{\bibnamefont{Champel}, \bibfnamefont{T.}}, and
  \bibinfo{author}{\bibfnamefont{M.}~\bibnamefont{Eschrig}},
  \bibinfo{year}{2005}{\natexlab{a}}, \bibinfo{journal}{cond-mat/0504198} .

\bibitem[{\citenamefont{Champel and Eschrig}(2005{\natexlab{b}})}]{eschrig05}
\bibinfo{author}{\bibnamefont{Champel}, \bibfnamefont{T.}}, and
  \bibinfo{author}{\bibfnamefont{M.}~\bibnamefont{Eschrig}},
  \bibinfo{year}{2005}{\natexlab{b}}, \bibinfo{journal}{Phys. Rev. B}
  \textbf{\bibinfo{volume}{71}}, \bibinfo{pages}{220506}.

\bibitem[{\citenamefont{Chandrasekhar}(1962)}]{chandra}
\bibinfo{author}{\bibnamefont{Chandrasekhar}, \bibfnamefont{B.~S.}},
  \bibinfo{year}{1962}, \bibinfo{journal}{Appl. Phys. Lett.}
  \textbf{\bibinfo{volume}{1}}, \bibinfo{pages}{7}.

\bibitem[{\citenamefont{Charlat} \emph{et~al.}(1996)\citenamefont{Charlat,
  Courtois, Gandit, Mailly, Volkov, and Pannetier}}]{volkov_pannetier}
\bibinfo{author}{\bibnamefont{Charlat}, \bibfnamefont{P.}},
  \bibinfo{author}{\bibfnamefont{H.}~\bibnamefont{Courtois}},
  \bibinfo{author}{\bibfnamefont{P.}~\bibnamefont{Gandit}},
  \bibinfo{author}{\bibfnamefont{D.}~\bibnamefont{Mailly}},
  \bibinfo{author}{\bibfnamefont{A.~F.} \bibnamefont{Volkov}}, and
  \bibinfo{author}{\bibfnamefont{B.}~\bibnamefont{Pannetier}},
  \bibinfo{year}{1996}, \bibinfo{journal}{Phys. Rev. Lett.}
  \textbf{\bibinfo{volume}{77}}, \bibinfo{pages}{4950}.

\bibitem[{\citenamefont{Chien and Chandrasekhar}(1999)}]{chien1}
\bibinfo{author}{\bibnamefont{Chien}, \bibfnamefont{C.~J.}}, and
  \bibinfo{author}{\bibfnamefont{V.}~\bibnamefont{Chandrasekhar}},
  \bibinfo{year}{1999}, \bibinfo{journal}{Phys. Rev. B}
  \textbf{\bibinfo{volume}{60}}, \bibinfo{pages}{3655}.

\bibitem[{\citenamefont{Chien and Reich}(1999)}]{chien_reich}
\bibinfo{author}{\bibnamefont{Chien}, \bibfnamefont{C.~L.}}, and
  \bibinfo{author}{\bibfnamefont{D.~H.} \bibnamefont{Reich}},
  \bibinfo{year}{1999}, \bibinfo{journal}{J. Magn. Magn. Mater.}
  \textbf{\bibinfo{volume}{200}}, \bibinfo{pages}{83}.

\bibitem[{\citenamefont{Chtchelkatchev}
  \emph{et~al.}(2001)\citenamefont{Chtchelkatchev, Belzig, Nazarov, and
  Bruder}}]{nazarov_sfs}
\bibinfo{author}{\bibnamefont{Chtchelkatchev}, \bibfnamefont{N.~M.}},
  \bibinfo{author}{\bibfnamefont{W.}~\bibnamefont{Belzig}},
  \bibinfo{author}{\bibfnamefont{Y.}~\bibnamefont{Nazarov}}, and
  \bibinfo{author}{\bibfnamefont{C.}~\bibnamefont{Bruder}},
  \bibinfo{year}{2001}, \bibinfo{journal}{JETP Lett.}
  \textbf{\bibinfo{volume}{74}}, \bibinfo{pages}{323}.

\bibitem[{\citenamefont{Chtchelkatchev}
  \emph{et~al.}(2002)\citenamefont{Chtchelkatchev, W.Belzig, and
  C.Bruder}}]{chtchelkatchev}
\bibinfo{author}{\bibnamefont{Chtchelkatchev}, \bibfnamefont{N.~M.}},
  \bibinfo{author}{\bibnamefont{W.Belzig}}, and
  \bibinfo{author}{\bibnamefont{C.Bruder}}, \bibinfo{year}{2002},
  \bibinfo{journal}{JETP Lett.} \textbf{\bibinfo{volume}{75}},
  \bibinfo{pages}{646}.

\bibitem[{\citenamefont{Clogston}(1962)}]{clogston}
\bibinfo{author}{\bibnamefont{Clogston}, \bibfnamefont{A.~M.}},
  \bibinfo{year}{1962}, \bibinfo{journal}{Phys. Rev. Lett.}
  \textbf{\bibinfo{volume}{9}}, \bibinfo{pages}{266}.

\bibitem[{\citenamefont{Coleman}
  \emph{et~al.}(1993{\natexlab{a}})\citenamefont{Coleman, Miranda, and
  Tsvelik}}]{coleman93}
\bibinfo{author}{\bibnamefont{Coleman}, \bibfnamefont{P.}},
  \bibinfo{author}{\bibfnamefont{E.}~\bibnamefont{Miranda}}, and
  \bibinfo{author}{\bibfnamefont{A.}~\bibnamefont{Tsvelik}},
  \bibinfo{year}{1993}{\natexlab{a}}, \bibinfo{journal}{Phys. Rev. Lett.}
  \textbf{\bibinfo{volume}{70}}, \bibinfo{pages}{2960}.

\bibitem[{\citenamefont{Coleman}
  \emph{et~al.}(1993{\natexlab{b}})\citenamefont{Coleman, Miranda, and
  Tsvelik}}]{coleman94}
\bibinfo{author}{\bibnamefont{Coleman}, \bibfnamefont{P.}},
  \bibinfo{author}{\bibfnamefont{E.}~\bibnamefont{Miranda}}, and
  \bibinfo{author}{\bibfnamefont{A.}~\bibnamefont{Tsvelik}},
  \bibinfo{year}{1993}{\natexlab{b}}, \bibinfo{journal}{Phys. Rev. B}
  \textbf{\bibinfo{volume}{49}}, \bibinfo{pages}{8955}.

\bibitem[{\citenamefont{Coleman} \emph{et~al.}(1995)\citenamefont{Coleman,
  Miranda, and Tsvelik}}]{coleman95}
\bibinfo{author}{\bibnamefont{Coleman}, \bibfnamefont{P.}},
  \bibinfo{author}{\bibfnamefont{E.}~\bibnamefont{Miranda}}, and
  \bibinfo{author}{\bibfnamefont{A.}~\bibnamefont{Tsvelik}},
  \bibinfo{year}{1995}, \bibinfo{journal}{Phys. Rev. Lett.}
  \textbf{\bibinfo{volume}{74}}, \bibinfo{pages}{165}.

\bibitem[{\citenamefont{{de Gennes}}(1964)}]{de_gennes}
\bibinfo{author}{\bibnamefont{{de Gennes}}, \bibfnamefont{P.~G.}},
  \bibinfo{year}{1964}, \bibinfo{journal}{Rev. Mod. Phys.}
  \textbf{\bibinfo{volume}{36}}, \bibinfo{pages}{225}.

\bibitem[{\citenamefont{{de Gennes}}(1966)}]{degennes_book}
\bibinfo{author}{\bibnamefont{{de Gennes}}, \bibfnamefont{P.~G.}},
  \bibinfo{year}{1966}, \emph{\bibinfo{title}{Superconductivity of Metals and
  Alloys}} (\bibinfo{publisher}{Benjamin}, \bibinfo{address}{New York}).

\bibitem[{\citenamefont{{de Jong} and Beenakker}(1994)}]{dejong}
\bibinfo{author}{\bibnamefont{{de Jong}}, \bibfnamefont{M.~J.~M.}}, and
  \bibinfo{author}{\bibfnamefont{C.~W.~J.} \bibnamefont{Beenakker}},
  \bibinfo{year}{1994}, \bibinfo{journal}{Phys. Rev. Lett.}
  \textbf{\bibinfo{volume}{74}}, \bibinfo{pages}{1657}.

\bibitem[{\citenamefont{Demler} \emph{et~al.}(1997)\citenamefont{Demler,
  Arnold, and Beasley}}]{demler}
\bibinfo{author}{\bibnamefont{Demler}, \bibfnamefont{E.~A.}},
  \bibinfo{author}{\bibfnamefont{G.~B.} \bibnamefont{Arnold}}, and
  \bibinfo{author}{\bibfnamefont{M.~R.} \bibnamefont{Beasley}},
  \bibinfo{year}{1997}, \bibinfo{journal}{Phys. Rev. B}
  \textbf{\bibinfo{volume}{55}}, \bibinfo{pages}{15174}.

\bibitem[{\citenamefont{Deutscher}(2005)}]{deutscher_rev}
\bibinfo{author}{\bibnamefont{Deutscher}, \bibfnamefont{G.}},
  \bibinfo{year}{2005}, \bibinfo{journal}{Rev. Mod. Phys.}
  \textbf{\bibinfo{volume}{77}}, \bibinfo{pages}{109}.

\bibitem[{\citenamefont{Deutscher and {de Gennes}}(1969)}]{deutscher}
\bibinfo{author}{\bibnamefont{Deutscher}, \bibfnamefont{G.}}, and
  \bibinfo{author}{\bibfnamefont{P.~G.} \bibnamefont{{de Gennes}}},
  \bibinfo{year}{1969}, \emph{\bibinfo{title}{Superconductivity}}, volume
  \bibinfo{volume}{Vol. 2} (\bibinfo{publisher}{Marcel Dekker Inc.},
  \bibinfo{address}{New York}).

\bibitem[{\citenamefont{Dimoulas} \emph{et~al.}(1995)\citenamefont{Dimoulas,
  Heida, van Wees, and Klapwijk}}]{vanWees_95}
\bibinfo{author}{\bibnamefont{Dimoulas}, \bibfnamefont{A.}},
  \bibinfo{author}{\bibfnamefont{J.~P.} \bibnamefont{Heida}},
  \bibinfo{author}{\bibfnamefont{B.~J.} \bibnamefont{van Wees}}, and
  \bibinfo{author}{\bibfnamefont{T.~M.} \bibnamefont{Klapwijk}},
  \bibinfo{year}{1995}, \bibinfo{journal}{Phys. Rev. Lett.}
  \textbf{\bibinfo{volume}{74}}, \bibinfo{pages}{602}.

\bibitem[{\citenamefont{Dubonos} \emph{et~al.}(2002)\citenamefont{Dubonos,
  Geim, Novoselov, and Grigorieva1}}]{Geim}
\bibinfo{author}{\bibnamefont{Dubonos}, \bibfnamefont{S.~V.}},
  \bibinfo{author}{\bibfnamefont{A.~K.} \bibnamefont{Geim}},
  \bibinfo{author}{\bibfnamefont{K.~S.} \bibnamefont{Novoselov}}, and
  \bibinfo{author}{\bibfnamefont{I.~V.} \bibnamefont{Grigorieva1}},
  \bibinfo{year}{2002}, \bibinfo{journal}{Phys. Rev. B}
  \textbf{\bibinfo{volume}{65}}, \bibinfo{pages}{220513}.

\bibitem[{\citenamefont{Edelstein}(1989)}]{Edel89}
\bibinfo{author}{\bibnamefont{Edelstein}, \bibfnamefont{V.~M.}},
  \bibinfo{year}{1989}, \bibinfo{journal}{Sov. Phys. JETP}
  \textbf{\bibinfo{volume}{68}}, \bibinfo{pages}{1244}.

\bibitem[{\citenamefont{Edelstein}(2001)}]{Edel03}
\bibinfo{author}{\bibnamefont{Edelstein}, \bibfnamefont{V.~M.}},
  \bibinfo{year}{2001}, \bibinfo{journal}{Phys. Rev. B}
  \textbf{\bibinfo{volume}{67}}, \bibinfo{pages}{153940205R}.

\bibitem[{\citenamefont{Eilenberger}(1968)}]{eilenberger}
\bibinfo{author}{\bibnamefont{Eilenberger}, \bibfnamefont{G.}},
  \bibinfo{year}{1968}, \bibinfo{journal}{Z. Phys.}
  \textbf{\bibinfo{volume}{214}}, \bibinfo{pages}{195}.

\bibitem[{\citenamefont{Eliashberg}(1971)}]{eliashberg}
\bibinfo{author}{\bibnamefont{Eliashberg}, \bibfnamefont{G.~M.}},
  \bibinfo{year}{1971}, \bibinfo{journal}{Sov. Phys. JETP}
  \textbf{\bibinfo{volume}{34}}, \bibinfo{pages}{668}.

\bibitem[{\citenamefont{Eremin} \emph{et~al.}(2004)\citenamefont{Eremin,
  Manske, Ovchinnikov, and F.Annett}}]{annett}
\bibinfo{author}{\bibnamefont{Eremin}, \bibfnamefont{I.}},
  \bibinfo{author}{\bibfnamefont{D.}~\bibnamefont{Manske}},
  \bibinfo{author}{\bibfnamefont{S.~G.} \bibnamefont{Ovchinnikov}}, and
  \bibinfo{author}{\bibfnamefont{J.}~\bibnamefont{F.Annett}},
  \bibinfo{year}{2004}, \bibinfo{journal}{Ann. Phys. (Berlin)}
  \textbf{\bibinfo{volume}{13}}, \bibinfo{pages}{149}.

\bibitem[{\citenamefont{Eschrig}(2000)}]{eschrig00}
\bibinfo{author}{\bibnamefont{Eschrig}, \bibfnamefont{M.}},
  \bibinfo{year}{2000}, \bibinfo{journal}{Phys. Rev. B}
  \textbf{\bibinfo{volume}{61}}, \bibinfo{pages}{9061}.

\bibitem[{\citenamefont{Eschrig} \emph{et~al.}(2003)\citenamefont{Eschrig,
  Kopu, Cuevas, and Sch\"on}}]{schoen_half}
\bibinfo{author}{\bibnamefont{Eschrig}, \bibfnamefont{M.}},
  \bibinfo{author}{\bibfnamefont{J.}~\bibnamefont{Kopu}},
  \bibinfo{author}{\bibfnamefont{J.~C.} \bibnamefont{Cuevas}}, and
  \bibinfo{author}{\bibfnamefont{G.}~\bibnamefont{Sch\"on}},
  \bibinfo{year}{2003}, \bibinfo{journal}{Phys. Rev. Lett.}
  \textbf{\bibinfo{volume}{90}}, \bibinfo{pages}{137003}.

\bibitem[{\citenamefont{Falko} \emph{et~al.}(1999)\citenamefont{Falko, Volkov,
  and Lambert}}]{falko_volkov1}
\bibinfo{author}{\bibnamefont{Falko}, \bibfnamefont{V.~I.}},
  \bibinfo{author}{\bibfnamefont{A.~F.} \bibnamefont{Volkov}}, and
  \bibinfo{author}{\bibfnamefont{C.~J.} \bibnamefont{Lambert}},
  \bibinfo{year}{1999}, \bibinfo{journal}{Phys. Rev. B}
  \textbf{\bibinfo{volume}{60}}, \bibinfo{pages}{15394}.

\bibitem[{\citenamefont{Fazio and Lucheroni}(1999)}]{fazio}
\bibinfo{author}{\bibnamefont{Fazio}, \bibfnamefont{R.}}, and
  \bibinfo{author}{\bibfnamefont{C.}~\bibnamefont{Lucheroni}},
  \bibinfo{year}{1999}, \bibinfo{journal}{Europhys. Lett.}
  \textbf{\bibinfo{volume}{45}}, \bibinfo{pages}{707}.

\bibitem[{\citenamefont{Fogelstr\"{o}m}(2000)}]{fogel00}
\bibinfo{author}{\bibnamefont{Fogelstr\"{o}m}, \bibfnamefont{M.}},
  \bibinfo{year}{2000}, \bibinfo{journal}{Phys. Rev. B}
  \textbf{\bibinfo{volume}{62}}, \bibinfo{pages}{11812}.

\bibitem[{\citenamefont{Fominov} \emph{et~al.}(2002)\citenamefont{Fominov,
  Chtchelkatchev, and Golubov}}]{golubov_tc}
\bibinfo{author}{\bibnamefont{Fominov}, \bibfnamefont{Y.~V.}},
  \bibinfo{author}{\bibfnamefont{N.~M.} \bibnamefont{Chtchelkatchev}}, and
  \bibinfo{author}{\bibfnamefont{A.~A.} \bibnamefont{Golubov}},
  \bibinfo{year}{2002}, \bibinfo{journal}{Phys. Rev. B}
  \textbf{\bibinfo{volume}{66}}, \bibinfo{pages}{014507}.

\bibitem[{\citenamefont{Fominov} \emph{et~al.}(2003)\citenamefont{Fominov,
  Golubov, and Kupriyanov}}]{golubov_triplet}
\bibinfo{author}{\bibnamefont{Fominov}, \bibfnamefont{Y.~V.}},
  \bibinfo{author}{\bibfnamefont{A.~A.} \bibnamefont{Golubov}}, and
  \bibinfo{author}{\bibfnamefont{M.~Y.} \bibnamefont{Kupriyanov}},
  \bibinfo{year}{2003}, \bibinfo{journal}{JETP Lett.}
  \textbf{\bibinfo{volume}{77}}, \bibinfo{pages}{510}.

\bibitem[{\citenamefont{Fulde and Ferrell}(1965)}]{fulde_state}
\bibinfo{author}{\bibnamefont{Fulde}, \bibfnamefont{P.}}, and
  \bibinfo{author}{\bibfnamefont{R.~A.} \bibnamefont{Ferrell}},
  \bibinfo{year}{1965}, \bibinfo{journal}{Phys. Rev.}
  \textbf{\bibinfo{volume}{135}}, \bibinfo{pages}{550}.

\bibitem[{\citenamefont{Galaktionov and Zaikin}(2002)}]{zaikin_b}
\bibinfo{author}{\bibnamefont{Galaktionov}, \bibfnamefont{A.~V.}}, and
  \bibinfo{author}{\bibfnamefont{A.~D.} \bibnamefont{Zaikin}},
  \bibinfo{year}{2002}, \bibinfo{journal}{Phys. Rev. B}
  \textbf{\bibinfo{volume}{65}}, \bibinfo{pages}{184507}.

\bibitem[{\citenamefont{Garifullin}
  \emph{et~al.}(2002)\citenamefont{Garifullin, Tikhonov, Garif'yanov,
  Fattakhov, Theis-Broehl, Westerholt, and Zabel}}]{garifullin}
\bibinfo{author}{\bibnamefont{Garifullin}, \bibfnamefont{I.~A.}},
  \bibinfo{author}{\bibfnamefont{D.~A.} \bibnamefont{Tikhonov}},
  \bibinfo{author}{\bibfnamefont{N.~N.} \bibnamefont{Garif'yanov}},
  \bibinfo{author}{\bibfnamefont{M.~Z.} \bibnamefont{Fattakhov}},
  \bibinfo{author}{\bibfnamefont{K.}~\bibnamefont{Theis-Broehl}},
  \bibinfo{author}{\bibfnamefont{K.}~\bibnamefont{Westerholt}}, and
  \bibinfo{author}{\bibfnamefont{H.}~\bibnamefont{Zabel}},
  \bibinfo{year}{2002}, \bibinfo{journal}{Appl. Magn. Reson.}
  \textbf{\bibinfo{volume}{22}}, \bibinfo{pages}{439}.

\bibitem[{\citenamefont{Ginzburg and Landau}(1950)}]{GL}
\bibinfo{author}{\bibnamefont{Ginzburg}, \bibfnamefont{V.~L.}}, and
  \bibinfo{author}{\bibfnamefont{L.~D.} \bibnamefont{Landau}},
  \bibinfo{year}{1950}, \bibinfo{journal}{JETP} \textbf{\bibinfo{volume}{20}},
  \bibinfo{pages}{1064}.

\bibitem[{\citenamefont{Giroud} \emph{et~al.}(1998)\citenamefont{Giroud,
  Courtois, Hasselbach, Mailly, and Pannetier}}]{pannetier}
\bibinfo{author}{\bibnamefont{Giroud}, \bibfnamefont{M.}},
  \bibinfo{author}{\bibfnamefont{H.}~\bibnamefont{Courtois}},
  \bibinfo{author}{\bibfnamefont{K.}~\bibnamefont{Hasselbach}},
  \bibinfo{author}{\bibfnamefont{D.}~\bibnamefont{Mailly}}, and
  \bibinfo{author}{\bibfnamefont{B.}~\bibnamefont{Pannetier}},
  \bibinfo{year}{1998}, \bibinfo{journal}{Phys. Rev. B}
  \textbf{\bibinfo{volume}{58}}, \bibinfo{pages}{11872}.

\bibitem[{\citenamefont{Giroud} \emph{et~al.}(2003)\citenamefont{Giroud,
  Hasselbach, H.Courtois, Mailly, and Pannetier}}]{giroud03}
\bibinfo{author}{\bibnamefont{Giroud}, \bibfnamefont{M.}},
  \bibinfo{author}{\bibfnamefont{K.}~\bibnamefont{Hasselbach}},
  \bibinfo{author}{\bibnamefont{H.Courtois}},
  \bibinfo{author}{\bibfnamefont{D.}~\bibnamefont{Mailly}}, and
  \bibinfo{author}{\bibfnamefont{B.}~\bibnamefont{Pannetier}},
  \bibinfo{year}{2003}, \bibinfo{journal}{Eur. Phys. J. B}
  \textbf{\bibinfo{volume}{31}}, \bibinfo{pages}{103}.

\bibitem[{\citenamefont{Golubov}(1999)}]{golubov}
\bibinfo{author}{\bibnamefont{Golubov}, \bibfnamefont{A.~A.}},
  \bibinfo{year}{1999}, \bibinfo{journal}{Physica C}
  \textbf{\bibinfo{volume}{326-327}}, \bibinfo{pages}{46}.

\bibitem[{\citenamefont{Golubov and Kupriyanov}(1996)}]{golubov_kupriyanov_96}
\bibinfo{author}{\bibnamefont{Golubov}, \bibfnamefont{A.~A.}}, and
  \bibinfo{author}{\bibfnamefont{M.~Y.} \bibnamefont{Kupriyanov}},
  \bibinfo{year}{1996}, \bibinfo{journal}{Physica C}
  \textbf{\bibinfo{volume}{259}}, \bibinfo{pages}{27}.

\bibitem[{\citenamefont{Golubov}
  \emph{et~al.}(2002{\natexlab{a}})\citenamefont{Golubov, Kupriyanov, and
  Fominov}}]{golubov_phi}
\bibinfo{author}{\bibnamefont{Golubov}, \bibfnamefont{A.~A.}},
  \bibinfo{author}{\bibfnamefont{M.~Y.} \bibnamefont{Kupriyanov}}, and
  \bibinfo{author}{\bibfnamefont{Y.~V.} \bibnamefont{Fominov}},
  \bibinfo{year}{2002}{\natexlab{a}}, \bibinfo{journal}{JETP Lett.}
  \textbf{\bibinfo{volume}{75}}, \bibinfo{pages}{588}.

\bibitem[{\citenamefont{Golubov}
  \emph{et~al.}(2002{\natexlab{b}})\citenamefont{Golubov, Kupriyanov, and
  Fominov}}]{golubov_sfifs}
\bibinfo{author}{\bibnamefont{Golubov}, \bibfnamefont{A.~A.}},
  \bibinfo{author}{\bibfnamefont{M.~Y.} \bibnamefont{Kupriyanov}}, and
  \bibinfo{author}{\bibfnamefont{Y.~V.} \bibnamefont{Fominov}},
  \bibinfo{year}{2002}{\natexlab{b}}, \bibinfo{journal}{JETP Lett.}
  \textbf{\bibinfo{volume}{75}}, \bibinfo{pages}{190}.

\bibitem[{\citenamefont{Golubov} \emph{et~al.}(2004)\citenamefont{Golubov,
  Kupriyanov, and Il'ichev}}]{golubov_rev}
\bibinfo{author}{\bibnamefont{Golubov}, \bibfnamefont{A.~A.}},
  \bibinfo{author}{\bibfnamefont{M.~Y.} \bibnamefont{Kupriyanov}}, and
  \bibinfo{author}{\bibfnamefont{E.}~\bibnamefont{Il'ichev}},
  \bibinfo{year}{2004}, \bibinfo{journal}{Rev. Mod. Phys.}
  \textbf{\bibinfo{volume}{76}}, \bibinfo{pages}{411}.

\bibitem[{\citenamefont{Golubov} \emph{et~al.}(2005)\citenamefont{Golubov,
  Kupriyanov, and Siegel}}]{golubov05}
\bibinfo{author}{\bibnamefont{Golubov}, \bibfnamefont{A.~A.}},
  \bibinfo{author}{\bibfnamefont{M.~Y.} \bibnamefont{Kupriyanov}}, and
  \bibinfo{author}{\bibfnamefont{M.}~\bibnamefont{Siegel}},
  \bibinfo{year}{2005}, \bibinfo{journal}{JETP Lett.}
  \textbf{\bibinfo{volume}{81}}, \bibinfo{pages}{180}.

\bibitem[{\citenamefont{Golubov} \emph{et~al.}(1997)\citenamefont{Golubov,
  Wilhelm, and Zaikin}}]{golubov_zaikin}
\bibinfo{author}{\bibnamefont{Golubov}, \bibfnamefont{A.~A.}},
  \bibinfo{author}{\bibfnamefont{F.~K.} \bibnamefont{Wilhelm}}, and
  \bibinfo{author}{\bibfnamefont{A.~D.} \bibnamefont{Zaikin}},
  \bibinfo{year}{1997}, \bibinfo{journal}{Phys. Rev. B}
  \textbf{\bibinfo{volume}{55}}, \bibinfo{pages}{1123}.

\bibitem[{\citenamefont{Gor'kov and Rashba}(2001)}]{GorkovRashba}
\bibinfo{author}{\bibnamefont{Gor'kov}, \bibfnamefont{L.~P.}}, and
  \bibinfo{author}{\bibfnamefont{E.~I.} \bibnamefont{Rashba}},
  \bibinfo{year}{2001}, \bibinfo{journal}{Phys. Rev. Lett.}
  \textbf{\bibinfo{volume}{87}}, \bibinfo{pages}{037004}.

\bibitem[{\citenamefont{Gor'kov and Rusinov}(1963)}]{GR}
\bibinfo{author}{\bibnamefont{Gor'kov}, \bibfnamefont{L.~P.}}, and
  \bibinfo{author}{\bibfnamefont{A.~I.} \bibnamefont{Rusinov}},
  \bibinfo{year}{1963}, \bibinfo{journal}{Zh. Eksp. Teor. Fiz.}
  \textbf{\bibinfo{volume}{46}}, \bibinfo{pages}{1363}, \bibinfo{note}{[Sov.
  Phys. JETP {\bf 19}, 922 (1964)}.

\bibitem[{\citenamefont{Gu} \emph{et~al.}(2002{\natexlab{a}})\citenamefont{Gu,
  You, Jiang, Pearson, Bazaliy, and Bader}}]{bader}
\bibinfo{author}{\bibnamefont{Gu}, \bibfnamefont{J.~Y.}},
  \bibinfo{author}{\bibfnamefont{C.-Y.} \bibnamefont{You}},
  \bibinfo{author}{\bibfnamefont{J.~S.} \bibnamefont{Jiang}},
  \bibinfo{author}{\bibfnamefont{J.}~\bibnamefont{Pearson}},
  \bibinfo{author}{\bibfnamefont{Y.~B.} \bibnamefont{Bazaliy}}, and
  \bibinfo{author}{\bibfnamefont{S.~D.} \bibnamefont{Bader}},
  \bibinfo{year}{2002}{\natexlab{a}}, \bibinfo{journal}{Phys. Rev. Lett.}
  \textbf{\bibinfo{volume}{89}}, \bibinfo{pages}{267001}.

\bibitem[{\citenamefont{Gu} \emph{et~al.}(2002{\natexlab{b}})\citenamefont{Gu,
  You, Jiang, Pearson, Bazaliy, and Bader}}]{argonne}
\bibinfo{author}{\bibnamefont{Gu}, \bibfnamefont{J.~Y.}},
  \bibinfo{author}{\bibfnamefont{C.-Y.} \bibnamefont{You}},
  \bibinfo{author}{\bibfnamefont{J.~S.} \bibnamefont{Jiang}},
  \bibinfo{author}{\bibfnamefont{J.}~\bibnamefont{Pearson}},
  \bibinfo{author}{\bibfnamefont{Y.~B.} \bibnamefont{Bazaliy}}, and
  \bibinfo{author}{\bibfnamefont{S.~D.} \bibnamefont{Bader}},
  \bibinfo{year}{2002}{\natexlab{b}}, \bibinfo{journal}{Phys. Rev. Lett.}
  \textbf{\bibinfo{volume}{89}}, \bibinfo{pages}{267001}.

\bibitem[{\citenamefont{Gubankov and Margolin}(1979)}]{gubankov}
\bibinfo{author}{\bibnamefont{Gubankov}, \bibfnamefont{V.~N.}}, and
  \bibinfo{author}{\bibfnamefont{N.~M.} \bibnamefont{Margolin}},
  \bibinfo{year}{1979}, \bibinfo{journal}{JETP Lett.}
  \textbf{\bibinfo{volume}{29}}, \bibinfo{pages}{673}.

\bibitem[{\citenamefont{Gu\'{e}ron}
  \emph{et~al.}(1996)\citenamefont{Gu\'{e}ron, Pothier, Birge, Esteve, and
  Devoret}}]{esteve96}
\bibinfo{author}{\bibnamefont{Gu\'{e}ron}, \bibfnamefont{S.}},
  \bibinfo{author}{\bibfnamefont{H.}~\bibnamefont{Pothier}},
  \bibinfo{author}{\bibfnamefont{N.~O.} \bibnamefont{Birge}},
  \bibinfo{author}{\bibfnamefont{D.}~\bibnamefont{Esteve}}, and
  \bibinfo{author}{\bibfnamefont{M.~H.} \bibnamefont{Devoret}},
  \bibinfo{year}{1996}, \bibinfo{journal}{Phys. Rev. Lett.}
  \textbf{\bibinfo{volume}{77}}, \bibinfo{pages}{3025}.

\bibitem[{\citenamefont{Guichard} \emph{et~al.}(2003)\citenamefont{Guichard,
  Bourgeois, Kontos, Lesueur, and Gandit}}]{kontos03}
\bibinfo{author}{\bibnamefont{Guichard}, \bibfnamefont{W.}},
  \bibinfo{author}{\bibfnamefont{M.~A.~O.} \bibnamefont{Bourgeois}},
  \bibinfo{author}{\bibfnamefont{T.}~\bibnamefont{Kontos}},
  \bibinfo{author}{\bibfnamefont{J.}~\bibnamefont{Lesueur}}, and
  \bibinfo{author}{\bibfnamefont{P.}~\bibnamefont{Gandit}},
  \bibinfo{year}{2003}, \bibinfo{journal}{Phys. Rev. Lett.}
  \textbf{\bibinfo{volume}{90}}, \bibinfo{pages}{167001}.

\bibitem[{\citenamefont{Gupta} \emph{et~al.}(2004)\citenamefont{Gupta,
  Crétinon, Moussy, Pannetier, and Courtois}}]{pannetier_dos_sn}
\bibinfo{author}{\bibnamefont{Gupta}, \bibfnamefont{A.~K.}},
  \bibinfo{author}{\bibfnamefont{L.}~\bibnamefont{Crétinon}},
  \bibinfo{author}{\bibfnamefont{N.}~\bibnamefont{Moussy}},
  \bibinfo{author}{\bibfnamefont{B.}~\bibnamefont{Pannetier}}, and
  \bibinfo{author}{\bibfnamefont{H.}~\bibnamefont{Courtois}},
  \bibinfo{year}{2004}, \bibinfo{journal}{Phys. Rev. B}
  \textbf{\bibinfo{volume}{69}}, \bibinfo{pages}{104514}.

\bibitem[{\citenamefont{Halterman and
  Valls}(2002{\natexlab{a}})}]{halterman_02}
\bibinfo{author}{\bibnamefont{Halterman}, \bibfnamefont{K.}}, and
  \bibinfo{author}{\bibfnamefont{O.~T.} \bibnamefont{Valls}},
  \bibinfo{year}{2002}{\natexlab{a}}, \bibinfo{journal}{Phys. Rev. B}
  \textbf{\bibinfo{volume}{66}}, \bibinfo{pages}{224516}.

\bibitem[{\citenamefont{Halterman and Valls}(2002{\natexlab{b}})}]{halterman}
\bibinfo{author}{\bibnamefont{Halterman}, \bibfnamefont{K.}}, and
  \bibinfo{author}{\bibfnamefont{O.~T.} \bibnamefont{Valls}},
  \bibinfo{year}{2002}{\natexlab{b}}, \bibinfo{journal}{Phys. Rev. B}
  \textbf{\bibinfo{volume}{65}}, \bibinfo{pages}{014509}.

\bibitem[{\citenamefont{Hashimoto}(2000)}]{hashimoto}
\bibinfo{author}{\bibnamefont{Hashimoto}, \bibfnamefont{K.}},
  \bibinfo{year}{2000}, \bibinfo{journal}{J. Phys. Soc. Jpn}
  \textbf{\bibinfo{volume}{69}}, \bibinfo{pages}{2229}.

\bibitem[{\citenamefont{Hauser} \emph{et~al.}(1963)\citenamefont{Hauser,
  Theurer, and Werthamer}}]{werthamer_sf}
\bibinfo{author}{\bibnamefont{Hauser}, \bibfnamefont{J.~J.}},
  \bibinfo{author}{\bibfnamefont{H.~C.} \bibnamefont{Theurer}}, and
  \bibinfo{author}{\bibfnamefont{N.~R.} \bibnamefont{Werthamer}},
  \bibinfo{year}{1963}, \bibinfo{journal}{Phys. Rev.}
  \textbf{\bibinfo{volume}{142}}, \bibinfo{pages}{118}.

\bibitem[{\citenamefont{Heikkil\"a}
  \emph{et~al.}(2000)\citenamefont{Heikkil\"a, Wilhelm, and Sch\"on}}]{schoen}
\bibinfo{author}{\bibnamefont{Heikkil\"a}, \bibfnamefont{T.~T.}},
  \bibinfo{author}{\bibfnamefont{F.~K.} \bibnamefont{Wilhelm}}, and
  \bibinfo{author}{\bibfnamefont{G.}~\bibnamefont{Sch\"on}},
  \bibinfo{year}{2000}, \bibinfo{journal}{Europhys. Lett.}
  \textbf{\bibinfo{volume}{51}}, \bibinfo{pages}{434}.

\bibitem[{\citenamefont{Heslinga} \emph{et~al.}(1994)\citenamefont{Heslinga,
  Shafranjuk, van Kempen, and Klapwijk}}]{heslinga}
\bibinfo{author}{\bibnamefont{Heslinga}, \bibfnamefont{D.~R.}},
  \bibinfo{author}{\bibfnamefont{S.~E.} \bibnamefont{Shafranjuk}},
  \bibinfo{author}{\bibfnamefont{H.}~\bibnamefont{van Kempen}}, and
  \bibinfo{author}{\bibfnamefont{T.~M.} \bibnamefont{Klapwijk}},
  \bibinfo{year}{1994}, \bibinfo{journal}{Phys. Rev. B}
  \textbf{\bibinfo{volume}{49}}, \bibinfo{pages}{10484}.

\bibitem[{\citenamefont{Huertas-Hernando}
  \emph{et~al.}(2002)\citenamefont{Huertas-Hernando, Nazarov, and
  Belzig}}]{nazarov02}
\bibinfo{author}{\bibnamefont{Huertas-Hernando}, \bibfnamefont{D.}},
  \bibinfo{author}{\bibfnamefont{Y.~V.} \bibnamefont{Nazarov}}, and
  \bibinfo{author}{\bibfnamefont{W.}~\bibnamefont{Belzig}},
  \bibinfo{year}{2002}, \bibinfo{journal}{Phys. Rev. Lett.}
  \textbf{\bibinfo{volume}{88}}, \bibinfo{pages}{047003}.

\bibitem[{\citenamefont{Izyumov} \emph{et~al.}(2002)\citenamefont{Izyumov,
  Proshin, and Khusainov}}]{proshin}
\bibinfo{author}{\bibnamefont{Izyumov}, \bibfnamefont{Y.~A.}},
  \bibinfo{author}{\bibfnamefont{Y.~N.} \bibnamefont{Proshin}}, and
  \bibinfo{author}{\bibfnamefont{M.~G.} \bibnamefont{Khusainov}},
  \bibinfo{year}{2002}, \bibinfo{journal}{Usp. Fiz. Nauk}
  \textbf{\bibinfo{volume}{172}}, \bibinfo{pages}{113}.

\bibitem[{\citenamefont{Jedema} \emph{et~al.}(1999)\citenamefont{Jedema, van
  Wees, Hoving, Filip, and Klapwijk}}]{vanWees_99}
\bibinfo{author}{\bibnamefont{Jedema}, \bibfnamefont{F.~J.}},
  \bibinfo{author}{\bibfnamefont{B.~J.} \bibnamefont{van Wees}},
  \bibinfo{author}{\bibfnamefont{B.~H.} \bibnamefont{Hoving}},
  \bibinfo{author}{\bibfnamefont{A.~T.} \bibnamefont{Filip}}, and
  \bibinfo{author}{\bibfnamefont{T.~M.} \bibnamefont{Klapwijk}},
  \bibinfo{year}{1999}, \bibinfo{journal}{Phys. Rev. B}
  \textbf{\bibinfo{volume}{60}}, \bibinfo{pages}{16549}.

\bibitem[{\citenamefont{Jiang} \emph{et~al.}(1995)\citenamefont{Jiang,
  Davidovi\'{c}, Reich, and Chien}}]{jiang}
\bibinfo{author}{\bibnamefont{Jiang}, \bibfnamefont{J.~S.}},
  \bibinfo{author}{\bibfnamefont{D.}~\bibnamefont{Davidovi\'{c}}},
  \bibinfo{author}{\bibfnamefont{D.}~\bibnamefont{Reich}}, and
  \bibinfo{author}{\bibfnamefont{C.~L.} \bibnamefont{Chien}},
  \bibinfo{year}{1995}, \bibinfo{journal}{Phys. Rev. Lett.}
  \textbf{\bibinfo{volume}{74}}, \bibinfo{pages}{314}.

\bibitem[{\citenamefont{Kadigrobov}
  \emph{et~al.}(2001)\citenamefont{Kadigrobov, Skehter, and
  Jonson}}]{kadigrobov}
\bibinfo{author}{\bibnamefont{Kadigrobov}, \bibfnamefont{A.}},
  \bibinfo{author}{\bibfnamefont{R.~I.} \bibnamefont{Skehter}}, and
  \bibinfo{author}{\bibfnamefont{M.}~\bibnamefont{Jonson}},
  \bibinfo{year}{2001}, \bibinfo{journal}{Europhys. Lett.}
  \textbf{\bibinfo{volume}{54}}, \bibinfo{pages}{394}.

\bibitem[{\citenamefont{Karchev} \emph{et~al.}(2001)\citenamefont{Karchev,
  Blagoev, Bedell, and Littlewood}}]{littlewood}
\bibinfo{author}{\bibnamefont{Karchev}, \bibfnamefont{N.~I.}},
  \bibinfo{author}{\bibfnamefont{K.~B.} \bibnamefont{Blagoev}},
  \bibinfo{author}{\bibfnamefont{K.~S.} \bibnamefont{Bedell}}, and
  \bibinfo{author}{\bibfnamefont{P.~B.} \bibnamefont{Littlewood}},
  \bibinfo{year}{2001}, \bibinfo{journal}{Phys. Rev. Lett.}
  \textbf{\bibinfo{volume}{86}}, \bibinfo{pages}{846}.

\bibitem[{\citenamefont{Kharitonov}
  \emph{et~al.}(2005)\citenamefont{Kharitonov, A.F.Volkov, and
  K.B.Efetov}}]{kharitonov05}
\bibinfo{author}{\bibnamefont{Kharitonov}, \bibfnamefont{M.}},
  \bibinfo{author}{\bibnamefont{A.F.Volkov}}, and
  \bibinfo{author}{\bibnamefont{K.B.Efetov}}, \bibinfo{year}{2005},
  \bibinfo{journal}{to be published} .

\bibitem[{\citenamefont{Khusainov and Proshin}(1997)}]{khusainov}
\bibinfo{author}{\bibnamefont{Khusainov}, \bibfnamefont{M.~G.}}, and
  \bibinfo{author}{\bibfnamefont{Y.~N.} \bibnamefont{Proshin}},
  \bibinfo{year}{1997}, \bibinfo{journal}{Phys. Rev. B}
  \textbf{\bibinfo{volume}{56}}, \bibinfo{pages}{R14283}.

\bibitem[{\citenamefont{Kirckpatrick and Belitz}(1992)}]{kirkpatrick92}
\bibinfo{author}{\bibnamefont{Kirckpatrick}, \bibfnamefont{T.~R.}}, and
  \bibinfo{author}{\bibfnamefont{D.}~\bibnamefont{Belitz}},
  \bibinfo{year}{1992}, \bibinfo{journal}{Phys. Rev. B}
  \textbf{\bibinfo{volume}{46}}, \bibinfo{pages}{8393}.

\bibitem[{\citenamefont{Kirkpatrick and Belitz}(1991)}]{kirkpatrick91}
\bibinfo{author}{\bibnamefont{Kirkpatrick}, \bibfnamefont{T.~R.}}, and
  \bibinfo{author}{\bibfnamefont{D.}~\bibnamefont{Belitz}},
  \bibinfo{year}{1991}, \bibinfo{journal}{Phys. Rev. Lett.}
  \textbf{\bibinfo{volume}{66}}, \bibinfo{pages}{1536}.

\bibitem[{\citenamefont{Kontos} \emph{et~al.}(2002)\citenamefont{Kontos,
  Aprili, Lesueur, Genet, Stephanidis, and Boursier}}]{kontos02}
\bibinfo{author}{\bibnamefont{Kontos}, \bibfnamefont{T.}},
  \bibinfo{author}{\bibfnamefont{M.}~\bibnamefont{Aprili}},
  \bibinfo{author}{\bibfnamefont{J.}~\bibnamefont{Lesueur}},
  \bibinfo{author}{\bibfnamefont{F.}~\bibnamefont{Genet}},
  \bibinfo{author}{\bibfnamefont{B.}~\bibnamefont{Stephanidis}}, and
  \bibinfo{author}{\bibfnamefont{R.}~\bibnamefont{Boursier}},
  \bibinfo{year}{2002}, \bibinfo{journal}{Phys. Rev. Lett.}
  \textbf{\bibinfo{volume}{89}}, \bibinfo{pages}{137007}.

\bibitem[{\citenamefont{Kontos} \emph{et~al.}(2001)\citenamefont{Kontos,
  Aprili, Lesueur, and Grison}}]{kontos01}
\bibinfo{author}{\bibnamefont{Kontos}, \bibfnamefont{T.}},
  \bibinfo{author}{\bibfnamefont{M.}~\bibnamefont{Aprili}},
  \bibinfo{author}{\bibfnamefont{J.}~\bibnamefont{Lesueur}}, and
  \bibinfo{author}{\bibfnamefont{X.}~\bibnamefont{Grison}},
  \bibinfo{year}{2001}, \bibinfo{journal}{Phys. Rev. Lett.}
  \textbf{\bibinfo{volume}{86}}, \bibinfo{pages}{304}.

\bibitem[{\citenamefont{Kopnin}(2001)}]{kopnin}
\bibinfo{author}{\bibnamefont{Kopnin}, \bibfnamefont{N.~B.}},
  \bibinfo{year}{2001}, \emph{\bibinfo{title}{Theory of Nonequilibrium
  Superconductivity}} (\bibinfo{publisher}{Claredon Press-Oxford},
  \bibinfo{address}{Oxford}).

\bibitem[{\citenamefont{Kopu} \emph{et~al.}(2004)\citenamefont{Kopu, Eschrig,
  Cuevas, , and Fogelström}}]{kopu}
\bibinfo{author}{\bibnamefont{Kopu}, \bibfnamefont{J.}},
  \bibinfo{author}{\bibfnamefont{M.}~\bibnamefont{Eschrig}},
  \bibinfo{author}{\bibfnamefont{J.~C.} \bibnamefont{Cuevas}}, , and
  \bibinfo{author}{\bibfnamefont{M.}~\bibnamefont{Fogelström}},
  \bibinfo{year}{2004}, \bibinfo{journal}{Phys. Rev. B}
  \textbf{\bibinfo{volume}{69}}, \bibinfo{pages}{094501}.

\bibitem[{\citenamefont{Kouh and Valles}(2003)}]{kouh}
\bibinfo{author}{\bibnamefont{Kouh}, \bibfnamefont{T.}}, and
  \bibinfo{author}{\bibfnamefont{J.}~\bibnamefont{Valles}},
  \bibinfo{year}{2003}, \bibinfo{journal}{Phys. Rev. B}
  \textbf{\bibinfo{volume}{67}}, \bibinfo{pages}{140506}.

\bibitem[{\citenamefont{Krawiec} \emph{et~al.}(2004)\citenamefont{Krawiec,
  Gy\"{o}rffy, and Annett}}]{annett04}
\bibinfo{author}{\bibnamefont{Krawiec}, \bibfnamefont{M.}},
  \bibinfo{author}{\bibfnamefont{B.~L.} \bibnamefont{Gy\"{o}rffy}}, and
  \bibinfo{author}{\bibfnamefont{J.~F.} \bibnamefont{Annett}},
  \bibinfo{year}{2004}, \bibinfo{journal}{Phys. Rev. B}
  \textbf{\bibinfo{volume}{70}}, \bibinfo{pages}{134519}.

\bibitem[{\citenamefont{Krivoruchko and
  Koshina}(2001{\natexlab{a}})}]{krivoruchko1}
\bibinfo{author}{\bibnamefont{Krivoruchko}, \bibfnamefont{V.~N.}}, and
  \bibinfo{author}{\bibfnamefont{E.~A.} \bibnamefont{Koshina}},
  \bibinfo{year}{2001}{\natexlab{a}}, \bibinfo{journal}{Phys. Rev. B}
  \textbf{\bibinfo{volume}{63}}, \bibinfo{pages}{224515}.

\bibitem[{\citenamefont{Krivoruchko and
  Koshina}(2001{\natexlab{b}})}]{krivoruchko2}
\bibinfo{author}{\bibnamefont{Krivoruchko}, \bibfnamefont{V.~N.}}, and
  \bibinfo{author}{\bibfnamefont{E.~A.} \bibnamefont{Koshina}},
  \bibinfo{year}{2001}{\natexlab{b}}, \bibinfo{journal}{Phys. Rev. B}
  \textbf{\bibinfo{volume}{64}}, \bibinfo{pages}{172511}.

\bibitem[{\citenamefont{Krivoruchko and Koshina}(2002)}]{krivoruchko3}
\bibinfo{author}{\bibnamefont{Krivoruchko}, \bibfnamefont{V.~N.}}, and
  \bibinfo{author}{\bibfnamefont{E.~A.} \bibnamefont{Koshina}},
  \bibinfo{year}{2002}, \bibinfo{journal}{Phys. Rev. B}
  \textbf{\bibinfo{volume}{66}}, \bibinfo{pages}{014521}.

\bibitem[{\citenamefont{Kulic and Endres}(2000)}]{kulic_endres}
\bibinfo{author}{\bibnamefont{Kulic}, \bibfnamefont{M.~L.}}, and
  \bibinfo{author}{\bibfnamefont{M.}~\bibnamefont{Endres}},
  \bibinfo{year}{2000}, \bibinfo{journal}{Phys. Rev. B}
  \textbf{\bibinfo{volume}{62}}, \bibinfo{pages}{11846}.

\bibitem[{\citenamefont{Kulic and Kulic}(2001)}]{kulic_kulic}
\bibinfo{author}{\bibnamefont{Kulic}, \bibfnamefont{M.~L.}}, and
  \bibinfo{author}{\bibfnamefont{I.~M.} \bibnamefont{Kulic}},
  \bibinfo{year}{2001}, \bibinfo{journal}{Phys. Rev. B}
  \textbf{\bibinfo{volume}{63}}, \bibinfo{pages}{104503}.

\bibitem[{\citenamefont{Kulik and Yanson.}(1970)}]{kulik_book}
\bibinfo{author}{\bibnamefont{Kulik}, \bibfnamefont{I.~O.}}, and
  \bibinfo{author}{\bibfnamefont{I.~K.} \bibnamefont{Yanson.}},
  \bibinfo{year}{1970}, \emph{\bibinfo{title}{The Josephson Effect in
  Superconducting Tunneling Structures}} (\bibinfo{publisher}{Nauka Publ.House,
  Moscow}).

\bibitem[{\citenamefont{Kuprianov and Lukichev}(1988)}]{kuprianov}
\bibinfo{author}{\bibnamefont{Kuprianov}, \bibfnamefont{M.~Y.}}, and
  \bibinfo{author}{\bibfnamefont{V.~F.} \bibnamefont{Lukichev}},
  \bibinfo{year}{1988}, \bibinfo{journal}{Sov. Phys. JETP}
  \textbf{\bibinfo{volume}{67}}, \bibinfo{pages}{1163}.

\bibitem[{\citenamefont{Lambert and Raimondi}(1998)}]{lambert_rev}
\bibinfo{author}{\bibnamefont{Lambert}, \bibfnamefont{C.}}, and
  \bibinfo{author}{\bibfnamefont{R.}~\bibnamefont{Raimondi}},
  \bibinfo{year}{1998}, \bibinfo{journal}{J. Phys. Cond. Matt.}
  \textbf{\bibinfo{volume}{10}}, \bibinfo{pages}{901}.

\bibitem[{\citenamefont{Lambert} \emph{et~al.}(1997)\citenamefont{Lambert,
  Raimondi, Sweeney, and Volkov}}]{lambert_bc}
\bibinfo{author}{\bibnamefont{Lambert}, \bibfnamefont{C.}},
  \bibinfo{author}{\bibfnamefont{R.}~\bibnamefont{Raimondi}},
  \bibinfo{author}{\bibfnamefont{V.}~\bibnamefont{Sweeney}}, and
  \bibinfo{author}{\bibfnamefont{A.~F.} \bibnamefont{Volkov}},
  \bibinfo{year}{1997}, \bibinfo{journal}{Phys. Rev. B}
  \textbf{\bibinfo{volume}{55}}, \bibinfo{pages}{6015}.

\bibitem[{\citenamefont{Larkin}(1965)}]{larkin_triplet}
\bibinfo{author}{\bibnamefont{Larkin}, \bibfnamefont{A.~I.}},
  \bibinfo{year}{1965}, \bibinfo{journal}{JETP Lett.}
  \textbf{\bibinfo{volume}{2}}, \bibinfo{pages}{130}.

\bibitem[{\citenamefont{Larkin and Ovchinikov}(1964)}]{lo_state}
\bibinfo{author}{\bibnamefont{Larkin}, \bibfnamefont{A.~I.}}, and
  \bibinfo{author}{\bibfnamefont{Y.~N.} \bibnamefont{Ovchinikov}},
  \bibinfo{year}{1964}, \bibinfo{journal}{Zh. Eksp. Teor. Fiz.}
  \textbf{\bibinfo{volume}{47}}, \bibinfo{pages}{1136}, \bibinfo{note}{[JETP
  {\bf 20}, 762 (1965)]}.

\bibitem[{\citenamefont{Larkin and Ovchinnikov}(1968)}]{larkin_ovch}
\bibinfo{author}{\bibnamefont{Larkin}, \bibfnamefont{A.~I.}}, and
  \bibinfo{author}{\bibfnamefont{Y.~N.} \bibnamefont{Ovchinnikov}},
  \bibinfo{year}{1968}, \bibinfo{journal}{Zh. Eksp. Teor. Fiz.}
  \textbf{\bibinfo{volume}{55}}, \bibinfo{pages}{2262}, \bibinfo{note}{[JETP
  {\bf 28}, 1200 (1969)]}.

\bibitem[{\citenamefont{Larkin and Ovchinnikov}(1984)}]{lo_book}
\bibinfo{author}{\bibnamefont{Larkin}, \bibfnamefont{A.~I.}}, and
  \bibinfo{author}{\bibfnamefont{Y.~N.} \bibnamefont{Ovchinnikov}},
  \bibinfo{year}{1984}, \emph{\bibinfo{title}{Nonequilibrium
  Superconductivity}} (\bibinfo{publisher}{Elservier},
  \bibinfo{address}{Amsterdam}), p. \bibinfo{pages}{530}.

\bibitem[{\citenamefont{Lawrence and
  Giordano}(1996{\natexlab{a}})}]{Giordano96}
\bibinfo{author}{\bibnamefont{Lawrence}, \bibfnamefont{M.~D.}}, and
  \bibinfo{author}{\bibfnamefont{N.}~\bibnamefont{Giordano}},
  \bibinfo{year}{1996}{\natexlab{a}}, \bibinfo{journal}{J. Phys. Cond. Matter}
  \textbf{\bibinfo{volume}{8}}, \bibinfo{pages}{563}.

\bibitem[{\citenamefont{Lawrence and
  Giordano}(1996{\natexlab{b}})}]{Giordano99}
\bibinfo{author}{\bibnamefont{Lawrence}, \bibfnamefont{M.~D.}}, and
  \bibinfo{author}{\bibfnamefont{N.}~\bibnamefont{Giordano}},
  \bibinfo{year}{1996}{\natexlab{b}}, \bibinfo{journal}{J. Phys. Cond. Matter}
  \textbf{\bibinfo{volume}{11}}, \bibinfo{pages}{1089}.

\bibitem[{\citenamefont{Lazar} \emph{et~al.}(2000)\citenamefont{Lazar,
  Westerholt, Zabel, Tagirov, Goryunov, Garif'yanov, and Garifullin}}]{lazar}
\bibinfo{author}{\bibnamefont{Lazar}, \bibfnamefont{L.}},
  \bibinfo{author}{\bibfnamefont{K.}~\bibnamefont{Westerholt}},
  \bibinfo{author}{\bibfnamefont{H.}~\bibnamefont{Zabel}},
  \bibinfo{author}{\bibfnamefont{L.~R.} \bibnamefont{Tagirov}},
  \bibinfo{author}{\bibfnamefont{Y.~V.} \bibnamefont{Goryunov}},
  \bibinfo{author}{\bibfnamefont{N.~N.} \bibnamefont{Garif'yanov}}, and
  \bibinfo{author}{\bibfnamefont{I.~A.} \bibnamefont{Garifullin}},
  \bibinfo{year}{2000}, \bibinfo{journal}{Phys. Rev. B}
  \textbf{\bibinfo{volume}{61}}, \bibinfo{pages}{3711}.

\bibitem[{\citenamefont{Legget}(1975)}]{legget}
\bibinfo{author}{\bibnamefont{Legget}, \bibfnamefont{A.~J.}},
  \bibinfo{year}{1975}, \bibinfo{journal}{Rev. Mod. Phys.}
  \textbf{\bibinfo{volume}{47}}, \bibinfo{pages}{331}.

\bibitem[{\citenamefont{Li} \emph{et~al.}(2002)\citenamefont{Li, Zheng, Xing,
  Sun, , and Dong}}]{li}
\bibinfo{author}{\bibnamefont{Li}, \bibfnamefont{X.}},
  \bibinfo{author}{\bibfnamefont{Z.}~\bibnamefont{Zheng}},
  \bibinfo{author}{\bibfnamefont{D.~Y.} \bibnamefont{Xing}},
  \bibinfo{author}{\bibfnamefont{G.}~\bibnamefont{Sun}}, , and
  \bibinfo{author}{\bibfnamefont{Z.}~\bibnamefont{Dong}}, \bibinfo{year}{2002},
  \bibinfo{journal}{Phys. Rev. B} \textbf{\bibinfo{volume}{65}},
  \bibinfo{pages}{134507}.

\bibitem[{\citenamefont{Likharev}(1979)}]{likharev}
\bibinfo{author}{\bibnamefont{Likharev}, \bibfnamefont{K.~K.}},
  \bibinfo{year}{1979}, \bibinfo{journal}{Rev. Mod. Phys.}
  \textbf{\bibinfo{volume}{51}}, \bibinfo{pages}{101}.

\bibitem[{\citenamefont{Lodder and Yu.V.Nazarov}(1998)}]{lodder}
\bibinfo{author}{\bibnamefont{Lodder}, \bibfnamefont{A.}}, and
  \bibinfo{author}{\bibnamefont{Yu.V.Nazarov}}, \bibinfo{year}{1998},
  \bibinfo{journal}{Phys. Rev. B} \textbf{\bibinfo{volume}{59}},
  \bibinfo{pages}{5783}.

\bibitem[{\citenamefont{Luetkens} \emph{et~al.}(2003)\citenamefont{Luetkens,
  Korecki, Morenzoni, Prokscha, Birke, Glückler, Khasanov, Klauss, Slezak,
  Suter, Forgan, Niedermayer} \emph{et~al.}}]{muon}
\bibinfo{author}{\bibnamefont{Luetkens}, \bibfnamefont{H.}},
  \bibinfo{author}{\bibfnamefont{J.}~\bibnamefont{Korecki}},
  \bibinfo{author}{\bibfnamefont{E.}~\bibnamefont{Morenzoni}},
  \bibinfo{author}{\bibfnamefont{T.}~\bibnamefont{Prokscha}},
  \bibinfo{author}{\bibfnamefont{M.}~\bibnamefont{Birke}},
  \bibinfo{author}{\bibfnamefont{H.}~\bibnamefont{Glückler}},
  \bibinfo{author}{\bibfnamefont{R.}~\bibnamefont{Khasanov}},
  \bibinfo{author}{\bibfnamefont{H.-H.} \bibnamefont{Klauss}},
  \bibinfo{author}{\bibfnamefont{T.}~\bibnamefont{Slezak}},
  \bibinfo{author}{\bibfnamefont{A.}~\bibnamefont{Suter}},
  \bibinfo{author}{\bibfnamefont{E.~M.} \bibnamefont{Forgan}},
  \bibinfo{author}{\bibfnamefont{C.}~\bibnamefont{Niedermayer}}, \emph{et~al.},
  \bibinfo{year}{2003}, \bibinfo{journal}{Phys. Rev. Lett.}
  \textbf{\bibinfo{volume}{91}}, \bibinfo{pages}{017204}.

\bibitem[{\citenamefont{Lyuksyutov and Pokrovsky}(2004)}]{pokrovsky}
\bibinfo{author}{\bibnamefont{Lyuksyutov}, \bibfnamefont{I.~F.}}, and
  \bibinfo{author}{\bibfnamefont{V.}~\bibnamefont{Pokrovsky}},
  \bibinfo{year}{2004}, \bibinfo{title}{Ferromagnet-superconductor hybrids},
  \bibinfo{note}{cond-mat/0409137}.

\bibitem[{\citenamefont{Maeno} \emph{et~al.}(1994)\citenamefont{Maeno,
  Hashimoto, Yoshida, Nishizaki, Fujita, Bednorz, and Lichtenberg}}]{maeno}
\bibinfo{author}{\bibnamefont{Maeno}, \bibfnamefont{Y.}},
  \bibinfo{author}{\bibfnamefont{H.}~\bibnamefont{Hashimoto}},
  \bibinfo{author}{\bibfnamefont{K.}~\bibnamefont{Yoshida}},
  \bibinfo{author}{\bibfnamefont{S.}~\bibnamefont{Nishizaki}},
  \bibinfo{author}{\bibfnamefont{T.}~\bibnamefont{Fujita}},
  \bibinfo{author}{\bibfnamefont{J.~G.} \bibnamefont{Bednorz}}, and
  \bibinfo{author}{\bibfnamefont{F.}~\bibnamefont{Lichtenberg}},
  \bibinfo{year}{1994}, \bibinfo{journal}{Nature}
  \textbf{\bibinfo{volume}{372}}, \bibinfo{pages}{532}.

\bibitem[{\citenamefont{Maki}(1968)}]{maki}
\bibinfo{author}{\bibnamefont{Maki}, \bibfnamefont{K.}}, \bibinfo{year}{1968},
  \bibinfo{journal}{Progr. Theoret. Phys.} \textbf{\bibinfo{volume}{39}},
  \bibinfo{pages}{897}.

\bibitem[{\citenamefont{Maki}(1969)}]{maki_book}
\bibinfo{author}{\bibnamefont{Maki}, \bibfnamefont{K.}}, \bibinfo{year}{1969},
  \emph{\bibinfo{title}{Superconductivity}} (\bibinfo{publisher}{Dekker, New
  York}).

\bibitem[{\citenamefont{McCann} \emph{et~al.}(2000)\citenamefont{McCann, Falko,
  Volkov, and Lambert}}]{mccann00}
\bibinfo{author}{\bibnamefont{McCann}, \bibfnamefont{E.}},
  \bibinfo{author}{\bibfnamefont{V.~I.} \bibnamefont{Falko}},
  \bibinfo{author}{\bibfnamefont{A.~F.} \bibnamefont{Volkov}}, and
  \bibinfo{author}{\bibfnamefont{C.~J.} \bibnamefont{Lambert}},
  \bibinfo{year}{2000}, \bibinfo{journal}{Phys. Rev. B}
  \textbf{\bibinfo{volume}{62}}, \bibinfo{pages}{6015}.

\bibitem[{\citenamefont{McMillan}(1968)}]{mcmillan}
\bibinfo{author}{\bibnamefont{McMillan}, \bibfnamefont{W.~L.}},
  \bibinfo{year}{1968}, \bibinfo{journal}{Phys. Rev.}
  \textbf{\bibinfo{volume}{175}}, \bibinfo{pages}{537}.

\bibitem[{\citenamefont{M\'{e}lin}(2001)}]{melin01}
\bibinfo{author}{\bibnamefont{M\'{e}lin}, \bibfnamefont{R.}},
  \bibinfo{year}{2001}, \bibinfo{journal}{J. Phys.: Condens. Matter}
  \textbf{\bibinfo{volume}{13}}, \bibinfo{pages}{6445}.

\bibitem[{\citenamefont{M\'{e}lin and Feinberg}(2004)}]{melin04_2}
\bibinfo{author}{\bibnamefont{M\'{e}lin}, \bibfnamefont{R.}}, and
  \bibinfo{author}{\bibfnamefont{D.}~\bibnamefont{Feinberg}},
  \bibinfo{year}{2004}, \bibinfo{journal}{Phys. Rev. B}
  \textbf{\bibinfo{volume}{70}}, \bibinfo{pages}{174509}.

\bibitem[{\citenamefont{M\'{e}lin and Peysson}(2003)}]{melin03}
\bibinfo{author}{\bibnamefont{M\'{e}lin}, \bibfnamefont{R.}}, and
  \bibinfo{author}{\bibfnamefont{S.}~\bibnamefont{Peysson}},
  \bibinfo{year}{2003}, \bibinfo{journal}{Phys. Rev. B}
  \textbf{\bibinfo{volume}{68}}, \bibinfo{pages}{174515}.

\bibitem[{\citenamefont{Melsen} \emph{et~al.}(1996)\citenamefont{Melsen,
  P.W.Brouwer, K.M.Frahm, and Beenakker}}]{beenakker96}
\bibinfo{author}{\bibnamefont{Melsen}, \bibfnamefont{J.~A.}},
  \bibinfo{author}{\bibnamefont{P.W.Brouwer}},
  \bibinfo{author}{\bibnamefont{K.M.Frahm}}, and
  \bibinfo{author}{\bibfnamefont{C.}~\bibnamefont{Beenakker}},
  \bibinfo{year}{1996}, \bibinfo{journal}{Europhys.Lett.}
  \textbf{\bibinfo{volume}{35}}, \bibinfo{pages}{7}.

\bibitem[{\citenamefont{Mercaldo} \emph{et~al.}(1996)\citenamefont{Mercaldo,
  Affanasio, Coccorese, Maritato, Prischepa, and Salvato}}]{mercaldo}
\bibinfo{author}{\bibnamefont{Mercaldo}, \bibfnamefont{V.}},
  \bibinfo{author}{\bibfnamefont{C.}~\bibnamefont{Affanasio}},
  \bibinfo{author}{\bibfnamefont{C.}~\bibnamefont{Coccorese}},
  \bibinfo{author}{\bibfnamefont{L.}~\bibnamefont{Maritato}},
  \bibinfo{author}{\bibfnamefont{S.~L.} \bibnamefont{Prischepa}}, and
  \bibinfo{author}{\bibfnamefont{M.}~\bibnamefont{Salvato}},
  \bibinfo{year}{1996}, \bibinfo{journal}{Phys. Rev. B}
  \textbf{\bibinfo{volume}{53}}, \bibinfo{pages}{14040}.

\bibitem[{\citenamefont{Millis} \emph{et~al.}(1988)\citenamefont{Millis,
  D.Rainer, and Sauls}}]{millis}
\bibinfo{author}{\bibnamefont{Millis}, \bibfnamefont{A.}},
  \bibinfo{author}{\bibnamefont{D.Rainer}}, and
  \bibinfo{author}{\bibfnamefont{J.~A.} \bibnamefont{Sauls}},
  \bibinfo{year}{1988}, \bibinfo{journal}{Phys. Rev. B}
  \textbf{\bibinfo{volume}{38}}, \bibinfo{pages}{4504}.

\bibitem[{\citenamefont{Mineev and Samokhin}(1999)}]{mineev}
\bibinfo{author}{\bibnamefont{Mineev}, \bibfnamefont{V.~P.}}, and
  \bibinfo{author}{\bibfnamefont{K.~V.} \bibnamefont{Samokhin}},
  \bibinfo{year}{1999}, \emph{\bibinfo{title}{Introdction to Unconventional
  Superconductivity}} (\bibinfo{publisher}{Gordon and Breach},
  \bibinfo{address}{Amsterdam}).

\bibitem[{\citenamefont{M\"{u}hge} \emph{et~al.}(1996)\citenamefont{M\"{u}hge,
  Garif'yanov, Goryunov, Khaliullin, Tagirov, Westerholt, Garifullin, and
  Zabel}}]{muehge_prl}
\bibinfo{author}{\bibnamefont{M\"{u}hge}, \bibfnamefont{T.}},
  \bibinfo{author}{\bibfnamefont{N.}~\bibnamefont{Garif'yanov}},
  \bibinfo{author}{\bibfnamefont{Y.~V.} \bibnamefont{Goryunov}},
  \bibinfo{author}{\bibfnamefont{G.~G.} \bibnamefont{Khaliullin}},
  \bibinfo{author}{\bibfnamefont{L.~R.} \bibnamefont{Tagirov}},
  \bibinfo{author}{\bibfnamefont{K.}~\bibnamefont{Westerholt}},
  \bibinfo{author}{\bibfnamefont{I.~A.} \bibnamefont{Garifullin}}, and
  \bibinfo{author}{\bibfnamefont{H.}~\bibnamefont{Zabel}},
  \bibinfo{year}{1996}, \bibinfo{journal}{Phys. Rev. Lett.}
  \textbf{\bibinfo{volume}{77}}, \bibinfo{pages}{1857}.

\bibitem[{\citenamefont{M\"{u}hge} \emph{et~al.}(1998)\citenamefont{M\"{u}hge,
  Garif'yanov, Goryunov, Theis-Br\"{o}hl, Westerholt, Garifullin, and
  Zabel}}]{muehge}
\bibinfo{author}{\bibnamefont{M\"{u}hge}, \bibfnamefont{T.}},
  \bibinfo{author}{\bibfnamefont{N.}~\bibnamefont{Garif'yanov}},
  \bibinfo{author}{\bibfnamefont{Y.~V.} \bibnamefont{Goryunov}},
  \bibinfo{author}{\bibfnamefont{K.}~\bibnamefont{Theis-Br\"{o}hl}},
  \bibinfo{author}{\bibfnamefont{K.}~\bibnamefont{Westerholt}},
  \bibinfo{author}{\bibfnamefont{I.~A.} \bibnamefont{Garifullin}}, and
  \bibinfo{author}{\bibfnamefont{H.}~\bibnamefont{Zabel}},
  \bibinfo{year}{1998}, \bibinfo{journal}{Physica C}
  \textbf{\bibinfo{volume}{296}}, \bibinfo{pages}{325}.

\bibitem[{\citenamefont{Nambu}(1960)}]{Nambu}
\bibinfo{author}{\bibnamefont{Nambu}, \bibfnamefont{Y.}}, \bibinfo{year}{1960},
  \bibinfo{journal}{Phys. Rev.} \textbf{\bibinfo{volume}{117}},
  \bibinfo{pages}{648}.

\bibitem[{\citenamefont{Nazarov}(1999)}]{nazarov_bc}
\bibinfo{author}{\bibnamefont{Nazarov}, \bibfnamefont{Y.~V.}},
  \bibinfo{year}{1999}, \bibinfo{journal}{Superlattices Microstructures}
  \textbf{\bibinfo{volume}{25}}, \bibinfo{pages}{1221}.

\bibitem[{\citenamefont{Nazarov and Stoof}(1996)}]{nazarov_stoof}
\bibinfo{author}{\bibnamefont{Nazarov}, \bibfnamefont{Y.~V.}}, and
  \bibinfo{author}{\bibfnamefont{T.~H.} \bibnamefont{Stoof}},
  \bibinfo{year}{1996}, \bibinfo{journal}{Phys. Rev. Lett.}
  \textbf{\bibinfo{volume}{76}}, \bibinfo{pages}{823}.

\bibitem[{\citenamefont{Nugent} \emph{et~al.}(2004)\citenamefont{Nugent,
  Sosnin, and Petrashov}}]{petrashov04}
\bibinfo{author}{\bibnamefont{Nugent}, \bibfnamefont{P.}},
  \bibinfo{author}{\bibfnamefont{I.}~\bibnamefont{Sosnin}}, and
  \bibinfo{author}{\bibfnamefont{V.~T.} \bibnamefont{Petrashov}},
  \bibinfo{year}{2004}, \bibinfo{journal}{J. Phys.: Condens. Matter}
  \textbf{\bibinfo{volume}{16}}, \bibinfo{pages}{L509}.

\bibitem[{\citenamefont{Obiand} \emph{et~al.}(1999)\citenamefont{Obiand, Ikebe,
  Kubo, and Fujimori}}]{obi}
\bibinfo{author}{\bibnamefont{Obiand}, \bibfnamefont{Y.}},
  \bibinfo{author}{\bibfnamefont{M.}~\bibnamefont{Ikebe}},
  \bibinfo{author}{\bibfnamefont{T.}~\bibnamefont{Kubo}}, and
  \bibinfo{author}{\bibfnamefont{H.}~\bibnamefont{Fujimori}},
  \bibinfo{year}{1999}, \bibinfo{journal}{Physica C}
  \textbf{\bibinfo{volume}{317-318}}, \bibinfo{pages}{149}.

\bibitem[{\citenamefont{Ogrin} \emph{et~al.}(2000)\citenamefont{Ogrin, Lee,
  Hillier, Mitchell, and Shen}}]{ogrin}
\bibinfo{author}{\bibnamefont{Ogrin}, \bibfnamefont{F.~Y.}},
  \bibinfo{author}{\bibfnamefont{S.~L.} \bibnamefont{Lee}},
  \bibinfo{author}{\bibfnamefont{A.~D.} \bibnamefont{Hillier}},
  \bibinfo{author}{\bibfnamefont{A.}~\bibnamefont{Mitchell}}, and
  \bibinfo{author}{\bibfnamefont{T.-H.} \bibnamefont{Shen}},
  \bibinfo{year}{2000}, \bibinfo{journal}{Phys. Rev. B}
  \textbf{\bibinfo{volume}{62}}, \bibinfo{pages}{6021}.

\bibitem[{\citenamefont{Oh} \emph{et~al.}(2000)\citenamefont{Oh, Kim, Youm, and
  Beasley}}]{beasley}
\bibinfo{author}{\bibnamefont{Oh}, \bibfnamefont{S.}},
  \bibinfo{author}{\bibfnamefont{Y.-H.} \bibnamefont{Kim}},
  \bibinfo{author}{\bibfnamefont{D.}~\bibnamefont{Youm}}, and
  \bibinfo{author}{\bibfnamefont{M.~R.} \bibnamefont{Beasley}},
  \bibinfo{year}{2000}, \bibinfo{journal}{Phys. Rev. B}
  \textbf{\bibinfo{volume}{63}}, \bibinfo{pages}{052501}.

\bibitem[{\citenamefont{Ostrovsky} \emph{et~al.}(2001)\citenamefont{Ostrovsky,
  Skvortsov, and Feigel'man}}]{ostrovsky}
\bibinfo{author}{\bibnamefont{Ostrovsky}, \bibfnamefont{P.~M.}},
  \bibinfo{author}{\bibfnamefont{M.~A.} \bibnamefont{Skvortsov}}, and
  \bibinfo{author}{\bibfnamefont{M.~V.} \bibnamefont{Feigel'man}},
  \bibinfo{year}{2001}, \bibinfo{journal}{Phys. Rev. Lett.}
  \textbf{\bibinfo{volume}{87}}, \bibinfo{pages}{027002}.

\bibitem[{\citenamefont{Petrashov} \emph{et~al.}(1995)\citenamefont{Petrashov,
  Antonov, Delsing, and Claeson}}]{petrashov_sn}
\bibinfo{author}{\bibnamefont{Petrashov}, \bibfnamefont{V.~T.}},
  \bibinfo{author}{\bibfnamefont{V.~N.} \bibnamefont{Antonov}},
  \bibinfo{author}{\bibfnamefont{P.}~\bibnamefont{Delsing}}, and
  \bibinfo{author}{\bibfnamefont{T.}~\bibnamefont{Claeson}},
  \bibinfo{year}{1995}, \bibinfo{journal}{Phys. Rev. Lett.}
  \textbf{\bibinfo{volume}{74}}, \bibinfo{pages}{5268}.

\bibitem[{\citenamefont{Petrashov} \emph{et~al.}(1999)\citenamefont{Petrashov,
  Sosnin, Cox, Parsons, and Troadec}}]{petrashov}
\bibinfo{author}{\bibnamefont{Petrashov}, \bibfnamefont{V.~T.}},
  \bibinfo{author}{\bibfnamefont{I.~A.} \bibnamefont{Sosnin}},
  \bibinfo{author}{\bibfnamefont{I.}~\bibnamefont{Cox}},
  \bibinfo{author}{\bibfnamefont{A.}~\bibnamefont{Parsons}}, and
  \bibinfo{author}{\bibfnamefont{C.}~\bibnamefont{Troadec}},
  \bibinfo{year}{1999}, \bibinfo{journal}{Phys. Rev. Lett.}
  \textbf{\bibinfo{volume}{83}}, \bibinfo{pages}{3281}.

\bibitem[{\citenamefont{Pilgram} \emph{et~al.}(2000)\citenamefont{Pilgram,
  Belzig, and Bruder}}]{bruder}
\bibinfo{author}{\bibnamefont{Pilgram}, \bibfnamefont{S.}},
  \bibinfo{author}{\bibfnamefont{W.}~\bibnamefont{Belzig}}, and
  \bibinfo{author}{\bibfnamefont{C.}~\bibnamefont{Bruder}},
  \bibinfo{year}{2000}, \bibinfo{journal}{Phys. Rev. B}
  \textbf{\bibinfo{volume}{62}}, \bibinfo{pages}{12462}.

\bibitem[{\citenamefont{Pothier} \emph{et~al.}(1994)\citenamefont{Pothier,
  Gueron, Esteve, and Devoret}}]{pothier}
\bibinfo{author}{\bibnamefont{Pothier}, \bibfnamefont{H.}},
  \bibinfo{author}{\bibfnamefont{S.}~\bibnamefont{Gueron}},
  \bibinfo{author}{\bibfnamefont{D.}~\bibnamefont{Esteve}}, and
  \bibinfo{author}{\bibfnamefont{M.~M.} \bibnamefont{Devoret}},
  \bibinfo{year}{1994}, \bibinfo{journal}{Phys. Rev. Lett.}
  \textbf{\bibinfo{volume}{73}}, \bibinfo{pages}{2488}.

\bibitem[{\citenamefont{Proshin} \emph{et~al.}(2001)\citenamefont{Proshin,
  Izyumov, and Khusainov}}]{proshin01}
\bibinfo{author}{\bibnamefont{Proshin}, \bibfnamefont{Y.~N.}},
  \bibinfo{author}{\bibfnamefont{Y.~A.} \bibnamefont{Izyumov}}, and
  \bibinfo{author}{\bibfnamefont{M.~G.} \bibnamefont{Khusainov}},
  \bibinfo{year}{2001}, \bibinfo{journal}{Phys. Rev. B}
  \textbf{\bibinfo{volume}{64}}, \bibinfo{pages}{064522}.

\bibitem[{\citenamefont{Proshin and Khusainov}(1998)}]{proshin98}
\bibinfo{author}{\bibnamefont{Proshin}, \bibfnamefont{Y.~N.}}, and
  \bibinfo{author}{\bibfnamefont{M.~G.} \bibnamefont{Khusainov}},
  \bibinfo{year}{1998}, \bibinfo{journal}{Sov. Phys. JETP}
  \textbf{\bibinfo{volume}{86}}, \bibinfo{pages}{930}.

\bibitem[{\citenamefont{Proshin and Khusainov}(1999)}]{proshin99}
\bibinfo{author}{\bibnamefont{Proshin}, \bibfnamefont{Y.~N.}}, and
  \bibinfo{author}{\bibfnamefont{M.~G.} \bibnamefont{Khusainov}},
  \bibinfo{year}{1999}, \bibinfo{journal}{Sov. Phys. JETP}
  \textbf{\bibinfo{volume}{89}}, \bibinfo{pages}{1021}.

\bibitem[{\citenamefont{Quirion} \emph{et~al.}(2002)\citenamefont{Quirion,
  C.Hoffmann, F.Lefloch, and M.Sanquer}}]{Sanquer02}
\bibinfo{author}{\bibnamefont{Quirion}, \bibfnamefont{D.}},
  \bibinfo{author}{\bibnamefont{C.Hoffmann}},
  \bibinfo{author}{\bibnamefont{F.Lefloch}}, and
  \bibinfo{author}{\bibnamefont{M.Sanquer}}, \bibinfo{year}{2002},
  \bibinfo{journal}{Phys. Rev. B} \textbf{\bibinfo{volume}{60}},
  \bibinfo{pages}{100508}.

\bibitem[{\citenamefont{Radovic} \emph{et~al.}(1991)\citenamefont{Radovic,
  Dobrosavljevic-Grujic, Buzdin, and Clem}}]{radovic2}
\bibinfo{author}{\bibnamefont{Radovic}, \bibfnamefont{Z.}},
  \bibinfo{author}{\bibfnamefont{L.}~\bibnamefont{Dobrosavljevic-Grujic}},
  \bibinfo{author}{\bibfnamefont{A.~I.} \bibnamefont{Buzdin}}, and
  \bibinfo{author}{\bibfnamefont{J.~R.} \bibnamefont{Clem}},
  \bibinfo{year}{1991}, \bibinfo{journal}{Phys. Rev. B}
  \textbf{\bibinfo{volume}{44}}, \bibinfo{pages}{759}.

\bibitem[{\citenamefont{Radovic} \emph{et~al.}(2003)\citenamefont{Radovic,
  Lazarides, and Flytzanis}}]{radovic03}
\bibinfo{author}{\bibnamefont{Radovic}, \bibfnamefont{Z.}},
  \bibinfo{author}{\bibfnamefont{N.}~\bibnamefont{Lazarides}}, and
  \bibinfo{author}{\bibfnamefont{N.}~\bibnamefont{Flytzanis}},
  \bibinfo{year}{2003}, \bibinfo{journal}{Phys. Rev. B}
  \textbf{\bibinfo{volume}{68}}, \bibinfo{pages}{014501}.

\bibitem[{\citenamefont{Rammer and Smith}(1986)}]{rammer}
\bibinfo{author}{\bibnamefont{Rammer}, \bibfnamefont{J.}}, and
  \bibinfo{author}{\bibfnamefont{H.}~\bibnamefont{Smith}},
  \bibinfo{year}{1986}, \bibinfo{journal}{Rev. Mod. Phys.}
  \textbf{\bibinfo{volume}{58}}, \bibinfo{pages}{323}.

\bibitem[{\citenamefont{Rashba}(1960)}]{rashba}
\bibinfo{author}{\bibnamefont{Rashba}, \bibfnamefont{E.}},
  \bibinfo{year}{1960}, \bibinfo{journal}{Sov. Phys. Solid State}
  \textbf{\bibinfo{volume}{2}}, \bibinfo{pages}{1109}.

\bibitem[{\citenamefont{Reymond} \emph{et~al.}(2000)\citenamefont{Reymond,
  SanGiorgio, Beasley, Kim, Kim, and Char}}]{beasley05}
\bibinfo{author}{\bibnamefont{Reymond}, \bibfnamefont{S.}},
  \bibinfo{author}{\bibfnamefont{P.}~\bibnamefont{SanGiorgio}},
  \bibinfo{author}{\bibfnamefont{M.}~\bibnamefont{Beasley}},
  \bibinfo{author}{\bibfnamefont{J.}~\bibnamefont{Kim}},
  \bibinfo{author}{\bibfnamefont{T.}~\bibnamefont{Kim}}, and
  \bibinfo{author}{\bibfnamefont{K.}~\bibnamefont{Char}}, \bibinfo{year}{2000},
  \bibinfo{journal}{cond-mat/0504739} .

\bibitem[{\citenamefont{Rusanov} \emph{et~al.}(2004)\citenamefont{Rusanov,
  Hesselberth, and Aarts}}]{rusanov}
\bibinfo{author}{\bibnamefont{Rusanov}, \bibfnamefont{A.~Y.}},
  \bibinfo{author}{\bibfnamefont{M.}~\bibnamefont{Hesselberth}}, and
  \bibinfo{author}{\bibfnamefont{J.}~\bibnamefont{Aarts}},
  \bibinfo{year}{2004}, \bibinfo{journal}{Phys. Rev. Lett.}
  \textbf{\bibinfo{volume}{93}}, \bibinfo{pages}{057002}.

\bibitem[{\citenamefont{Rusinov}(1969)}]{rusinov}
\bibinfo{author}{\bibnamefont{Rusinov}, \bibfnamefont{A.~I.}},
  \bibinfo{year}{1969}, \bibinfo{journal}{Sov. Phys. JETP Lett.}
  \textbf{\bibinfo{volume}{9}}, \bibinfo{pages}{85}.

\bibitem[{\citenamefont{Ryazanov} \emph{et~al.}(2001)\citenamefont{Ryazanov,
  Oboznov, Rusanov, Veretennikov, Golubov, and Aarts}}]{ryazanov}
\bibinfo{author}{\bibnamefont{Ryazanov}, \bibfnamefont{V.~V.}},
  \bibinfo{author}{\bibfnamefont{V.~A.} \bibnamefont{Oboznov}},
  \bibinfo{author}{\bibfnamefont{A.~Y.} \bibnamefont{Rusanov}},
  \bibinfo{author}{\bibfnamefont{A.~V.} \bibnamefont{Veretennikov}},
  \bibinfo{author}{\bibfnamefont{A.~A.} \bibnamefont{Golubov}}, and
  \bibinfo{author}{\bibfnamefont{J.}~\bibnamefont{Aarts}},
  \bibinfo{year}{2001}, \bibinfo{journal}{Phys. Rev. Lett.}
  \textbf{\bibinfo{volume}{86}}, \bibinfo{pages}{2427}.

\bibitem[{\citenamefont{Saint-James}(1964)}]{sjames}
\bibinfo{author}{\bibnamefont{Saint-James}, \bibfnamefont{D.}},
  \bibinfo{year}{1964}, \bibinfo{journal}{J. Phys. (Paris)}
  \textbf{\bibinfo{volume}{25}}, \bibinfo{pages}{899}.

\bibitem[{\citenamefont{Sakurai}(1970)}]{sakurai}
\bibinfo{author}{\bibnamefont{Sakurai}, \bibfnamefont{A.}},
  \bibinfo{year}{1970}, \bibinfo{journal}{Prog. Theor. Phys.}
  \textbf{\bibinfo{volume}{44}}, \bibinfo{pages}{1472}.

\bibitem[{\citenamefont{Salkola} \emph{et~al.}(1997)\citenamefont{Salkola,
  Balatsky, and Schrieffer}}]{schrieffer}
\bibinfo{author}{\bibnamefont{Salkola}, \bibfnamefont{M.~I.}},
  \bibinfo{author}{\bibfnamefont{A.~V.} \bibnamefont{Balatsky}}, and
  \bibinfo{author}{\bibfnamefont{J.~R.} \bibnamefont{Schrieffer}},
  \bibinfo{year}{1997}, \bibinfo{journal}{Phys. Rev. B}
  \textbf{\bibinfo{volume}{55}}, \bibinfo{pages}{12648}.

\bibitem[{\citenamefont{Sarma}(1963)}]{sarma}
\bibinfo{author}{\bibnamefont{Sarma}, \bibfnamefont{G.}}, \bibinfo{year}{1963},
  \bibinfo{journal}{J. Phys. Chem. Solids} \textbf{\bibinfo{volume}{24}},
  \bibinfo{pages}{1029}.

\bibitem[{\citenamefont{Sefrioui} \emph{et~al.}(2003)\citenamefont{Sefrioui,
  Arias, Pe$\tilde{\rm n}$a, Villegas, Varela, Prieto, Le\'{o}n, Martinez, and
  Santamaria}}]{pena}
\bibinfo{author}{\bibnamefont{Sefrioui}, \bibfnamefont{Z.}},
  \bibinfo{author}{\bibfnamefont{D.}~\bibnamefont{Arias}},
  \bibinfo{author}{\bibfnamefont{V.}~\bibnamefont{Pe$\tilde{\rm n}$a}},
  \bibinfo{author}{\bibfnamefont{J.~E.} \bibnamefont{Villegas}},
  \bibinfo{author}{\bibfnamefont{M.}~\bibnamefont{Varela}},
  \bibinfo{author}{\bibfnamefont{P.}~\bibnamefont{Prieto}},
  \bibinfo{author}{\bibfnamefont{C.}~\bibnamefont{Le\'{o}n}},
  \bibinfo{author}{\bibfnamefont{J.~L.} \bibnamefont{Martinez}}, and
  \bibinfo{author}{\bibfnamefont{J.}~\bibnamefont{Santamaria}},
  \bibinfo{year}{2003}, \bibinfo{journal}{Phys. Rev. B}
  \textbf{\bibinfo{volume}{67}}, \bibinfo{pages}{214511}.

\bibitem[{\citenamefont{Sellier} \emph{et~al.}(2004)\citenamefont{Sellier,
  Baraduc, Lefloch, and Calemczuk}}]{sellier}
\bibinfo{author}{\bibnamefont{Sellier}, \bibfnamefont{H.}},
  \bibinfo{author}{\bibfnamefont{C.}~\bibnamefont{Baraduc}},
  \bibinfo{author}{\bibfnamefont{F.}~\bibnamefont{Lefloch}}, and
  \bibinfo{author}{\bibfnamefont{R.}~\bibnamefont{Calemczuk}},
  \bibinfo{year}{2004}, \bibinfo{journal}{Phys. Rev. Lett.}
  \textbf{\bibinfo{volume}{92}}, \bibinfo{pages}{257005}.

\bibitem[{\citenamefont{Serene and Reiner}(1983)}]{serene}
\bibinfo{author}{\bibnamefont{Serene}, \bibfnamefont{J.~W.}}, and
  \bibinfo{author}{\bibfnamefont{D.}~\bibnamefont{Reiner}},
  \bibinfo{year}{1983}, \bibinfo{journal}{Phys. Reports}
  \textbf{\bibinfo{volume}{101}}, \bibinfo{pages}{222}.

\bibitem[{\citenamefont{Shapira} \emph{et~al.}(2000)\citenamefont{Shapira,
  Linfield, Lambert, Serviour, Volkov, and Zaitsev}}]{shapira}
\bibinfo{author}{\bibnamefont{Shapira}, \bibfnamefont{S.}},
  \bibinfo{author}{\bibfnamefont{E.~H.} \bibnamefont{Linfield}},
  \bibinfo{author}{\bibfnamefont{C.~J.} \bibnamefont{Lambert}},
  \bibinfo{author}{\bibfnamefont{R.}~\bibnamefont{Serviour}},
  \bibinfo{author}{\bibfnamefont{A.~F.} \bibnamefont{Volkov}}, and
  \bibinfo{author}{\bibfnamefont{A.~V.} \bibnamefont{Zaitsev}},
  \bibinfo{year}{2000}, \bibinfo{journal}{Phys. Rev. Lett.}
  \textbf{\bibinfo{volume}{84}}, \bibinfo{pages}{159}.

\bibitem[{\citenamefont{Shelankov and Ozana}(2000)}]{shelankov}
\bibinfo{author}{\bibnamefont{Shelankov}, \bibfnamefont{A.}}, and
  \bibinfo{author}{\bibfnamefont{M.}~\bibnamefont{Ozana}},
  \bibinfo{year}{2000}, \bibinfo{journal}{Phys. Rev. B}
  \textbf{\bibinfo{volume}{61}}, \bibinfo{pages}{7077}.

\bibitem[{\citenamefont{Shen} \emph{et~al.}(2003)\citenamefont{Shen, Zheng,
  Liu, and Xing}}]{shen}
\bibinfo{author}{\bibnamefont{Shen}, \bibfnamefont{R.}},
  \bibinfo{author}{\bibfnamefont{Z.~M.} \bibnamefont{Zheng}},
  \bibinfo{author}{\bibfnamefont{S.}~\bibnamefont{Liu}}, and
  \bibinfo{author}{\bibfnamefont{D.~Y.} \bibnamefont{Xing}},
  \bibinfo{year}{2003}, \bibinfo{journal}{Phys. Rev. B}
  \textbf{\bibinfo{volume}{67}}, \bibinfo{pages}{024514}.

\bibitem[{\citenamefont{Shiba}(1968)}]{shiba}
\bibinfo{author}{\bibnamefont{Shiba}, \bibfnamefont{H.}}, \bibinfo{year}{1968},
  \bibinfo{journal}{Prog. Theor. Phys.} \textbf{\bibinfo{volume}{40}},
  \bibinfo{pages}{435}.

\bibitem[{\citenamefont{Stahn} \emph{et~al.}(2005)\citenamefont{Stahn,
  Chakhalian, Niedermayer, Hoppler, Gutberlet, Voigt, Treubel, Habermeier,
  Cristiani, Keimer, and Bernhard}}]{bernhard}
\bibinfo{author}{\bibnamefont{Stahn}, \bibfnamefont{J.}},
  \bibinfo{author}{\bibfnamefont{J.}~\bibnamefont{Chakhalian}},
  \bibinfo{author}{\bibfnamefont{C.}~\bibnamefont{Niedermayer}},
  \bibinfo{author}{\bibfnamefont{J.}~\bibnamefont{Hoppler}},
  \bibinfo{author}{\bibfnamefont{T.}~\bibnamefont{Gutberlet}},
  \bibinfo{author}{\bibfnamefont{J.}~\bibnamefont{Voigt}},
  \bibinfo{author}{\bibfnamefont{F.}~\bibnamefont{Treubel}},
  \bibinfo{author}{\bibfnamefont{H.-U.} \bibnamefont{Habermeier}},
  \bibinfo{author}{\bibfnamefont{G.}~\bibnamefont{Cristiani}},
  \bibinfo{author}{\bibfnamefont{B.}~\bibnamefont{Keimer}}, and
  \bibinfo{author}{\bibfnamefont{C.}~\bibnamefont{Bernhard}},
  \bibinfo{year}{2005}, \bibinfo{journal}{Phys. Rev. B}
  \textbf{\bibinfo{volume}{71}}, \bibinfo{pages}{140509}.

\bibitem[{\citenamefont{Strunk} \emph{et~al.}(1994)\citenamefont{Strunk,
  S\"urgers, Paschen, and v.~L\"ohneysen}}]{strunk}
\bibinfo{author}{\bibnamefont{Strunk}, \bibfnamefont{C.}},
  \bibinfo{author}{\bibfnamefont{C.}~\bibnamefont{S\"urgers}},
  \bibinfo{author}{\bibfnamefont{U.}~\bibnamefont{Paschen}}, and
  \bibinfo{author}{\bibfnamefont{H.}~\bibnamefont{v.~L\"ohneysen}},
  \bibinfo{year}{1994}, \bibinfo{journal}{Phys. Rev. B}
  \textbf{\bibinfo{volume}{49}}, \bibinfo{pages}{4053}.

\bibitem[{\citenamefont{Taddei} \emph{et~al.}(2001)\citenamefont{Taddei,
  Sanvito, and Lambert}}]{lambert01}
\bibinfo{author}{\bibnamefont{Taddei}, \bibfnamefont{F.}},
  \bibinfo{author}{\bibfnamefont{S.}~\bibnamefont{Sanvito}}, and
  \bibinfo{author}{\bibfnamefont{C.~J.} \bibnamefont{Lambert}},
  \bibinfo{year}{2001}, \bibinfo{journal}{Phys. Rev. B}
  \textbf{\bibinfo{volume}{63}}, \bibinfo{pages}{012404}.

\bibitem[{\citenamefont{Tagirov}(1998)}]{tagirov_C}
\bibinfo{author}{\bibnamefont{Tagirov}, \bibfnamefont{L.~R.}},
  \bibinfo{year}{1998}, \bibinfo{journal}{Physica C}
  \textbf{\bibinfo{volume}{307}}, \bibinfo{pages}{145}.

\bibitem[{\citenamefont{Taras-Semchuk and Altland}(2001)}]{altland01}
\bibinfo{author}{\bibnamefont{Taras-Semchuk}, \bibfnamefont{D.}}, and
  \bibinfo{author}{\bibfnamefont{A.}~\bibnamefont{Altland}},
  \bibinfo{year}{2001}, \bibinfo{journal}{Phys. Rev. B}
  \textbf{\bibinfo{volume}{64}}, \bibinfo{pages}{014512}.

\bibitem[{\citenamefont{Tkachov} \emph{et~al.}(2002)\citenamefont{Tkachov,
  McCann, and Falko}}]{falko02}
\bibinfo{author}{\bibnamefont{Tkachov}, \bibfnamefont{G.}},
  \bibinfo{author}{\bibfnamefont{E.}~\bibnamefont{McCann}}, and
  \bibinfo{author}{\bibfnamefont{V.~I.} \bibnamefont{Falko}},
  \bibinfo{year}{2002}, \bibinfo{journal}{Phys. Rev. B}
  \textbf{\bibinfo{volume}{65}}, \bibinfo{pages}{024519}.

\bibitem[{\citenamefont{Tokuyasu} \emph{et~al.}(1988)\citenamefont{Tokuyasu,
  Sauls, and Rainer}}]{tokuyasu88}
\bibinfo{author}{\bibnamefont{Tokuyasu}, \bibfnamefont{T.}},
  \bibinfo{author}{\bibfnamefont{J.~A.} \bibnamefont{Sauls}}, and
  \bibinfo{author}{\bibfnamefont{D.}~\bibnamefont{Rainer}},
  \bibinfo{year}{1988}, \bibinfo{journal}{Phys. Rev. B}
  \textbf{\bibinfo{volume}{38}}, \bibinfo{pages}{8823}.

\bibitem[{\citenamefont{Tollis}(2004)}]{tollis}
\bibinfo{author}{\bibnamefont{Tollis}, \bibfnamefont{S.}},
  \bibinfo{year}{2004}, \bibinfo{journal}{Phys. Rev. B}
  \textbf{\bibinfo{volume}{69}}, \bibinfo{pages}{104532}.

\bibitem[{\citenamefont{Tollis} \emph{et~al.}(2005)\citenamefont{Tollis,
  Daumens, and Buzdin}}]{buzdin05}
\bibinfo{author}{\bibnamefont{Tollis}, \bibfnamefont{S.}},
  \bibinfo{author}{\bibfnamefont{M.}~\bibnamefont{Daumens}}, and
  \bibinfo{author}{\bibfnamefont{A.}~\bibnamefont{Buzdin}},
  \bibinfo{year}{2005}, \bibinfo{journal}{Phys. Rev. B}
  \textbf{\bibinfo{volume}{71}}, \bibinfo{pages}{024510}.

\bibitem[{\citenamefont{Toplicar and Finnemore}(1977)}]{toplicar}
\bibinfo{author}{\bibnamefont{Toplicar}, \bibfnamefont{J.~R.}}, and
  \bibinfo{author}{\bibfnamefont{D.~K.} \bibnamefont{Finnemore}},
  \bibinfo{year}{1977}, \bibinfo{journal}{Phys. Rev. B}
  \textbf{\bibinfo{volume}{16}}, \bibinfo{pages}{2072}.

\bibitem[{\citenamefont{Tsuei and Kirtley}(2003)}]{kirtley}
\bibinfo{author}{\bibnamefont{Tsuei}, \bibfnamefont{C.~C.}}, and
  \bibinfo{author}{\bibfnamefont{J.~R.} \bibnamefont{Kirtley}},
  \bibinfo{year}{2003}, \bibinfo{journal}{Rev. Mod. Phys.}
  \textbf{\bibinfo{volume}{72}}, \bibinfo{pages}{657}.

\bibitem[{\citenamefont{Usadel}(1970)}]{usadeleq}
\bibinfo{author}{\bibnamefont{Usadel}, \bibfnamefont{K.~L.}},
  \bibinfo{year}{1970}, \bibinfo{journal}{Phys. Rev. Lett.}
  \textbf{\bibinfo{volume}{25}}, \bibinfo{pages}{507}.

\bibitem[{\citenamefont{Vaks} \emph{et~al.}(1962)\citenamefont{Vaks, Galitskii,
  and Larkin}}]{vaks}
\bibinfo{author}{\bibnamefont{Vaks}, \bibfnamefont{V.~G.}},
  \bibinfo{author}{\bibfnamefont{V.~M.} \bibnamefont{Galitskii}}, and
  \bibinfo{author}{\bibfnamefont{A.~I.} \bibnamefont{Larkin}},
  \bibinfo{year}{1962}, \bibinfo{journal}{Sov. Phys. JETP}
  \textbf{\bibinfo{volume}{14}}, \bibinfo{pages}{1177}.

\bibitem[{\citenamefont{Velez} \emph{et~al.}(1999)\citenamefont{Velez, Cyrille,
  Kim, Vicent, and Schuller}}]{velez}
\bibinfo{author}{\bibnamefont{Velez}, \bibfnamefont{M.}},
  \bibinfo{author}{\bibfnamefont{M.~C.} \bibnamefont{Cyrille}},
  \bibinfo{author}{\bibfnamefont{S.}~\bibnamefont{Kim}},
  \bibinfo{author}{\bibfnamefont{J.~L.} \bibnamefont{Vicent}}, and
  \bibinfo{author}{\bibfnamefont{I.~K.} \bibnamefont{Schuller}},
  \bibinfo{year}{1999}, \bibinfo{journal}{Phys. Rev. B}
  \textbf{\bibinfo{volume}{59}}, \bibinfo{pages}{14659}.

\bibitem[{\citenamefont{Volkov}(1995)}]{volkov}
\bibinfo{author}{\bibnamefont{Volkov}, \bibfnamefont{A.~F.}},
  \bibinfo{year}{1995}, \bibinfo{journal}{Phys. Rev. Lett.}
  \textbf{\bibinfo{volume}{74}}, \bibinfo{pages}{4730}.

\bibitem[{\citenamefont{Volkov} \emph{et~al.}(1996)\citenamefont{Volkov,
  Allsopp, and Lambert}}]{volkov_allsopp}
\bibinfo{author}{\bibnamefont{Volkov}, \bibfnamefont{A.~F.}},
  \bibinfo{author}{\bibfnamefont{N.}~\bibnamefont{Allsopp}}, and
  \bibinfo{author}{\bibfnamefont{C.~J.} \bibnamefont{Lambert}},
  \bibinfo{year}{1996}, \bibinfo{journal}{J. Phys. Condens Matter}
  \textbf{\bibinfo{volume}{8}}, \bibinfo{pages}{45}.

\bibitem[{\citenamefont{Volkov and Anishchanka}(2004)}]{volkov04}
\bibinfo{author}{\bibnamefont{Volkov}, \bibfnamefont{A.~F.}}, and
  \bibinfo{author}{\bibfnamefont{A.}~\bibnamefont{Anishchanka}},
  \bibinfo{year}{2004}, \bibinfo{title}{Alternative mechanism of the
  sign-reversal effect in superconductor-ferromagnet-superconductor josephson
  junctions}, \bibinfo{note}{cond-mat/0407330}.

\bibitem[{\citenamefont{Volkov} \emph{et~al.}(2003)\citenamefont{Volkov,
  Bergeret, and Efetov}}]{BVE3}
\bibinfo{author}{\bibnamefont{Volkov}, \bibfnamefont{A.~F.}},
  \bibinfo{author}{\bibfnamefont{F.~S.} \bibnamefont{Bergeret}}, and
  \bibinfo{author}{\bibfnamefont{K.~B.} \bibnamefont{Efetov}},
  \bibinfo{year}{2003}, \bibinfo{journal}{Phys. Rev. Lett.}
  \textbf{\bibinfo{volume}{90}}, \bibinfo{pages}{117006}.

\bibitem[{\citenamefont{Volkov and Pavlovskii}(1996)}]{volkov_pavlo}
\bibinfo{author}{\bibnamefont{Volkov}, \bibfnamefont{A.~F.}}, and
  \bibinfo{author}{\bibfnamefont{V.~V.} \bibnamefont{Pavlovskii}},
  \bibinfo{year}{1996}, in \emph{\bibinfo{booktitle}{Proceedings of the XXXI
  Rencontres de Moriond}}, edited by
  \bibinfo{editor}{\bibfnamefont{T.}~\bibnamefont{Martin}},
  \bibinfo{editor}{\bibfnamefont{G.}~\bibnamefont{Montambaux}}, and
  \bibinfo{editor}{\bibnamefont{{J. Tran Thanh Van}}}
  (\bibinfo{publisher}{Frontiers}, \bibinfo{address}{France}).

\bibitem[{\citenamefont{Volkov} \emph{et~al.}(1993)\citenamefont{Volkov,
  Zaitsev, and Klapwijk}}]{VZK}
\bibinfo{author}{\bibnamefont{Volkov}, \bibfnamefont{A.~F.}},
  \bibinfo{author}{\bibfnamefont{A.~V.} \bibnamefont{Zaitsev}}, and
  \bibinfo{author}{\bibfnamefont{T.~M.} \bibnamefont{Klapwijk}},
  \bibinfo{year}{1993}, \bibinfo{journal}{Physica C}
  \textbf{\bibinfo{volume}{210}}, \bibinfo{pages}{21}.

\bibitem[{\citenamefont{Vollhardt and W\"olfle}(1990)}]{woelfle_book}
\bibinfo{author}{\bibnamefont{Vollhardt}, \bibfnamefont{D.}}, and
  \bibinfo{author}{\bibfnamefont{P.}~\bibnamefont{W\"olfle}},
  \bibinfo{year}{1990}, \emph{\bibinfo{title}{The superfluid phases of He 3}}
  (\bibinfo{publisher}{Taylor and Francis}, \bibinfo{address}{London,New York,
  Philadelphia}).

\bibitem[{\citenamefont{Wilhelm} \emph{et~al.}(1998)\citenamefont{Wilhelm,
  Sch\"on, and Zaikin}}]{wilhelm_schon}
\bibinfo{author}{\bibnamefont{Wilhelm}, \bibfnamefont{F.~K.}},
  \bibinfo{author}{\bibfnamefont{G.}~\bibnamefont{Sch\"on}}, and
  \bibinfo{author}{\bibfnamefont{A.~D.} \bibnamefont{Zaikin}},
  \bibinfo{year}{1998}, \bibinfo{journal}{Phys. Rev. Lett.}
  \textbf{\bibinfo{volume}{81}}, \bibinfo{pages}{1682}.

\bibitem[{\citenamefont{Wong} \emph{et~al.}(1986)\citenamefont{Wong, Jin, Yang,
  Ketterson, and Hillard}}]{wong}
\bibinfo{author}{\bibnamefont{Wong}, \bibfnamefont{H.~K.}},
  \bibinfo{author}{\bibfnamefont{B.}~\bibnamefont{Jin}},
  \bibinfo{author}{\bibfnamefont{H.~Q.} \bibnamefont{Yang}},
  \bibinfo{author}{\bibfnamefont{J.~B.} \bibnamefont{Ketterson}}, and
  \bibinfo{author}{\bibfnamefont{J.~E.} \bibnamefont{Hillard}},
  \bibinfo{year}{1986}, \bibinfo{journal}{J. Low Temp. Phys.}
  \textbf{\bibinfo{volume}{63}}, \bibinfo{pages}{307}.

\bibitem[{\citenamefont{Xia} \emph{et~al.}(2002)\citenamefont{Xia, Kelly,
  Bauer, and Turek}}]{bauer02}
\bibinfo{author}{\bibnamefont{Xia}, \bibfnamefont{K.}},
  \bibinfo{author}{\bibfnamefont{P.~J.} \bibnamefont{Kelly}},
  \bibinfo{author}{\bibfnamefont{G.~E.~W.} \bibnamefont{Bauer}}, and
  \bibinfo{author}{\bibfnamefont{I.}~\bibnamefont{Turek}},
  \bibinfo{year}{2002}, \bibinfo{journal}{Phys. Rev. Lett.}
  \textbf{\bibinfo{volume}{89}}, \bibinfo{pages}{166603}.

\bibitem[{\citenamefont{Yip}(1998)}]{yip}
\bibinfo{author}{\bibnamefont{Yip}, \bibfnamefont{S.~K.}},
  \bibinfo{year}{1998}, \bibinfo{journal}{Phys. Rev. B}
  \textbf{\bibinfo{volume}{58}}, \bibinfo{pages}{5803}.

\bibitem[{\citenamefont{You} \emph{et~al.}(2004)\citenamefont{You, Bazaliy, Gu,
  Oh, and Bader}}]{you}
\bibinfo{author}{\bibnamefont{You}, \bibfnamefont{C.-Y.}},
  \bibinfo{author}{\bibfnamefont{Y.~B.} \bibnamefont{Bazaliy}},
  \bibinfo{author}{\bibfnamefont{J.~Y.} \bibnamefont{Gu}},
  \bibinfo{author}{\bibfnamefont{S.-J.} \bibnamefont{Oh}}, and
  \bibinfo{author}{\bibfnamefont{L.~M. L. S.~D.} \bibnamefont{Bader}},
  \bibinfo{year}{2004}, \bibinfo{journal}{Phys. Rev. B}
  \textbf{\bibinfo{volume}{70}}, \bibinfo{pages}{014505}.

\bibitem[{\citenamefont{Zaitsev}(1984)}]{zaitsev}
\bibinfo{author}{\bibnamefont{Zaitsev}, \bibfnamefont{A.~V.}},
  \bibinfo{year}{1984}, \bibinfo{journal}{Zh. Eksp. Teor. Fiz.}
  \textbf{\bibinfo{volume}{86}}, \bibinfo{pages}{1742}, \bibinfo{note}{[JETP
  {\bf 59}, 1015 (1984)]}.

\bibitem[{\citenamefont{Zaitsev}(1990)}]{SubgapZ}
\bibinfo{author}{\bibnamefont{Zaitsev}, \bibfnamefont{A.~V.}},
  \bibinfo{year}{1990}, \bibinfo{journal}{JETP Lett.}
  \textbf{\bibinfo{volume}{51}}, \bibinfo{pages}{35}.

\bibitem[{\citenamefont{Zareyan} \emph{et~al.}(2001)\citenamefont{Zareyan,
  Belzig, and Nazarov}}]{nazarov_dos}
\bibinfo{author}{\bibnamefont{Zareyan}, \bibfnamefont{M.}},
  \bibinfo{author}{\bibfnamefont{W.}~\bibnamefont{Belzig}}, and
  \bibinfo{author}{\bibfnamefont{Y.~V.} \bibnamefont{Nazarov}},
  \bibinfo{year}{2001}, \bibinfo{journal}{Phys. Rev. Lett.}
  \textbf{\bibinfo{volume}{86}}, \bibinfo{pages}{308}.

\bibitem[{\citenamefont{Zyuzin} \emph{et~al.}(2003)\citenamefont{Zyuzin,
  Spivak, and Hruska}}]{zyuzin}
\bibinfo{author}{\bibnamefont{Zyuzin}, \bibfnamefont{A.~Y.}},
  \bibinfo{author}{\bibfnamefont{B.}~\bibnamefont{Spivak}}, and
  \bibinfo{author}{\bibfnamefont{M.}~\bibnamefont{Hruska}},
  \bibinfo{year}{2003}, \bibinfo{journal}{Europhys. Lett.}
  \textbf{\bibinfo{volume}{62}}, \bibinfo{pages}{97}.

\end{thebibliography}


\end{document}